\RequirePackage{tikz,pgfplots}
\documentclass[sn-mathphys,Numbered, iicol, a4paper, 10pt]{sn-jnl}

\usepackage{amsmath}
\pgfplotsset{compat=newest}
\pgfkeys{/pgf/number format/.cd,1000 sep={}}
\usetikzlibrary{arrows.meta}
\usetikzlibrary{decorations.markings}
\usetikzlibrary{plotmarks}
\usepgfplotslibrary{groupplots}

\usepackage{graphicx}%
\usepackage{multirow}%
\usepackage{amsmath,amssymb,amsfonts}%
\usepackage{amsthm}%
\usepackage{mathrsfs}%
\usepackage[title]{appendix}%
\usepackage{xcolor}%
\usepackage{textcomp}%
\usepackage{manyfoot}%
\usepackage{booktabs}%
\usepackage{algorithm}%
\usepackage{algorithmicx}%
\usepackage{algpseudocode}%
\usepackage{listings}%

\RequirePackage{xspace}
\usepackage{siunitx}
\usepackage{hyperref}
\usepackage{orcidlink}
\RequirePackage[backend=biber,sorting=none]{biblatex}

\usepackage{caption} 
\usepackage{rotating}

\usepackage{soul}
\usepackage{fancyhdr}
\begin{document}

\title{Impact of Beam-Beam Effects on Absolute Luminosity Calibrations 
at the CERN Large Hadron Collider}


\author{A. Babaev$^1$ \orcidlink{0000-0001-8876-3886}} 
\author{T. Barklow$^2$ \orcidlink{0000-0002-7709-037X}}
\author{O. Karacheban$^{3,4}$ \orcidlink{0000-0002-2785-3762}}
\author{W. Kozanecki$^5$ \orcidlink{0000-0001-6226-8385}}
\author{I. Kralik$^6$ \orcidlink{0000-0001-6441-9300}}
\author{A. Mehta$^7$ \orcidlink{0000-0002-0433-4484}}
\author{G. Pasztor$^7$ \orcidlink{0000-0003-0707-9762}}
\author*{T. Pieloni$^8$ \orcidlink{0000-0002-3218-0048}} 
\email{Tatiana.Pieloni@epfl.ch}
\author{D. Stickland$^9$ \orcidlink{0000-0003-4702-8820}}
\author{C. Tambasco$^8$}
\author{R. Tomas$^4$ \orcidlink{0000-0002-9857-1703}}
\author{J. Wa\'{n}czyk$^{4,8}$ \orcidlink{0000-0002-8562-1863}}

\affil[1]{Tomsk Polytechnic University, Tomsk, Russia}

\affil[2]{SLAC National Accelerator Laboratory, Stanford, CA, USA}

\affil[3]{Rutgers, The State University of New Jersey, Piscataway, NJ, USA}
\affil[4]{CERN, Geneva, Switzerland}
\affil[5]{IRFU-CEA, Universit\'{e} Paris-Saclay, 91191 Gif-sur-Yvette, France}
\affil[6]{Institute of Experimental Physics SAS, Watsonova 47, 04001 Kosice, Slovak Republic}
\affil[7]{MTA-ELTE Lendület CMS Particle and Nuclear Physics Group, Eötvös Loránd University (ELTE), Budapest, Hungary}
\affil[8]{Laboratory of  Particle Accelerator Physics, EPFL, Lausanne, Switzerland}
\affil[9]{ Princeton University, Princeton, NJ, USA}

\abstract{
At the Large Hadron Collider (LHC), absolute luminosity calibrations obtained by the van der Meer (\vdM) method are affected by the mutual electromagnetic interaction of the two beams. The colliding bunches experience relative orbit shifts, as well as optical distortions akin to the dynamic-$\beta$ effect, that both depend on the transverse beam separation and must therefore be corrected for when deriving the absolute luminosity scale. In the \vdM regime, the beam-beam parameter is small enough that the orbit shift can be calculated analytically. The dynamic-$\beta$ corrections to the luminometer calibrations, however, had until the end of Run 2 been estimated in the linear approximation only. In this report, the influence of beam-beam effects on the \vdM-based luminosity scale is quantified, together with the associated systematic uncertainties, by means of simulations that fully take into account the non-linearity of the beam-beam force, as well as the resulting non-Gaussian distortions of the transverse beam distributions. Two independent multiparticle simulations, one limited to the weak-strong approximation and one that models strong-strong effects in a self-consistent manner, are found in excellent agreement; both predict a percent-level shift of the absolute $pp$-luminosity values  with respect to those assumed until recently in the physics publications of the LHC experiments. These results also provide guidance regarding further studies aimed at reducing the beam--beam-related systematic uncertainty on beam-beam corrections to absolute luminosity calibrations by the van der Meer method.}


\newcommand{\main}{.}
%
\newcommand {\EDC}[1]{{\em[#1]}}
%
\newcommand{\ie}{{i.e.}\xspace}
\newcommand{\eg}{{e.g.}\xspace}
\newcommand{\vs}{{\it vs.}~}
\newcommand{\figFromRef}  [1]  {Figure reproduced from Ref.~\cite{#1}}
%
\newcommand{\vdM}{{\em vdM}\xspace}
\newcommand{\svis}{\ensuremath{\,\sigma_\mathrm{vis}}\xspace}
\newcommand{\svisz}{\ensuremath{\,\sigma^0_\mathrm{vis}}\xspace}
\newcommand{\svisNF}{\ensuremath{\,\sigma_\mathrm{vis2D}}\xspace}
\newcommand{\svisNFz}{\ensuremath{\,\sigma^0_\mathrm{vis2D}}\xspace}
\newcommand{\svisc}{\ensuremath{\,\sigma^c_\mathrm{vis}}\xspace}
\newcommand{\sviscOpt}{\ensuremath{\,\sigma^{c\mathrm{, Opt}}_\mathrm{vis}}\xspace}
\newcommand{\muvis}{\ensuremath{\,\mu_\mathrm{vis}}}
\newcommand{\muvismax}{\ensuremath{\mu_{\mathrm{vis, pk}}}}
\newcommand{\muz}{\ensuremath{\,\mu^0}}
\newcommand{\muc}{\ensuremath{\,\mu^c}}
\newcommand{\Lum} {\ensuremath{\mathcal{L}}\xspace}
\newcommand{\Lumc} {\ensuremath{\mathcal{L}^c}\xspace}
\newcommand{\Lumu} {\ensuremath{\mathcal{L}^u}\xspace}
\newcommand{\LoLz} {\ensuremath{[\Lum / \Lumz]}\xspace}
\newcommand{\LoLzF} {\ensuremath{[\Lum / \Lumz]_\mathrm{Full\,BB}}\xspace}
\newcommand{\LoLzOrb} {\ensuremath{[\Lum / \Lumz]_\mathrm{Orb}}\xspace}
\newcommand{\LoLzOpt} {\ensuremath{[\Lum / \Lumz]_\mathrm{Opt}}\xspace}
\newcommand{\LoLzFD} {\ensuremath{[\Lum / \Lumz]_\mathrm{Full\,BB}(\Delta)}\xspace}
\newcommand{\LoLzOrbD} {\ensuremath{[\Lum / \Lumz]_\mathrm{Orb}(\Delta)}\xspace}
\newcommand{\LoLzOptD} {\ensuremath{[\Lum / \Lumz]_\mathrm{Opt}(\Delta)}\xspace}
\newcommand{\NSIP}  {\ensuremath{N_{\mathrm {NSIP}}}\xspace}
\newcommand{\Lnobb}  {\ensuremath{\Lum_{\mathrm {no-bb}}}\xspace}
\newcommand{\Lumz} {\ensuremath{\mathcal{L}^0}\xspace}
\newcommand{\LoLu} {\ensuremath{[\Lum / \Lumu]}\xspace}
\newcommand{\LuoLz} {\ensuremath{[\Lumu / \Lumz]}\xspace}
\newcommand{\LoLuF} {\ensuremath{[\Lum / \Lumu]_\mathrm{Full\,BB}}\xspace}
\newcommand{\LoLnobb} {\ensuremath{[\Lum / \Lnobb]}\xspace}
\newcommand{\LuGeq} {\ensuremath{\Lum^u_\mathrm{Geq}}\xspace}
\newcommand{\cms} {\ensuremath{\mathrm{cm}^{-2}\, {\mathrm s}^{-1}}\xspace}

\newcommand{\bst}{\ensuremath{\,\beta^*}\xspace}
\newcommand{\Qxy}{\ensuremath{\text{Q}_{x,y}}\xspace}
\newcommand{\Qx}{\ensuremath{\text{Q}_x}\xspace}
\newcommand{\Qy}{\ensuremath{\text{Q}_y}\xspace}
\newcommand{\qx}{\ensuremath{\text{q}_x}\xspace}
\newcommand{\qy}{\ensuremath{\text{q}_y}\xspace}
\newcommand{\DQmIP}{\ensuremath{\Delta \text{Q}_\mathrm{mIP}}\xspace}
\newcommand{\CSx}{\ensuremath{\Sigma_x}\xspace}
\newcommand{\CSy}{\ensuremath{\Sigma_y}\xspace}
\newcommand{\CSxz}{\ensuremath{\Sigma^0_x}\xspace}
\newcommand{\CSyz}{\ensuremath{\Sigma^0_y}\xspace}
\newcommand{\CSR}{\ensuremath{\Sigma_R}\xspace}
\newcommand{\CSxc}{\ensuremath{\Sigma^c_x}\xspace}
\newcommand{\CSyc}{\ensuremath{\Sigma^c_y}\xspace}
\newcommand{\CSxcOrb}{\ensuremath{\Sigma^{c,\mathrm{Orb}}_x}\xspace}
\newcommand{\CSycOrb}{\ensuremath{\Sigma^{c,\mathrm{Orb}}_y}\xspace}
\newcommand{\CSxcOpt}{\ensuremath{\Sigma^{c,\mathrm{Opt}}_x}\xspace}
\newcommand{\CSycOpt}{\ensuremath{\Sigma^{c,\mathrm{Opt}}_y}\xspace}
\newcommand{\CSxu}{\ensuremath{\Sigma^u_x}\xspace}
\newcommand{\CSxuGeq}{\ensuremath{\Sigma^u_{x\mathrm{,Geq}}}\xspace}
\newcommand{\CSyu}{\ensuremath{\Sigma^u_y}\xspace}
\newcommand{\CSyuGeq}{\ensuremath{\Sigma^u_{y\mathrm{,Geq}}}\xspace}
\newcommand{\CSP}{\ensuremath{[\Sigma _x  \Sigma_y]_\mathrm{2D}}\xspace}
\newcommand{\CSPz}{\ensuremath{[\Sigma _x  \Sigma_y]^0_\mathrm{2D}}\xspace}
\newcommand{\sigSG}{\ensuremath{\,\sigma^0_\mathrm{SG}}\xspace}
\newcommand{\xiR}{\ensuremath{\xi_R}\xspace}
\newcommand{\sigR}{\ensuremath{\sigma_R}\xspace}
\newcommand{\xiAv}{\ensuremath{ \langle \xi \rangle}\xspace}
\newcommand{\xibar}{\ensuremath{\bar{\xi}}\xspace}
%




\maketitle
\tableofcontents
%

\section{Introduction}
\label{sec:Intro}

The determination of the absolute scale of the luminosity delivered to the ALICE, ATLAS, CMS and LHCb experiments at the LHC relies almost entirely~\cite{bib:LumR} on the van der Meer (\vdM) method~\cite{bib:vdm, bib:Rubbia}. Luminosity calibrations, which must be performed under specially tailored beam conditions, use beam-separation scans, also known as \vdM\ scans, to relate the collision rate measured by a given luminometer to the absolute luminosity inferred from directly measured beam parameters. The proportionality between the measured rate and the absolute luminosity is expressed in terms of a ``visible'' cross-section denoted \svis, which is specific to the luminometer and the counting method considered.

\par
The accuracy requirements on \svis are driven by the physics program. At the LHC, the comparison of the most precisely measured cross-sections with the corresponding theoretical predictions of the Standard model provide some of the most stringent tests of higher-order calculations; they also put strong constraints on parton-density distributions. In particular, the experimental uncertainty affecting the fiducial cross-sections for inclusive vector-boson production ($pp \rightarrow Z + X$, $pp \rightarrow W + X$) is totally dominated~\cite{bib:ATL_WZ8TeV, bib:CMS_Z2016, bib:CMS_W2016} by the systematic uncertainty in the integrated luminosity, that in these three publications ranges from 1.8 to 2.5\%; reducing that uncertainty by a factor of three would make it comparable to the combination of all other experimental uncertainties, and significantly improve the sensitivity of the associated Standard-Model tests. For more experimentally challenging processes such as $t \bar t$ production, the overall measurement uncertainty remains significantly impacted by that on the luminosity~\cite{bib:ATL_ttbR2, bib:CMS_tlXR2}, even with the recently achieved sub-percent uncertainty on the Run-2 integrated luminosity~\cite{bib:ATLR2Lum}. At the HL-LHC, the measurement of the absolute Higgs couplings drives the accuracy specifications. Projections of the experimental and theoretical uncertainties that should be achievable by then led to set a 1\% goal on the overall integrated-luminosity uncertainty~\cite{bib:CYR2019007, bib:BRIL_TDR}. Since the latter is known to receive comparable contributions from \vdM calibrations, from rate-related instrumental non-linearities, and from long-term luminometer stability~\cite{bib:ATLR2Lum}, the corresponding target for \vdM uncertainties at the HL-LHC has to remain significantly smaller than this goal; it has been set to approximately 0.6\% by both the ATLAS and the CMS~\cite{bib:BRIL_TDR} Collaboration.

\par
The  beam parameters used to determine the absolute luminosity during \vdM\ scans are the particle populations and transverse sizes associated with each colliding-bunch pair. During a beam-separation scan, the mutual electromagnetic interaction between two opposing bunches induces separation-dependent orbit shifts, as well as variations in the transverse size and shape of each bunch, thereby distorting the luminosity-scan curves from which the beam-overlap integrals, or equivalently the convolved transverse beam sizes, are extracted. Depending on the beam conditions under which the scans are performed, the magnitude of the resulting calibration bias -- if left uncorrected -- represents a significant fraction of, or can even exceed, the \svis\ systematic-uncertainty budget. 



\par
The methodology first developed in Refs.~\cite{Herr, bib:ATL2011Lum} to quantify the impact of beam-beam effects on the absolute luminosity scale has been adopted since 2013 by all LHC Collaborations. With as input the beam separation dialed-in at each scan step, plus the measured bunch currents and convolved beam sizes, the orbit shift induced by beam-beam deflections was calculated analytically~\cite{Bassetti}, and the optical distortion associated with the dynamic-$\beta$ effect was evaluated in the linear approximation~\cite{Herr} using the MAD-X package~\cite{MADX}. The two effects impacted the luminosity scale in opposite ways, resulting at the time in a net upwards correction to \svis\ of 1--1.5\% for vdM calibrations at $\sqrt{s} = 13$\,TeV.

\par
 A recent reevaluation of this methodology, using a new beam-beam simulation package specifically developed for \vdM-scan studies, revealed that the linear approximation used in the MAD-X simulation results in a significant underestimate of the optical-distortion correction~\cite{Balagura_2021}.
These findings motivated the studies reported in the present paper, which is organized as follows.

\par
After a brief overview of the \vdM-calibration methodology at the LHC and of the impact thereon of beam-beam effects (Sec.\,\ref{sec:lumCalMeth}), the relevant simulation tools are presented in Sec.~\ref{sec:bbSimCodes}: the MAD-X package used in the original implementation~\cite{Herr, bib:ATL2011Lum}; the B*B package presented in Ref.~\cite{Balagura_2021}, that models beam-beam dynamics in the transverse plane under the weak-strong approximation; and the long-established COMBI~\cite{COMBICODE} multiparticle code, that is more CPU-intensive but can simulate beam-beam effects in the strong-strong regime with the optional inclusion of longitudinal dynamics. 
Following a systematic cross-validation of the latter two packages, B*B is used in Sec.~\ref{sec:bbImpact} to develop an easy-to-use parameterization of beam-beam corrections to \vdM\ scans in the limit of round, initially Gaussian bunches of equal brightness\footnote{There exist multiple definitions of the concept of {\em beam brightness}, depending on the type of beam and the application considered~\cite{bib:AccPhysHandbook}.  In the context of the present paper, the term {\em brightness} refers to the particle density in transverse phase space. For round beams, it is defined as $\mathcal{B} = n_p / \epsilon_N$, where $n_p$  is the bunch population and $ \epsilon_N$ the normalized transverse emittance.} 
that collide at a single interaction point (IP) with zero crossing angle. Deviations of the colliding-bunch configuration from this idealized limit: non-Gaussian tails, elliptical transverse bunch profiles, non-zero crossing angle, collisions at multiple IPs, or unequal-brightness beams, are then either accounted for using simulation-based adjustments to the idealized parameterization, or found to be small enough to be treated as a systematic uncertainty. These and other sources of systematic uncertainty, such as tune or \bst\ settings, that may affect the beam-beam correction in the \vdM\ regime  are consolidated in Sec.~\ref{sec:bbsysts}. An overall summary and a brief outlook are offered in Sec.~\ref{sec:Concl}.
\section{Luminosity-calibration methodology at the LHC}
\label{sec:lumCalMeth}

The \vdM-scan formalism~\cite{bib:vdm, bib:Rubbia} that underpins the determination of the absolute luminosity scale at the CERN ISR, RHIC and the LHC is summarized, for the simplest case, in Sec.\,\ref{subsec:vdMPrinc} below; generalizations of this formalism can be found in Refs.~\cite{Balagura_2021, bib:BalagNIM, bib:cai2000lum}. At the LHC, because of both instrumental and accelerator-physics reasons~\cite{bib:LumR}, \vdM\ scans are not performed during normal physics operation, but rather under dedicated beam conditions (Sec.\,\ref{subsec:beamParms}). Their two fundamental ingredients, the transverse beam separation and the measured collision rate, are both affected by the beam-beam interaction (Sec.\,\ref{subsec:bbBias}), at a  level that is significant on the scale of the precision goals outlined in Sec.\,\ref{sec:Intro}.

	\subsection{Absolute luminosity scale from measured beam parameters}
	\label{subsec:vdMPrinc}
	
\begin{figure}[b!]
\centering
\includegraphics[width=0.47\textwidth]{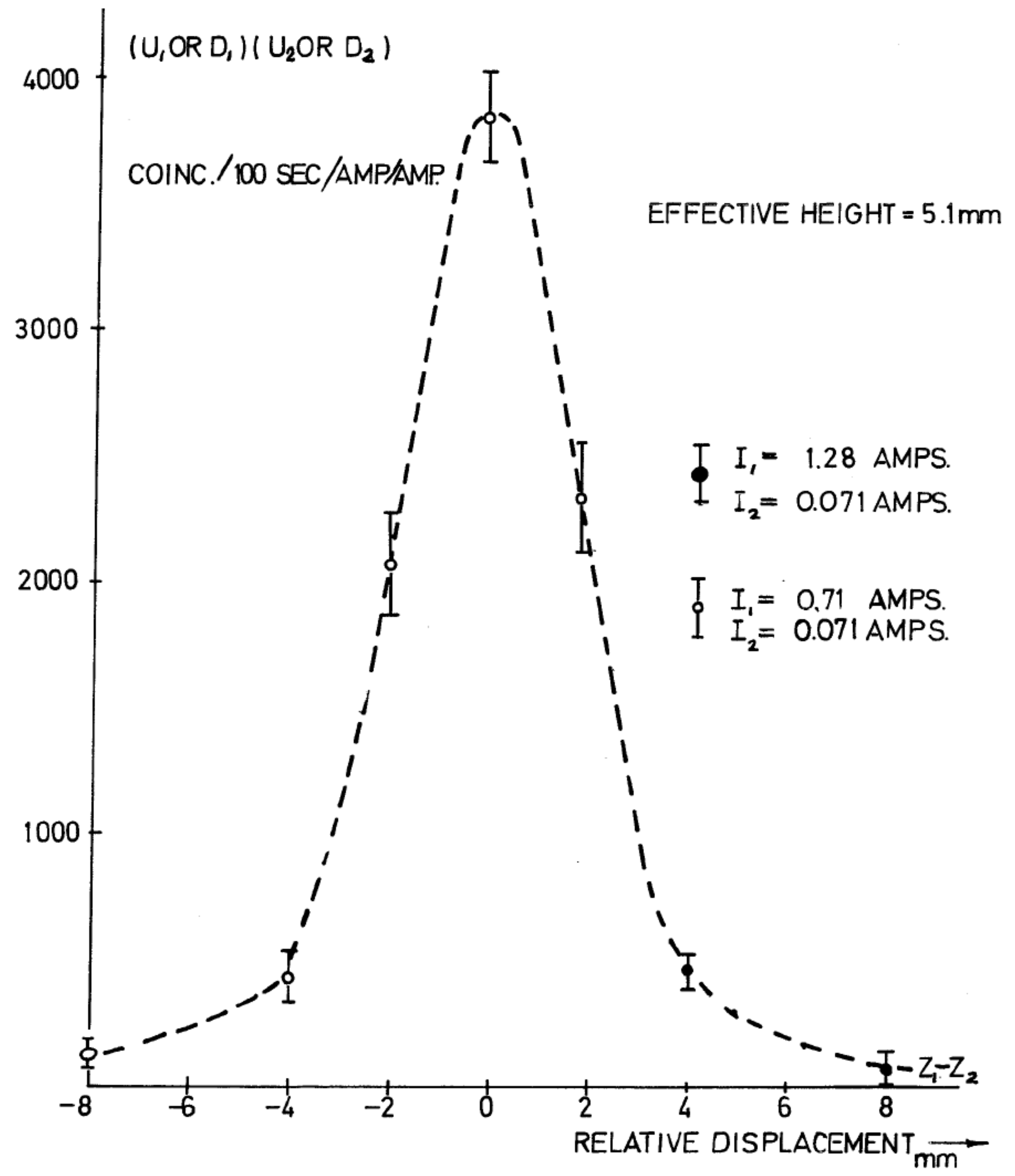}
\includegraphics[width=0.50\textwidth]{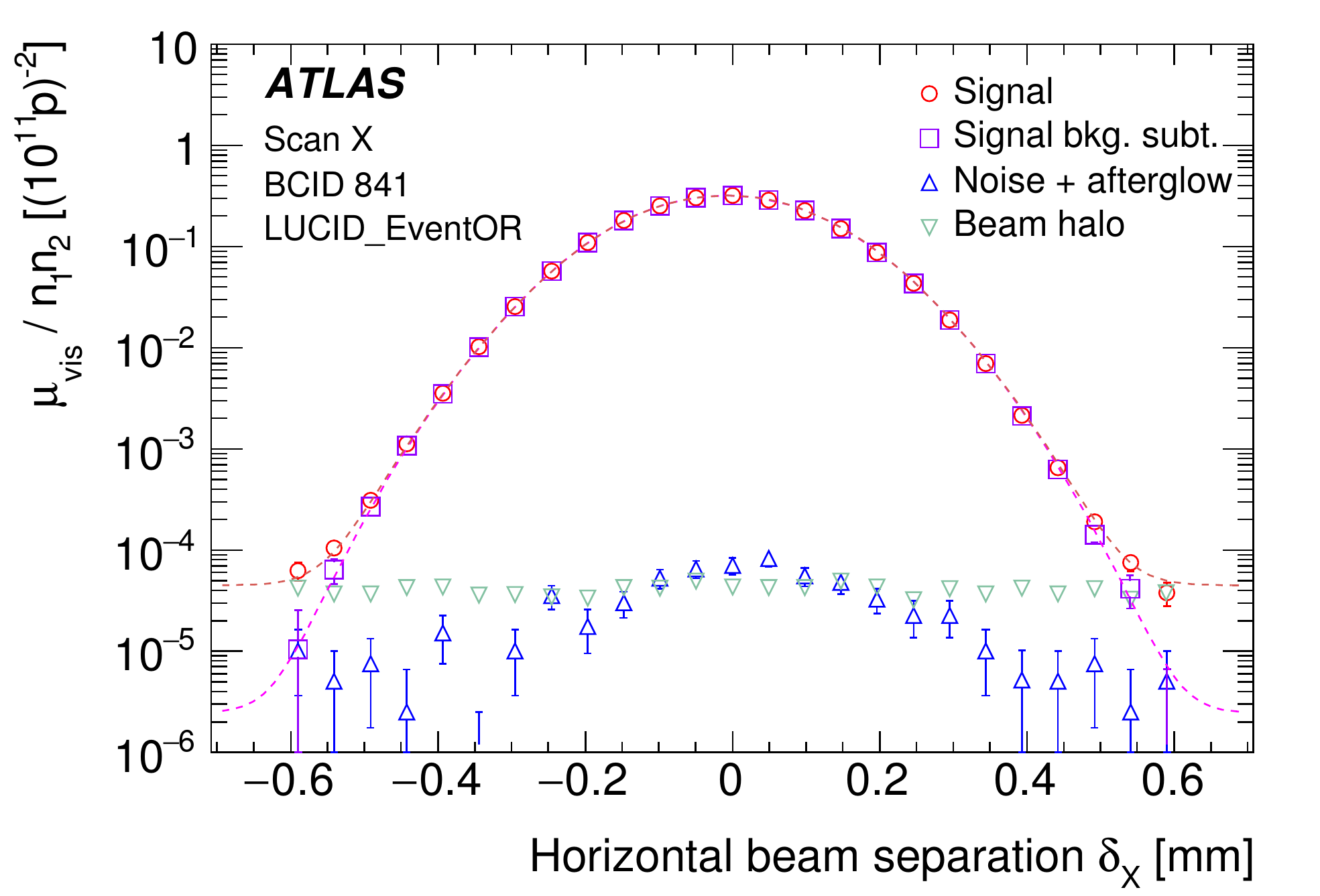}
\caption{Top: van der Meer method at the CERN ISR. Shown is the monitor rate $R_y(\delta_y)$ as a function of the relative vertical separation $\delta_y$ of the two beams (\figFromRef{bib:NP71-6}, \copyright\ CERN). 
Bottom: beam-separation dependence of the  visible interaction rate measured at the ATLAS IP
during a horizontal \vdM\ scan in $pp$ collisions at $\sqrt{s}=8$\,TeV (LHC fill 3311), before (red circles) and after (purple squares) noise and background subtraction. The subtracted contributions are shown as triangles. The scan curve is fitted to a Gaussian function multiplied by a sixth-order polynomial, plus a constant (\figFromRef{bib:ATL2012Lum}, \copyright\ CERN, 
CC-BY-4.0 license).}
\label{fig:LscanCurves} 
\end{figure}

In terms of colliding-beam parameters, the bunch luminosity ${\mathcal L}_{\mathrm b}$ is given by
\begin{equation}
{\mathcal L}_{\mathrm b} =  f_{\mathrm r}\, n_1 n_2\, \int {\hat{\rho} _1 (x,y)}\, \hat{\rho} _2(x,y)\,\mathrm{d}x\,\mathrm{d}y~,
\label{eqn:lumi}
\end{equation}
where the opposing bunches are assumed to collide head-on and with zero crossing angle, $f_{\mathrm r}$ is the LHC revolution frequency, $n_1 n_2$ is the bunch-population product and $\hat{\rho}_{1(2)}(x,y)$ is the normalized particle density in the transverse ($x$--$y$) plane of beam 1 (2) at the IP.\footnote{This paper uses a right-handed coordinate system with its origin at the nominal IP, and the $z$-axis along the direction of LHC beam 1; the latter circulates in the clockwise direction when the LHC rings are viewed from above. The $x$-axis points from the center of the LHC ring to the IP, and the $y$-axis points upwards.}
With the standard assumption that the particle densities can be factorized into independent horizontal and vertical component distributions,
$\hat{\rho}(x,y)=\rho_x(x)\,\rho_y(y)$, Eq.~(\ref{eqn:lumi}) can be rewritten as
\begin{equation}
{\mathcal L}_{\mathrm b} =  f_{\mathrm r}\, n_1 n_2 \,\Omega _x (\rho _{x1},\rho _{x2}) \,\Omega _y (\rho _{y1},\rho _{y2})~,
\label{eqn:lumi1}
\end{equation}
where \[ \Omega _x (\rho _{x1},\rho _{x2} ) = \int {\rho _{x1} (x)\,\rho _{x2} (x)\,\mathrm{d}x}\] 
is the beam-overlap integral in the $x$ direction (with an analogous definition in the $y$ direction). 
In the method proposed by van der Meer~\cite{bib:vdm} at the ISR (Fig.~\ref{fig:LscanCurves}, top), the overlap integral (for example in the $y$ direction) can be calculated as
\begin{equation}
\Omega _y (\rho _{y1},\rho _{y2}) = \frac{{R_y (0)}}{{\int {R_y (\delta_y)\,\mathrm{d}\delta_y} }}~,
\label{eqn:vdm}
\end{equation}
where $R_y(\delta_y)$ is the collision rate, or equivalently the luminosity in arbitrary units, measured during a vertical scan at the time the two beams are separated vertically by the distance $\delta_y$.  Because the collision rate $R_y(\delta_y)$ is normalized to that at zero separation $R_y(0)$, any quantity proportional to the luminosity can be substituted in Eq.~(\ref{eqn:vdm}) in place of $R$.

\par
Defining the vertical convolved bunch size $\Sigma_y$ \cite{bib:LumR} as
\begin{equation}
\Sigma _y  = \frac{1}{{\sqrt {2\pi } }}\frac{{\int {R_y (\delta_y)\,\mathrm{d}\delta_y} }}{{R_y (0)}}~,
\label{eqn:caps}
\end{equation}
and similarly for $\Sigma_x$, the bunch luminosity in Eq.~(\ref{eqn:lumi1}) can be rewritten as
\begin{equation}
{\mathcal L}_{\mathrm b} = \frac{ f_{\mathrm r} n_1 n_2 }{{2\pi \Sigma _x \Sigma _y}}~,
\label{eqn:lumifin}
\end{equation}
which allows the absolute bunch luminosity at zero separation to be determined from the revolution frequency
$f_{\mathrm r}$, the bunch-population product $n_1 n_2$,  and the product
$\Sigma_x \Sigma_y$ which is measured directly during a pair of orthogonal scans. 
\par
If the transverse density profile of each beam $B$ ($B = 1, 2$) can be described by a single Gaussian of width $\sigma_{iB}$ ($i = x, y$), the convolved widths are given by
\begin{equation}
\label{eqn:capSigDef}
\Sigma_i = \sqrt{\sigma_{i1}^2 + \sigma_{i2}^2} = \sqrt{\beta^*_{i1} \epsilon_{i1}  +  \beta^*_{i2} \epsilon_{i2} } 
\end{equation}
where $\epsilon_{iB}$ is the geometrical emittance of beam $B$ in plane $i$, and $\beta^*_{iB} $ is the corresponding value of the $\beta$ function at the IP\footnote{Throughout most of the present paper, and unless explicitly specified otherwise, the IP $\beta$-function is implicitly assumed to be the same for the two beams and in the two planes: $\beta^*_{x1} = \beta^*_{y1} =\beta^*_{x2} =\beta^*_{y2} = \bst$.}.
In such a case, the beam-separation dependence of the collision rate is given by
\begin{equation}
\label{eqn:LvsDelta}
R_i(\delta_i) = R_i(0)~e^{- \frac {\delta_i^2}{2 \Sigma_i^2}}\, .
\end{equation}
The luminosity curve $R_i(\delta_i)$ is also Gaussian, and $\Sigma _i $ coincides with the standard deviation of that distribution. It is important to note, however, that the \vdM method does not rely on any particular functional form of $R_i(\delta_i)$: the quantities $\Sigma_x$ and $\Sigma_y$ can be determined for any observed luminosity curve
from Eq.~(\ref{eqn:caps}) and used with Eq.~(\ref{eqn:lumifin}) to determine the absolute luminosity
at $\delta_x = \delta_y = 0$.
\par
In the more general case where the factorization assumption breaks down, \ie when the particle densities cannot be factorized into a product of uncorrelated $x$ and $y$ components, Eq.~(\ref{eqn:lumi1}) no longer holds, and a single pair of horizontal and vertical scans is no longer sufficient to measure the overlap integral in Eq.~(\ref{eqn:lumi}). One must then generalize the formalism to the two-dimensional case~\cite{bib:Rubbia}, and scan over a  grid in the ($\delta_x,\delta_y$) beam-separation space  to measure the product of the convolved bunch widths~\cite{bib:LumR,Balagura_2021}:
\begin{equation}
[\Sigma _x  \Sigma_y]= \frac{1}{{2\pi}}\frac{{\int {R_{x,y} (\delta_x,\delta_y)\,\mathrm{d}\delta_x\,\mathrm{d}\delta_y}}}{{R_{x,y} (0,0)}}~.
\label{eq:CapSNonFact}
\end{equation}
Here the square brackets highlight the fact that in the presence of non-factorization, the quantity $[\Sigma _x  \Sigma_y]$ can no longer be broken down into a product of two independent quantities. Equation~(\ref{eqn:lumifin}), however, remains formally unaffected, as do Eqs.~(\ref{eqn:defmu})--(\ref{eqn:sigmaVis}) below.

\par
In terms of luminometer observables, the bunch luminosity can be written as
\begin{equation}
\mathcal{L}_{\mathrm b} = \frac{\muvis \, f_{\mathrm r} } {\svis}\,,
\label{eqn:defmu}
\end{equation}
where \muvis\ is the average number of inelastic collisions per bunch crossing detected by the luminometer considered, and \svis\ is the associated visible cross-section. Since \muvis\ is a directly measurable quantity,  the calibration of the absolute luminosity scale amounts to determining the visible cross-section \svis. Equating the absolute luminosity computed from beam parameters using Eq.~(\ref{eqn:lumifin}) to that measured according to Eq.~(\ref{eqn:defmu}), yields:
\begin{equation}
\svis =\muvismax \frac{2\pi\, \Sigma_x \Sigma_y}{n_1 n_2}~,
\label{eqn:sigmaVis}
\end{equation}
where \muvismax\ is the visible interaction rate per bunch crossing reported at the peak of the scan curve (Fig.~\ref{fig:LscanCurves}, bottom). Equation~(\ref{eqn:sigmaVis}) provides a direct calibration of the visible cross-section \svis\ in terms of the peak visible interaction rate \muvismax, the product of the convolved bunch widths $\Sigma_x \Sigma_y$, and the bunch-population product $n_1 n_2$.

\par
In the presence of a significant crossing angle, the formalism becomes more involved~\cite{Balagura_2021, bib:BalagNIM, Herr:2003em}. A non-zero crossing angle in either the horizontal or the vertical plane widens the corresponding luminosity-scan curve by the so-called geometrical factor $F$:
\begin{equation}
F = \frac{   \sqrt{  1 + \tan^2 \, \theta_c/2 ~ (\sigma_{z1}^2+\sigma_{z2}^2) /(\sigma_{c1}^2+\sigma_{c2}^2)  }  } {\cos \,\theta_c/2} .
\label{eqn:geomFact}
\end{equation}
Here $\theta_c$ is the full crossing angle,  $\sigma_{zB}$ ($B = 1, 2$) are the RMS bunch lengths of beams 1 and 2, and $\sigma_{cB}$ the transverse single-beam sizes in the crossing plane. The peak luminosity is reduced by the same factor. The corresponding increase in the measured value of $\Sigma_x$ or $\Sigma_y$ is exactly compensated by the decrease in \muvismax, so that Eqs.~(\ref{eqn:caps})--(\ref{eqn:sigmaVis}) remain valid, and no correction for the crossing angle is needed in the determination of \svis.

		\subsection{Beam conditions during van der Meer scans}
		\label{subsec:beamParms}

The strength of the beam-beam interaction is traditionally quantified by the {\em linear beam-beam parameter}, defined as~\cite{bib:GenXiDef, bib:AccPhysHandbook}:
\begin{equation}
\label{eqn:xiGDef}
\xi_{x2} = \frac{n_1 \,  r_0 \, Z_1\, Z_2\,\beta^*_{x2}}{2 \pi \, A_{\mathrm{ion},2}\, \gamma_2 \, \sigma_{x1} (\sigma_{x1} + \sigma_{y1})}
\end{equation}	
Here $\xi_{x2}$ is the horizontal beam-beam parameter experienced by beam 2 (B2), the ``witness beam''; $n_1$ is the bunch population of beam 1 (B1), the ``source beam'';  $r_0 = e^2/4 \pi \epsilon_0 m_p c^2$ is the classical radius of the proton;  $A_{\mathrm{ion},B}$ and $Z_B$ ($B = 1,2$) are the atomic mass number and charge number of the beam-$B$ particle type (proton or fully stripped ion); $\beta^*_{x2}$ is the value of the B2 horizontal $\beta$ function at the IP; $\gamma_2$ is the relativistic factor of the B2 particles, and $ \sigma_{x1}$ ($ \sigma_{y1}$) is the horizontal (vertical) transverse RMS size of B1. Formulas for the other beam and the other plane are obtained by interchanging B1 and B2, and/or $x$ and $y$. 

\par
For the most frequent case of $pp$ collisions, Equation~(\ref{eqn:xiGDef}) takes the more familiar form:
\begin{equation}
\xi_{x2} = \frac{n_1 \,  r_0 \, \beta^*_{x2}}{2 \pi \, \gamma_2 \, \sigma_{x1} (\sigma_{x1} + \sigma_{y1})} \, .
\nonumber
\end{equation}		
In the case of equally populated, equally sized round beams,  this expression becomes much simpler:
\begin{equation}
\label{eqn:xiRDef}
\xi = \frac{n r_0 \bst}{4 \pi \gamma \sigma_0^2}
\end{equation}
where $n$ is the bunch population, and $\sigma_0 = \sqrt{\epsilon \bst}$ is the nominal RMS beam size at the IP.

\par
In 2018, during high-luminosity physics running in proton-proton ($pp$) mode, the LHC collided up to  2544 bunches with typical initial intensities of $1.1\times 10^{11}$\,$p$/bunch, grouped in trains of 36 to 144 bunches with  a minimum interbunch spacing of 25\,ns. Beams crossed with a half-angle $\theta_c/2$ of $\pm 130\, \mu$rad in order to mitigate the impact of the long-range beam-beam interaction at parasitic crossings. At the start of stable beams, the emittance was typically 2\,$\mu$m$\cdot$rad \cite{bib:R2LHCconfig}, the single-bunch luminosity around $7.8\times 10^{30}\,\cms$ at IP1 and IP5, and the total luminosity close to $1.9\times 10^{34}\,\cms$ at each of these two IPs. These values correspond to a pile-up parameter $\mu$ of around 55 inelastic $pp$ collisions per bunch crossing, and to a bunch-averaged, head-on beam-beam parameter $\langle \xi \rangle$ of approximately 0.005. The brightness, however,  varied significantly along the bunch string, occasionally resulting in $\xi$ values as high as 0.007 for some of the bunches.

\par 
In contrast, during $pp$ \vdM\ scans (Table~\ref{tab:vdMBeamParms}), the injected emittance is deliberately blown up and the bunch population significantly lowered, in order to reduce the impact of beam-beam effects as well as minimize the unbunched-beam fraction and the intensity of satellite bunches. The bunches are isolated rather than in trains, and their number is limited to 152 at most, in order to eliminate parasitic crossings and collide with zero crossing angle in the interaction regions where the beam-line layout so permits. The $\beta$ function at the IP is increased such as to bring the pile-up parameter $\mu$ down to around 0.5, \ie in a regime where luminometers are free of instrumental non-linearities; this carries the additional advantage that it significantly increases the transverse luminous size, allowing a more precise measurement of its beam-separation dependence~\cite{bib:LumR, bib:ATL2012Lum}.

\begin{table*} [htb]
\centering
\begin{tabular}{c|ccc} 
Parameter 							& Typical scans							& Reference		\\ 
									& (LHC Run 2)							& parameter set	\\
\hline
Beam energy $E_B$ [TeV]				& 6.5									&  3.5		    	\\
Nominal tune settings \Qx/\Qy	            		& 64.31/59.32							& 64.31/59.32		\\
Normalized emittance $\epsilon_N$ [$\mu$m$\cdot$rad] 				
									&  2.2 -- 3.5							&  4.00 	    		\\  
IP1,5: \bst [m]							& 19.2 								& 1.50		    	\\
IP2/8: \bst [m]							& 19.2 / 24.0							& -			    	\\	
IP1,5: $\theta_c/2$ [$\mu$rad]				& 0 									& 0			    	\\
IP2: $\theta_c/2$ [$\mu$rad]				& 70 -- 195 ($y$)						& -			    	\\
IP8: $\theta_c/2$ [$\mu$rad]				& 450 -- 550 ($x$)					    	& -			    	\\
Transverse convolved bunch sizes $\CSx \approx \CSy$ [$\mu$m]				
									& 110 -- 160							& 56.7			\\
Bunch spacing [ns] 						& $ \ge 525$ 							& -			    	\\  
Number of colliding bunches 				& 30 -- 124 							& 1   		    		\\  
Bunch population $n_p$ [$10^{11}$] 			& 0.7 -- 1.0 							&  0.85		    	\\  
Linear beam-beam parameter $\xi$ [per IP] 	& 0.0022 -- 0.0056						& 0.0026			\\
Typical bunch luminosity @ IP1, 5 [$\cms$]	            			
									& $3 - 8 \times 10^{28} $					& $4 \times 10^{29}$	\\
Typical total luminosity @ IP1, 5 [$\cms$]	        & $2.5 - 9 \times 10^{30}$					& $4 \times 10^{29}$	\\

\end{tabular}
\caption{Beam conditions during \vdM\ scans at the LHC. Middle column: typical parameters during the 2015-2018 $pp$ scan sessions at $\sqrt{s}=13$\,TeV. Right column: reference parameters used for the cross-validation of the beam-beam codes described in Sec.\,\ref{sec:bbSimCodes}; the parameter values are chosen to match those used in the initial, MAD-X based simulations of Ref.~\cite{Herr}, which were representative of 2011 \vdM\ scans at IP1 and IP5. In the absence of the beam-beam interaction, the simulated bunches are assumed round in the transverse plane, of equal radius, and strictly Gaussian in all three dimensions.}
\label{tab:vdMBeamParms}
\end{table*}

	\subsection{Beam--beam-induced biases and their correction}
	\label{subsec:bbBias}
	
	The mutual electromagnetic interaction between colliding bunches shifts their orbits, and therefore modifies their transverse separation at the IP (Sec.\,\ref{subsubsec:orbShift}); it also distorts their transverse density distributions (Sec.\,\ref{subsubsec:optclDist}). These two effects depend on the {\em nominal separation} $\Delta$  dialed-in at each scan step. Their combination impacts both the normalized integrals (\CSx, \CSy) and the peak (\muvismax) of the luminosity-scan curves used in determining the absolute luminosity scale. The strategy for correcting the resulting biases is outlined in Sec.\,\ref{subsubsec:bbCorStrtgy}; its detailed implementation is developed in later chapters, in particular in Secs.~\ref{subsubsec:CorImplement} and \ref {subsubsec:mIPparam}.

		\subsubsection{Orbit shift}
		\label{subsubsec:orbShift}
	
		When two positively charged  bunches collide with a non-zero impact parameter, they experience a mutually repulsive angular kick equivalent to that of a dipole located at the collision point, the strength of which depends on the beam separation. In the round-beam limit and for $pp$ collisions, the angular kick experienced by a B2 bunch during a horizontal beam-separation scan is given by~\cite{Bambade}:
\begin{equation}
\label{eqn:deflAng}
\theta_{2x}  = \frac{2 r_0 n_1}{\gamma_2} \frac{\delta_x}{\delta}   \left[ \frac{1 - e^{-\delta^2/ 2 \CSR^2}}{\delta} \right] ,
\end{equation}
and similarly for a vertical scan. Here 
$\delta_x$ (resp. $\delta$) is the horizontal (resp. total) beam separation, and $\CSR = \CSx = \CSy$ is the transverse convolved beam size. This formalism has been extended by Bassetti and Erskine~\cite{Bassetti} and by Ziemann~\cite{Ziemann} to the case of elliptical beams.

\begin{figure}[b!]
\centering
\includegraphics[width=0.5\textwidth]{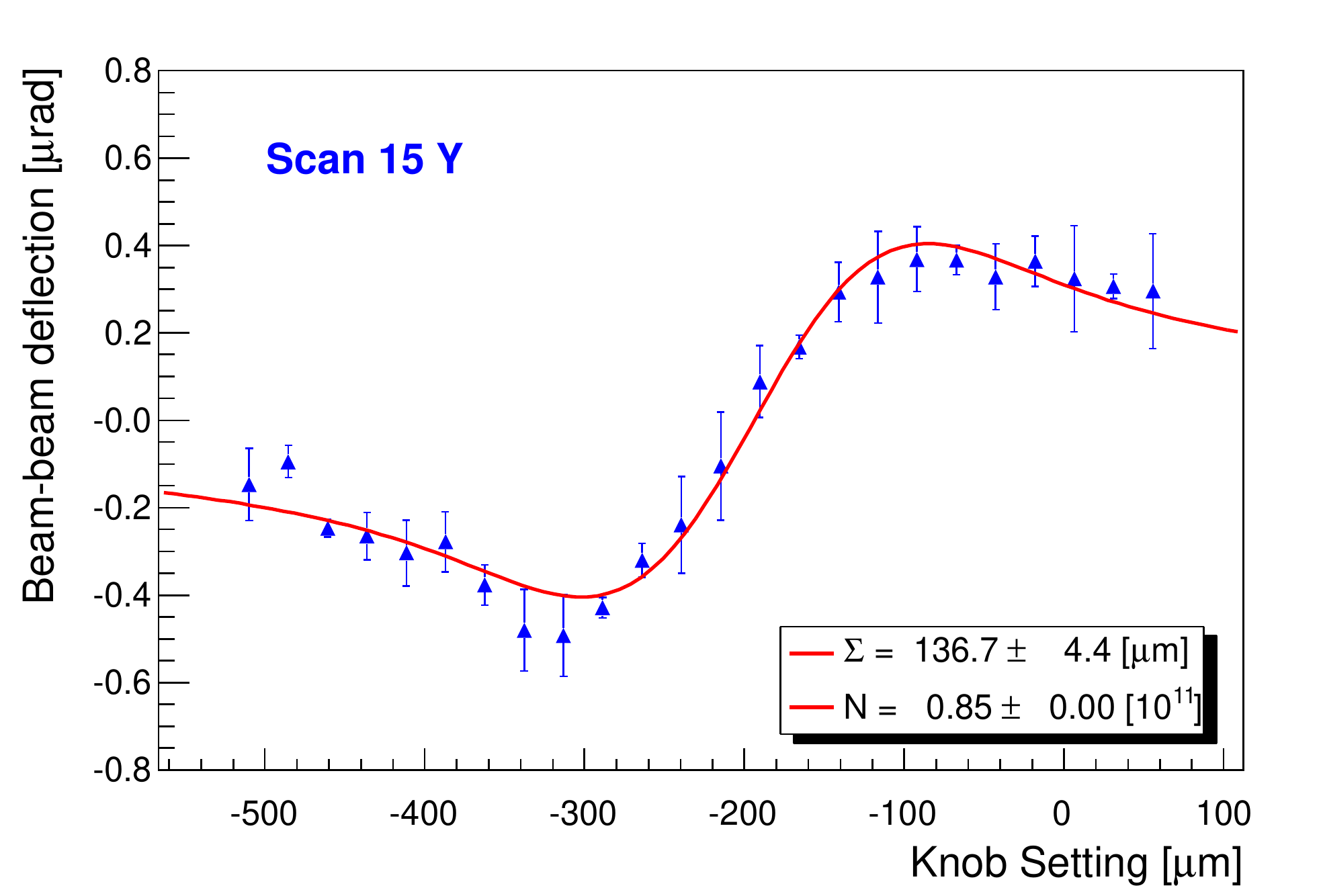}
\caption{Total beam-beam deflection angle (B1-B2)  as a function of the nominal separation $\Delta$ (denoted here by ``knob setting''), during a vertical vdM scan at the ATLAS IP in $pp$ collisions at $\sqrt{s}=8$\,TeV (LHC fill 3316). For each beam separately, the deflection angle is obtained from the difference between the outgoing- and incoming-beam angles measured by BPMs located in the LHC arcs outside the closed-orbit bump used for the scans. The zero of the horizontal axis is arbitrary, since only relative beam displacements matter. The vertical convolved beam size, denoted here by $\Sigma$, is extracted from a fit to Eq.~(\ref{eqn:deflAng}), shown by the red curve, with as input the bunch-averaged population $\mathrm{N}$ (\figFromRef{Jorg}, \copyright\ CERN)
}
\label{fig:bbDefl_LHC} 
\end{figure}

\par
This angular deflection, first observed at the Stanford Linear Collider in $e^+ e^-$ collisions~\cite{Bambade}, can be measured using beam-position monitors (BPMs) installed both upstream and downstream of the IP, as illustrated in Fig.\,\ref{fig:bbDefl_LHC} for the LHC~\cite{Jorg}. 

\par
In a circular collider, the beam-beam angular kick experienced by each beam $B$ ($B = 1, 2$) in the $i$-plane ($i = x, y$) results in a shift of its position at the IP, given by~\cite{bib:AccPhysics_Lee}
\begin{equation}
\label{eqn:orbShift}
\delta^{bb}_{iB} = \frac{\beta^*_{iB}}{2\,\tan (\pi Q_i)} ~ \theta_{iB} ,
\end{equation}
where \bst is the value of the $\beta$ function at the IP and $Q_i$ is the betatron tune.
The actual beam separation at the IP $\delta_i$ therefore differs slightly from the nominal separation $\Delta_i$:
\begin{equation}
\label{eqn:bbCSep}
\delta_i = \Delta_i + \delta^{bb}_i ,
\end{equation}
where the beam--beam-induced change in beam separation, hereafter denoted by ``orbit shift'', is given by $ \delta^{bb}_i = \delta^{bb}_{i1} - \delta^{bb}_{i2}$. Since $\delta^{bb}_{iB}$ is, to first order, proportional to the corresponding beam-beam parameter $\xi_{iB}$ (Eqs.~(\ref{eqn:xiGDef}) and (\ref{eqn:deflAng})), the orbit shift varies from one bunch to the next. The $\Delta$-dependence of $\delta^{bb}_i$ mirrors that displayed in  Fig.\,\ref{fig:bbDefl_LHC} for the deflection angle, with a peak-to-peak swing of $\pm (1\text{-}2 ) \,\mu$m under typical \vdM-scan conditions, to be compared to typical $\Sigma_i$ values in the 110--160\,$\mu$m range. The value of $ \delta^{bb}_i$ changes from scan step to scan step, thereby expanding in a non-linear fashion the beam-separation scale, i.e. the horizontal axis of scan curves such as that illustrated in Fig.\,\ref{fig:LscanCurves}b. As a result, the overlap integral of Eq.\,(\ref{eqn:caps}) increases by typically 0.7--1.4\% per plane, corresponding to a positive correction to \svis\ in the 1.4--2.8\% range. 

\par
In practice, the correction for the beam-beam orbit shift is implemented as follows. At each scan step and for each colliding-bunch pair separately, $\delta^{bb}_i$ is calculated using the Bassetti-Erskine formula~\cite{Bassetti}, with as input the measured bunch populations ($n_1, n_2$) and uncorrected convolved bunch widths (\CSx, \CSy), as well as the beam energy, the \bst setting and  the tunes. The separation-dependent collision rate $R_i(\delta_i)$ is then integrated according to Eq.\,(\ref{eqn:caps}) to obtain the beam-beam corrected values \CSxcOrb and \CSycOrb, using the beam-beam corrected separation $\delta$ (Eq.\,(\ref{eqn:bbCSep}))  instead of the nominal separation $\Delta$. The agreement of this simple analytical procedure with the predictions of self-consistent multi-particle simulations will be addressed in Sec.\,\ref{subsec:XValidtn}.

		\subsubsection{Optical distortions}
		\label{subsubsec:optclDist}

Not only does the electromagnetic field of the B1 bunch deflect the B2 bunch as a whole: it also acts as a non-linear lens that perturbs the trajectory of the individual particles in that bunch, thereby modifying the transverse density distribution of both bunches in a separation-dependent manner.

\begin{figure}[b!]
\centering
\includegraphics[width=0.45\textwidth]{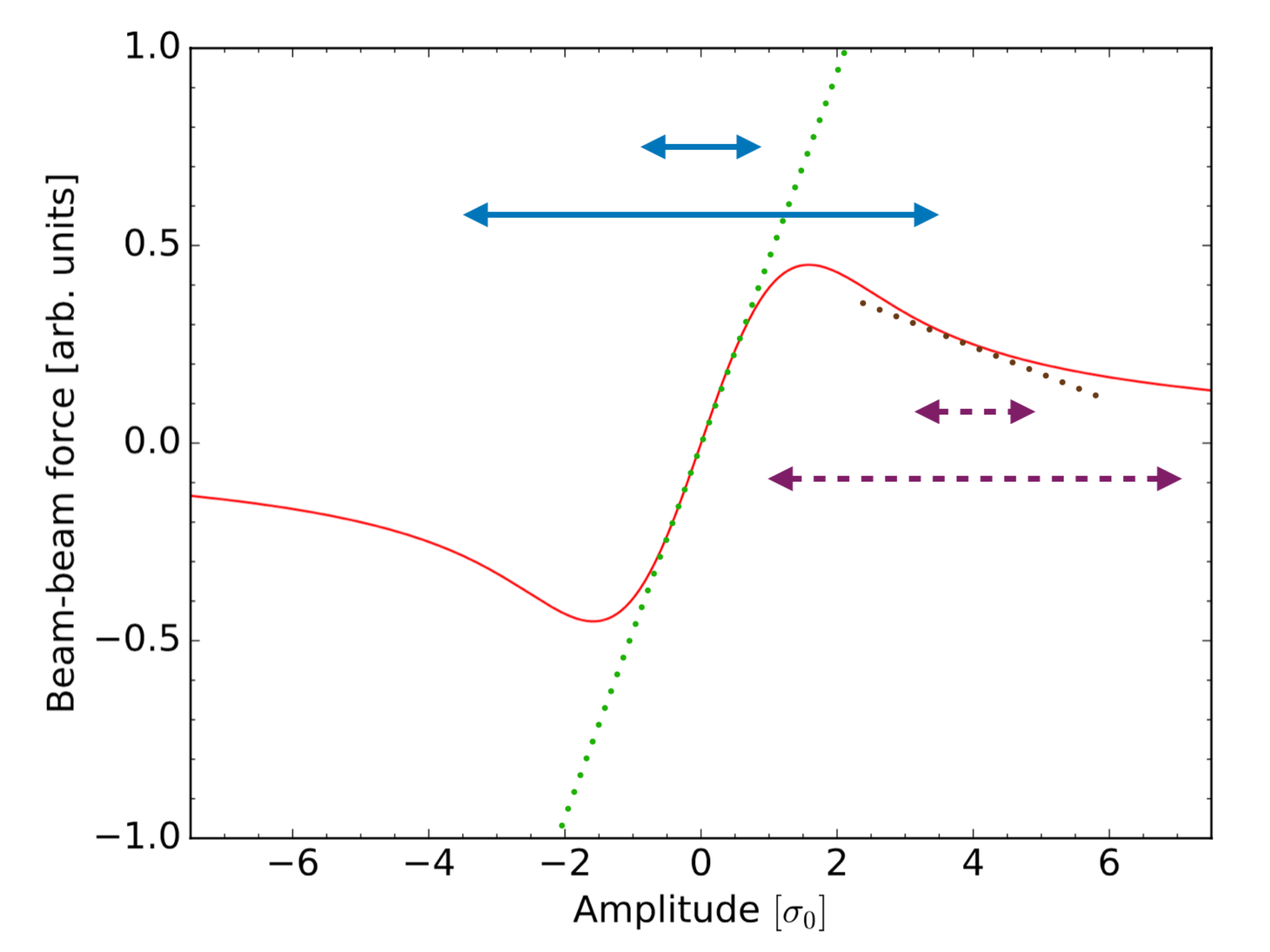}
\caption{Beam-beam force exerted by the B1 (B2) source bunch as a whole (red curve), as a function of the betatron amplitude of a B2 (B1) test particle, for equally sized round beams. The amplitude is in units of the RMS beam size. The green and brown lines correspond to the linear component of the force at amplitudes of zero and $4\sigma$  respectively, and are akin to the effect of a quadrupole. The short (long) double arrows illustrate the range of transverse kicks experienced by a small (large) amplitude particle for beams either in head-on collision (solid blue lines), or transversely separated by $4\sigma$ (dashed magenta lines).
}
\label{fig:bbForce} 
\end{figure}

\par
For small-amplitude particles and beams in head-on collision (Fig.~\ref{fig:bbForce}, short blue arrows), the force is rather linear and resembles that of a quadrupole (dotted green line), resulting in a tune shift proportional to the beam-beam parameter $\xi$ and in the subsequent dynamic-$\beta$ effect~\cite{Herr}. This ``quadrupole strength'' is proportional to the derivative of the beam-beam force; it is largest, and repulsive, for small-amplitude particles, changes sign around $1.6\sigma$, and becomes weakly attractive at larger amplitude (dotted brown line).

\par
If for simplicity one assumes that for a given beam separation, all particles are subject to the same quadrupolar-like force (the strength and sign of which depend on the beam separation), then the value of \bst at the scanning IP is modulated by the linear component of the beam-beam force. This results in a modulation of the transverse beam size, and therefore in a beam--separation-dependent modulation of the actual luminosity; however the actual shapes of the transverse density distributions projected on the $x$ and $y$ axes remain unaffected by the quadrupolar-like force. This is the approximation that was adopted in the first implementation of the optical-distortion correction~\cite{Herr, bib:ATL2011Lum}, and that will be further discussed in Sec.\,\ref{subsec:MADXdescr}.

\par
While for small amplitudes (short arrows) the force remains approximately linear, at amplitudes larger than $1\sigma$ (long arrows) it includes significant non-linear contributions. Large-amplitude particles, therefore, experience a tune shift and a $\beta$-beating that depend both on the particle amplitude (short \vs long arrows)~\cite{BBCAS}, and on the beam separation (blue \vs magenta arrows). The resulting optical distortions include not only a change in optical magnification as in Ref.~\cite{Herr}, but also distortions of the shape of the transverse density distributions. Describing their beam-separation dependence requires numerical simulations, that are detailed in Sec.\,\ref{sec:bbSimCodes}.

\par
In practice, the correction for optical distortions is implemented as follows. Separately for each scan step in a horizontal and vertical \vdM-scan pair, the {\it luminosity-bias factor}  \LoLzOpt associated with beam--beam-induced optical distortions is extracted, as a function of the nominal separation $\Delta_i$, from one of the multiparticle simulations described in Sec.\,\ref{sec:bbSimCodes}. Here \Lum refers to the luminosity that would be measured in the presence of beam-beam optical-distortion effects, and \Lumz is the corresponding luminosity if beam-beam effects were turned off altogether, all other conditions remaining unchanged. The quantity \Lumz is dubbed the {\em nominal luminosity}; it is  akin to ``Monte Carlo truth'', and  is accessible in the simulation only. The beam-beam corrected collision rate $R^c_i(\delta_i)$ is then computed by dividing the measured collision rate by this simulation-based luminosity-bias factor:
\begin{equation}
\label{eqn:corPrinc}
R^c_i(\delta_i) = \frac{R_i(\Delta_i)}{[\Lum(\Delta_i)/\Lumz(\Delta_i)]_\mathrm{Opt}} 
\end{equation}
and used instead of $R_i(\delta_i)$ in computing the beam-beam corrected convolved bunch sizes \CSxcOpt and \CSycOpt (Eq.\,(\ref{eqn:caps})), the peak rate $\muc = \muvismax \, R^c_i(0)/ R_i(0)$, and from these quantities the visible cross-section \sviscOpt (Eq.\,(\ref{eqn:sigmaVis})).

		\subsubsection{Beam-beam correction strategy}
		\label{subsubsec:bbCorStrtgy}

Conceptually, the principle of the beam-beam correction to \vdM\ calibrations is to determine the visible cross-section \svisc from the convolved bunch sizes and peak collision rates corrected both for the orbit shift (Sec.~\ref{subsubsec:orbShift}) and for optical distortions (Sec.~\ref{subsubsec:optclDist}), \ie corrected to the values (\CSxc, \CSyc, \muc) that  these observables would take if the beam-beam interaction could be turned off during the scan. The quantity \svisc is then the proportionality constant that translates a measured visible interaction rate \muvis\ into the corresponding bunch luminosity $\mathcal{L}_{\mathrm b}$. It is important to note that even though the actual luminosity is always modified by the beam-beam interaction, including during head-on collisions typical of routine physics running,  beam-beam {\em corrections} are only needed {\em during scans}, basically because the beam-separation dependence of the beam-beam effects distorts the \vdM-scan curves. Once \svisc has been determined as specified above, it can always be used to translate the measured collision rate into luminosity units, irrespective of the extent to which this collision rate has been enhanced by the beam-beam interaction.

\par
In practice, orbit-shift and optical-distortion corrections must be applied on a bunch-by-bunch basis, with as inputs the measured bunch populations ($n_1, n_2$) and uncorrected convolved bunch widths (\CSx, \CSy), as well as the beam energy, the \bst setting and  the tunes. In many cases, these corrections can be extracted from a one-time parameterization of the simulation results in terms of the bunch-specific beam-beam parameter value and of the nominal tunes \Qx, \Qy. This procedure avoids the repeated use of CPU-intensive, time-consuming multiparticle simulations; it is detailed in Sec.\,\ref{sec:bbImpact}.
\section{Beam-beam simulation codes}
\label{sec:bbSimCodes}

Beam-beam corrections to \vdM\ calibrations were originally based on MAD-X (Sec.\,\ref{subsec:MADXdescr}). Since then, only the zero-separation case has proven amenable to analytical treatment (Sec.\,\ref{subsec:Analytical}). Beam--separation-dependent effects have been investigated using two independent multiparticle codes dubbed B*B (Sec.\,\ref{subsec:BsBDescr}) and COMBI (Sec.\,\ref{subsec:COMBIDescr}), that have been extensively cross-validated (Sec.\,\ref{subsec:XValidtn}).

	\subsection{Linear approximation with MAD-X}
	\label{subsec:MADXdescr}

Since the linear part of the beam-beam force is similar to a quadrupolar field (Fig.\,\ref{fig:bbForce}), one expects the beam-beam interaction to contribute additional focusing or defocusing, thereby shifting the tunes and affecting the optical functions all around the ring, including at the IP itself: this is known as the ``dynamic-$\beta$'' effect. In the specialized case of equally sized round beams colliding head-on at a single location, the resulting change in the $\beta$ function at the IP is given by~\cite{Herr}:
\begin{equation}
\label{eqn:dynBeta}
\frac{\bst}{\beta^*_0} = \frac{1}{\sqrt{1  -  4 \pi \xi \cot (2 \pi \mathrm{Q})  - 4 \pi^2 \xi^2}}
\end{equation}
where $\beta^*_0$ is the value of the unperturbed $\beta$ function at the IP, \bst  its value in the presence of the beam-beam interaction and Q the tune. The beam-beam parameter $\xi$ is proportional to the derivative of the beam-beam force.\footnote{The sign of the linear term in the denominator of Eq.~(\ref{eqn:dynBeta}) is flipped compared to that in Eq.~(3) of Ref.~\cite{Herr}. This is because the latter follows the convention that $\xi$ is negative (positive) for equal- (opposite-) charge beams. In the present paper, in contrast, $\xi$ is positive by definition (Eq.~(\ref{eqn:xiGDef})), and therefore Eq.~(\ref{eqn:dynBeta}) is adjusted so as to be applicable to equal-charge beams.}
Equation\,(\ref{eqn:dynBeta}) implies that the beam-beam induced change in the IP $\beta$ function, and therefore in the IP beam-size squared and in the luminosity, depends only on $\xi$ and on Q. In addition, this dynamic-$\beta$ effect is, to first order, proportional to $\xi$. This is the physical motivation underlying the parameterized-correction approach developed in Sec.\,\ref{sec:bbImpact}.

\par
In Ref.~\cite{Herr}, the general-purpose optics code MAD-X~\cite{MADX} was used to model the dynamic-$\beta$ effect during simulated \vdM\ scans. In this software package, beam-beam elements can be inserted at one or several IPs, and their impact on the tunes and on the single-particle optical functions computed as a function of the beam separation $\Delta$. The procedure effectively assumes that for a given beam separation, all particles in the bunch experience the same beam-beam kick, equal to that applied to a zero-amplitude particle for that particular value of $\Delta$. Unperturbed bunches are implicitly supposed to be strictly Gaussian, and to remain so in the presence of the beam-beam interaction: only the change in optical magnification between the LHC arcs and the IP is accounted for in this method.
	
\par
\begin{figure}[t!]
\centering
\includegraphics[width=0.48\textwidth]{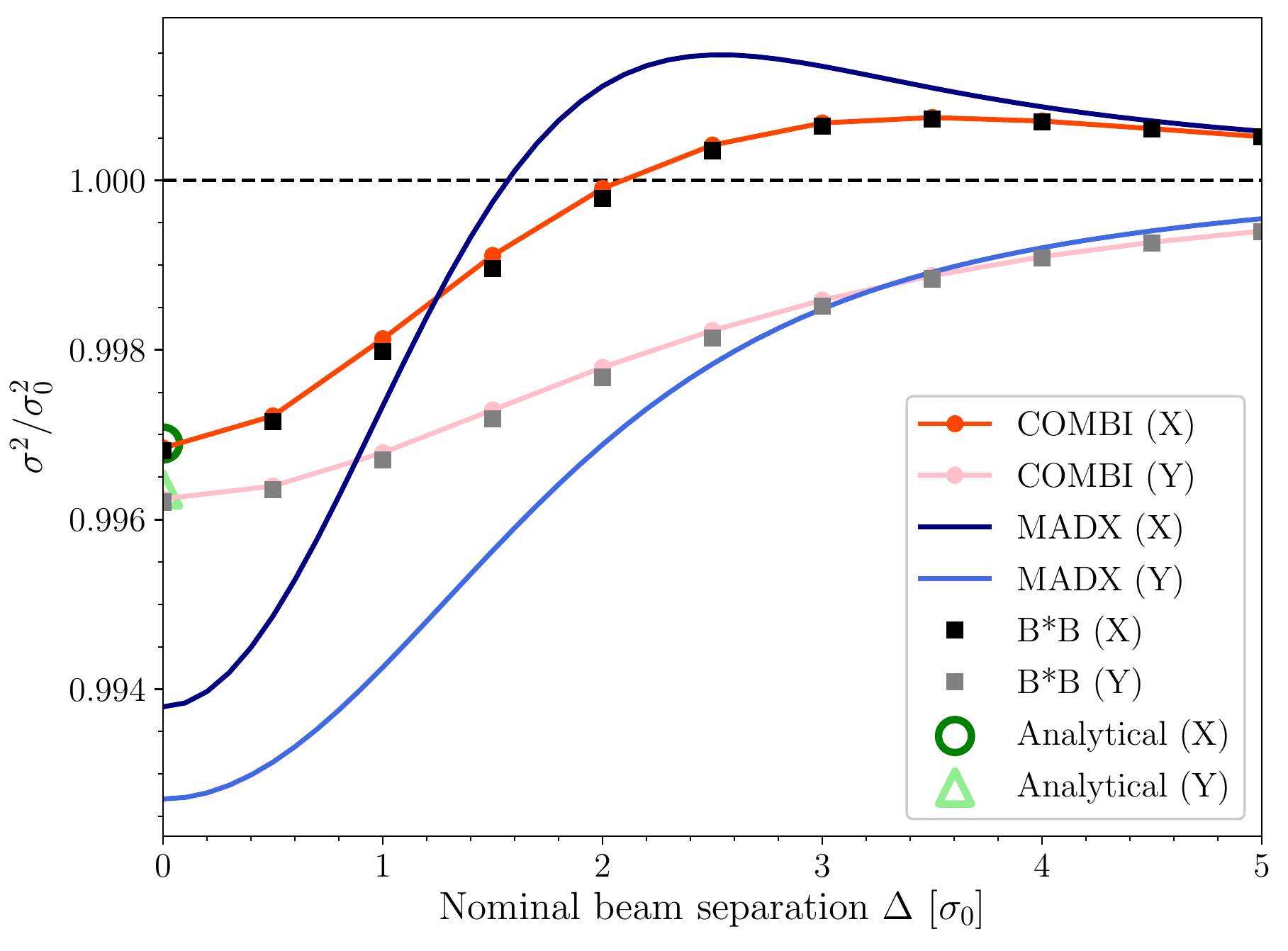}
\caption{Beam-separation dependence of the transverse RMS beam sizes during a simulated horizontal \vdM\ scan. The unperturbed beams are assumed to be round and perfectly Gaussian; their parameters are listed in the right column of Table\,\ref{tab:vdMBeamParms}. The vertical axis is the ratio squared of the actual beam size $\sigma$ to its unperturbed value $\sigma_0$; the horizontal axis is the nominal beam separation in units of $\sigma_0$. The MAD-X calculation of the single-particle dynamic-$\beta$ effect (blue curves) is compared to the results of the B*B (square markers) and COMBI (red curves) multiparticle codes.
The dark- and light-colored curves and markers illustrate the evolution, during the horizontal scan, of the horizontal and the vertical beam size respectively. The green circular and triangular markers display the predictions of Eq.~(\ref{eq:deltasig}), that only apply at zero beam separation.
}
\label{fig: beta/beta0 H scan} 
\vspace*{-0.3cm}
\end{figure}

This study was carried out for the reference parameter set listed in Table\,\ref{tab:vdMBeamParms}.  The corrections were then adapted to the  beam conditions of different \vdM\ sessions, using the assumption that $\Delta \beta/\beta^*_0 = \bst / \beta^*_0 - 1$ scales linearly with the value of $\xi$ inferred from the beam parameters measured during each scan. The simulated beam-separation dependence of the transverse beam size squared, \ie the value of $(\Lum / \Lumz)^{-1}$,  is illustrated in Fig.\,\ref{fig: beta/beta0 H scan}. Under head-on conditions ($\Delta = 0$), the intrinsically {\em defocusing} beam-beam force results in a $\sim 0.7$\% {\em reduction} of the beam size squared at the IP, \ie to an {\em increase} of the luminosity \Lum. This apparent contradiction results from the numerical value of the fractional tunes the LHC optics was designed for (Eq.~(\ref{eqn:dynBeta})). As $\Delta$ increases, the tune shift and the dynamic-$\beta$ effect weaken, change sign  (in the scanning plane only) around $\Delta/\sigma^0 \sim 1.6$, then peak and finally vanish asymptotically at very large separation. The change in optical magnification at $\Delta = 0$ is slightly different in the $x$ and $y$ planes, because the corresponding fractional tunes \qx and \qy differ by design; the difference in beam-separation dependence also reflects the fact that the vertical kick never changes sign during a horizontal scan, while the horizontal kick does.

\par
Historically, the optical-distortion correction to \svis\ predicted by MAD-X under typical Run-1 and Run-2 \vdM\ conditions lay in the 0.2--0.4\% range, much smaller than that associated with the orbit effect. For a long time, therefore, it was considered small enough that even if imperfect, it remained sufficiently accurate in view of the systematic uncertainty assigned at the time to the overall beam-beam correction, as documented \eg in Refs.~\cite{bib:ATL2011Lum,bib:ATL2012Lum}. 

\par
In hindsight, however, the limitations of applying the MAD-X approach to beam-beam corrections may not have been fully appreciated. Since at zero beam separation, the slope of the beam-beam force is steepest for zero-amplitude particles (Fig.\,\ref{fig:bbForce}, dotted green line), and since in MAD-X all particles in the witness bunch are assumed to experience the same linearized force, the dynamic-$\beta$ effect predicted by MAD-X at $\Delta =0$ is likely to be an overestimate; this is confirmed analytically below. The situation is reversed at large beam separation. When the beams are separated by (for instance) $\Delta = 4\,\sigma_0$,  the derivative of the force at large amplitude (say 3 to $4\,\sigma_0$) is typically larger (dotted green line) than at small amplitude (dotted brown line), suggesting that MAD-X underestimates the optical distortions at large beam separation. This conjecture, however, can only be confirmed using multiparticle simulations.

\subsection{Analytical estimate of optical distortions at zero beam separation}
\label{subsec:Analytical}

The particle phase space at any location $s$ along a storage ring is described by an ellipse, the shape of which depends on $s$ in a manner described by the well known Courant-Snyder parameters $\alpha$, $\beta$ and $\gamma$ \cite{bib:AccPhysHandbook}:
\begin{eqnarray*}
J_{u} = \frac{1}{2} (\gamma_{u} u^2 + 2 \alpha_{u} u p_{u} + \beta_u p_{u}^2 )\ .
\end{eqnarray*}
Here $u =x, y$ is the deviation of the single-particle orbit from the reference trajectory,  $p_u = du/ds$, and $J_{u}$ is the action variable that represents the invariant of the motion for each single particle when the reference energy is not changing. The particle position in transverse phase space is fully described by the action $J_u$ and by the corresponding phase variable $\phi_{u}$ defined as:
\begin{eqnarray}
\tan{\phi_u} = - \beta_{u} \frac{p_{u}}{u} - \alpha_{u}\ .
\label{Angle}
\end{eqnarray}
In addition to  distorting the closed orbit, the beam-beam interaction acts as a non-linear electromagnetic lens that includes a quadrupolar term; the latter affects the optical functions $\beta$ and the betatron tunes of the individual particles~\cite{BBCAS, Chao}. During a \vdM\ scan and for each colliding bunch, the quadrupolar component of the electromagnetic field of the opposing bunch changes as a function of the relative transverse separations $\delta_i$ ($i = x, y$) between the beams centroids, and of the single-particle transverse actions $J_{x}$ and $J_{y}$ ~\cite{Chao,Chao1}. This results in a shift in the betatron tune, the so-called detuning with amplitude, given by~\cite{AmpDetuning}:
\begin{align*}													    
\Delta \Qx(J_x,J_y) = - \xi_{x}  \int_0^{\infty}\frac{1}{(1+t)^2} ~ e^{-\frac{J_x+J_y}{2\epsilon (1+t)}}  \times 	\nonumber  \\
\left[I_0\left(\frac{J_x}{2\epsilon (1+t)}\right)  - I_1\left(\frac{J_x}{2\epsilon (1+t)}\right) \right]   \times   \nonumber  \\
I_0\left(\frac{J_y}{2\epsilon  (1+t)}\right)    \mathrm{d}t	   				  
\end{align*}
for the horizontal dimension. Here  $\xi_{x}$ is the beam-beam parameter, $I_0$ and $I_1$ are modified Bessel functions of the first kind, $\epsilon$ is the emittance (assumed equal in $x$ and $y$), and $t$ is a bound variable. This integral can only be solved when one action is zero, obtaining (for $J_y = 0$):
\begin{eqnarray}
\Delta \Qx (J_x)  	&=& - \xi_x  \frac{2 \epsilon}{J_x}  \left[1  - I_0 \left( \frac{J_x}{2 \epsilon} \right) \, e^{-\frac{J_x}{2 \epsilon} } \right] \,	 
                                \label{tunespread}  \\
\Delta \Qy (J_x)  	&=& - \xi_x    \left[I_0 \left( \frac{J_x}{2 \epsilon} \right) 
                                + I_1\left( \frac{J_x}{2 \epsilon} \right)  \right] \, e^{-\frac{J_x}{2 \epsilon} } \, . \nonumber	      
\end{eqnarray}
For a zero-amplitude particle ($J_x = J_y =0$) and beams in head-on collision ($\delta_x = \delta_y = 0$), the change in tune is equal to the beam-beam parameter $ \xi_x$.

\par
Head-on beam-beam $\beta$-beating can be derived from the tune shift\footnote{The overall minus sign in Eqs.~(\ref{tunespread}) and (\ref{eq:beta}) expresses the fact that when the charges of the colliding bunches have the same sign, the tune shift is negative, even though the beam-beam parameter remains positive by definition (Eq.~(\ref{eqn:xiGDef})).} in Eq.~(\ref{tunespread}) as:

\begin{eqnarray}
\frac{\Delta \beta_x}{\beta_x}(J_x,J_y) &=& - \frac{\Delta \Qx(J_x,J_y)}{\xi_x}  \cdot  \frac{\Delta \beta_0}{\beta_x}  
\label{eq:beta}
\end{eqnarray}
where 
\begin{equation}
\frac{\Delta \beta_0}{\beta_x}= \frac{\Delta \beta_x}{\beta_x}(0,0)
\nonumber
\end{equation}
is the linear $\beta$-beating or, equivalently, the $\beta$-beating at $J_x=J_y=0$. 

\par
In the zero-separation case, the $\beta$-beating averaged over the particle distribution (assumed Gaussian) can be calculated as:
\begin{eqnarray}
\overline{\frac{\Delta \beta}{\beta_x}}  
& = 		& \int_0^{\infty} dJ_x\int_0^{\infty}dJ_y \frac{\Delta \beta_x}{\beta_x}(J_x,J_y) e^{-J_x-J_y}\nonumber \\
& \approx	&0.633~ \frac{\Delta \beta_0}{\beta_x}\, .
\nonumber
\end{eqnarray}
%
%
%
For bunches colliding with zero transverse separation, therefore, the particle-action distribution is modified such that the average beam-beam beating is reduced to about $ 63\%$ of the single-particle estimate computed in Ref.~\cite{Herr}. The corresponding impact on the head-on luminosity can be derived analytically, as follows.

\par
The RMS single-beam size $\sigma_x = \sqrt{\beta_x \, \epsilon}$, including the action-dependent, beam--beam-induced  $\beta$-beating given by Eq.~(\ref{eq:beta}), is computed using the following relation:
\begin{eqnarray*}
x^2 	& = & 2\beta_x J_x \left(1+\frac{\Delta\beta_x}{\beta_x}\right)\cos^2\phi_x =\nonumber \\
  	& = & \beta_x J_x \left(1+\frac{\Delta\beta_x}{\beta_x}\right)(1+\cos2\phi_x)  \, .
\end{eqnarray*}
Assuming Gaussian particle-density distributions, this yields%
\begin{align*}
&\sigma_x^2 = \int_0^{\infty} dJ_x\int_0^{\infty}dJ_y\int_0^{2\pi} \frac{d\phi}{2\pi}\ x^2 e^{-J_x-J_y}  \nonumber \\ & =\beta_x \left(1 + \int_0^{\infty} dJ_x\int_0^{\infty}dJ_y e^{-J_x-J_y} J_x\frac{\Delta \beta_x}{\beta_x}(J_x,J_y)\right) 
\end{align*}
where, for simplicity, we  used an emittance value of $\epsilon=1$. The triple integral has an exact solution:
\begin{eqnarray}
\sigma_x^2 &=&   \beta_x\left(1+   \frac{1}{2}  \frac{\Delta \beta_0}{\beta_x}  \right)\, . 
\label{eq:deltasig}
\end{eqnarray}
This  calculation implies that in head-on collisions, beam--beam-induced linear $\beta$-beating of magnitude ${\Delta \beta_0}/{\beta}$ in both the horizontal and the vertical plane, causes a relative luminosity change of
\begin{equation}
\frac{\Delta L}{L}  \approx -   \frac{1}{2}  \frac{\Delta \beta_0}{\beta} \, .
\label{eq:deltaL}
\end{equation}
For collisions with zero transverse separation, in other words, the beam-beam interaction changes the luminosity by only half of what is expected from linear $\beta$-beating: this is consistent with the COMBI predictions displayed in Fig.~\ref{fig: beta/beta0 H scan}. \

\par
Using the reference parameter set in Table~\ref{tab:vdMBeamParms}, for instance, the beam--beam-induced linear $\beta$-beating amounts to about -0.6\%, leading to a head-on luminosity enhancement of about 0.3\%. During high-luminosity physics running typical of LHC Run 2, the effect is computed to be two to three times larger; it is both $\xi$- and tune-dependent (see Eq.~(\ref{eqn:dynBeta})), and therefore its magnitude changes as beam conditions evolve.

\par
When beams are transversely separated, analytical computations become impossible, forcing one to resort to numerical simulations such as those described below.

	\subsection{Weak-strong limit: B*B}
	\label{subsec:BsBDescr}
	
The B*B package~\cite{Balagura_2021} is a multiparticle-simulation program developed specifically for assessing beam-beam biases in \vdM scans, that was optimized for speed by adopting several simplifying assumptions. It aims at predicting the corresponding beam-beam corrections with better than $0.1\%$ accuracy (for a given set of input parameters), so as not to contribute significantly to the overall \vdM-calibration uncertainty. The code is written in C++; it can be used as a standalone application, or as a library available in C, C++, Python and R.

\par
The initial transverse particle density distributions $\hat{\rho} _B (x,y)$, where $B = s, w$ refers to either the source ($s$) or the witness ($w$) bunch, are modeled by a (linear combination of) two-dimensional elliptical Gaussian(s). Since during typical \vdM\ scans at LHC, beam--beam-induced bunch-shape deformations remain small, they are taken as negligible when computing the electromagnetic field of the source bunch, which therefore remains unperturbed as a function of the transverse beam separation. This makes it possible to precompute this field over a two-dimensional grid in the transverse plane at initialization time, and to use only fast interpolations on all subsequent machine turns. The opposing, ``witness''  bunch is represented by a set of $\mathcal{O}(1000)$ macroparticles, that are transported around the ring using linear maps, with as input the nominal ($x$, $y$) phase advance between consecutive collision points. 
The B*B simulation, therefore, falls in the ``weak-strong'' category in that it models the transverse deformation of the density distribution of the witness bunch ($\hat{\rho} _w \to \hat{\rho}_w \, + \, \delta \hat{\rho}_w$) caused by the electromagnetic field of an unperturbed source bunch, the density distribution of which remains unaffected. Effects such as coherent bunch oscillations, or the distortion of the shape (and therefore of the field) of the source bunch induced by the deformation of the witness bunch, are therefore implicitly neglected.

\par
The macroparticles are selected from a two-dimensional ($x$, $y$) grid of betatron amplitudes, and assigned weights that are precalculated from the initial Gaussian density distribution $\hat{\rho}_w$. Their initial  betatron phase is chosen randomly; the phase then samples the full $[0,2\pi]$ interval over the next $\mathcal{O}(100-1000)$ turns or so.

\par
The overlap integral of the perturbed and unperturbed bunches (Eq.~(\ref{eqn:lumi}))  is computed as the sum over macroparticles, of the source-bunch density $\hat{\rho}_s (x_w, y_w)$ evaluated at the current location $ (x_w, y_w)$ of the macroparticle considered, and multiplied by the weight of that same macroparticle.  If the populations and initial transverse-density distributions of the two colliding bunches are identical ($n_1 = n_2$ and $\hat{\rho} _1 (x,y) = \hat{\rho} _2 (x,y)$), the simulation needs to be run once only; in the presence of any initial B1-B2 asymmetry, it needs to be run twice, with the roles of the source and the witness bunch swapped between the two beams. The overall beam-beam bias affecting the overlap integral is approximated by
\begin{align}
  \int  \left( \hat{\rho}_1+\delta\hat{\rho}_1 \right)  \left(\hat{\rho}_2+\delta\hat{\rho}_2\right) \mathrm{d}x\,\mathrm{d}y 
  - \int \hat{\rho}_1 \hat{\rho}_2 \, \mathrm{d}x \, \mathrm{d}y  									\nonumber	\\
  \approx   
  \int  \left( \delta\hat{\rho}_1 \cdot \hat{\rho}_2   +  \hat{\rho}_1 \cdot \delta\hat{\rho}_2  \right) \mathrm{d}x \, \mathrm{d}y,
   \nonumber
\end{align}
up to some physical constants and where the second-order term $\int  \delta\hat{\rho}_1 \, \delta\hat{\rho}_2 \, \mathrm{d}x \, \mathrm{d}y$ is neglected.

\par
The uncertainties arising from the manner in which the beam-beam force is switched on in the calculation (gradually or instantaneously), from the granularity of the simulation (finite number of macroparticles, largest sampled transverse amplitude, random choice of the initial phases), and from other simulation-control parameters such as the number of accelerator turns, are discussed in detail in Ref.~\cite{Balagura_2021} and found to lie well below $0.1\%$. 

\par
Finally, even though the density distributions $\hat{\rho} _{s,w}$ are only two-dimensional, and therefore represent the projection of the full six-dimensional distribution onto a plane perpendicular to the beam axis, the B*B package is capable of simulating the {\em geometrical} effects that arise in the presence of a non-zero crossing angle in a plane of arbitrary orientation~\cite{Balagura_2021}. Simulating {\em longitudinal dynamics}, however, and in particular the potential impact of a finite crossing angle on beam-beam corrections to the \vdM\ calibration, requires a fully six-dimensional treatment such as that outlined below.

\subsection{Strong-strong model: COMBI}
\label{subsec:COMBIDescr}

The COMBI code \cite{COMBICODE} has been developed over the years for simulating, in a self-consistent manner, the coherent  beam-beam interaction between multiple bunches coupled by head-on and/or long-range beam-beam encounters~\cite{combitatiana, pielonithesis, combiWT}.  It includes a first level of parallelization based on the Message Passing Interface (MPI)~\cite{jonesWT}, and a second one sharing several CPUs per node using OpenMPI~\cite{buffatthesis, openmpiweb}. 

\par
The code has been optimized to handle simultaneously multiple bunches at several interaction points, thereby allowing flexible collision patterns. The circumference of each accelerator ring is modeled by a number of equally spaced slots that define the possible bunch positions. At each location one can assign an action (e.g. head-on or long-range beam-beam interaction, linear or non-linear magnetic element, Landau octupole, linear or non-linear map,...) that will be executed when a bunch is present. The actions corresponding to head-on or long-range beam-beam interactions require one bunch from each beam in order to be performed. Each macroparticle is tracked individually under the effect of the preassigned actions, in either four or six dimensions. In the LHC arcs, the macroparticles are transported by applying a linear transfer map to their coordinates, using phase advances precomputed by MADX with, as input, the nominal optical configuration used during \vdM sessions. 

\par
In B*B, the transverse density distribution of the source bunch, and therefore its electromagnetic field, remain unchanged from turn to turn; only the witness bunch contains macroparticles, that are tracked over thousands of turns in the transverse plane. Coherent effects, therefore, cannot be modeled, and neither can longitudinal dynamics. COMBI, in contrast, describes the two partners in a colliding-bunch pair as independent sets of macroparticles. In this paper, the initial, unperturbed density distributions are uncoupled single Gaussians by default; however, arbitrary distributions, such as a linear combination of Gaussians, can be used instead. 

\par
The beam-beam interaction can be described by different models: a four-dimensional Gaussian lens \cite{pielonithesis, combiWT}, a six-dimensional Gaussian lens \cite{Hirata, Laurent}, or a field computed from the actual charge distributions by the HFMM method \cite{HFMM}. Multiple beam-beam encounters, and therefore the evolution of bunch parameters such as emittance or transverse barycenter position, as well as coherent beam-beam effects, are treated in a self-consistent manner: the particle trajectories affected by the beam-beam interaction modify the density distribution of the corresponding bunch, and the fields produced by both partners are updated turn by turn from these modified distributions. Whether these fields are estimated in the Gaussian approximation, or by the HFMM method, yields effectively identical results for the full range of beam-beam parameter values considered in this paper (see Appendix \ref{sec:nonGausFieldImpact}).

\par
At a given IP, the luminosity per colliding-bunch pair can be computed either analytically, using the Gaussian formalism of Ref.~\cite{Herr:2003em}, or by evaluating numerically the actual overlap integral of the two colliding distributions. The first method evaluates the luminosity from Eqs.\,(\ref{eqn:lumifin})--(\ref{eqn:LvsDelta}) by substituting the single-beam sizes $\sigma_{iB}$ with the RMS transverse widths of the macroparticle distributions, and the separation $\delta$ with the distance between the barycenters of these distributions. In the second method, the overlap integral of the macroparticle distributions is computed using functionalities developed specifically for this purpose. This is the approach adopted throughout this paper, except where explicitly specified otherwise; the numerical-integration procedure and the associated convergence studies are detailed in Appendix \ref{sec:Integration}.

\subsection{Cross-validation of simulation codes}
\label{subsec:XValidtn}

The full impact of the beam-beam interaction on the beam-separation dependence of the luminosity-bias factor \LoLzFD can be broken down as follows.

\par
The impact of the orbit shift  can be expressed either in terms of the beam-beam induced distortion of the {\em actual} beam separation $\delta_i$, as discussed in Sec.\,\ref{subsubsec:XValid_orb} below, or in terms of an orbit-related luminosity-bias factor, denoted by \LoLzOrbD for a given {\em nominal} separation $\Delta$. In physically intuitive terms, and in the context of a weak-strong model such as B*B, the orbit shift results from applying to all particles in the witness bunch the same electromagnetic kick, computed from the field produced by the source bunch as a whole and averaged over all particles in the witness bunch. Since this is tantamount to a dipole kick, only the orbit of the witness bunch is affected; the size and shape of the macroparticle distribution remain invariant as $|\Delta|$ increases.

\par
The impact of the optical distortions is quantified in terms of the luminosity-bias factor \LoLzOptD first introduced in Sec.\,\ref{subsubsec:optclDist}. It results from the combination of beam-separation and amplitude-dependent $\beta$-beating effects that modulate the RMS transverse beam size (Sec.\,\ref{subsubsec:XValid_bBeat}), and of $\Delta$-dependent bunch-shape distortions that affect the beam-beam overlap integral (Sec.\,\ref{subsubsec:XValid_ovlp}).

\par
The B*B and COMBI packages have been mutually benchmarked, and their results compared to those obtained either analytically (where possible) or using MADX. The cross-package comparisons of the beam-beam induced orbit shift and of the predicted optical distortions are based on the reference parameter set of Table~\ref{tab:vdMBeamParms}, and assume that in the absence of the beam-beam interaction the transverse density distributions are strictly Gaussian. The consistency of B*B and COMBI results at higher beam-beam parameters is quantified in Sec.\,\ref{subsubsec:XValid_chsi}. All the results presented in this Section assume in addition that the beams collide only at the IP where the \vdM\ scan is taking place; multiple-IP effects will be discussed in Sec.\,\ref{subsec:multiIP}.

\subsubsection{Beam-beam-induced orbit shift}
\label{subsubsec:XValid_orb}

\begin{figure}[h!]
\centering
\includegraphics[width=0.48\textwidth]{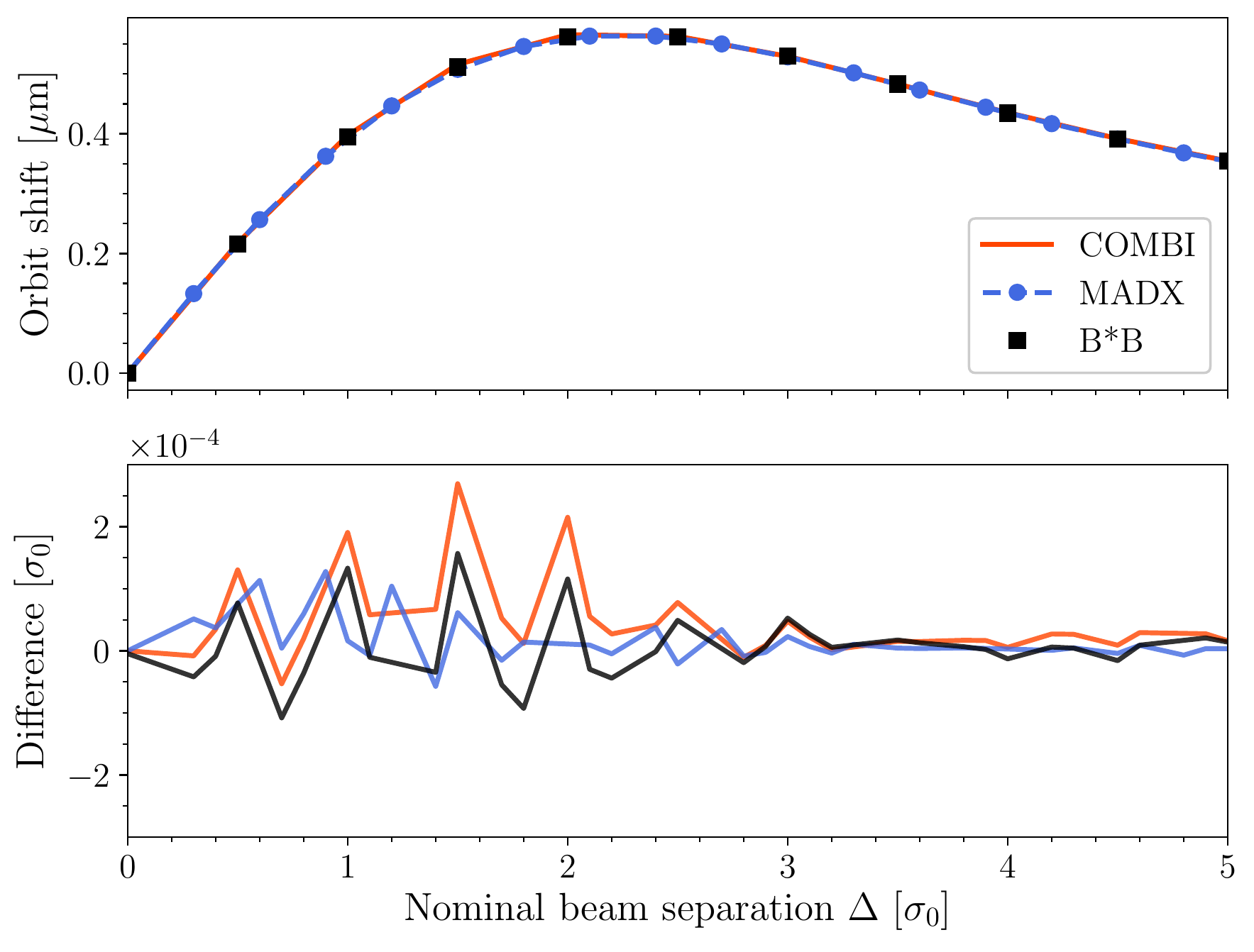}
\caption{Top: bunch-centroid displacement $\delta^{bb}_{iB}$ due to the beam-beam deflection during a simulated vdM scan, as predicted by COMBI, MAD-X and B*B. The bunch parameters are listed in the right column of Table \ref{tab:vdMBeamParms}. The horizontal axis is the nominal beam separation in units of the unperturbed transverse beam size $\sigma_0$. Bottom: difference between the simulated orbit shift and that calculated analytically using Eqs.\,(\ref{eqn:deflAng})-(\ref{eqn:orbShift}), in units of $\sigma_0$. The wiggles reflect the combination of a finite sampling step ($0.2\,\sigma_0$) and of numerical noise.
}
\label{fig:Orbit} 
\end{figure}

Figure~\ref{fig:Orbit} displays the beam--beam-induced, single-beam orbit shift ($\delta^{bb}_{iB}$) at the IP, as simulated by MADX, B*B and COMBI, and compares it to the analytical prediction. The three models agree extremely well among themselves, and their results are indistinguishable from those of the analytical prediction. 

\par
During a luminosity-calibration session, the bunches typically collide at more than one IP (Sec.~\ref{subsec:multiIP}); however performing simultaneous beam-separation scans at different IPs is carefully avoided. Under these conditions and to an excellent approximation, the orbit shifts associated with slightly misaligned collisions at a non-scanning IP remain static during a beam-separation scan at another IP. 
Their impact on the beam separation at the scanning IP is therefore expected to remain constant during the scan, and has been neglected in the simulations reported in this paper.

\subsubsection{Impact of amplitude-dependent $\beta$-beating effects in the Gaussian-bunch approximation}
\label{subsubsec:XValid_bBeat}

The beam-separation dependence of the transverse RMS beam-size ratio (squared for easier interpretation in terms of luminosity) is presented in Fig.~\ref{fig: beta/beta0 H scan}. The results of B*B and COMBI agree to better than $2\times 10^{-4}$.

\par
At $\Delta = 0$, both predict a decrease in beam-size squared (or equivalently an increase in head-on luminosity) about half of that obtained using MAD-X. This is remarkably consistent with the analytical predictions of Eqs.~(\ref{eq:deltasig}) and (\ref{eq:deltaL}): in MAD-X, all particles in the witness bunch are subject to the same quadrupole-like force as the zero-amplitude particle, while in B*B and COMBI, the slope of the beam-beam force  decreases and even changes sign as the betatron amplitude increases (Fig.~\ref{fig:bbForce}), resulting in a smaller overall change in optical demagnification at the IP.

\par
Amplitude detuning also explains the different $\Delta$-dependence of the beam sizes, with that in MAD-X being more pronounced than in B*B and COMBI. At very large beam separation, where the electromagnetic force becomes similar to that of a distant, point-like charge, all three curves tend towards a common asymptote. As for the different evolution of the horizontal and vertical beam sizes during a  scan, it simply reflects the tune-dependence apparent in, for instance, Eq.~(\ref{eqn:dynBeta}).

\par
If one assumes that the optical distortions discussed above modify the transverse RMS bunch sizes at the IP without significantly affecting their initial, purely Gaussian shape, Eqs.~(\ref{eqn:lumifin})--(\ref{eqn:LvsDelta}) apply. The combination of the beam-beam induced orbit shift and of the optical distortions then yields the following expression for the  luminosity-bias factor during a beam-separation scan in plane $i$ ($i = x,y$):
\begin{align}
\frac{\Lum}{\Lumz} (\Delta_i) & = &	\frac{\Sigma^0_x \Sigma^0_y}{\Sigma_x (\Delta_i) \Sigma_y(\Delta_i)} 	
 							e^{- \frac{1}{2}  [(\delta_i/\Sigma_i)^2 - (\Delta_i/\Sigma^0_i)^2]}		
\label{eq:LbBeat} \\
						& = &	\frac{(\sigma^0)^2}{\sigma_x (\Delta_i) \sigma_y(\Delta_i)} 	
 							e^{- \frac{1}{4}  [(\delta_i / \sigma_i)^2 - (\Delta_i /  \sigma^0)^2]}	
\label{eq:LbBeatR}					
\end{align}
where $\sigma_{x,y}$, $\Sigma_{x,y}$, \Lum (resp. $\sigma_0$, $\Sigma^0_{x,y}$,  $\Lumz$) are the RMS single-beam sizes, the inverse of the overlap integrals and the luminosity in the presence (resp. absence) of the beam-beam interaction.

\par
\begin{figure}[b!]
\centering
\includegraphics[width=0.48\textwidth]{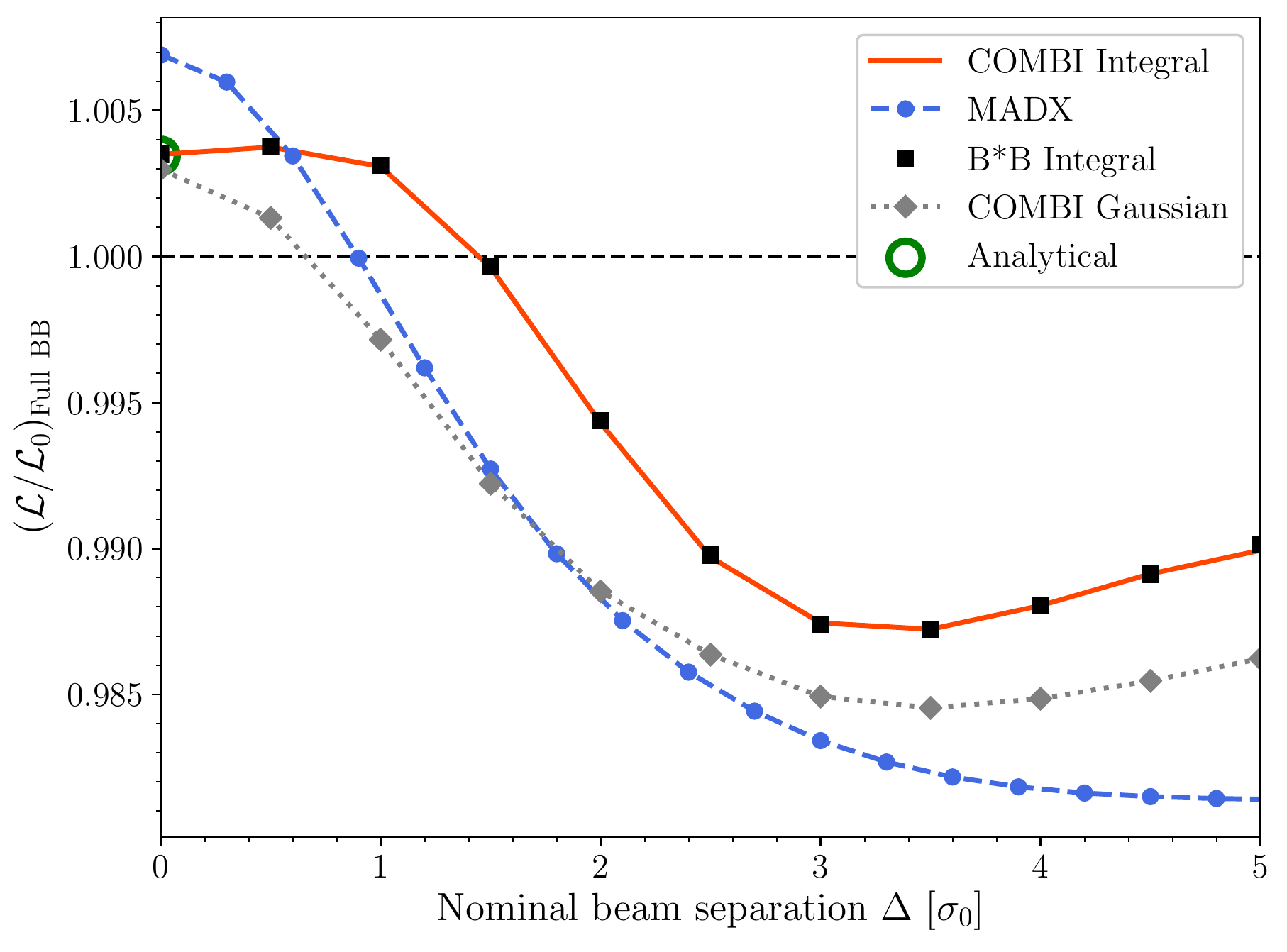}
\caption{Beam-separation dependence of the luminosity-bias factor \LoLzF during a simulated horizontal \vdM\ scan. The unperturbed beams are assumed to be round and perfectly Gaussian; their parameters are listed in the right column of Table~\ref{tab:vdMBeamParms}. The  horizontal axis is the nominal beam separation in units of the unperturbed transverse beam size $\sigma_0$. The MAD-X (dashed blue) and COMBI (dotted grey) curves are valid only in the Gaussian-bunch approximation (Sec.\,\ref{subsubsec:XValid_bBeat}); the COMBI (red curve) and the B*B markers (black squares) are obtained from the overlap integrals of the multiparticle distributions (Sec.\,\ref{subsubsec:XValid_ovlp}). The green circular marker displays the prediction of Eq.~(\ref{eq:deltaL}), that only applies at zero beam separation. 
}
\label{fig:L/L0} 
\end{figure}

Combining Eq.~(\ref{eq:LbBeatR}) with the beam-size ratios shown in Fig.~\ref{fig: beta/beta0 H scan} yields the beam-separation dependence of the luminosity-bias factor $\LoLz (\Delta)$ in the Gaussian-bunch approximation, as illustrated in Fig.\,\ref{fig:L/L0} for MAD-X (dashed blue) and COMBI (dotted grey). These curves are not simply the inverse of the beam-size ratios displayed in Fig.~\ref{fig: beta/beta0 H scan}, because they include in addition the impact, at a given nominal separation, of the beam-beam induced orbit shift. At zero and moderate separation ($\Delta / \sigma_0 < 1$). the luminosity bias $\LoLz -1$ is positive and dominated by the dynamic-$\beta$ effect; as the separation increases, the orbit shift, which is represented by the exponential term in Eq.~(\ref{eq:LbBeat}),  progressively takes over.

\subsubsection{Impact of optical distortions on the beam-beam overlap integral}
\label{subsubsec:XValid_ovlp}

If the transverse-density distributions are sufficiently distorted by the non-linearity of the beam-beam force, the Gaussian-bunch approximation encapsulated in Eqs.~(\ref{eq:LbBeat})--(\ref{eq:LbBeatR}) is no longer valid, and the luminosity bias must be calculated numerically from the beam--separation-dependent overlap integrals of the macroparticle distributions. To this effect, an integrator module, detailed in Appendix \ref{sec:Integration}, has been developed for COMBI; B*B provides an equivalent functionality~\cite{Balagura_2021}. The resulting beam-separation dependence of the  luminosity-bias factor is shown by the red curve and the black markers in Fig.~\ref{fig:L/L0}. At each simulated scan step (indicated by the markers), B*B and COMBI agree to better than $2\times 10^{-4}$; the difference is not systematic, but fluctuates around zero. The difference between the Gaussian-bunch approximation and the numerically calculated overlap integral demonstrates that the non-Gaussian tails (or more precisely non-Gaussian deviations from the originally Gaussian shape), that are induced by the beam-beam interaction, have a significant impact on the luminosity as soon as $\Delta / \sigma_0 \ge 0.2$. The excellent agreement between the two multiparticle simulations also demonstrates that strong-strong beam-beam effects, that are modeled by COMBI but not by B*B, remain negligible in the low beam-beam parameter regime considered here.

\par
Since the predicted orbit effect is identical in all simulations (Fig.\,\ref{fig:Orbit}), the differences between MAD-X on the one hand (Fig.\,\ref{fig:L/L0}, dashed blue curve), and B*B/COMBI on the other (black markers/red curve), is entirely associated with optical distortions. At zero separation, the difference is entirely explained by amplitude detuning (Eq.~(\ref{eq:deltaL})); once the separation increases, beam--beam-induced non-Gaussian tails play a growing role, as illustrated by the difference between the red and grey curves.\footnote{The very small difference, at zero separation, between the grey and red curves in Fig.\,\ref{fig:L/L0} may grow when the bunches collide not only at the scanning IP, but also at additional IPs (Sec.~\ref{subsec:multiIP}). In this case, and depending on the phase advance between IPs, beam--beam-induced non-Gaussian tails at these additional IPs may contribute noticeably to the overlap integral at the scanning IP, even for zero beam separation.}       

\par
\begin{figure}[b!]
\centering
\includegraphics[width=0.48\textwidth]{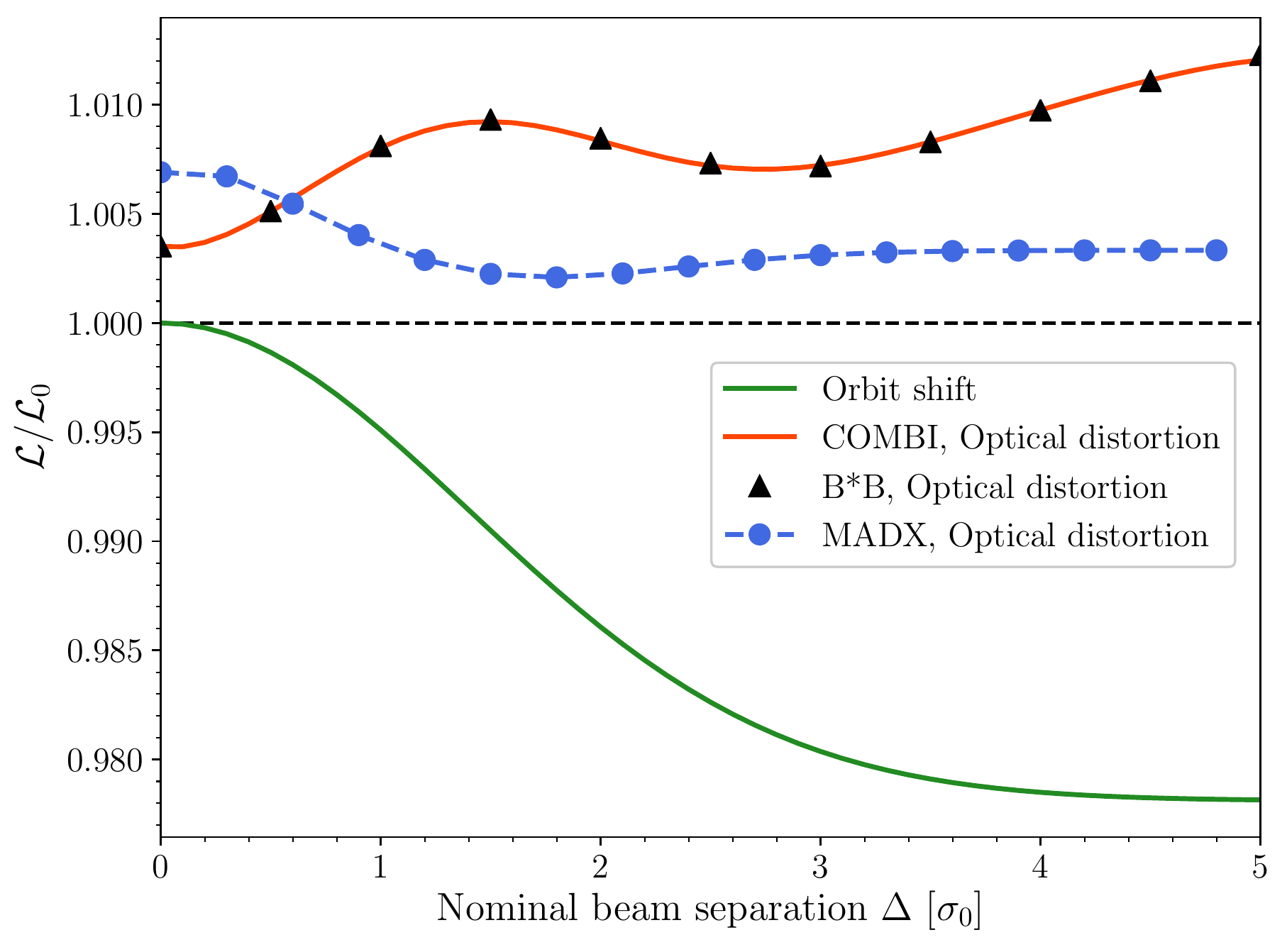}
\caption{Beam-separation dependence, during a simulated horizontal \vdM\ scan, of the luminosity-bias factor \LoLzOpt associated with the optical distortions, and of that induced by the beam-beam orbit effect (\LoLzOrb). The unperturbed beams are assumed to be round and perfectly Gaussian; their parameters are listed in the right column of Table~\ref{tab:vdMBeamParms}. The  horizontal axis is the nominal beam separation in units of the unperturbed transverse beam size $\sigma_0$. The MAD-X (dashed blue line with markers), weak-strong B*B (black triangles) and strong-strong COMBI (red curve) optical-distortion curves are obtained by dividing the corresponding full beam-beam bias factor shown in Fig.\,\ref{fig:L/L0} by that predicted analytically for the orbit effect alone (green curve).
}
\label{fig:L/L0_OrbOpt} 
\end{figure}

In the context of the beam-beam correction strategy outlined in Sec.~\ref{subsec:bbBias}, it is convenient to separate the full luminosity bias presented in Fig.\,\ref{fig:L/L0} into its optical-distortion and orbit-shift components. The luminosity-bias factor associated with optical distortions and denoted by \LoLzOpt is defined as the ratio, scan step by scan step, of the full luminosity-bias factor \LoLzF and of the bias factor \LoLzOrb associated with the orbit shift:
\begin{equation}
\LoLzOptD = \frac{\LoLzFD} {\LoLzOrbD}
\label{eqn:OptFulOrbdef}
\end{equation}
Its beam-separation dependence is presented in Fig.\,\ref{fig:L/L0_OrbOpt}. While MAD-X overestimates the dynamic-$\beta$ effect at zero separation, it strongly underestimates the optical distortions at medium and large beam separations. Since all models predict the same orbit effect, and since the optical distortions and the orbit effect impact the overlap integrals in opposite ways, their mutual cancellation is stronger in B*B and COMBI than in MAD-X, resulting in a net overall beam-beam correction of significantly smaller magnitude. This will be discussed quantitatively in Sec.~\ref{sec:bbImpact}.

\par
The beam-separation dependence of the optical-distortion factor \LoLzOpt predicted by B*B and COMBI (Fig.~\ref{fig:L/L0_OrbOpt}) exhibits a characteristic $S$-shape or ``wiggle'', the amplitude of which increases with $\xi$ (Fig.~\ref{fig:L/L0VsXi}) and with the fractional tunes (Fig.~\ref{fig:Opt-ll0-Q-dependence}). The beam separation where the wiggle amplitude exhibits either a peak ($\Delta \sim 1.6\, \sigma_0$) or a dip ($\Delta \sim 3\, \sigma_0$), is determined by the shape of the particle-density distribution. This is because the latter dictates both the dependence of the beam-beam force on the distance to the center of the source bunch, and the transverse distribution of the witness particles in the opposing bunch; an example will be offered in Sec.~\ref{subsec:nonGausImpact} (Fig.~\ref{fig:LoLzOpt_SGDG}). 

\par
Some intuitive insight into the beam-separation dependence of  \LoLzOpt can be gained by conceptually separating what occurs in the Gaussian core of the witness bunch from what happens to the large-amplitude particles in its non-Gaussian tails. The grey curve in Fig.~\ref{fig:L/L0} reflects the evolution, during the scan, of the transverse RMS bunch sizes (Fig.~\ref{fig: beta/beta0 H scan}); in a sense, therefore, it represents the impact of beam-beam effects on the overlap of the Gaussian component of the particle distribution. The difference between the red and grey curves, in contrast, reflects the beam-separation dependence of the luminosity contribution of the non-Gaussian transverse tails; the latter dominate the overall upward trend of the red curve in Fig.~\ref{fig:L/L0_OrbOpt}, and are responsible for most of its $S$-shape. A quantitative understanding of these features, however, is accessible neither to analytical calculations nor to simple physical arguments. It can only be obtained from simulations, because it requires taking into account, at each scan step, the interplay between:
\begin{itemize}
\item
the actual beam separation;
\item
the transverse bunch shapes;
\item
the non-linearity of the beam-beam force;
\item
the combined separation- and amplitude-dependence of the force exerted by the source bunch as a whole on those  particles in the witness bunch that significantly contribute to the luminosity at the scan step considered. 
\end{itemize}

\subsubsection{Beam-parameter dependence of beam-beam corrections}
\label{subsubsec:XValid_chsi}

\begin{figure}[htb]
\centering
\includegraphics[width=0.48\textwidth]{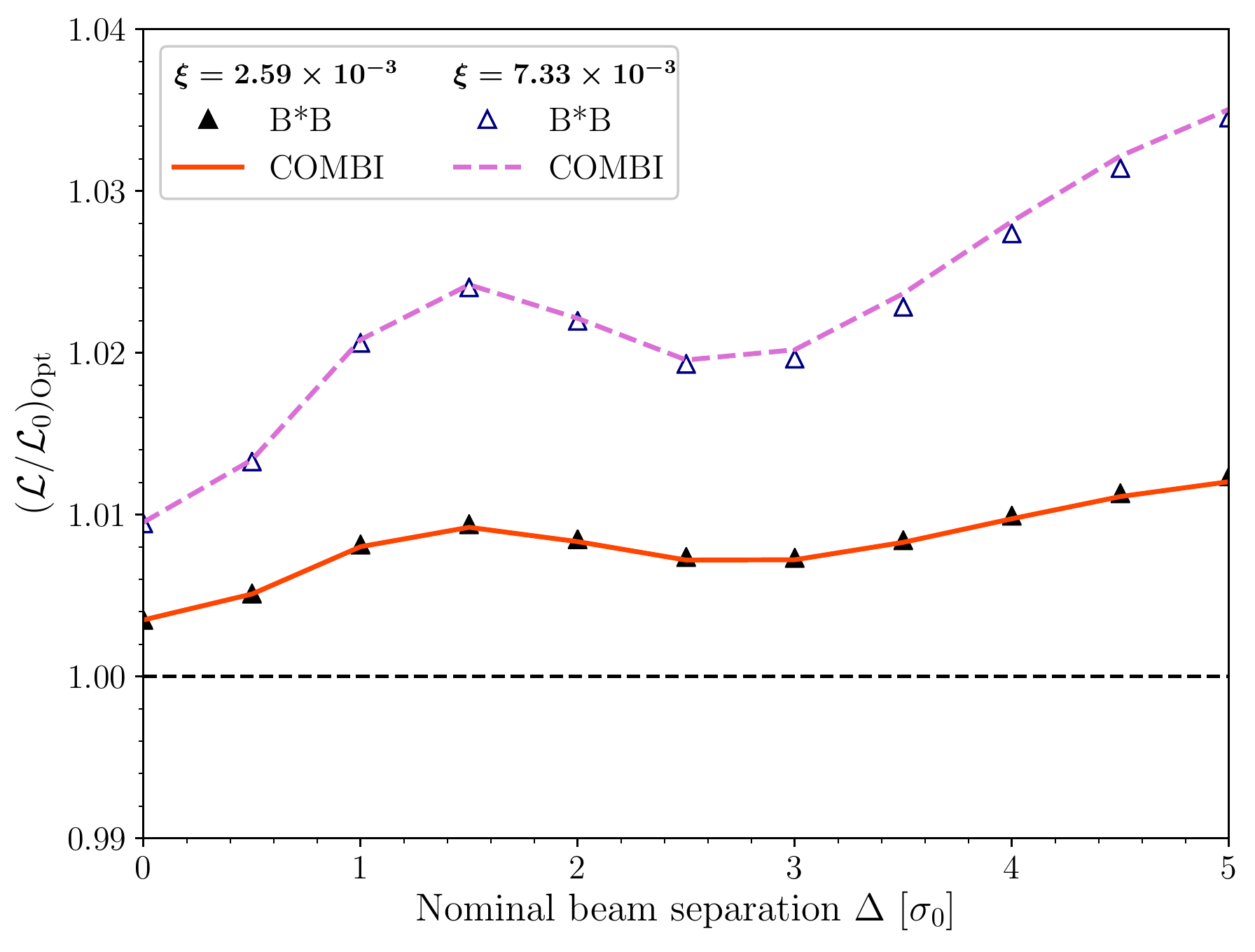}
\caption{Beam-separation dependence, during a simulated horizontal \vdM\ scan, of the luminosity-bias factor \LoLzOpt associated with optical distortions, for  the \vdM\ reference parameter set ($\xi = 2.59 \times 10^{-3}$, solid red curve and black filled triangles), and for a beam-beam parameter typical of routine physics running  ($\xi = 7.33 \times 10^{-3}$, dashed purple curve and blue open triangles). The COMBI (B*B) results are shown by the  straight-line segments (markers). The  horizontal axis is the nominal beam separation in units of the unperturbed transverse beam size $\sigma_0$.
}
\label{fig:L/L0VsXi} 
\end{figure}

The results in Figs.~\ref{fig:L/L0} and \ref{fig:L/L0_OrbOpt} were obtained using the reference parameter set in the right column of Table~\ref{tab:vdMBeamParms}. This corresponds to a beam-beam parameter $\xi = 2.59 \times 10^{-3}$,  at the low end of the $\xi$ range explored during \vdM\ scans at $\sqrt{s}=13$\,TeV.  The cross-validation of B*B and COMBI was therefore repeated with a larger bunch current and a smaller emittance, both  typical of routine physics running and corresponding to a beam-beam parameter value well beyond the \vdM\ range. The beam-separation dependence of the optical-distortion luminosity-bias factor at these two $\xi$ values is presented in Fig.~\ref{fig:L/L0VsXi}. While in perfect agreement at low $\xi$, the two codes exhibit hints of a small but systematic difference in the high-$\xi$ regime when the beam separation becomes large enough. This discrepancy is attributed to the fact that B*B is intrinsically a weak-strong model, and therefore cannot account for coherent bunch oscillations, contrarily to COMBI. This interpretation is confirmed by the comparison of the tune spectra predicted by the two packages: even in the vdM regime, a $\pi$-mode peak is apparent in the COMBI spectrum, but is missing (as it should be) from the B*B spectrum.
\par
Quantitatively, it turns out that throughout the \vdM regime and even in the high-$\xi$ case considered here, the discrepancy is of no practical significance. Since it manifests itself only in the tails of the beams ($\Delta > 2\sigma_0$), where the particle density is low, the difference, between B*B and COMBI, in beam--beam-induced distortions has only a very small impact on the integral of the scan curves, \ie on the perturbed transverse convolved beam sizes (Eq.~(\ref{eqn:caps})). This can be quantified using the methodology that will be presented in Sec.~\ref{subsubsec:FoMs}. For the high-$\xi$ setting illustrated in Fig.~\ref{fig:L/L0VsXi}, the difference between B*B and COMBI translates into a systematic uncertainty of 0.04\% on the absolute luminosity scale. In typical \vdM scans, where $\xi$ is lower by a factor of about 1.5, the inconsistency becomes negligible; there the two simulation codes remain in more than adequate agreement, validating the use of the less resource-intensive B*B package for parameterizing beam-beam corrections to \vdM\ calibrations.

\section{Calculated impact of beam-beam dynamics on luminosity calibrations}
\label{sec:bbImpact}

	\subsection{Methodology}
	\label{subsec:bbCorMeth}

	The beam-beam biases affecting a \vdM\ calibration are accounted for by correcting the luminosity-scan curves, one bunch pair at a time, according to the procedure outlined in Sec.~\ref{subsec:bbBias}. While the orbit shift (Sec.~\ref{subsubsec:orbShift}) can be calculated analytically, correcting for optical distortions (Sec.~\ref{subsubsec:optclDist}) requires the knowledge of the beam-separation dependence of the luminosity-bias factor \LoLzOpt, such as that illustrated in Fig.~\ref{fig:L/L0VsXi}. 
The latter can be obtained by running B*B or COMBI with, as input:
\begin{itemize}
 \item
 the interaction-region (IR) configuration (beam energy, nominal or measured \bst and crossing-angle values);
 \item
 the unperturbed tunes \Qx and \Qy (or equivalently the unperturbed fractional tunes \qx and \qy), \ie the values of the horizontal and vertical tunes with the beam-beam interaction switched off at the IP where the simulated scans are taking place. Physically, these correspond to the tune values that would be measured, for the bunch pair under study, before the beams are brought into collision at the scanning IP considered;
\item
 the measured parameters of the bunch pair under study: bunch population, horizontal and vertical beam sizes or emittances, as well as the bunch length in case of a non-zero nominal crossing angle.
\end{itemize}	

\par
Since \vdM\ calibrations must be performed on a bunch-by-bunch basis, with possibly over 100 colliding-bunch pairs and typically five to ten $x$-$y$ scan pairs per scan session and per IP, the approach sketched above can become unwieldy from the computing viewpoint. This motivated the development of a much lighter technique, based on the fact that at least in the \vdM\ regime and under some simplifying assumptions, beam-beam biases, to a very good approximation, scale almost linearly with the beam-beam parameter $\xi$. This makes it possible to construct a simple polynomial parameterization of the luminosity-bias curves, that is extracted from B*B or COMBI simulations over a grid in beam-parameter space and is applicable to most cases of practical interest for $pp$ \vdM\ calibrations at the LHC. The impact of violating the underlying assumptions is accounted for either by a contribution to the systematic uncertainty that is associated with the beam-beam correction procedure, or, in the case of multi-IP effects, by a simulation-guided adjustment to the parameterized correction. 

\par
Some cases do not lend themselves to the parameterization technique, such as off-axis \vdM scans, diagonal scans, or \vdM\ calibrations performed with a large crossing angle (which is unavoidable at the LHCb IP).

\par
An off-axis \vdM scan is a horizontal (or vertical) beam-separation scan where the beams are partially separated in the non-scanning plane, \ie in the vertical (or horizontal) direction. A diagonal scan is one in which the beams are scanned transversely along an inclined straight line in the $x$-$y$ plane, rather than only along either the $x$ or the $y$ axis. Such ``generalized'', one-dimensional \vdM scans are sometimes used to measure and correct for non-factorization effects~\cite{bib:ATL2012Lum, bib:CMS2018Lum, bib:CMS-LUM-18-001}. While the orbit-shift correction can still be calculated analytically using a more general form of the Bassetti-Erskine formula, the increased dimensionality of the parameter space makes it impractical to invest in general enough and precise enough a parameterization of optical-distortion effects. In such a case, deriving the corrections from a large set of fully simulated scans, labor- and CPU-intensive as it may be, appears more tractable by comparison.

\par
Allowing non-zero crossing angles also increases the dimensionality of the parameter space, with similar implications. Therefore, except where explicitly stated otherwise, the results offered in the remainder of this report are restricted to the zero (or moderate) crossing-angle case, which covers the needs of the  ATLAS and CMS experiments (as well as those of  ALICE, at the cost of a slight increase in systematic uncertainty).

\par 
This Section is organized as follows. The parameterization approach mentioned above is described in Sec.~\ref{subsec:corProcRefConf}. The associated ``fully symmetric Gaussian-beam configuration'' assumes that in the absence of any beam-beam interaction, the colliding bunches in beam 1 and beam 2:
\begin{itemize}
 \item
 can be modeled by factorizable transverse-density distributions, and exhibit a single-Gaussian profile in all three dimensions.  The impact of violating this assumption is evaluated in Sec.~\ref{subsec:nonGausImpact};
 \item 
are round in the transverse plane ($\sigma_{xB} = \sigma_{yB}, B = 1,2$). The impact of violating this assumption is evaluated in Sec.~\ref{subsec:EllipImpact};
 \item
 intersect at zero crossing angle. The impact of violating this assumption is evaluated in Sec.~\ref{subsec:XingAngImpact};
\item
collide only at the IP where beam-separation scans are performed.  The impact of violating this assumption is evaluated in Sec.~\ref{subsec:multiIP};
\item
are beam-beam symmetric, \ie equally populated ($n_1 = n_2$), of the same transverse size ($\sigma_{i,1} = \sigma_{i,2}$, $i = x, y$), and therefore subject to the same beam-beam parameter ($\xi_{i,1} = \xi_{i,2}, i= x,y$). The impact of violating these assumptions is evaluated in Sec.~\ref{subsec:ImbalImpact}.
\end{itemize}

	\subsection{Beam-beam correction procedure in the fully symmetric Gaussian-beam configuration}
	\label{subsec:corProcRefConf}

The parametrization strategy relies on the fact that at fixed tunes, beam-beam induced biases to the visible cross-section depend only on the beam-beam parameter $\xi$ (Sec.~\ref{subsubsec:scalHypVal}). The simulated $\xi$- and tune-dependence of the luminosity-bias functions (Sec.~\ref{subsubsec:xiNTuneDep}) leads to parameterizing them by second-order polynomials that can be used, in the context of the beam-beam correction procedure of \vdM calibrations, as a proxy for full-fledged B*B or COMBI simulations (Sec.~\ref{subsubsec:CorImplement}). Combining these parameterized bias functions with hypothetical \vdM-scan curves devoid of fitting biases and of step-to-step fluctuations provides robust and intuitive insight into the magnitude and beam-conditions dependence of beam-beam corrections to luminosity calibrations (Sec.~\ref{subsubsec:FoMs}).

\subsubsection{Validation of the scaling hypothesis}
\label{subsubsec:scalHypVal}

To characterize the scaling properties (or lack thereof) of beam-beam biases during vdM scans, pairs of horizontal and vertical beam-separation scans were generated under the assumptions above using B*B.\footnote{In view of the consistency of  the B*B and COMBI results demonstrated in Sec.~\ref{sec:bbSimCodes}, the B*B package was chosen for practical reasons, first and foremost because it is less computationally expensive.}
The beam-beam parameter spanned the range $\xi \sim 0.002 - 0.008$, thereby covering from \vdM\ scans with low-brightness proton beams at the low end, to conditions slightly above routine physics running at the high end. The input beam energies, bunch populations, emittance and \bst\ values were chosen to be representative of \vdM\ scans during LHC Runs 1 and 2 (Sec.~\ref{subsec:beamParms}); also included was a high-$\xi$ setting representative of Run-2 high-luminosity physics running. The unperturbed fractional tunes were constrained to satisfy $\qy = \qx+0.010$ so as to reflect routine LHC collision settings. 

\par
\begin{figure}
    \centering
      \includegraphics[width=0.5\textwidth]{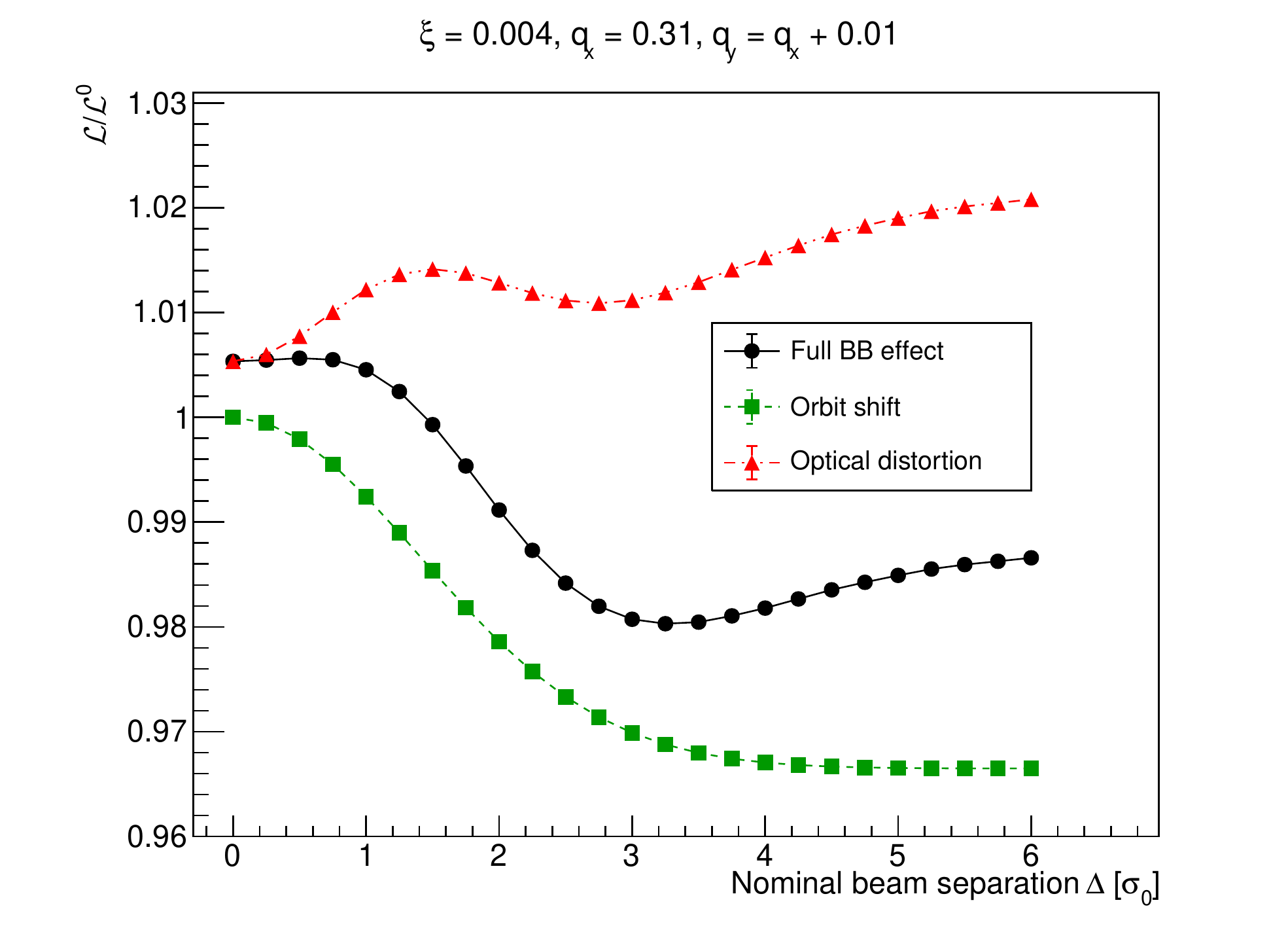}
    \caption{Beam-separation dependence, during a simulated horizontal \vdM\ scan, of the luminosity-bias factor associated with the beam-beam orbit shift only (green squares), the optical distortions only (red triangles), and their combination (black circles). The  horizontal axis is the nominal beam separation in units of the unperturbed transverse beam size $\sigma_0$. The beams satisfy the assumptions listed in Sec.~\ref{subsec:bbCorMeth}. The input unperturbed-tune and beam-beam parameter values are indicated at the top; the other relevant parameters are $E_B=6.5$\,TeV, $\bst=19.2$\,m and  $\epsilon_{N}= 2.6\,\mu$m$\cdot$rad. The points are the results of the simulation; the lines are intended to guide the eye.
}
    \label{fig:ll0-3-curves}
    \vspace*{-0.3cm}
\end{figure}

 The primary output of the simulation is the beam-separation dependence of the luminosity-bias factor \LoLzF. An example is presented in Fig.~\ref{fig:ll0-3-curves}, with  a value of $\xi$ chosen to lie roughly in the middle of the \vdM\ range (Table~\ref{tab:vdMBeamParms}). The orbit-shift bias factor \LoLzOrb (green squares) is computed analytically:
\begin{equation}
    \label{eqn:gausorbll0}
    \LoLzOrbD  = e^{- \frac{1}{2}  [(\delta_i/\CSR)^2 - (\Delta/\CSR)^2]}
\end{equation}
where $\Delta$ is the nominal separation in the scanning plane, $\delta_{i}$ is computed using Eqs.~(\ref{eqn:deflAng}) to  (\ref{eqn:bbCSep}), and $\CSR = \sqrt{\CSxz\, \CSyz}$ is the unperturbed, round beam-equivalent convolved beam size that in this particular case satisfies $\CSR = \CSxz=\CSyz=\sqrt{2} \, \sigma_0$. The optical-distortion bias factor \LoLzOpt (red triangles) can then be calculated using Eq.~(\ref{eqn:OptFulOrbdef}).

\par
For given values of \qx and \qy, beam--beam-induced distortions scale with the beam-beam parameter, in the sense that they depend only on $\xi$. This can be proven mathematically, both for the orbit shift (Eqs.~(\ref{eqn:deflAng})-(\ref{eqn:orbShift})) and  for the dynamic-$\beta$ effect at zero separation (Eq.~(\ref{eqn:dynBeta})). That this remains true at any beam separation, within the constraints of the fully symmetric beam configuration and over the $\xi$ range specified above, can only be demonstrated by simulation. To this effect, luminosity-bias curves such as those presented in Fig.~\ref{fig:ll0-3-curves} were compared for different combinations of bunch populations, beam energies, \bst\ and emittance values that all correspond to a given value of $\xi$; they were found to be identical within the statistics of the simulation. 
\begin{figure}
    \centering
    \includegraphics[width=0.5\textwidth]{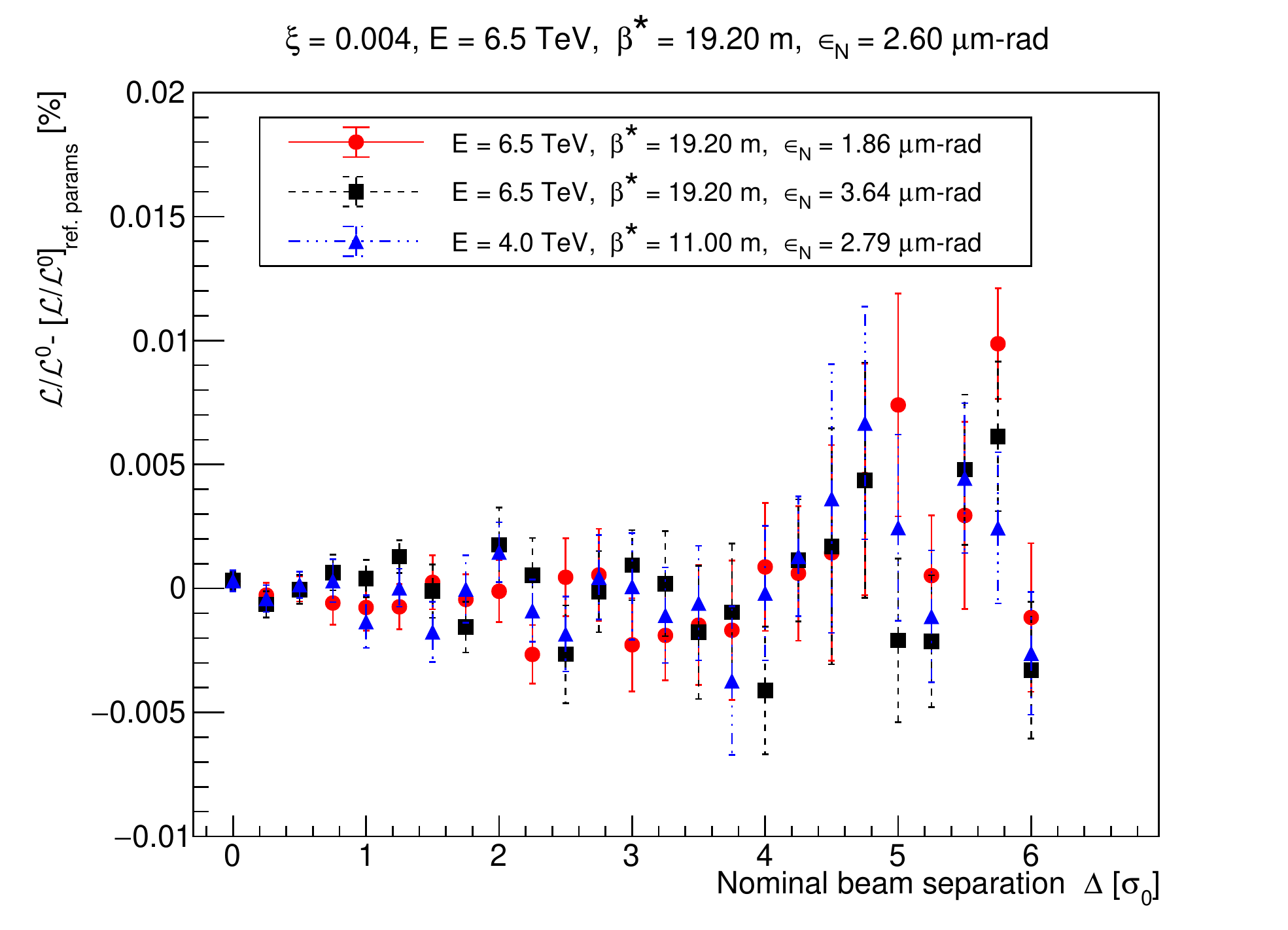}
    \caption{Beam-separation dependence, during simulated horizontal \vdM\ scans, of the difference between the luminosity-bias factor (full beam-beam effect) calculated for the collision parameters shown in the legend, and that computed using the reference parameters (Fig.~\ref{fig:ll0-3-curves}, black circles) listed at the top of the present figure. For each parameter set, the bunch intensity is adjusted such that the value of $\xi$ remains the same. The error bars are explained in the text.
}    
\label{fig:ll0-difference}
\vspace*{-0.3cm}
\end{figure}
An example is presented in Fig.~\ref{fig:ll0-difference}, which displays the difference, at each scan step, between the luminosity-bias factor computed for three distinct sets of beam parameters, and that associated with the parameters used in Fig.~\ref{fig:ll0-3-curves}; all four parameter sets correspond to $\xi = 0.004$. The error bars represent the statistical uncertainty associated with the randomization of initial conditions in the B*B code \cite{Balagura_2021}; they are computed as the error on the mean over eight different runs, with different random seeds, for each beam separation and each parameter set. The differences in luminosity-bias values between the four beam-parameter sets and the reference set are statistically consistent with zero, and never exceed $10^{-4}$. The exercise was repeated for a range of $\xi$ values, leading to the conclusion that at fixed \Qx and \Qy, the luminosity-bias curves indeed depend only on $\xi$, to better than $10^{-4}$ on the absolute luminosity scale.

\subsubsection{Beam-beam parameter and tune dependence of the optical-distortion correction}
\label{subsubsec:xiNTuneDep}  
    
\begin{figure}[h!]
    \centering
    \includegraphics[width=0.5\textwidth]{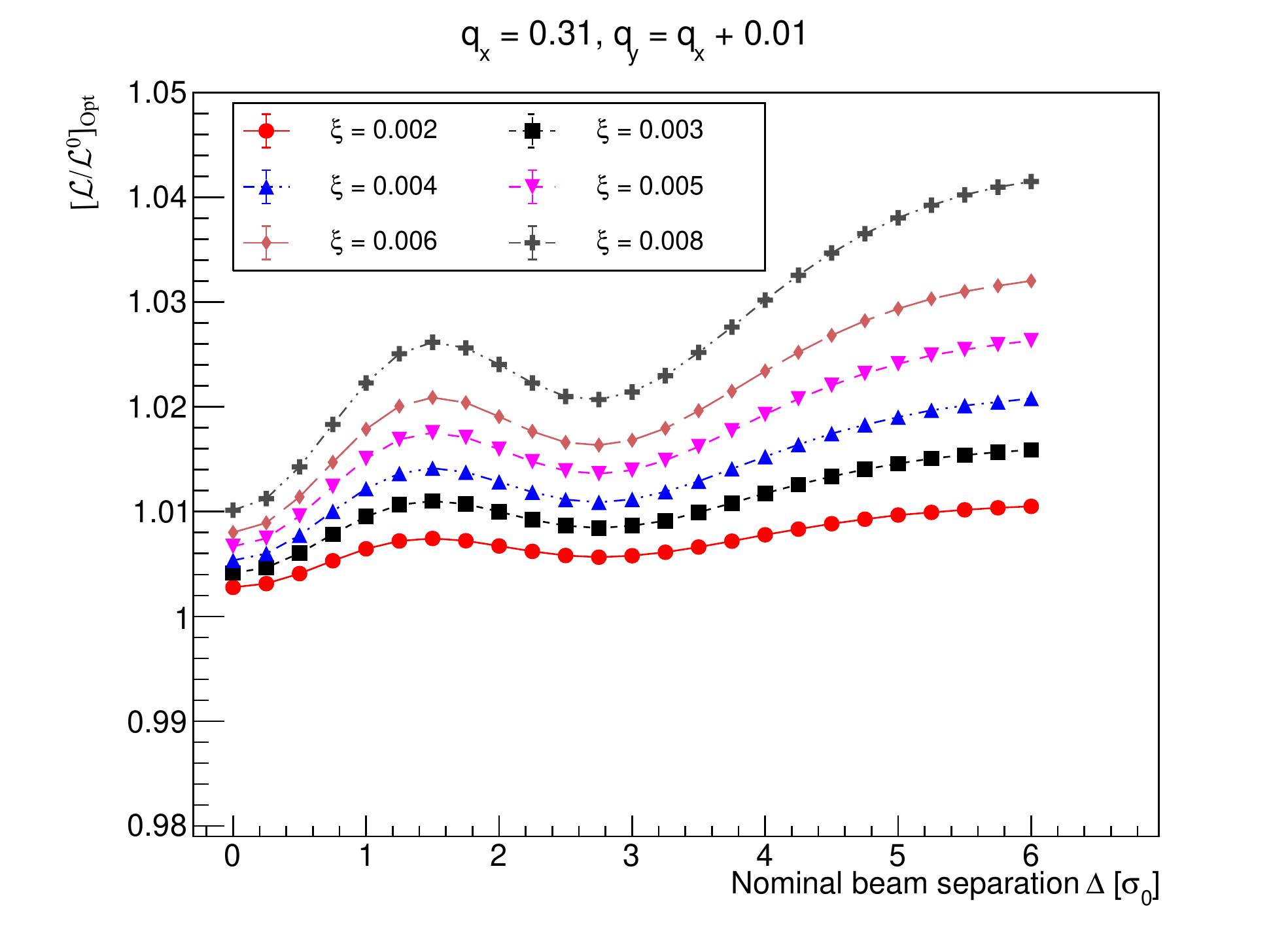}
    \caption{Beam-separation dependence, during simulated horizontal \vdM\ scans, of the luminosity-bias factor  \LoLzOpt associated with optical distortions, for several values of the beam-beam parameter $\xi$. The  horizontal axis is the nominal beam separation in units of the unperturbed transverse beam size $\sigma_0$. The beams satisfy the assumptions listed in Sec.~\ref{subsec:bbCorMeth}. The input unperturbed fractional-tune values are indicated at the top. The points are the results of the simulation; the lines are intended to guide the eye.
}
    \label{fig:FoM-ll0-xi-dependence}
    \vspace*{-0.3cm}
\end{figure}

The combined beam-separation and $\xi$ dependence of the luminosity-bias factor \LoLzOpt is illustrated in Fig.~\ref{fig:FoM-ll0-xi-dependence}. The simulations shown span the full beam-beam parameter range ($2.04\times 10^{-3} \le \xi \le 7.83\times 10^{-3}$), and are carried out at the nominal LHC tune settings ($\Qx=64.31$, $\Qy=59.32$). A similar $\xi$-dependence is observed for  \LoLzF and \LoLzOrb (not shown). The beam-beam parameter dependence of all three variables at fixed nominal separation is found to be well modeled by a second-order polynomial of $\xi$, the coefficients of which depend on the nominal separation.
    
\par
Since both the orbit shift (Eq.~(\ref{eqn:orbShift})) and the dynamic-$\beta$ effect (Eq.~(\ref{eqn:dynBeta})) depend on the unperturbed tunes, and as part of investigating the systematic uncertainties associated with beam-beam corrections, their sensitivity to the input tune values was characterized by extending the above-described simulations to an area in the (\qx, \qy) plane that encompasses the full range of operational tune settings and of beam-beam tune shifts expected during \vdM sessions. Horizontal and vertical beam-separation scans were simulated, for several values of $\xi$, over a grid\footnote{For technical reasons associated with the numerical evaluation of overlap integrals in B*B~\cite{Balagura_2021}, the input tune values had to be chosen such that $\gcd(10000 \, \qx,10000 \, \qy)=1$, where gcd refers to the greatest common divider. This avoids integration over closed curves that may lead to unexpected systematic effects.} 
bounded by  $0.2975 < \qx < 0.3100$, with the vertical unperturbed fractional tune set to $\qy = \qx + \delta_\mathrm{q}$ where $\delta_\mathrm{q} = 0.0075, 0.0100, 0.0125$.

\par
\begin{figure}[h!]
    \centering
    \includegraphics[width=0.5\textwidth]{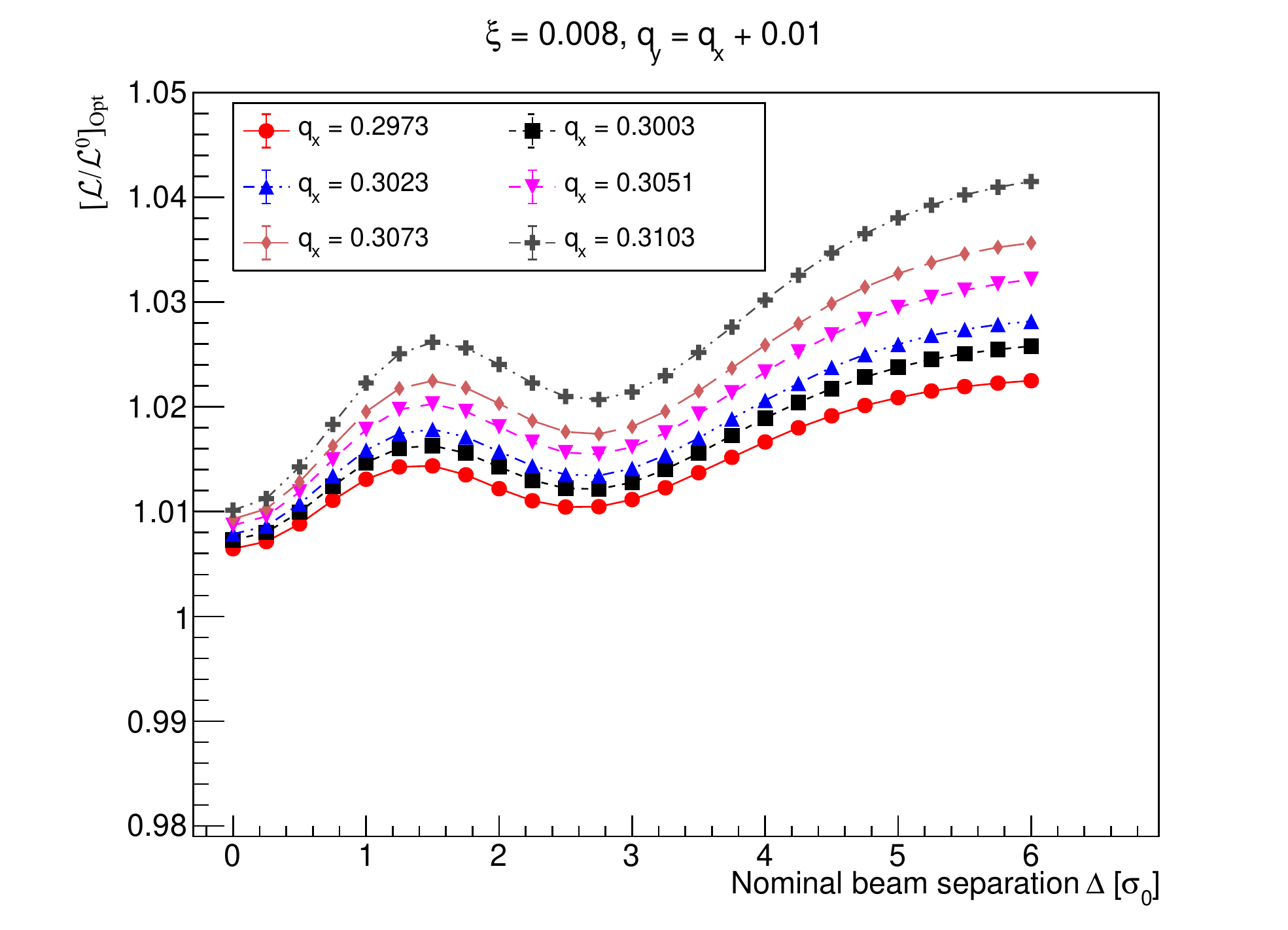}
    \caption{Beam-separation dependence, during simulated horizontal \vdM\ scans, of the luminosity-bias factor  \LoLzOpt associated with optical distortions, for several values of the unperturbed fractional tunes (\qx, \qy). The  horizontal axis is the nominal beam separation in units of the unperturbed transverse beam size $\sigma_0$. The beams satisfy the assumptions listed in Sec.~\ref{subsec:bbCorMeth}. The input $\xi$ value is indicated at the top. For each curve, the value of the horizontal tune is indicated in the legend; the value of the vertical tune is set to $\qy = \qx + 0.01$. The points are the results of the simulation; the lines are intended to guide the eye.
}
    \label{fig:Opt-ll0-Q-dependence}
\vspace*{-0.5cm}
\end{figure}
\begin{figure}[h!]
    \centering
    \includegraphics[width=0.5\textwidth]{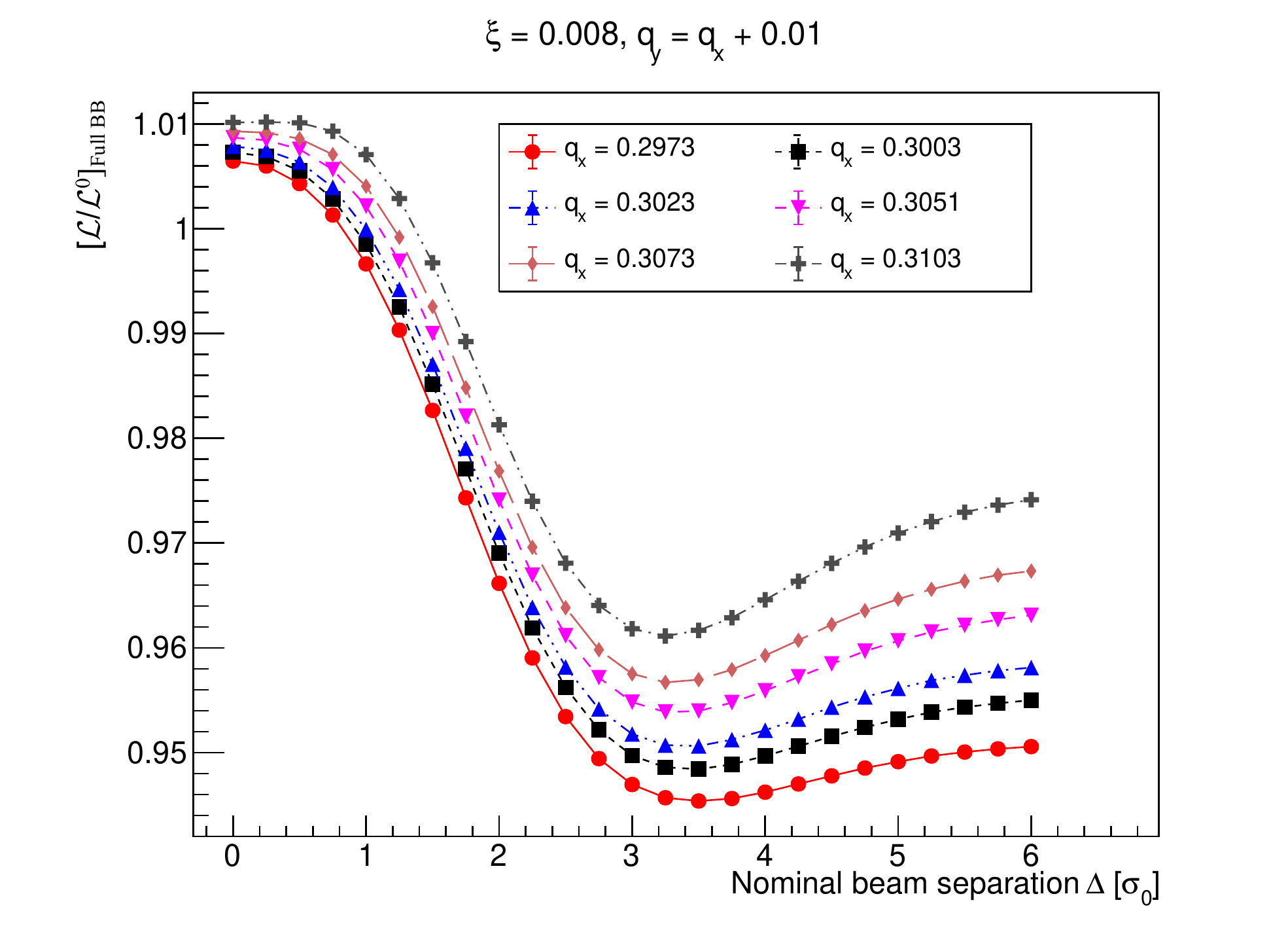}
    \caption{Beam-separation dependence, during simulated horizontal \vdM\ scans, of the luminosity-bias factor  \LoLzF associated with the full beam-beam effect, for several values of the unperturbed  fractional tunes (\qx, \qy). The  horizontal axis is the nominal beam separation in units of the unperturbed transverse beam size $\sigma_0$. The beams satisfy the assumptions listed in Sec.~\ref{subsec:bbCorMeth}. The input $\xi$ value is indicated at the top. For each curve, the value of the horizontal tune is indicated in the legend; the value of the vertical tune is set to $\qy = \qx + 0.01$. The points are the results of the simulation; the lines are intended to guide the eye.
}
    \label{fig:FullBB-ll0-Q-dependence}
    \vspace*{-0.6cm}
\end{figure}

The combined beam-separation and tune dependence of the luminosity-bias factor \LoLzOpt is illustrated in Fig.~\ref{fig:Opt-ll0-Q-dependence} for $\xi=7.83\times 10^{-3}$. For a given nominal separation, the lower the tunes, the more they approach the quarter integer, and therefore the smaller (\ie the closer to unity) the optical-distortion bias factor becomes~\cite{Herr}. Similarly, the lower the tunes, the smaller (\ie the further away from unity) the orbit-shift bias factor \LoLzOrb (not shown), albeit with a weaker tune dependence. As a result, the tune dependence of the full beam--beam-bias factor \LoLzF is slightly steeper than that associated with optical distortions only: the lower the tunes, the faster \LoLzF  drops in magnitude (Fig.~\ref{fig:FullBB-ll0-Q-dependence}), and therefore the larger the overall beam-beam correction to the luminosity scale. In addition, the lower the tunes, the less the optical distortions contribute to the overall beam-beam bias to \svis.


\subsubsection{Practical implementation of parameterized beam-beam corrections}
\label {subsubsec:CorImplement}

    Concretely, given a measured \vdM-scan curve such as that displayed in Fig.~\ref{fig:LscanCurves} (bottom), the beam-beam correction procedure can be implemented as follows.
 \begin{enumerate}
 	\item
	At each scan step in the $i$ scanning plane ($i = x, y$), correct the nominal separation $\Delta_i$  for the deflection-induced separation shift $\delta^{bb}_i$, as detailed in Sec.~\ref{subsubsec:orbShift}, using Eqs.~(\ref{eqn:deflAng})--(\ref{eqn:bbCSep}).
	\item
		At each scan step in the $i$ scanning plane, correct the measured rate $R_i(\Delta_i)$ for the optical-distortion bias as detailed in Sec.~\ref{subsubsec:optclDist}, using Eq.~(\ref{eqn:corPrinc}). The beam-separation dependence  of the luminosity-bias factor \LoLzOpt is either extracted directly from a set of B*B or COMBI simulations, or, if the assumptions listed in Sec.~\ref{subsec:bbCorMeth} can be considered valid, simply computed using the polynomial parameterization described below. The latter is valid only for on-axis, one-dimensional \vdM scans with zero crossing angle; other configurations, and in particular off-axis one-dimensional scans as well as two-dimensional grid scans in the $(\delta_x, \delta_y)$ beam-separation plane, require dedicated simulations.
\end{enumerate}
    
\par
The simulations described in Sec.~\ref{subsubsec:xiNTuneDep} were used to parameterize the optical-distortion luminosity-bias factor \LoLzOpt in bins of normalized nominal separation over the range $0 \le \Delta_i/\sigma_0 \le 6$, separately for $x$ and $y$ scans, by polynomials of the form  
\begin{eqnarray}
   [ \Lum / \Lumz (\xiR, \qx, \qy)]_k	
    			&=	& p_{0k}  										\nonumber		\\
                         	&+	& p_{1k}\, \xiR + p_{2k}\, \qx + p_{3k}\, \qy 			\nonumber		\\
                         	&+	& p_{4k}\, \xiR^{2} + p_{5k}\, \qx^{2} + p_{6k}\, \qy^{2} 	\nonumber	 	\\
                         	&+	& p_{7k} \xiR \qx + p_{8k}\, \xiR\, \qy 				\nonumber		\\
			&+	&p_{9k}\, \qx\, \qy\, .							\label{eqn:ll0_xiqxqy}
    \end{eqnarray}
Here the index $k$ ($k =$ 1, 25) labels the nominal-separation bin in the scanning plane considered. The round-beam equivalent beam-beam parameter \xiR is defined as
\begin{equation}
\label{eqn:xiRDef2}
\xiR = \frac{n r_0 \bst}{4 \pi \gamma \sigR^2}\, ,
\end{equation}
where the single-beam transverse RMS beam size $\sigma_0$ in Eq.~(\ref{eqn:xiRDef}) is approximated by its measured\footnote{It will be shown in Sec.~\ref{subsubsec:FoMs} that in the \vdM regime, beam-beam effects, if left uncorrected, lead to a percent-level underestimate of \CSx and \CSy, and therefore of \sigR. The use of the measured transverse beam size -- as opposed to that of the experimentally inaccessible unperturbed beam size -- in estimating \xiR therefore leads to a slight overestimate of the beam-beam parameter input to the simulation. However, since the overall beam-beam bias on the effective cross-section is typically less than 1\% (Fig.~\ref{fig:FoM-sigvis}, black curve), a sub-percent overestimate of \xiR biases the overall beam-beam {\em correction} to \svis by 1--1.5\% of itself. This can usually be neglected in view of the much larger fractional systematic uncertainties detailed in Sec.~\ref{sec:bbsysts}. Alternatively, the correction can be iterated upon, by using as input to the second iteration a \xiR value based on the corrected effective single-beam size obtained from the first iteration.}
round-beam equivalent $\sigR = \sqrt{\CSx \CSy/2}$, thereby partially accounting for a potential ellipticity of the beams at the IP. The input fractional-tune values (\qx, \qy) are those of the unperturbed tunes defined in Sec.~\ref{subsec:bbCorMeth}. In the case where bunches collide only at the IP where the scan is taking place, \ie remain fully separated, transversely and/or longitudinally, at the other three IPs, these unperturbed tunes are identical to the nominal LHC tunes, or equivalently to the values one would measure before the beams are put in collision at the scanning IP. If, however, some bunches also collide at one or more  other IPs, the unperturbed tunes input to the parameterization for those bunches must take into account the beam-beam tune shift arising from collisions at non-scanning IPs; a prescription to this effect is offered in Sec.~\ref{subsec:multiIP}.

\par
The parameterization above, that amounts to a second-order Taylor expansion in \xiR, \qx and \qy, was found sufficient to approximate the exact simulation results to better than $10^{-3}$ on \LoLz over most of the grid of simulated points in ($\xi$, \qx, \qy, $\Delta$) space.. The function defined by Eq.~(\ref{eqn:ll0_xiqxqy}) is linear in all 10 parameters $p_{lk}$. In each nominal-separation bin $k$, the set of parameters $p_{0k}, \dots, p_{9k}$ is determined by a weighted linear least-square fit. A smooth dependence of \LoLz on the nominal separation, for given values of \xiR, \qx and \qy, can be achieved by, for instance, cubic spline interpolation between adjacent separation bins. Parameterizations based on the same functional form were also constructed for the orbit-shift and full beam-beam bias factors  \LoLzOrb and \LoLzF, and achieved similar numerical accuracy.

\par
Tabulated values of the parameters $p_{lk}$ in separation steps of $0.25$ in $\Delta/\sigma_{R}$, for both vertical and horizontal scans and over the full range of \xiR and tune values defined earlier, have been made available to all LHC experimental Collaborations. The optical-distortion parameters, which are the most useful, are documented in Appendix~\ref{subsec:OptCorrTables}, and are  publicly accessible in computer-readable form~\cite{bib:OptCorrTables}.

\subsubsection{Separation-integrated estimates of beam-beam corrections to vdM calibrations}
\label {subsubsec:FoMs}

The beam-beam correction procedure detailed in the preceding Section is designed to be applied to measured \vdM-scan curves, one scan step at a time. One then extracts the beam-beam corrected peak rate and convolved beam sizes needed to calculate the visible cross-section (Eq.~(\ref{eqn:sigmaVis})) using a carefully chosen fit function. More often than not, the latter must diverge from a perfect Gaussian in order to faithfully model the data over the full beam-separation range.

\par
In order to provide consistency checks on this intricate analysis chain, as well as a physically intuitive, albeit approximative, breakdown of the impact of individual beam-beam effects on the luminosity calibration under study, the same procedure can be applied to an  $x$-$y$ pair of hypothetical beam-separation scan curves, that suffer from no statistical fluctuations and that, in the $\xiR \rightarrow 0$ limit would be perfectly Gaussian:
\begin{equation}
  \Lumz(\Delta)  \propto \frac{1}{4\pi\sigma_0^{2}}  \exp\left( -\frac{\Delta^{2}}{4\sigma_0^{2}} \right) \, .
\nonumber
\end{equation}
Here and in the remainder of this paper, the ``0'' index indicates 
a ``nominal''
quantity, \ie one for which beam-beam effects have been fully turned off in the simulation.

\par
At a given nominal separation $\Delta_i$ in scanning plane $i$ ($i = x, y$), and with the beams transversely centered on each other in the non-scanning plane, the luminosity in the presence of, say, the full beam-beam effect is then calculated as
\begin{equation}
\Lum (\Delta_i) = \Lumz(\Delta_i) \times \LoLzF(\Delta_i)
\label{eqn:LumD}
\end{equation}
with $\LoLzF(\Delta_i)$ provided by the parametrization detailed in Sec.~\ref{subsubsec:CorImplement}, with input parameters representative of the actual beam conditions during the scans under study. The impact of either the orbit shift or the optical distortions alone can be evaluated by substituting  $\LoLzOrb(\Delta_i)$ or  $\LoLzOpt(\Delta_i)$ for $\LoLzF(\Delta_i)$ in Eq.~(\ref{eqn:LumD}).

\par
One then defines  ``figures of merit'' (FoMs) that characterize the impact of beam-beam effects on \vdM observables.
\begin{enumerate}
    \item 
    The peak-rate bias factor characterizes the dynamic-$\beta$ effect at zero beam separation:
    \begin{equation}
    \mu_{pk}/\mu_{pk}^{0} = \frac{\Lum(\Delta_i = 0)}{\Lumz(\Delta_i = 0)} ,
\label{eqn:FoM-mupk}
    \end{equation}
    and is sensitive to optical distortions only.
    \item  
    The beam-size bias factors characterize, in each plane, the beam-beam impact on the convolved transverse beam size:
    \begin{equation}
    \Sigma_{i}/\Sigma_{i}^{0}~ (i=x,y),
    \label{eqn:SoSz}
    \end{equation}    
    where the overlap integrals
    \begin{eqnarray}
    \Sigma_{i} 		&=
    				& \frac{1}{\sqrt{2 \pi}}~\frac{\int \Lum(\Delta_{i}) \,\mathrm{d}\Delta_{i}}{\Lum(\Delta_i=0)} 			\label{eqn:FoM-Sigma}\\
    \Sigma_{i}^{0} 	&=
    				&\frac{1}{\sqrt{2 \pi}}~ \frac{\int \Lumz(\Delta_{i}) \,\mathrm{d}\Delta_{i}}{\Lumz(\Delta_i=0)}		\label{eqn:FoM-Sigma0}		
    \end{eqnarray}
    are calculated numerically, with their integrands smoothed by cubic-spline interpolation.    
    \item  
    The \svis-bias factor characterizes the beam-beam impact on the absolute luminosity scale, and is given by the product of the previous three FoMs:
 \begin{eqnarray}
    \svis/\svisz  &=	& ( \mu_{pk}/\mu_{pk}^{0} ) 					\nonumber \\ 
    			     &	& \times  (\Sigma_{x}/\Sigma_{x}^{0}) \times  (\Sigma_{y}/\Sigma_{y}^{0})\, .
    \label{eqn:FoM-svis}
    \end{eqnarray}
 \end{enumerate}
 
\par
\begin{figure}
    \centering
    \includegraphics[width=0.5\textwidth]{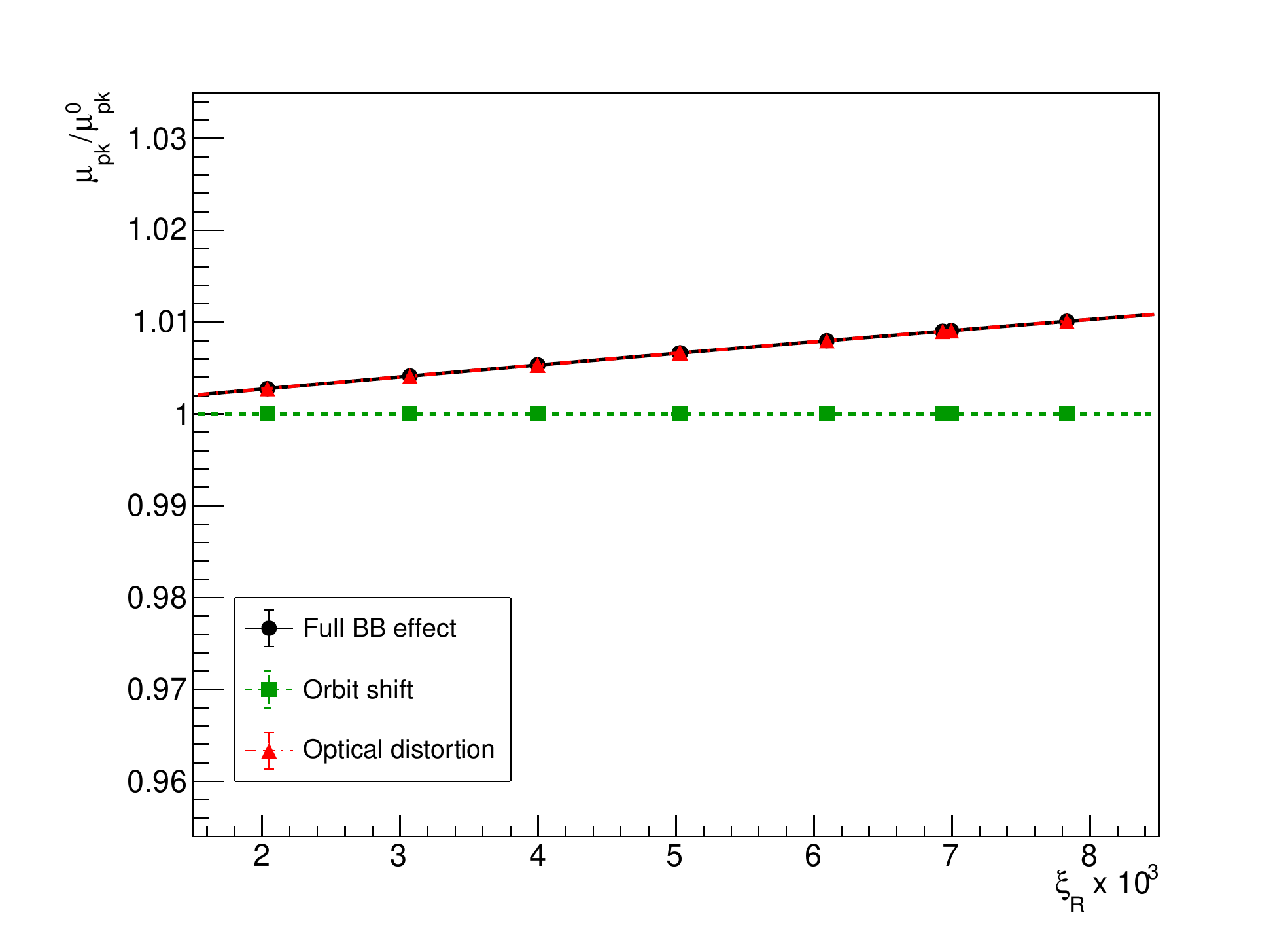}
    \caption{Dependence of the peak-rate bias factor on the beam-beam parameter, with beams colliding at the scanning IP only. The unperturbed tunes are fixed at the nominal LHC values (\Qx = 64.31, \Qy = 59.32). The symbols display the simulation results; the curves are described in the text. Since the orbit shift does not impact the peak-rate bias factor, the red and black symbols and curves lie exactly on top of each other.
}
    \label{fig:FoM-mu}
    \vspace*{-0.3cm}
\end{figure}

As an illustration, Fig.~\ref{fig:FoM-mu} displays the beam-beam parameter dependence of the peak-rate bias factor for the full beam-beam effect (black circles), the orbit shift only (green squares), and the optical distortion only (red triangles). Since at zero separation there is no orbit shift, only optical distortions matter in this case. 

\par
\begin{figure}
    \centering
    \includegraphics[width=0.5\textwidth]{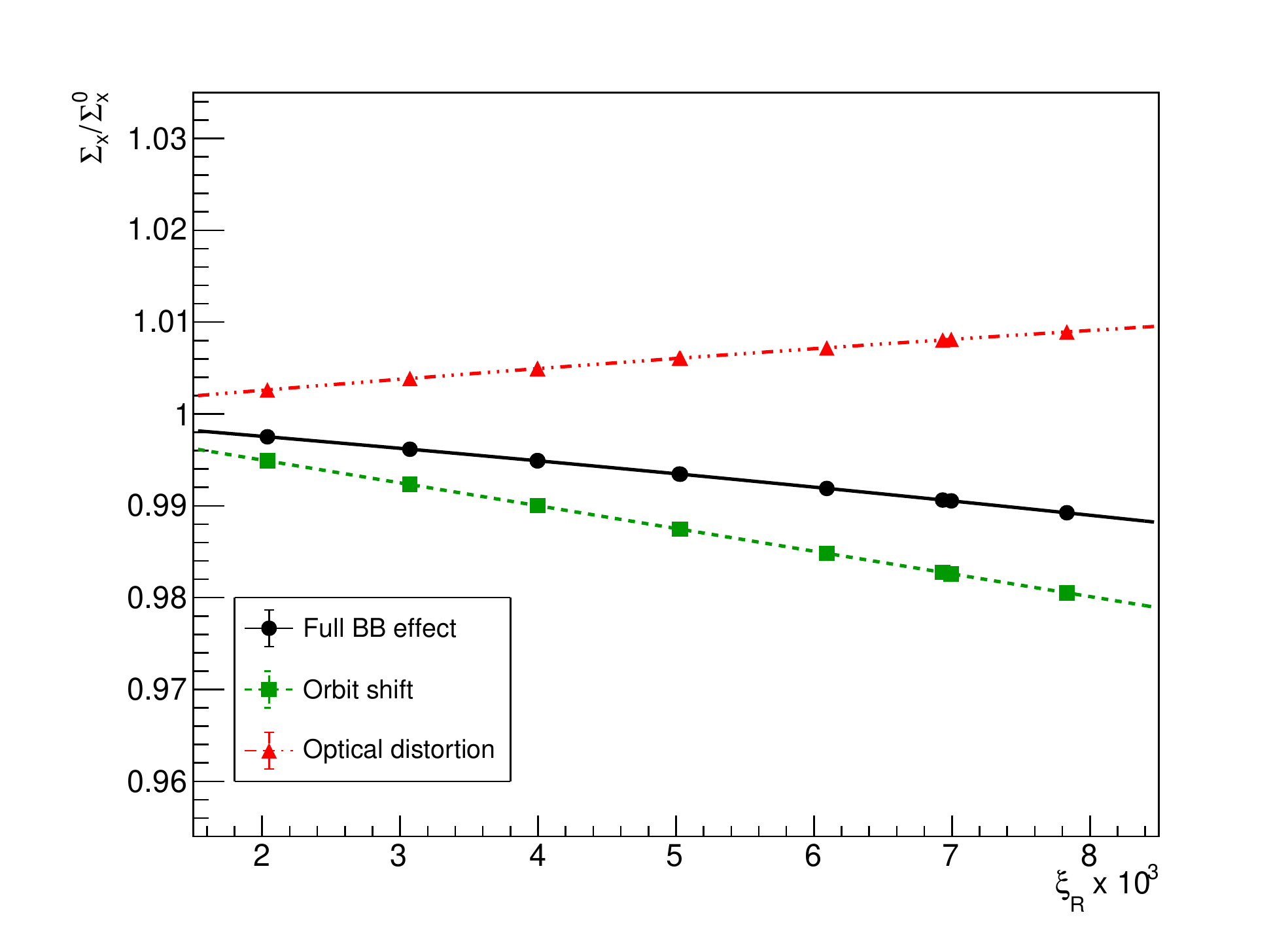}
    \caption{Dependence of the horizontal beam-size bias factor on the beam-beam parameter, with beams colliding at the scanning IP only. The unperturbed tunes are fixed at the nominal LHC values (\Qx = 64.31, \Qy = 59.32). The symbols display the simulation results; the curves are described in the text.
}    
    \label{fig:FoM-capsx}
    \vspace*{-0.3cm}
\end{figure}

The evolution of the horizontal beam-size bias factor is presented in Fig. \ref{fig:FoM-capsx}. While orbit shifts, if left uncorrected, result in an underestimate of \CSx and \CSy, optical distortions have the opposite effect, resulting in a partial cancellation of the bias. 

\par
\begin{figure}
    \centering
    \includegraphics[width=0.5\textwidth]{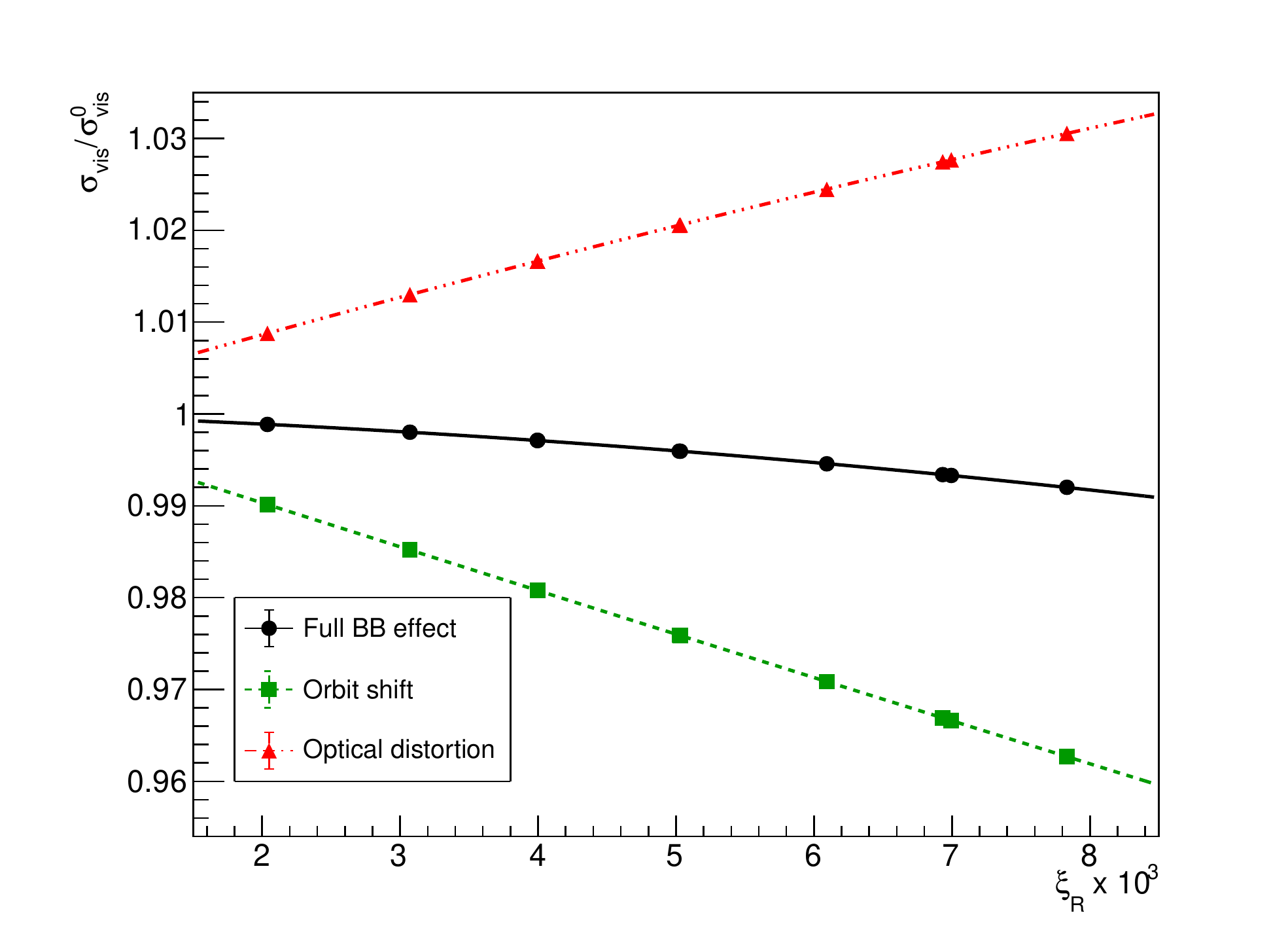}
    \caption{Dependence of the \svis bias factor on the beam-beam parameter, with beams colliding at the scanning IP only. The unperturbed tunes are fixed at the nominal LHC values (\Qx = 64.31, \Qy = 59.32). The symbols display the simulation results; the curves are described in the text.
}
    \label{fig:FoM-sigvis}
    \vspace*{-0.3cm}
\end{figure}

This is even more apparent in the visible-cross-section bias curves (Fig. \ref{fig:FoM-sigvis}). While the full beam-beam effect on the visible cross section remains below $1\%$ over the whole $\xi$ range (black circles), individual contributions from the orbit shift (green squares) and the optical distortion (red triangles) can each lead to a $3-4\%$ bias, a clear demonstration of how much overcorrection can be expected from too approximate a treatment of optical distortions.

\par	
In Figs.~\ref{fig:FoM-mu}--\ref{fig:FoM-sigvis}, the curves are fits to quadratic functions of \xiR:
\begin{equation}
    f(\xiR)_j = p_{0j} + p_{1j} \xiR + p_{2j}\xiR^{2} \label{eqn:quadxifun} ,
\end{equation}
where $j$ labels the type of correction (orbit shift only, optical distortion only, or full beam-beam). Even though the linear term is clearly dominant, a quadratic term is needed to describe the evolution of all three FoMs to satisfactory precision over the full range of beam-beam parameter values.

\par
 The impact of beam-beam effects on the absolute-luminosity scale can be expressed equivalently by either the \svis bias factor $\svis / \svisz$, the visible cross-section bias $\svis / \svisz-1$ (typically a fraction of a percent), or a multiplicative correction factor $(\svis / \svisz)^{-1}$ to the raw visible cross-section. This FoM approach, that lends itself to a simple polynomial parameterization in terms of \xiR, \qx and \qy, offers the advantage that its results are easy to interpret, as well as insensitive both to the quality of the fit to the measured scan curves and to rate fluctuations, from one scan step to the next, due for instance to counting statistics or beam-position jitter. This technique also simplifies the evaluation of some of the systematic uncertainties discussed in the remainder of this chapter. It is, however, not recommended for determining the central value of the beam-beam correction on real data, since it ignores the deviations of the actual scan curves from a perfect Gaussian.

	\subsection{Impact of non-Gaussian unperturbed transverse beam profiles}
	\label{subsec:nonGausImpact}
	
	All the simulation results presented in this report so far make the explicit assumption that in the absence of the beam-beam interaction at the scanning IP, the unperturbed transverse beam profiles, \ie the particle density functions $\hat{\rho} _B (x,y)$ ($B = 1,2$) in Eq.~(\ref{eqn:lumi}), can be perfectly modeled by the uncorrelated $x$-$y$ product of two single, one-dimensional Gaussians. Should this not be the case, the beam-separation dependence of the luminosity-bias curves will deviate from that presented in Sec.~\ref{subsec:corProcRefConf}. This is due both to the modified spatial dependence of the field generated by the source bunch, and to the fact that the fraction of particles in the witness bunch that experience a given electromagnetic kick is different from that in the pure-Gaussian case.
	
\par
Both B*B and COMBI accept as input unperturbed transverse-density distributions, functional forms that are more general than a single Gaussian. In order to  assess the potential impact, on beam-beam corrections, of non-Gaussian beams, a realistic model of $\hat{\rho} _B (x,y)$ is needed, or at least one that represents the ``worst-case'' deviation from the ideal Gaussian shape while remaining representative of actual beam conditions during \vdM-calibration sessions at the LHC (Sec.~\ref{subsubsec:2GBunchParms}). Taking into account potentially non-factorizable unperturbed density distributions requires a minor generalization of the beam-beam correction formalism (Sec.~\ref{subsubsec:2GFrmlsm}). While the influence of non-Gaussian tails on the beam-separation dependence of the luminosity-bias factors appears sizeable, the resulting bias on the beam-beam corrections to the visible cross-section remains moderate enough to be treated as a systematic uncertainty (Sec.~\ref{subsubsec:2GImpact}).

		\subsubsection{Single-bunch models}
		\label {subsubsec:2GBunchParms}


In a hadron collider, transverse-density distributions cannot be calculated from first principles, and existing beam-profile monitors, such as wire scanners or synchrotron-light telescopes, 
have too limited a dynamic range to provide sufficiently precise beam-tail measurements.
The only remaining experimental handle, therefore, is that provided by non-factorization analyses such as those described in, for instance, Refs.~\cite{bib:ATLR2Lum, bib:ATL2012Lum, bib:LHCbLumPap2, bib:CMS201516Lum, bib:ALICE2016-18Lum}. In this approach, the  bunch-density distributions are modeled by the sum of two or three three-dimensional Gaussians. A simultaneous fit to the beam-separation dependence, during a \vdM-scan pair, not only of the luminosity but also of the position, size, shape and orientation of the luminous region, makes it possible to estimate the single-bunch parameters of each colliding-bunch pair.

\par
In order to quantify the largest plausible impact of non-Gaussian beam profiles on the beam-beam corrections calculated in Sec.~\ref{subsec:corProcRefConf},  single-bunch parameters were extracted from the results of non-factorization analyses of  a 2012, a 2017 and a 2018 \vdM session. The former is representative of Run-1 \vdM conditions at the ATLAS IP, up to and including July 2012~\cite{bib:ATL2012Lum}, the latter two of Run-2 \vdM sessions at the CMS IP, for which ``bunch tailoring''  in the injector chain~\cite{bib:ATL2012Lum, bib:injBunchPrep} significantly reduced non-Gaussian tails. In both cases, a ''worst-case''  parameter set (in terms of deviations from a perfectly Gaussian shape) was selected from the fitted parameters of the analyzed colliding-bunch pairs. In order to separate the impact of non-Gaussian profiles discussed in this Section, from that of elliptical beams (Sec.~\ref{subsec:EllipImpact}) and of beam-beam imbalance (Sec.~\ref{subsec:ImbalImpact}), as well as to maximise the sensitivity of the study, in each parameter set the most non-Gaussian transverse profile is assigned to both planes and both beams.

\par
In a first step, the particle density distribution input to B*B is chosen to be factorizable by construction, and modeled by the product of two uncorrelated double Gaussians:
\begin{align}  
&\hat{\rho} _B (x,y)		= \rho_{xB}(x) \, \rho_{yB}(y)									\label{eqn:FDGparmtn}						\\
&					= 		\frac{1}{\sqrt{2 \pi}}\, \left[	\frac{w_n}{\sigma_n} \exp \left(-\frac{x^2}{2 \sigma^2_n} \right) 
												+ \frac{w_w}{\sigma_w} \exp \left(-\frac{x^2}{2 \sigma^2_w} \right)  \right]	 \nonumber	\\
&					\times  	\frac{1}{\sqrt{2 \pi}}\, \left[ 	\frac{w_n}{\sigma_n} \exp \left(-\frac{y^2}{2 \sigma^2_n} \right) 
												+ \frac{w_w}{\sigma_w} \exp \left(-\frac{y^2}{2 \sigma^2_w} \right)  \right]	 \nonumber	
\end{align}
where $B = 1$ or 2, the labels ``$n$'' and ``$w$'' refer to the narrow and wide components of the distribution, and their relative population is constrained by $w_{w} = 1-w_{n}$. The functional form reflects the assumptions of unperturbed round beams and of equal particle-density distributions ($\hat{\rho} _1 (x,y) = \hat{\rho} _2 (x,y)$). The three parameter sets input to the B*B simulation are listed in Table~\ref{tab:2gModels}. The 2012 (2017) parameter set clearly results in the most (the least) non-Gaussian shape, as evidenced qualitatively by the combination of the largest (smallest) beam-size ratio $\sigma_w / \sigma_n$ and the largest (smallest) weight of the wide component $w_w$, and as quantified by the single-beam kurtosis computed from the parameters listed in the Table.\footnote{Deviations of a distribution from the strictly Gaussian shape can be characterized by its kurtosis. This statistic is defined as $\Gamma_2 = m_4/m^2_2 - 3$, where $m_4$ and $m_2$ are, respectively, the fourth and the second moment of the distribution~\cite{bib:PDG2022}. The kurtosis is zero for a Gaussian, positive for a leptokurtic distribution with longer tails, and negative for a platykurtic distribution with tails that fall off more quickly than those of a Gaussian.}

\begin{table*} [htb]
\centering
\begin{tabular}{c|ccc} 
Date of \vdM session 			& July 2012	&	June 2018					& July 2017		\\
LHC fill number 				& 2855		&	6868						& 6016			\\ 		
\hline
$E_B$ [TeV]					& 4.0			&	6.5						&  6.5			\\
$n_p$ [$10^{11}$ p/bunch] 		& 1.1			& 	0.85						&  0.84			\\  
\bst [m]						& 11.0		&	19.2						& 19.2			\\
 \qx , \qy						& 0.31, 0.32	&	0.31, 0.32					& 0.31, 0.32		\\
$\sigma_n$ [$\mu$m]			& 57.9		& 	85.0						& 84.0			\\
$\sigma_w$ [$\mu$m]			& 115		&	125						& 116			\\
$\sigma_w / \sigma_n$			& 1.99		&	1.47						& 1.38			\\
$w_n$						& 0.634		& 	0.670					& 0.840			\\
$w_w = 1 - w_n$				& 0.366		&	0.330					& 0.160			\\
Single-beam kurtosis $\Gamma_2$ 	& 1.40		&	0.47						& 0.25			

\end{tabular}
\caption{Single-bunch parameters representative of strongly non-Gaussian single-beam profiles during ATLAS and CMS $pp$ \vdM scans in 2012, 2018 and 2017. The bunches are assumed to be round and, within each parameter set separately, beam-beam symmetric, \ie of equal brightness for beam 1 and beam 2.
}
\label{tab:2gModels}
\end{table*}

\par
The functional form chosen for $\hat{\rho} _B (x,y)$ makes it possible to compute analytically the unperturbed convolved transverse beam size $\Sigma^0$ from the values of $\sigma_n$, $\sigma_w$, $w_n$ and $w_w$, using Eq.~(\ref{eqn:caps}): these are listed in the top half of Table~\ref{tab:2gScanParms} (rows 3 to 6). The Gaussian-equivalent single-beam size \sigSG, \ie the transverse R.M.S. width of perfectly Gaussian bunches that would yield the same unperturbed convolved beam size as the non-Gaussian bunches studied here, is given by $\sigSG = \Sigma^0 / \sqrt{2}$ (Eq.~(\ref{eqn:capSigDef})). These definitions naturally lead to using the ``Gaussian-equivalent'' beam-beam parameter $\xi$, inferred from \sigSG using Eq.~(\ref{eqn:xiRDef}), as the common metric in which to quantify the impact of non-Gaussian tails.

\begin{table*} [htb]
\centering
\begin{tabular}{c|ccc} 
\hline
Date of \vdM session 				& July 2012	&	June 2018					& July 2017		\\
Single-beam kurtosis 				& 1.40		&	0.47						& 0.25			\\		
\hline \hline
\multicolumn{4}{c}{Factorizable density distribution: Eq.~(\ref{eqn:FDGparmtn})}  						\\
 \hline
 $\Sigma^0$  (from Eq.~(\ref{eqn:caps})) [$\mu$m]		
								& 107.1		&	137.3					& 125.4			\\	
$\sigSG = \Sigma^0 / \sqrt{2}$ [$\mu$m]		
								&   75.7		&	  97.1					&   88.7			\\
Gaussian-equivalent $\xi ~[10^{-3}]$ 		&     6.04		&	3.04						& 3.61			\\
\hline \hline
\multicolumn{4}{c}{Non-factorizable density distribution: Eq.~(\ref{eqn:NFDGparmtn})}  					\\
 \hline
 $\Sigma^0_F$  (from Eq.~(\ref{eqn:capsGen})) [$\mu$m]		
								& 100.4		&	134.9					& 124.4			\\	
$\sigSG = \Sigma^0_\text{F} / \sqrt{2}$ [$\mu$m]		
								&   71.0		&	  95.4					&   88.0			\\
Gaussian-equivalent $\xi ~[10^{-3}]$ 		&     6.88		&	    3.16					&     3.67			\\
\hline
\end{tabular}
\caption{Scan parameters representative of strongly non-Gaussian beam profiles during ATLAS and CMS $pp$ \vdM scans  in 2012, 2018 and 2017, calculated analytically from the single-bunch parameters listed in Table~\ref{tab:2gModels}.
}
\label{tab:2gScanParms}
\end{table*}

		\subsubsection{Beam--beam-correction formalism in the presence of non-factorization}
		\label {subsubsec:2GFrmlsm}

	In a second step, and in order to explore the potential impact of non-factorization on beam-beam corrections,  the simplest two-dimensional, non-factorizable double Gaussian:
\begin{align}  
\hat{\rho} _B (x,y)	=	\frac{w_{n}}{2 \pi \sigma^2_{n}}	~	\exp \left(-\frac{x^2+y^2}{2 \sigma^2_{n}} \right)	\nonumber	\\
				+ \frac{w_{w}}{2 \pi \sigma^2_{w}}	~	\exp \left(-\frac{x^2+y^2}{2 \sigma^2_{w}} \right)			
\label{eqn:NFDGparmtn}	
\end{align}
is chosen as an alternative density-distribution model to be input to B*B. Since by construction such a distribution cannot be expressed as the product of uncorrelated horizontal and vertical components, the concept of one-dimensional convolved beam size defined by Eq.~(\ref{eqn:caps}) ceases to be meaningful, and only the convolved--beam-size product \CSP introduced in Eq.~(\ref{eq:CapSNonFact}) remains amenable to physical interpretation. Accordingly, the two beam-size bias factors $\Sigma_{i}/\Sigma_{i}^{0}$  ($i = x, y)$ introduced in Eqs.~(\ref{eqn:SoSz})-(\ref{eqn:FoM-Sigma0}) must be replaced by the single quantity
\begin{eqnarray}
\frac{\CSP} {\CSPz}	&=	& \frac{\int {\Lum (\Delta_x,\Delta_y)\,\mathrm{d}\Delta_x\,\mathrm{d}\Delta_y}}
				 	 {\int {\Lumz(\Delta_x,\Delta_y)\,\mathrm{d}\Delta_x\,\mathrm{d}\Delta_y}}  		
																			\nonumber\\
				&	& \times \frac{\Lumz(0,0)}{\Lum(0,0)} \, ,		
\label{eqn:	FoM-CSP}	
\end{eqnarray}
where the scan and the integration are carried out over a grid in the two-dimensional nominal-separation space ($\Delta_x, \Delta_y$) that extends over $\pm 5 \sigSG$ in both directions.
The \svis-bias factor defined by Eq.~(\ref{eqn:FoM-svis}) then takes the more general form
\begin{eqnarray}
 \svisNF / \svisNFz	&  =	& \frac{\mu_{pk}} {\mu_{pk}^0}   \times  \frac{\CSP} {\CSPz} ,~~ 
 \label{eqn:FoM-svis2D}	
\end{eqnarray}
where the subscript ``$2D$'' indicates that the two-dimensional formalism introduced in Eq.~(\ref{eq:CapSNonFact}) is used throughout the analysis. 

\par
Using $\xi$ as a common metric for both the factorizable and the non-factorizable case also requires a generalization of the formula that expresses, in terms of single-beam parameters, an observable that is akin to the transverse convolved bunch size and that can be measured in a standard, one-dimensional \vdM scan. For instance, for a vertical scan with the beams centered on each other in the horizontal plane, the measured apparent convolved width is given by:
\begin{equation}
\Sigma_{y\text{F}} = \frac{1}{{\sqrt{2\pi}}}\frac{{\int {R_{x,y} (\delta_x =0,\delta_y)\,\mathrm{d}\delta_y}}}{{R_{x,y} (\delta_x =0,\delta_y =0)}}~,
\label{eqn:capsGen}
\end{equation}
which reduces to Eq.~(\ref{eqn:caps}) when $\hat{\rho} (x,y) =  \rho_x(x) \, \rho_y(y)$. The subscript ``F'' emphasizes that while measurable, this observable cannot be used in Eqs.~(\ref{eqn:lumifin}) and (\ref{eqn:sigmaVis}) unless the density distributions are factorizable. The importance of this distinction is illustrated by the significant differences between the top half (rows 3-6) and the bottom half (rows 7-10) of Table~\ref{tab:2gScanParms}: even though both sets of scan parameters are based on Table~\ref{tab:2gModels}, their values are significantly different in the two cases, because the non-linear $x$-$y$ correlation built into Eq.~(\ref{eqn:NFDGparmtn}) affects the beam-overlap integral (Eq.~(\ref{eqn:lumi})). These differences, the size of which appears correlated with the single-beam kurtosis, are indicative of the magnitude of the non-factorization biases that can be introduced, even in the absence of the beam-beam interaction, by applying the one-dimensional, factorizable FoM formalism of Sec.~\ref{subsubsec:FoMs} to non-factorizable beams. In contrast, the analysis strategy encapsulated in Eq.~(\ref{eqn:FoM-svis2D}) guarantees that non-factorizable correlations embedded in the unperturbed density distributions cancel in the \svisNF ratio; the latter, therefore, reflects the impact of beam-beam biases only. The study of non-factorization biases proper lies beyond the scope of the present report.

		\subsubsection{Impact of non-Gaussian tails on beam--beam-induced biases}
		\label {subsubsec:2GImpact}

\begin{figure}
    \centering
    \includegraphics[width=0.47\textwidth]{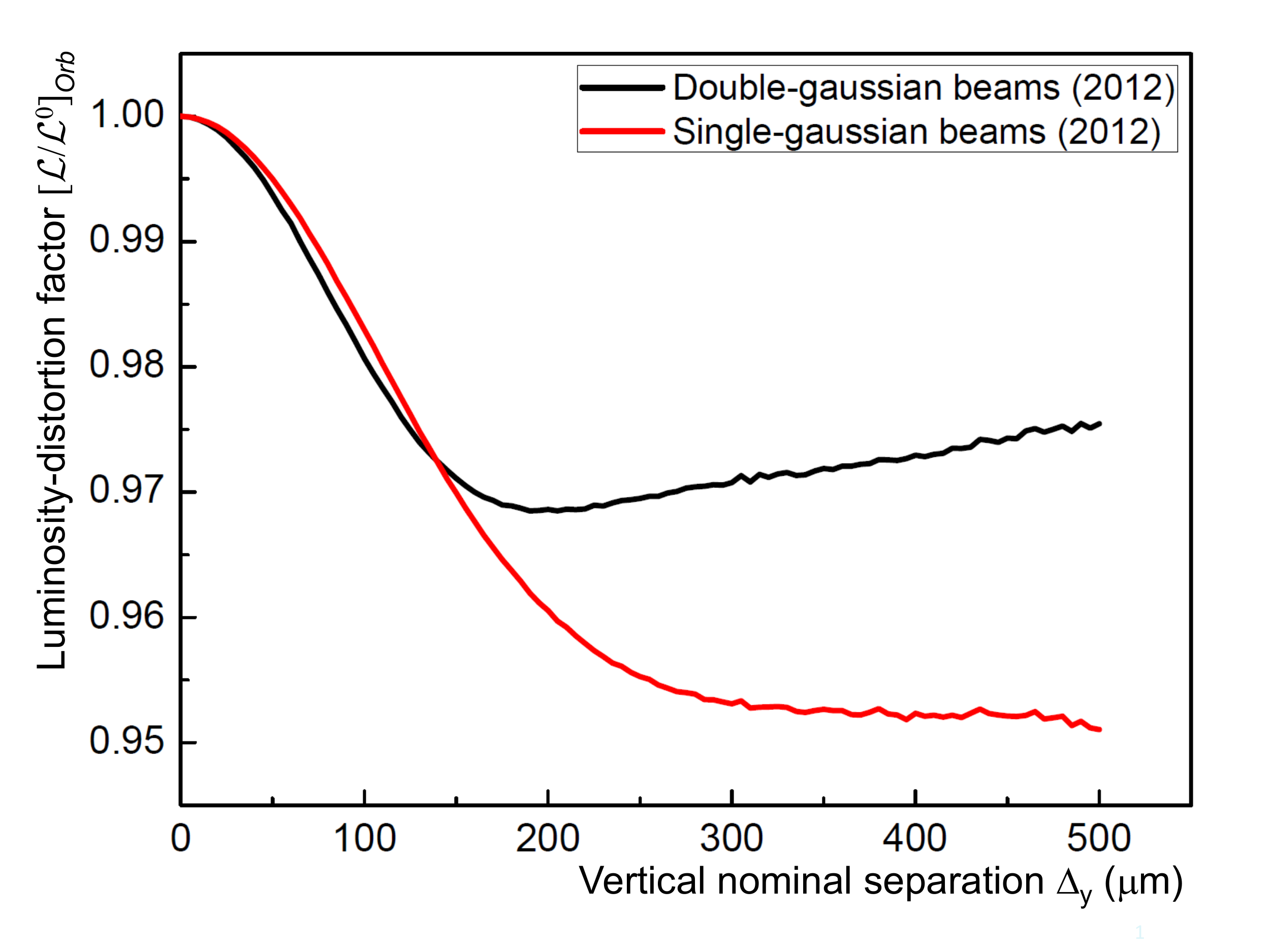}
    \caption{Beam-separation dependence, during simulated vertical \vdM\ scans, of the luminosity-bias factor \LoLzOrb associated with the deflection-induced orbit shift, for single-Gaussian (red) and factorizable double-Gaussian (black) bunch-density distributions that yield the same unperturbed convolved beam size. The horizontal axis is the nominal vertical beam separation. The bunch parameters are listed in the second column of Table~\ref{tab:2gModels} (LHC fill 2855). The beams satisfy the assumptions listed in Sec.~\ref{subsec:bbCorMeth}, except (in the case of the black curve) that regarding the single-Gaussian transverse profile. 
}
    \label{fig:LoLzOrb_SGDG}
\end{figure}

	In the case of factorizable density distributions, the impact of non-Gaussian tails on beam-beam corrections can be visualized by comparing the beam-separation dependence of the luminosity-bias factors \LoLz for a given set of non-Gaussian beam parameters, to that computed for perfectly Gaussian beams that in the absence of beam-beam effects, would yield the same value of the unperturbed convolved beam size $\Sigma^0$, and therefore the same head-on luminosity.
This is illustrated in Fig.~\ref{fig:LoLzOrb_SGDG} for the orbit-shift effect. In the single-Gaussian case (red curve), the steeper drop  of \LoLzOrb with increasing separation, and its asymptotic behavior for $\Delta_y \ge 300\,\mu$m, indicates that at such large separations ($\Delta_y > 4 \sigSG$), the number of particles contributing to the source field becomes negligible, so that the latter increasingly appears to the witness bunch as a point source. In the non-Gaussian case (black curve), the charge distribution is more spread out, and the interplay between the source-field distribution and the density distribution in the witness bunch produces a less intuitive separation dependence. 
	
\par
\begin{figure}
    \centering
    \includegraphics[width=0.47\textwidth]{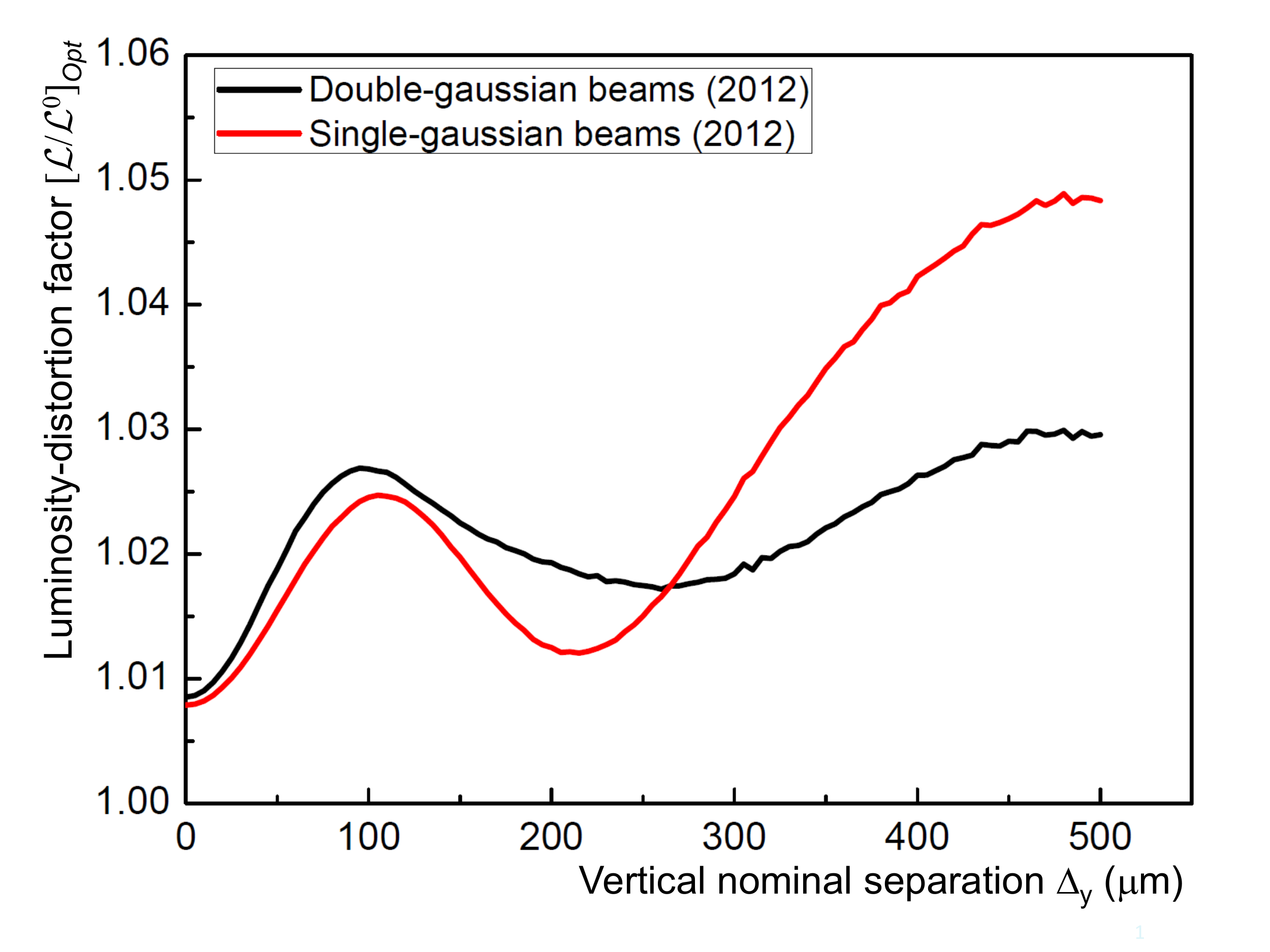}
    \caption{Beam-separation dependence, during simulated vertical \vdM\ scans, of the luminosity-bias factor \LoLzOpt associated with optical distortions, for single-Gaussian (red) and factorizable double-Gaussian (black) bunch-density distributions that yield the same unperturbed convolved beam size. The horizontal axis is the nominal vertical beam separation. The bunch parameters are listed in the second column of Table~\ref{tab:2gModels} (LHC fill 2855). The beams satisfy the assumptions listed in Sec.~\ref{subsec:bbCorMeth}, except (in the case of the black curve) that regarding the single-Gaussian transverse profile.     
}
    \label{fig:LoLzOpt_SGDG}
\end{figure}

	The situation is reversed for optical distortions (Fig.~\ref{fig:LoLzOpt_SGDG}):  the Gaussian beams are subject to larger optical distortions at large beam separation (red curve), presumably because in such configurations the tail particles in the witness bunch experience a more intense source-field gradient when close to the center of the source bunch. The tails of the non-Gaussian beams, being subject to a more diffuse source-field distribution, experience weaker gradients and correspondingly smaller optical distortions (black curve).

%
%
\begin{table*} [htb]
\centering
\begin{tabular}{cc|ccc} 
\hline
\multicolumn{2}{c|}{Date of \vdM session} 	& July 2012	&	June 2018		& July 2017		\\
\multicolumn{2}{c|}{Single-beam kurtosis} 	& 1.40		&	0.47			& 0.25			\\		
\hline \hline
\multicolumn{5}{c}{Factorizable density distribution: Eq.~(\ref{eqn:FDGparmtn})}  				\\
 \hline
 $\xi ~[10^{-3}]$		& 					&  6.03		&  3.05			& 3.61			\\
 \svisNF bias [\%]:	& DG				& -0.27		& -0.15			& -0.25			\\
				& SG				& -0.57		& -0.22			& -0.27			\\
				& DG$ - $SG			&  0.30		&  0.07			&  0.02			\\
\multicolumn{2}{c|}{DG$ - $SG scaled to $\xi = 5.6 \cdot 10^{-3}$}			
									&  0.28		&  0.13   			&  0.03   			\\
 \hline \hline
\multicolumn{5}{c}{Non-factorizable density distribution: Eq.~(\ref{eqn:NFDGparmtn})}  			\\
\hline
$\xi ~[10^{-3}]$		& 					&  6.88		&  3.16			&   3.67			\\
\svisNF bias [\%]:	& DG				& -0.50		& -0.17			& -0.23			\\
				& SG				& -0.64		& -0.21			& -0.25			\\
				& DG$ - $SG			&  0.14		&  0.04			&  0.02			\\
\multicolumn{2}{c|}{DG$ - $SG scaled to $\xi = 5.6 \cdot 10^{-3}$}			
									&  0.11		& 0.07   			&  0.03   			\\			
\hline
\end{tabular}
\centering
\caption{Visible cross-section bias associated with the full beam-beam effect, calculated using B*B, for double- (DG) and single- (SG) Gaussian beams yielding the same Gaussian-equivalent beam-beam parameter $\xi$, and difference between the two (DG$ - $SG). The unperturbed DG transverse-density distributions are described either by a factorizable or by a non-factorizable function; their parameters are listed in Table~\ref{tab:2gModels}. The bunches are assumed to collide at the scanning IP only.
}
\label{tab:2Gimpact}
\end{table*}

\par
In the case of non-factorizable distributions, a physically intuitive graphical representation of beam-beam biases becomes impractical, since \LoLz depends on both $\Delta_x$ and $\Delta_y$, and the curves in Figs.~\ref{fig:LoLzOrb_SGDG} and \ref{fig:LoLzOpt_SGDG} must  be replaced by surfaces. 

\par
The impact of non-Gaussian tails on the corresponding \vdM calibrations is best summarized by the \svis-bias factors presented in Table~\ref{tab:2Gimpact}. These are all computed using Eqs.~(\ref{eqn:	FoM-CSP})-(\ref{eqn:FoM-svis2D}), in order to ensure that factorizable and  non-factorizable models are treated identically; it has been verified that for the factorizable models, the one-dimensional FoM formalism detailed in Sec.~\ref{subsubsec:FoMs} yields the same results within 0.01\% of \svis, well within the expected accuracy of the overlap-integral calculation (Sec.~\ref{subsec:BsBDescr}). Listed in this table is the full beam-beam bias for either factorizable or non-factorizable double-Gaussian models, for the corresponding single-Gaussian model (that is factorizable by construction), as well as for their difference DG$-$SG. 
This difference is also listed after scaling it
linearly to a common, arbitrary reference value of $\xi = 5.6\cdot 10^{-3}$, that lies at the upper end of the beam-beam parameter range covered by Run-2 \vdM scans. This simple-minded procedure, inspired by the roughly linear dependence of the \svis bias factor on $\xi$ apparent in Fig.~\ref{fig:FoM-sigvis}, allows direct comparisons across configurations corresponding to different values of $\xi$.

\par
Globally, non-Gaussian tails modify the \svis bias by an amount that - for the models chosen here - varies from less than 10\% to about half of the bias itself. Predictably enough, the larger the single-beam kurtosis, the larger the relative impact of the tails. The non-factorizable models seemingly predict a smaller influence of the tails than their factorizable counterpart; this may however depend on the particular form of $x$-$y$ correlation embedded into the specific functional form chosen for $\hat{\rho} _B (x,y)$. A practical proposal  to translate the results listed in Table~\ref{tab:2Gimpact} into a realistic systematic uncertainty is offered in Sec.~\ref{subsubsec:nonGSyst}. Repeating the above studies using more refined single-beam models extracted from future \vdM sessions may help refine the understanding of such biases.


	\subsection{Impact of beam ellipticity at the interaction point}
	\label{subsec:EllipImpact}
	
	In the fully symmetric beam configuration (Sec.~\ref{subsec:bbCorMeth}), the beams are round at the IP, and the beam-beam interaction can be described by a single parameter $\xi$ (Eq.~(\ref{eqn:xiRDef})). In the more general case where the transverse beam profile is an upright\footnote{In view of the minor impact predicted by the simulations described in this section for upright (horizontal or vertical) elliptical beams, and of the small residual transverse coupling measured at the LHC IPs, the fully general case of beam ellipses tilted in the $x$-$y$ plane is not considered in the present paper.} ellipse, for instance because of unequal horizontal and vertical emittances or \bst values, the horizontal and vertical beam-beam parameters $\xi_x$ and $\xi_y$ (Eq.~(\ref{eqn:xiGDef})) no longer coincide, and the parametric approach described in Sec.~\ref{subsubsec:CorImplement} is, strictly speaking, no longer applicable.

\par
During \vdM scans at the LHC, however, the round-beam hypothesis is only moderately violated: the measured horizontal and vertical IP-$\beta$ functions typically differ by 10\% or less, the corresponding emittances by 30\% at most, and the increase in projected transverse beam size associated with a non-zero crossing angle at the ALICE and LHCb IPs does not exceed 15\%.

\par
\begin{figure}
    \centering
    \includegraphics[width=0.5\textwidth]{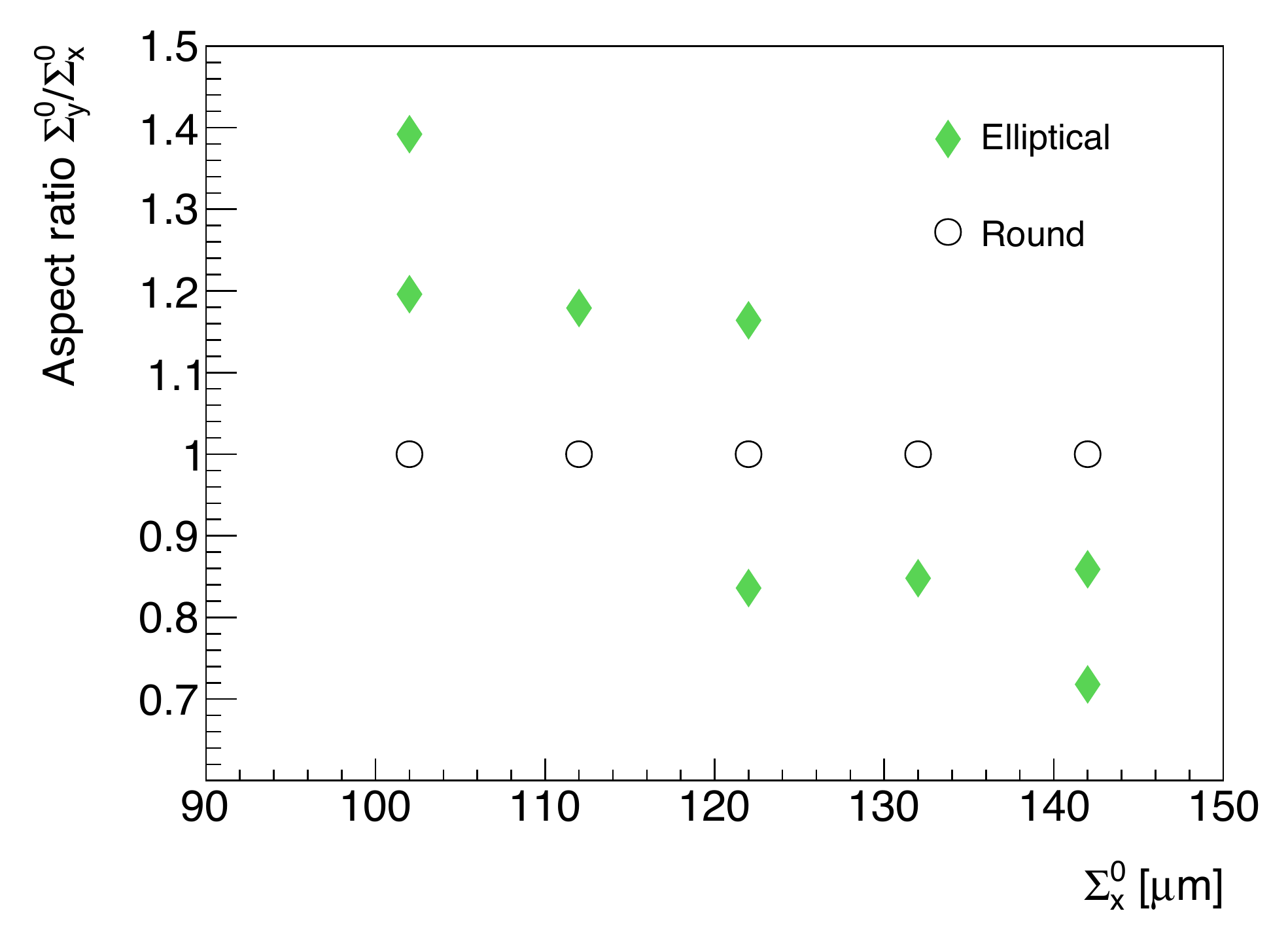}
    \caption{Grid of unperturbed transverse convolved beam sizes used to characterize the impact of beam ellipticity on beam-beam corrections, and their vertical/horizontal aspect ratio. The diamonds (circles) correspond to elliptical (round) beams. The beam energy, bunch intensity, IP-$\beta$ function and unperturbed tunes are fixed at, respectively, $E_B = 6.5$\,TeV, $n_p = 0.777 \times 10^{11}$ protons/bunch, $\bst = 19.2$\,m and $(\Qx, \Qy) = (64.31, 59.32)$.}
    \label{fig:AR-vs-CSx}
\end{figure}
\begin{figure}
    \centering
    \includegraphics[width=0.5\textwidth]{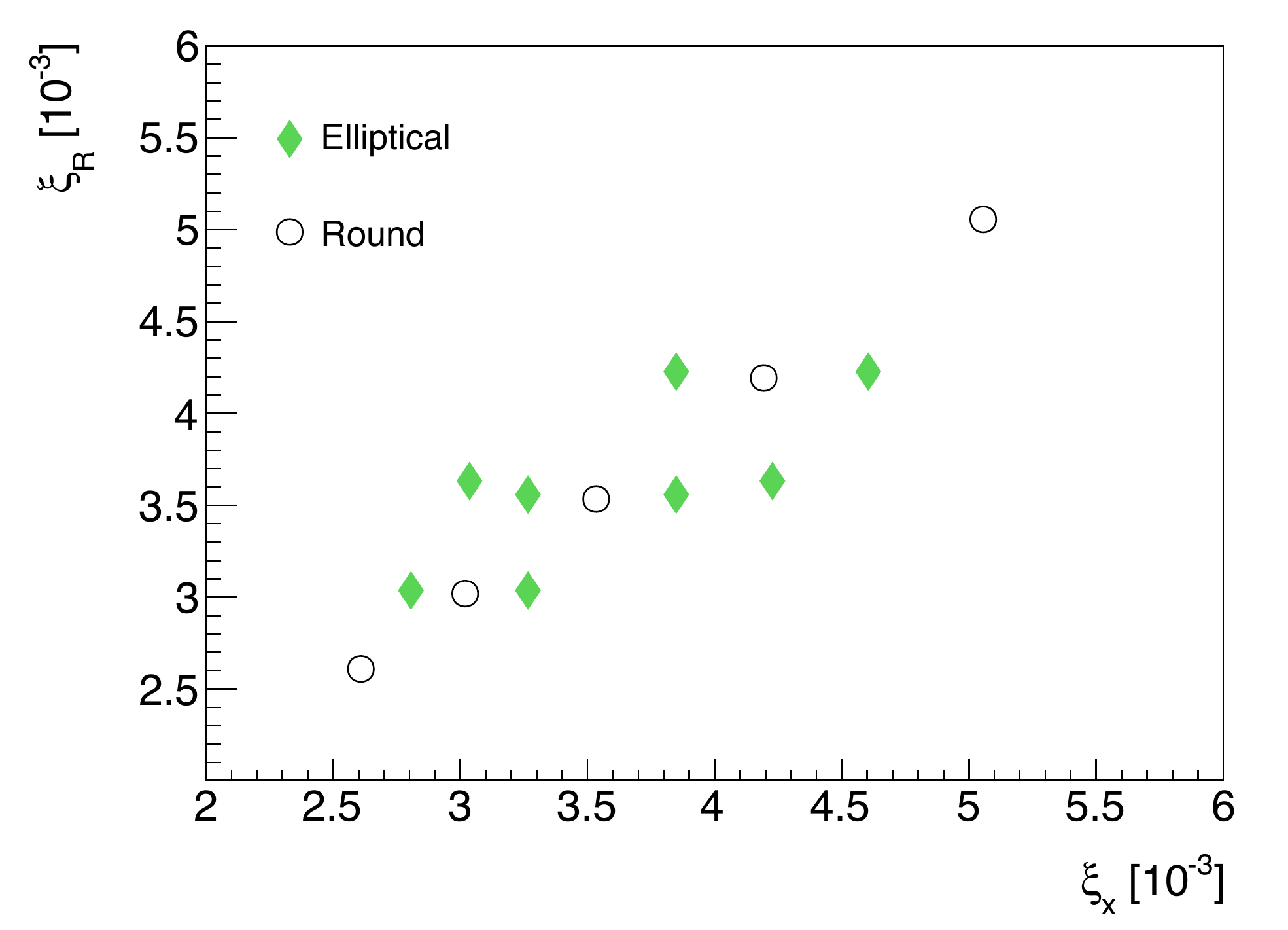}
    \caption{Grid of beam-beam parameter values used to characterize the impact of beam ellipticity on beam-beam corrections. The diamonds (circles) correspond to elliptical (round) beams. The horizontal axis shows the value of $\xi_x$ defined by Eq.~(\ref{eqn:xiGDef}), the vertical axis the corresponding round-beam equivalent parameter \xiR defined by Eq.~(\ref{eqn:xiRDef2}).}
    \label{fig:xiR-vs-xiX}
\end{figure}

To assess the impact of beam ellipticity on beam-beam corrections, B*B is used to simulate \vdM scans over a square grid in  (\CSxz, \CSyz) space (Fig.~\ref{fig:AR-vs-CSx}), with as input unperturbed beams that satisfy the assumptions listed in Sec.~\ref{subsec:bbCorMeth} except for the round-beam hypothesis. The chosen \CSxz range is slightly wider than that observed during Run-2 \vdM scans at $\sqrt{s} = 13$\,TeV; the vertical/horizontal aspect ratio \CSyz/\CSxz is varied from 0.7 to 1.4, corresponding to sampling the corresponding emittance ratio from 0.5 to 2.0, well beyond the range observed during \vdM-calibration sessions. The horizontal and vertical beam-beam parameters cover the range $2.6  \times 10^{-3} < \xi_x, \xi_y < 5.1 \times 10^{-3}$ (Fig.~\ref{fig:xiR-vs-xiX}). The grid is chosen such that for a constant value of the round-beam equivalent beam-beam parameter \xiR, the horizontal and vertical beam-beam parameters sample the full span allowed by the range of chosen \CSxz and aspect-ratio values. For example, the group of five points at $\xiR \approx 3.6 \times 10^{-3}$ in Fig.~\ref{fig:xiR-vs-xiX} corresponds to the five diagonal points in Fig.~\ref{fig:AR-vs-CSx}; the residual curvature is due to the fact that the grid is built from equally spaced (\CSxz, \CSyz), rather than ($\xi_x$, $\xi_y$), values.

\par
\begin{figure}
    \centering
    \includegraphics[width=0.5\textwidth]{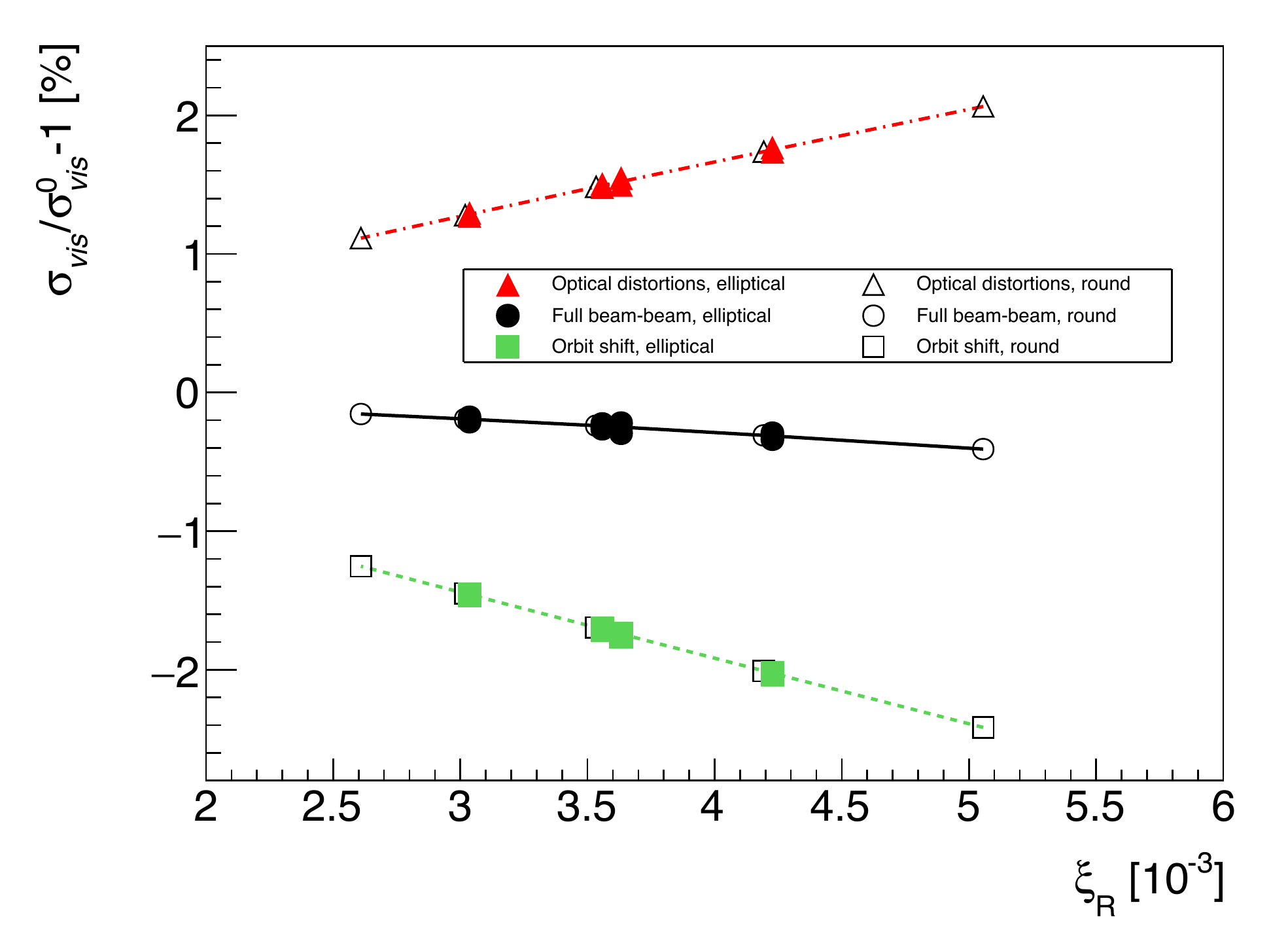}
    \caption{Dependence of the visible cross-section bias on the round-beam equivalent beam-beam parameter \xiR, for optical distortions only (red), orbit shift only (green), and full beam-beam effect (black). The filled (open) markers tag the results of individual B*B simulations assuming upright elliptical (round) beams; the orbit-shift bias is calculated analytically using the elliptical-beam formula~\cite{Bassetti}. The curves show the round-beam estimates based on the parameterization of Eqs.~(\ref{eqn:FoM-svis})-(\ref{eqn:quadxifun}). The points clustered around each \xiR value correspond to different settings of the \CSyz/\CSxz aspect ratio.}
    \label{fig:svisBias-vs-xiR}
\end{figure}
\begin{figure}
    \centering
    \includegraphics[width=0.5\textwidth]{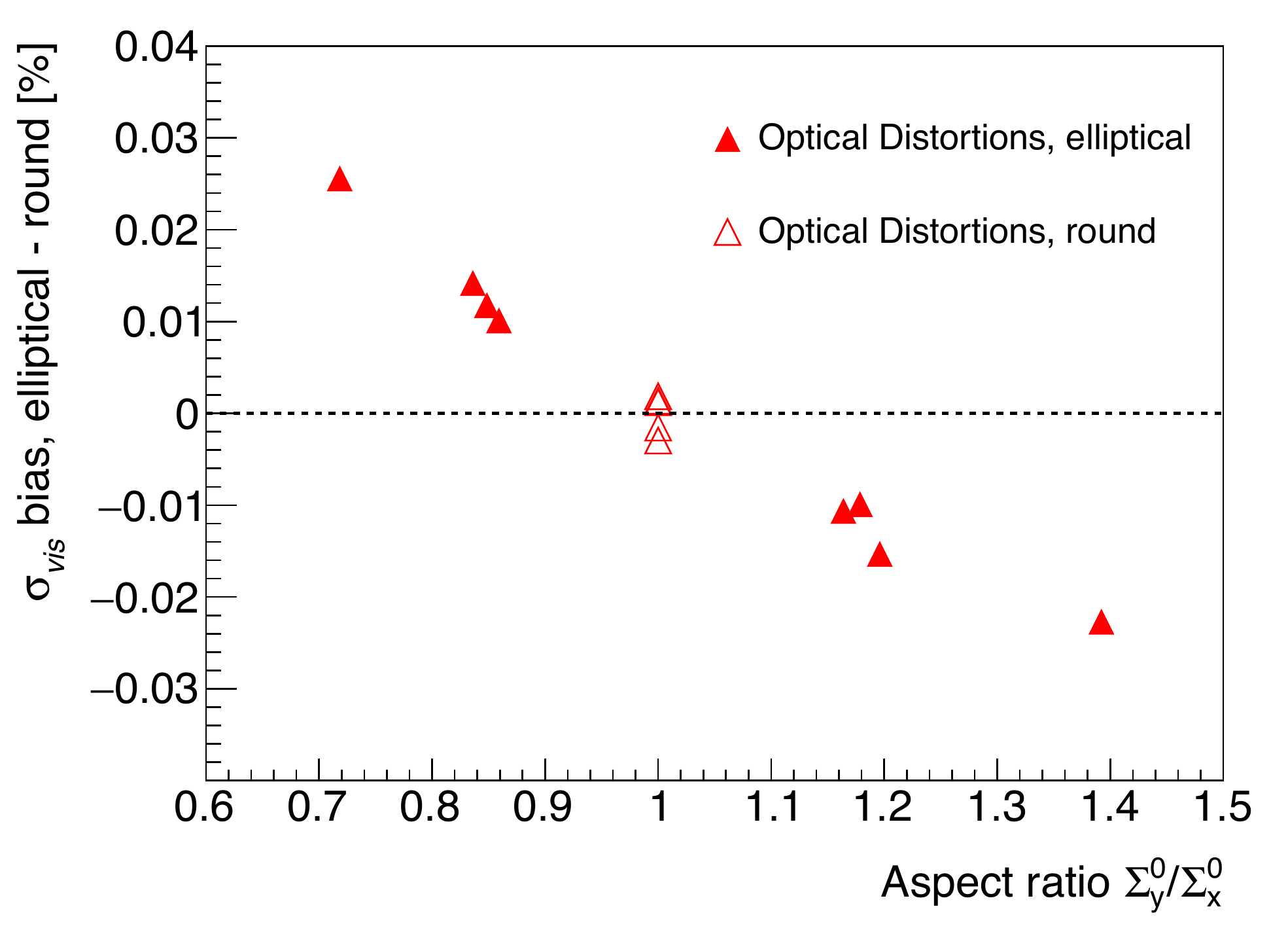}
    \caption{Aspect-ratio dependence of the difference in \svis bias associated with optical distortions, between the result of B*B elliptical-beam simulations, and the corresponding round-beam estimate based on the parameterization of Eqs.~(\ref{eqn:FoM-svis})-(\ref{eqn:quadxifun}). The filled (open) markers tag the results of individual simulations assuming upright elliptical (round) beams. The points clustered around each value of the aspect ratio correspond to different \xiR settings. The horizontal dashed line corresponds to no difference in \svis bias between round and elliptical beams. The spread of the round-beam simulation results (open triangles) reflects the numerical accuracy of the subtracted parameterization.
}
    \label{fig:svisBias_EmR-vs-AR}
\end{figure}

The resulting peak-rate, beam-size and visible cross-section bias factors (Eqs.~(\ref{eqn:FoM-mupk})--(\ref{eqn:FoM-svis})) for elliptical configurations are then compared to the same quantities calculated using the round-beam parameterization (Eq.~(\ref{eqn:quadxifun})) at the corresponding value of \xiR. The results are illustrated in Fig.~\ref{fig:svisBias-vs-xiR} for the visible cross-section. On the scale shown here, the bias ($\svis/\svisz -1$), and therefore the corresponding beam-beam correction, seem to depend on \xiR only; stated differently, the beam-beam-induced biases appear rather insensitive to the aspect ratio. Upon closer inspection however, all four FoMs do reveal an actual dependence on the \CSyz/\CSxz aspect ratio, albeit a weak one. Since the orbit-shift effect can be accurately calculated, even for elliptical beams, using the Bassetti-Erskine formalism~\cite{Bassetti}, only optical-distortion biases matter. The difference between the bias calculated by B*B for a given set of elliptical-beam parameters, and that extracted from the round-beam parameterization using the corresponding value of \xiR, is displayed in Fig.~\ref{fig:svisBias_EmR-vs-AR}. Its dependence on the vertical/horizontal aspect ratio is close to linear, and remains below 0.03\% over the full range of beam parameters considered in this study. Its dependence on \xiR is illustrated by the spread over the three parameter sets clustered at $\CSxz / \CSyz \approx$ 0.8 and 1.2. The beam-beam parameter values considered in this study, and that are associated with elliptical-beam configurations, span the range $3.0 \times 10^{-3} < \xiR < 4.2 \times 10^{-3}$\,: this is representative of, but does not quite cover, the full $\xi$ range listed in Table~\ref{tab:vdMBeamParms}. Even if the mean value of \xiR was doubled, however, the impact of potentially elliptical beam profiles on the beam-beam correction to \svis would still remain well below 1 permil, and can therefore be treated as a systematic uncertainty.

	\subsection{Impact of non-zero crossing angle}
	\label{subsec:XingAngImpact}
		
	In the presence of a significant crossing angle, the analysis becomes considerably more involved, and a full exposition thereof would exceed the scope of the present report. The discussion below is therefore focussed on assessing the beam-beam correction uncertainty associated with \vdM scans at zero nominal crossing angle, during which the actual, residual crossing angle does not exceed few tens of microradians: this covers most luminosity-calibration sessions at IP1 (ATLAS) and IP5 (CMS). The same treatment, albeit at the cost of a small additional uncertainty, remains applicable to scans at IP2 (ALICE) (and to some non-standard sessions at IPs 1 and 5) that involve nominal crossing angles $\theta_c$ in the 140--400\,$\mu$rad range. At IP8 however, the LHCb requirement for an unambiguous separation, in the crossing plane, of beam-gas vertices from beam 1 and beam 2, leads to $\theta_c$ settings of up to 1100\,$\mu$rad. Such large crossing angles would require simulation studies more extensive than can be reported here.
	
\par
In the absence of any beam-beam interaction, the geometric factor $F$ defined in Eq.~(\ref{eqn:geomFact}), and generalizations thereof~\cite{Balagura_2021, bib:BalagNIM}, provide an accurate description of the combined impact of a non-zero crossing angle and of a finite bunch length on the unperturbed apparent transverse beam size in the crossing plane, as well as of the beam-separation dependence of the unperturbed luminosity. 

\par
In the presence of the beam-beam interaction, the implications of a non-zero crossing-angle for the calculation of beam-beam biases fall in two conceptually distinct categories.
\begin{itemize}
\item
{\em Geometrical effects in the transverse plane} that depend on the crossing angle and on the bunch length, such as modifications of the transverse beam-beam overlap and the calculation of the effective strength of the source field experienced by the particles in the witness beam, are adequately modeled~\cite{Balagura_2021} by the B*B package and well reproduced by COMBI simulations. B*B simulations, however, are intrinsically limited to four dimensions (horizontal and vertical positions and angles), since the longitudinal variables are ``projected out'' and the associated dynamics implicitly neglected.
\item
{\em Longitudinal dynamics}, such as synchrotron motion or beam--beam-induced transverse-shape variations along the bunch, can be modeled by COMBI using the 6D beam-beam strong-strong model~\cite{Laurent}. This model divides particle distributions into slices along the length of each bunch. The kick experienced by each macroparticle in one beam is computed based on the statistical moments of the charge distributions in each slice of the opposing bunch; the computation is repeated for the macroparticles in the other beam. 
\end{itemize}
In order to properly account for longitudinal effects and assess their importance, COMBI is chosen here to characterize the interplay of crossing-angle and beam-beam effects. This first study is exploratory in nature and restricted to the \vdM regime. 

\par
As predicted analytically~\cite{Balagura_2021, bib:ABXingAng}, the magnitude of the orbit shift extracted from COMBI simulations for a purely horizontal crossing angle  of several hundred microradians, is found consistent with that computed using the Bassetti-Erskine formula with, as input:
\begin{itemize}
\item
in the crossing (horizontal) plane, the effective convolved beam size given by $\Sigma_{x,eff} = F \times \CSx$, where $F$ is the geometric factor defined in Eq.~(\ref{eqn:geomFact}) and \CSx is the horizontal convolved beam size at zero crossing angle, given by Eq.~(\ref{eqn:capSigDef});
\item
in the non-crossing (vertical) plane, the vertical convolved beam size \CSy.
\end{itemize}

\par
\begin{figure}
	    \centering
	    \includegraphics[width=0.48\textwidth]{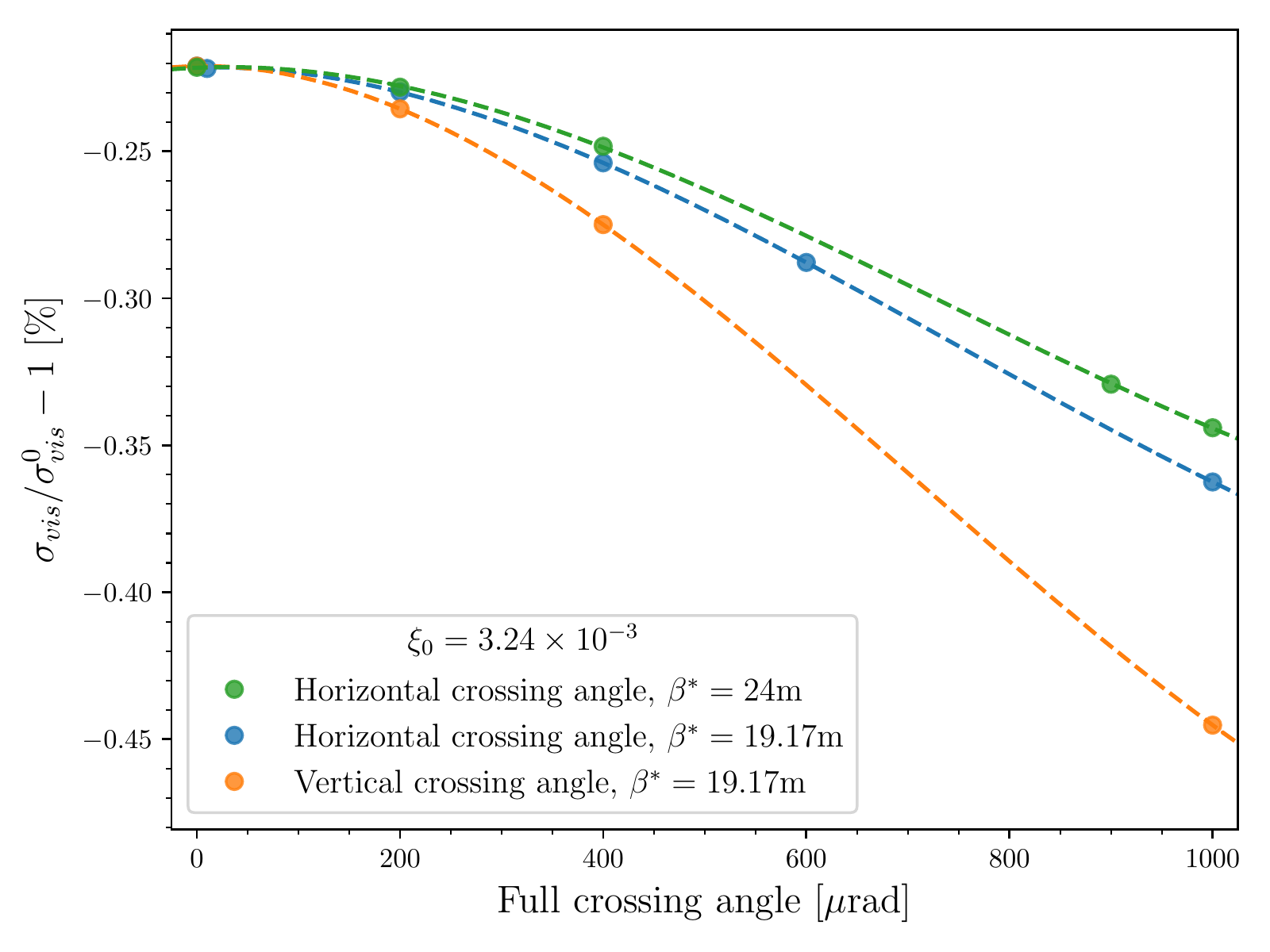}
	    \caption{Crossing-angle dependence of the visible cross-section bias predicted by COMBI strong-strong, 6-dimensional simulations, for a horizontal (blue) and a vertical (orange) crossing angle at \bst = 19.17\,m, and for a horizontal crossing angle at \bst = 24\,m (green). The beam-beam parameter value at zero crossing angle is indicated in the legend. The beam energy, bunch intensity, normalized emittance, bunch length and  tunes are fixed at, respectively, $E_B=6.5$\,TeV, $n_p =  0.783 \times 10^{11}$\,p/bunch,  $\epsilon_{N}= 2.95\, \mu$m$\cdot$rad, $\sigma_z = 7.5$\,cm, $\Qx = 64.31$, $\Qy = 59.32$, $\mathrm{Q_s} = 0.0023$. The bunches are assumed to collide only at the IP where the \vdM scans are performed. The points are the results of the simulation; the lines are meant to guide the eye.
}
	    \label{fig:sigVis_vsXA}
	\end{figure}

The dependence of the beam--beam-induced visible cross-section bias on the full crossing angle is illustrated in Fig.~\ref{fig:sigVis_vsXA} for a beam-beam parameter setting typical of Run-2 \vdM scans. For the configurations shown in this figure, the geometric factor spans the range $1.0 < F < 1.09$. Since this factor impacts both \svisz and \svis in an almost identical manner\footnote{In the ratio $\svis / \svisz$,  both numerator and denominator include a geometric factor. In the case of \svis, the transverse convolved beam size is in addition modified by the dynamic-$\beta$ effect, resulting in a 0.07\% increase in $F$ in the worst case ($\bst = 19$\,m, $\theta_c = 1000\, \mu$rad).}, it cancels in the crossing-angle dependence of the ratio $\svis / \svisz$. 
The absolute value of the \svis bias increases with the angle, even though the ``effective'' beam-beam parameter in the crossing plane (\ie that calculated from the effective transverse beam size) decreases. The difference between the horizontal- and vertical-crossing configurations is due to the difference between the $x$ and $y$ tunes. Also notable is the sensitivity to \bst: the crossing-angle dependence of the beam-beam bias is weaker for $\bst = 24$\,m than for $\bst = 19$\,m. The results are  insensitive to the assumed value of the synchrotron tune, as well as to the chromaticity of the ring lattice (reducing the chromaticity to zero from its nominal setting of 15 units impacts the results by less than 0.01 \%).  This suggests that the effect is mostly geometrical in nature, as confirmed by the qualitative agreement of the COMBI results with those of a preliminary B*B study that covered a similar parameter space.

\par
In practical terms, the impact of the crossing angle on the estimated beam-beam correction can be taken as the difference in visible cross-section bias between $\theta_c = 0$ and the actual crossing-angle value. For a nominally zero crossing angle, the actual angle, as measured by the BPM system or by K-modulation in the final-triplet quadrupoles, is of the order of $10\, \mu \mathrm{rad}$: the associated shift in the beam-beam bias, as read off Fig.~\ref{fig:sigVis_vsXA} at $\bst = 19$\,m, does not exceed 0.01\%. For a nominal crossing angle around $150\, \mu \mathrm{rad}$, which is typical of ALICE \vdM sessions, neglecting the crossing angle would underestimate the beam-beam correction by less than 0.02\%, a value small enough to be neglected. Crossing angles in excess of $500 \,\mu$rad, however, induce still-to-be understood biases of 0.1\% or more (Fig.~\ref{fig:sigVis_vsXA}), and may require further study.
 
	\subsection{Impact of multiple interaction points}
	\label{subsec:multiIP}

	\subsubsection{Motivation and methodology}
	\label{subsubsec:mIPmethod}

	All results presented in this paper so far assume that the beams collide at the scanning IP only, \ie that they are fully separated, either transversely or longitudinally, at the other three IPs of the LHC. In practice however, this is often not the case: to make optimal use of beam time during luminosity-calibration scans at, say, the CMS IP,  the experimental detectors at some or all of the other IPs typically collect collision data at fixed (most often zero) beam separation, thereby significantly affecting the unperturbed-tune spectra and therefore the magnitude of beam-beam induced biases.
	
\par

	 For a given colliding-bunch pair at a given scanning IP, the {\em number of non-scanning interaction points} \NSIP is defined as the average number of head-on collisions experienced, at interaction points {\em other} than the scanning IP, by the two members of the pair. 
The parameter $\NSIP$ ranges from zero when beams are fully separated at all but the scanning IP, to $\NSIP =3$ when both bunches in the pair collide at three other IPs; it can assume half-integer values when the two members of the pair experience different numbers of additional collisions. 	 

\par
\begin{figure}
	    \centering
	    \includegraphics[width=0.485\textwidth]{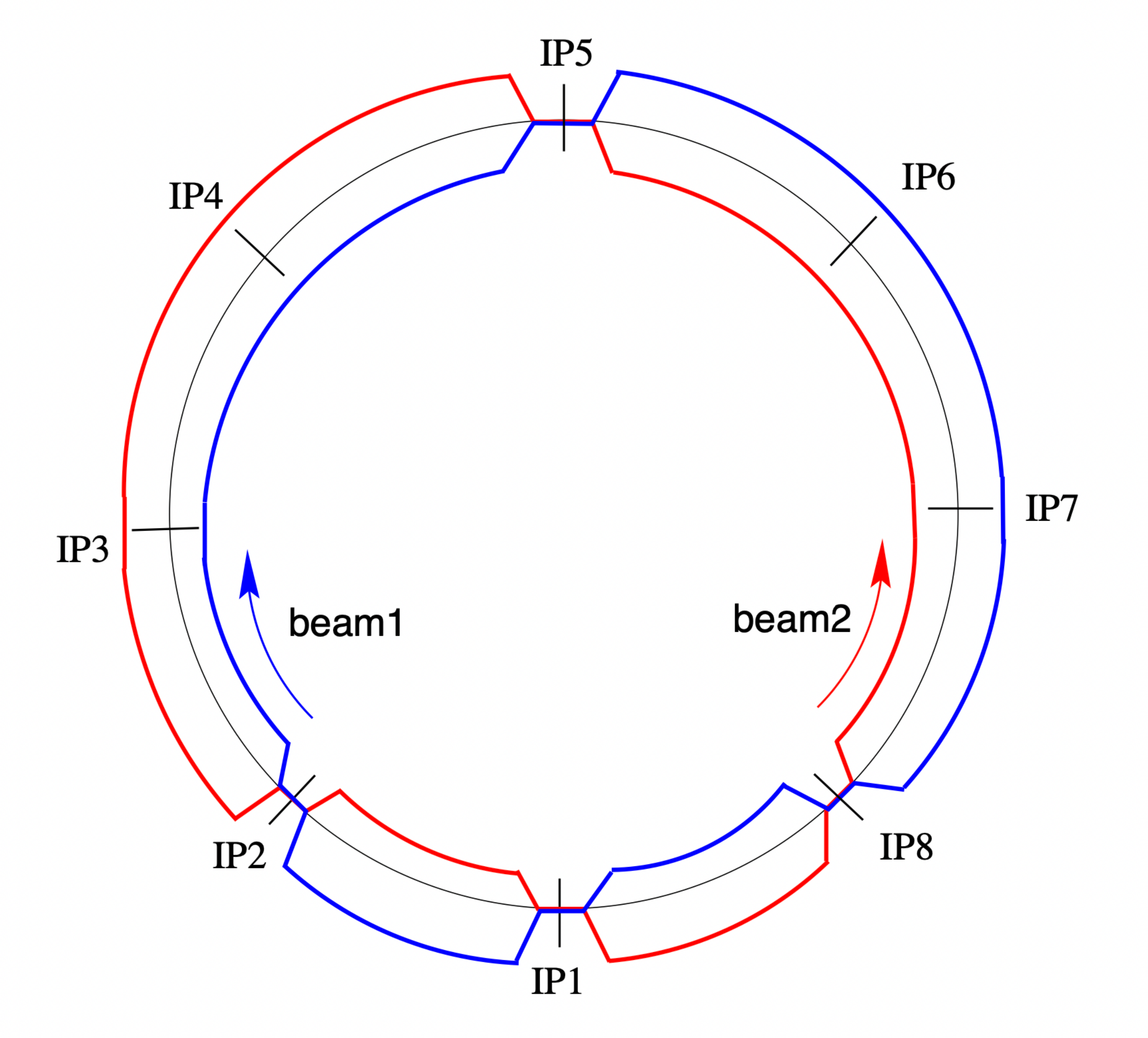}
	    \caption{Schematic layout of the LHC beams and interaction points. Interaction points 1, 2, 5 and 8 house, respectively, the ATLAS, ALICE, CMS and LHCb experiments. (\figFromRef{BBCAS}, \copyright\ CERN)}
	    \label{fig:LHCRingLayout}
	\end{figure}

Because IP1 (ATLAS) and IP5 (CMS) are separated by half a ring circumference, all bunches that collide in IP1 also collide in IP5, unless fully separated in the transverse plane at the non-scanning IP. Each member of such a colliding-bunch pair can optionally also collide in IP2 (ALICE) and/or IP8 (LHCb), both of which are located 1/8th of the ring away\footnote{Up to a longitudinal shift of 11.24\,m (30 RF buckets) at IP8, a requirement driven by the longitudinal asymmetry of the LHCb detector.} from IP1; however its collision ``partner'' at these IPs cannot be the same as that at IPs 1 and 5, because of the ring layout (Fig.~\ref{fig:LHCRingLayout}). Injected-bunch patterns are optimized, on a fill-by-fill basis, to achieve the desired sharing of collisions (or absence thereof) among the four experimental detectors. 

\par	 
	 The parameter \NSIP, therefore, depends on the scanning IP considered, and, for a given scanning IP, can vary along the bunch string in a manner that depends on the injected-bunch pattern. From 2010 to 2017, the most frequent configuration during $pp$ \vdM scans at the ATLAS and CMS IPs corresponded to $\NSIP=1$ at IPs 1 and 5, \ie to bunches that collide in IP1 and IP5 only; from 2018 onwards, all four scanning IPs had to deal with multiple non-scanning IPs, for reasons of operational efficiency. 

\par
	In order to characterize the impact of multiple collision points on beam-beam corrections to luminosity calibrations, COMBI was used to simulate \vdM scans at one IP (\eg IP1), with additional head-on collision points optionally inserted at one or more other IPs (in this example IP5, IP2 and/or IP8). The bunch parameters are listed in Table~\ref{tab:mIPparms}. The beam transport between consecutive IPs is modeled by linear maps that reflect the nominal phase advance in the LHC rings during the Run-2 \vdM\ sessions at $\sqrt{s} = 13$\,TeV (Table~\ref{tab:R2PhasAdv}), and head-on collisions are simulated at each non-scanning IP as described in Sec.~\ref{subsec:COMBIDescr}. The unperturbed beams satisfy the assumptions listed in Sec.~\ref{subsec:bbCorMeth}, except for the absence of additional collisions.

\begin{table} [htb]
\centering
\begin{tabular}{c|ccc} 
Parameter 							& Value				\\ 
\hline
Beam energy $E_B$ [TeV]				& 6.5					\\
Typical bunch population $n_p$ [$10^{11}$]	& 0.78				\\   
Bunches colliding						& 1 					\\  
$\epsilon_N$ [$\mu$m-rad] 				&  2.9				\\   
IP1,5: \bst [m]							& 19.2 				\\   
IP2/8: \bst [m]							& 19.2 / 24.0			\\	
$\theta_c$ [$\mu$rad]					& 0 					\\
$\CSxz  = \CSyz$ [$\mu$m]				& 128				\\ 
Nominal tune settings \Qx/\Qy 				& 64.31/59.32			\\
Typical $\xi$ [per IP] 						& 0.0032 				\\  
\end{tabular}
\caption{Beam conditions assumed for the multi-IP simulations described in Sec.~\ref{subsec:multiIP}. The crossing angle $\theta_c$ is set to 0 at all IPs so as to fully decouple the multi-IP study from potential longitudinal-dynamics effects.}
\label{tab:mIPparms}
\end{table}
\begin{table*} [htb]
\centering
\begin{tabular}{|c|c|c|c|c|c|} 	
\hline
\multicolumn{3}{|c|}{Beam 1}				&  \multicolumn{3}{c|}{Beam 2}	\\
\hline
From - to		& $\mu_\mathrm{x}^{{\mathrm B}1}~[2\pi]$	& $\mu_\mathrm{y}^{{\mathrm B}1}~[2\pi]$	&		
From - to		& $\mu_\mathrm{x}^{{\mathrm B}2}~[2\pi]$	& $\mu_\mathrm{y}^{{\mathrm B}2}~[2\pi]$	\\
\hline \hline
IP1 - IP5		& 31.9757					&   29.6486				&			
IP5 - IP1		& 31.9844					&   29.7613				\\
IP5 - IP1		& 32.3343					&   29.6714				& 
IP1 - IP5		& 32.3256					&   29.5587				\\
\hline
IP1 - IP2		& 8.2960					&    7.6692				&   
IP2 - IP1		& 8.2728					&    7.9577				\\
IP2 - IP5		& 23.6797					&  21.9794				&
IP5 - IP2		& 23.7116					&  21.8036				\\
IP5 - IP8		& 24.0891					&  21.3685				& 
IP8 - IP5		& 23.8146					&   21.9544				\\
IP8 - IP1		&   8.2452					&     8.3029				&  
IP1 - IP8 		&  8.5110					&     7.6042				\\
\hline\hline
Full turn		& 64.3100					&  59.3200				&
Full turn		& 64.3100					&  59.3200				\\
\hline
\end{tabular}
\caption{Horizontal ($\mu_\mathrm{x}$) and vertical ($\mu_\mathrm{y}$) betatron phase advance between consecutive collision points in the LHC Run-2 \vdM\ configuration for $pp$ collisions at $\sqrt{s} = 13$\,TeV. Beam 1 moves from the top to the bottom row, beam 2 from the bottom to the top row.
}
\label{tab:R2PhasAdv}
\end{table*}

\par

	This Section is organized as follows. Multi-IP collisions increase the beam-beam tune spread, thereby shifting the means of the unperturbed-tune spectra to lower values (Sec.~\ref{subsubsec:mIPtunes}). They also affect the peak-rate and beam-size bias factors
in a manner that depends on the phase advance between consecutive collision points around the rings, but that leaves the \svis bias almost totally insensitive to these phase-advance values. The analysis can be greatly simplified by an astute choice of the reference ``no beam-beam'' configuration (Sec.~\ref{subsubsec:LzVsLu}), up to  a small residual ambiguity that depends on the phase advance between the non-scanning IP(s) and that where the scans are taking place (Sec.~\ref{subsubsec:mIPambig}). This approach naturally leads to an effective parameterization of the impact of multi-IP collisions on  beam-beam corrections to \vdM\ calibrations (Sec.~\ref{subsubsec:mIPparam}). Candidate beam-beam correction strategies in the presence of multi-IP collisions are summarized in Sec.~\ref{subsubsec:mIPSmry}.


	\subsubsection{Impact of multiple interaction points on the tune spectra}
	\label{subsubsec:mIPtunes}
	 
\begin{figure}
	    \centering
	    \includegraphics[width=0.48\textwidth]{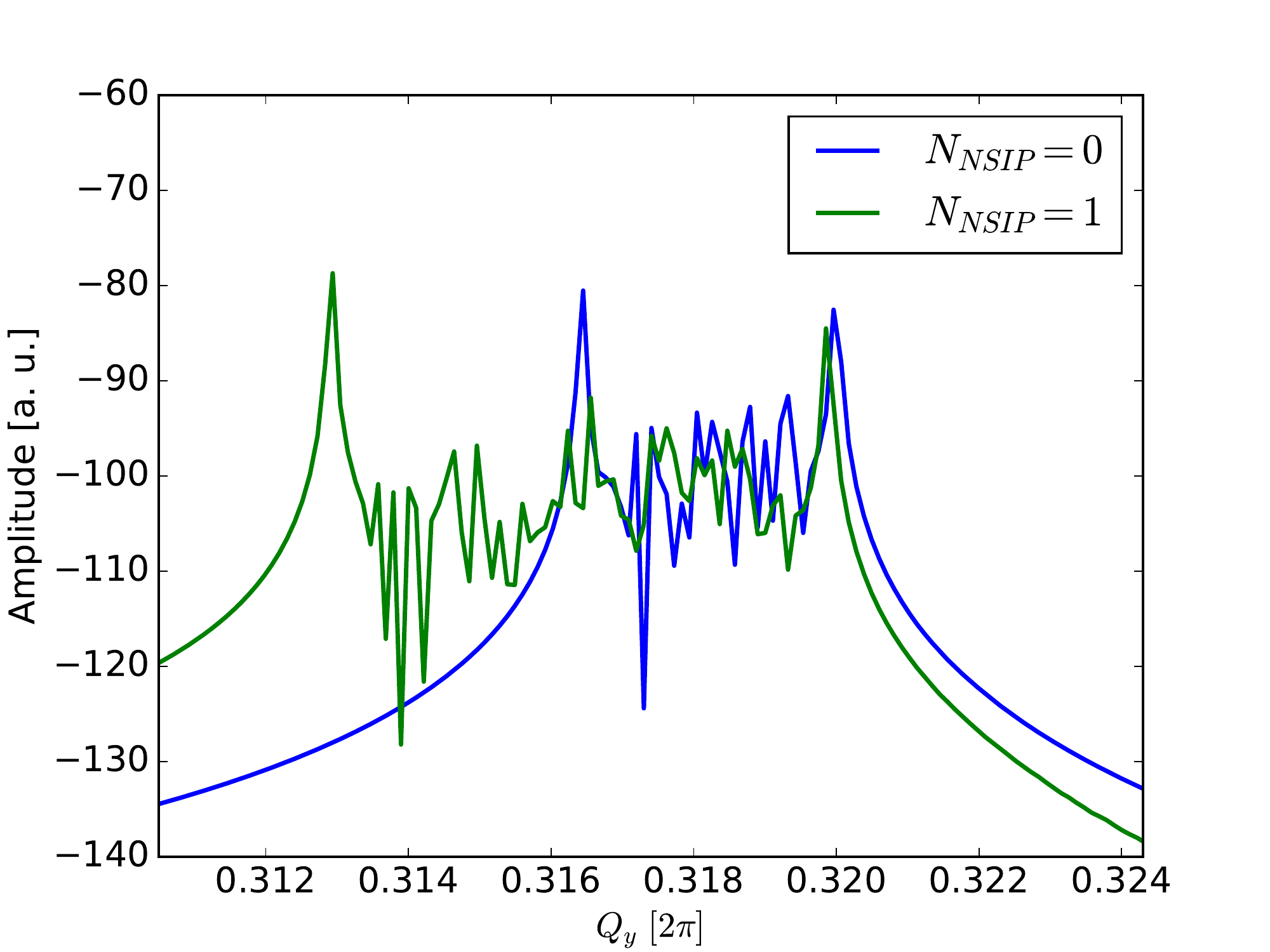}
	    \caption{Vertical coherent-tune spectra in collisions simulated using COMBI, for head-on collisions at the scanning IP only ($\NSIP =0$, blue curve), and in the presence of head-on collisions at one additional IP ($\NSIP =1$, green curve).
	     The beams satisfy the assumptions listed in Sec.~\ref{subsec:bbCorMeth}, except for the absence of additional collisions. The beam-beam parameter $\xi$ is set to $3.24 \times 10^{-3}$ per collision, and the tunes to their nominal value $(\Qx, \Qy)=(64.31, 59.32)$}
	    \label{fig:tuneSpectr}
	\end{figure}

A comparison of simulated vertical-tune spectra for $\NSIP =0$ and $\NSIP =1$ is presented in Fig.~\ref{fig:tuneSpectr}. The single-IP beam-beam parameter $\xi$ is chosen to be typical of \vdM scans in $pp$ collisions. For beams colliding at the scanning IP only ($\NSIP =0$), two narrow peaks appear: one near the nominal tune ($\qy \approx 0.320$), and the other near $\qy - Y*\xi$, where $Y \approx 1.1$ is the Yokoya factor in the soft-Gaussian approximation~\cite{Yokoya}. These peaks correspond to, respectively, the so-called $\sigma$ and $\pi$ modes of coherent bunch oscillation. When beams collide in addition at one non-scanning IP ($\NSIP =1$), the peak of the $\sigma$ mode is unaffected, but the beam-beam tune spread roughly doubles and the $\pi$ mode is shifted further down, near $\qy - 2Y*\xi$.  In both cases, the mean tunes lie roughly half-way between the $\sigma$ and $\pi$ peaks, and therefore differ by roughly $-\xi/2$ between the two configurations. Adding a second non-scanning IP results in a further widening of the tune spread, to roughly three times that for $\NSIP =0$, and in a correspondingly larger downward shift of the mean tune. 

	\subsubsection{Comparison of reference ``no beam-beam'' configurations in the presence of one non-scanning IP}
	\label{subsubsec:LzVsLu}

	In Secs.~\ref{sec:lumCalMeth}, \ref{sec:bbSimCodes} and \ref{subsec:bbCorMeth} to \ref{subsec:XingAngImpact}, it is implicitly assumed that the beams collide at the scanning IP only. The beam--beam-induced peak-rate, beam-size and luminosity-bias factors defined in Sec.~\ref{subsubsec:FoMs} are all expressed with respect to an unambiguous reference configuration that corresponds to ``no beam-beam anywhere'', in which the ``nominal '' reference variables ($\sigma_0$, $\muz_{pk}$, \CSxz, \CSyz, \Lumz, \svisz,...) are identified by a ``0'' index.

\par
	In the presence of collisions at non-scanning IPs, however, an ambiguity arises: should beam-beam-induced biases be referenced to a configuration in which the beam-beam interaction is switched off {\em at every single IP}, or to one in which the beam-beam interaction is switched off {\em at the scanning IP only}, but remains active at the non-scanning IPs (if any)? 
	
\par
	Confronting this question requires a more precise nomenclature than that used so far:
\begin{itemize}
\item
when the reference configuration is one in which the beam-beam interaction is switched off at all IPs, \ie if \vdM\ observables are corrected to ``no beam-beam anywhere'', the corresponding reference variables carry a ``0'' index as before. In particular, the reference luminosity \Lnobb at the scanning IP is equal to the nominal luminosity defined in Sec.~\ref{subsubsec:optclDist}  ($\Lnobb = \Lumz$). This is referred to below as the ``\Lumz normalization'';
\item
when the reference configuration is one in which \vdM\ observables are corrected to ``no beam-beam at the scanning IP only'' even when the bunches considered also collide at one or more other IPs, the corresponding reference variables carry a ``$u$'' index instead. In particular, the reference luminosity \Lnobb at the scanning IP is given by the {\em unperturbed luminosity} \Lumu (\Lnobb = \Lumu), \ie that which would be measured at the scanning IP if the beam-beam interaction could be turned off at that IP only. This quantity is accessible only in simulations; its beam-separation dependence is in general notably different from that of \Lumz. The same notation, and physical meaning, are applied to the other reference variables. This is referred to below as the ``\Lumu normalization'';
\item
in the limiting case of collisions at the scanning IP only ($\NSIP =0$), \Lumu and \Lumz are one and the same, as are all the other reference variables: there is no difference between ``nominal'' and ``unperturbed'' quantities.
\end{itemize}

\par
\begin{figure}
    \centering
    \includegraphics[width=0.48\textwidth]{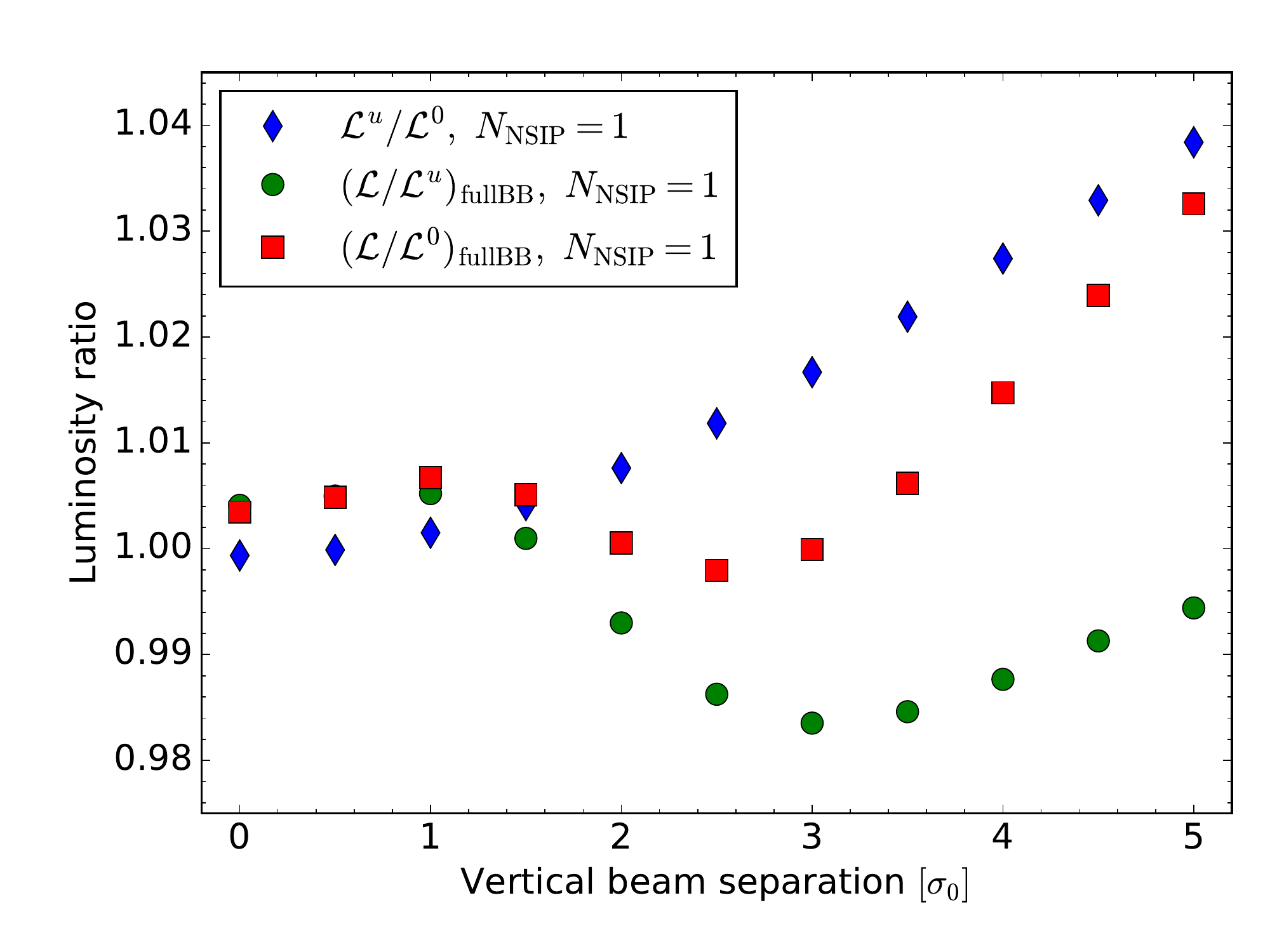}
    \caption{Beam-separation dependence, during a simulated vertical \vdM\ scan in the Run-2 configuration with one non-scanning IP,  of the luminosity-bias factors at the scanning IP associated with the unperturbed luminosity ($\Lumu/\Lumz$, blue lozenges), with  the full beam-beam effect in the \Lumz normalization (\LoLzF, red squares), and with the full beam-beam effect in the \Lumu normalization (\LoLuF, filled green circles). The  horizontal axis is the beam separation in units of the nominal transverse beam size $\sigma_0$.
}
    \label{fig:Lmu-o-L0_1wIP}
\end{figure}
\begin{figure}
    \centering
    \includegraphics[width=0.5\textwidth]{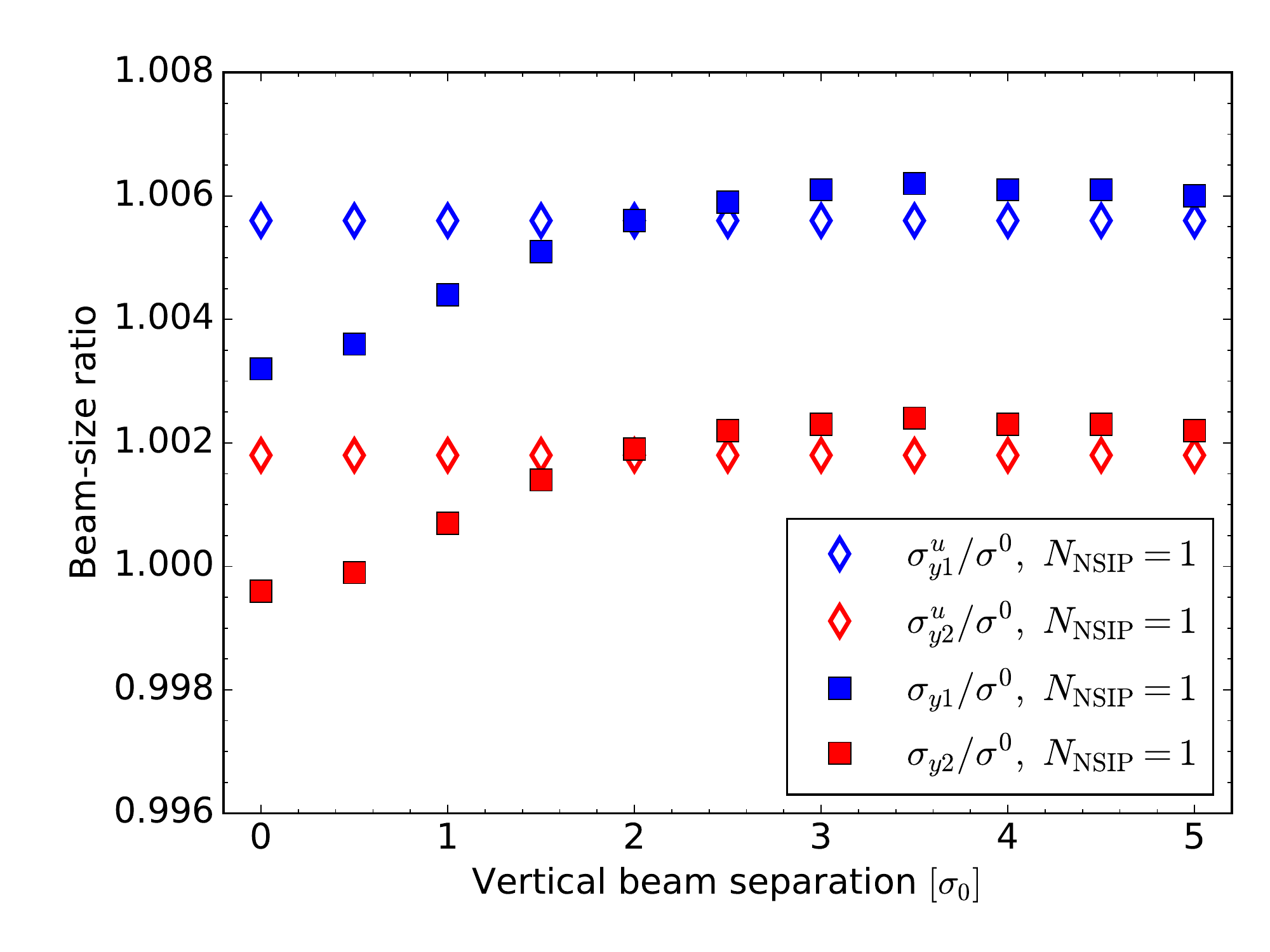}
    \caption{Beam-separation dependence, during a simulated vertical \vdM\ scan, of the beam--size-bias factors at the scanning IP associated with the full beam-beam effect ($\sigma_{yB}/\sigma^0$, filled squares) and of the corresponding unperturbed quantities ($\sigma^u_{yB}/\sigma^0$, open lozenges), separately for B = beam 1 (blue) and B = beam 2 (red), using the \Lumz normalization and in the presence of collisions at one non-scanning IP ($\NSIP =1$). The  horizontal axis is the beam separation in units of the nominal transverse beam size $\sigma_0$.}
    \label{fig:sYmu-o-s0_1wIP}
\end{figure}

The beam-separation dependence of the $\Lumu/\Lumz$ ratio, during a vertical scan at IP1 and in the presence of  head-on collisions at IP5, is presented in Fig.~\ref{fig:Lmu-o-L0_1wIP} (blue lozenges). Even though the unperturbed luminosity \Lumu is by definition unaffected by beam-beam effects at the scanning IP,  it exhibits a monotonic increase with beam separation when compared to the reference luminosity \Lumz. This seemingly counter-intuitive observation is explained by the fact that the horizontal axis in Fig.~\ref{fig:Lmu-o-L0_1wIP} is the {\em nominal} transverse beam size $\sigma^0$, not the {\em actual} beam size during the scan; equivalently, the actual relative separation, \ie that normalized to the actual unperturbed beam size, differs from the nominal relative separation, which is normalized to the nominal beam size.
The monotonic increase of the $\Lumu/\Lumz$ ratio therefore suggests that in this particular configuration, the collisions at IP5 result in an enlarged vertical convolved beam size at IP1 ($\CSyu > \CSyz$). This interpretation is confirmed by the evolution, during the scan, of the individual unperturbed single-beam sizes $\sigma^u_{yB}$ ($B = 1,2$) that is illustrated in Fig.~\ref{fig:sYmu-o-s0_1wIP} (open lozenges): both are larger than $\sigma_0$, they remain constant as a function of the beam separation at the scanning IP, and their difference reflects the fact that the vertical phase-advance values from IP1 to IP5 differ significantly between the two beams (Table~\ref{tab:R2PhasAdv}, top two rows).

\par
Also shown in Fig.~\ref{fig:Lmu-o-L0_1wIP} is the luminosity-bias factor \LoLzF in the \Lumz normalization (red squares), that displays an $s$-like shape superposed on the monotonic beam-separation dependence of the $\Lumu/\Lumz$ curve. The ratio of these two curves:
\begin{equation}
\LoLzF\, / \,(\Lumu/\Lumz) = \LoLuF
\nonumber
\end{equation}
represents the luminosity-bias factor in the \Lumu normalization in the presence of one non-scanning IP ; it is displayed as filled green circles, and is highly reminiscent of the black curve in Fig.~\ref{fig:ll0-3-curves}.
%

\par
\begin{figure}
    \centering
    \includegraphics[width=0.48\textwidth]{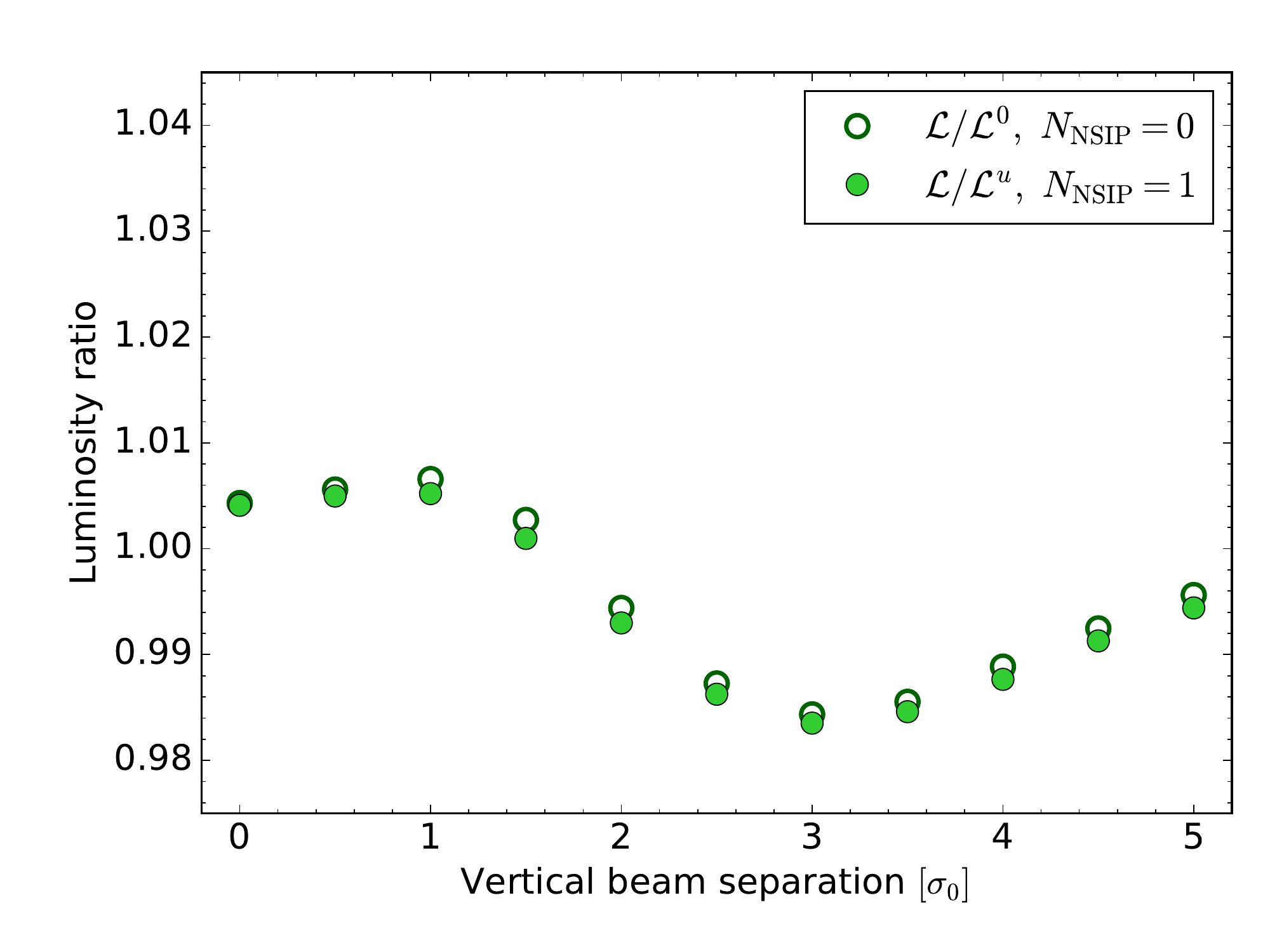}
    \caption{Beam-separation dependence, during a simulated vertical \vdM\ scan, of the luminosity-bias factors at the scanning IP associated with the full beam-beam effect, for $\NSIP = 0$ using the \Lumz normalization (open circles), and for $\NSIP = 1$ using the \Lumu normalization (filled green circles). The  horizontal axis is the beam separation in units of the nominal transverse beam size $\sigma_0$. The span of the vertical scale is identical to that of Fig.~\ref{fig:Lmu-o-L0_1wIP}.}
    \label{fig:Lm-o-Lu0_1n0wIP}
\end{figure}
\begin{figure}
    \centering
    \includegraphics[width=0.48\textwidth]{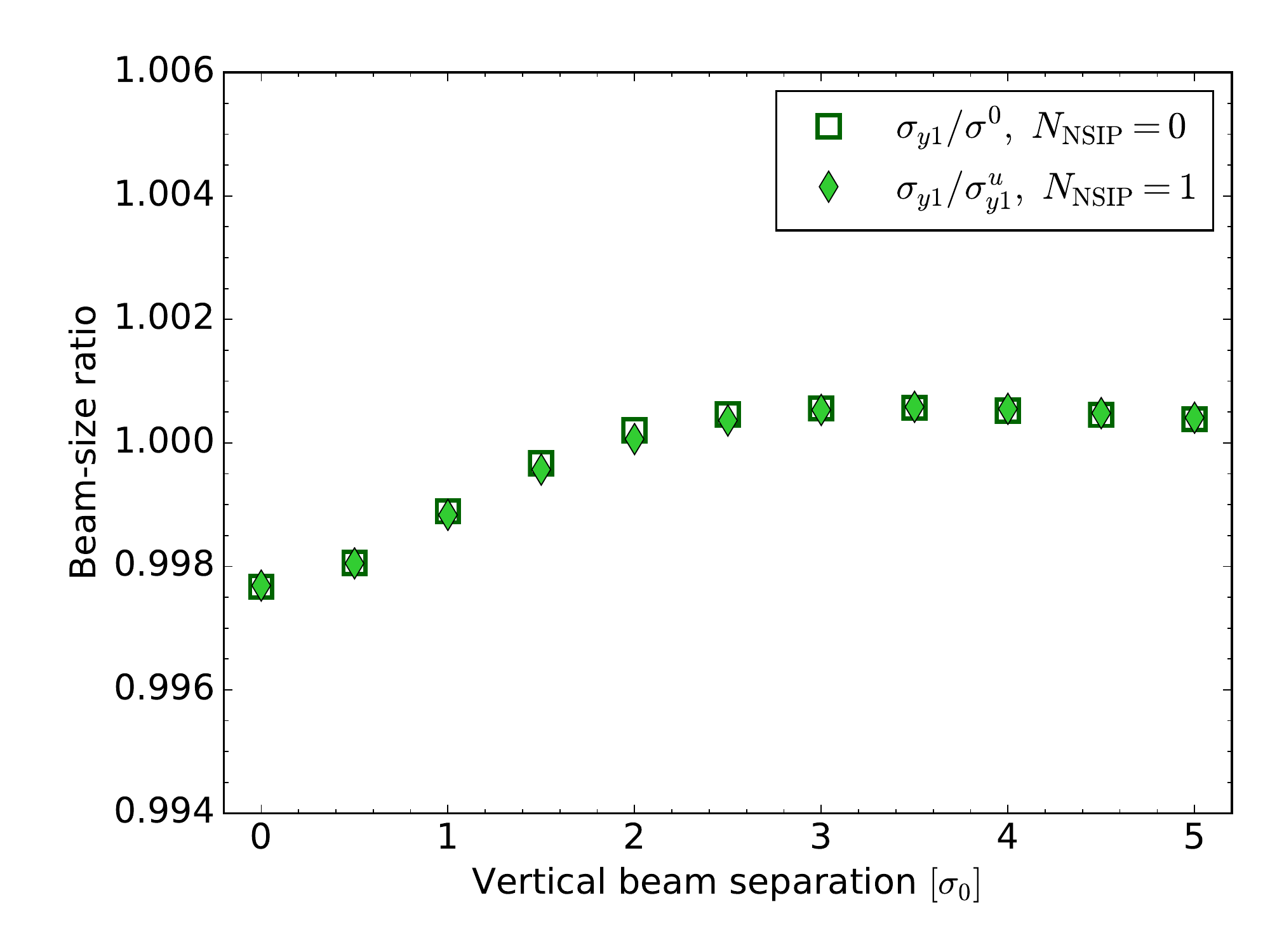}
    \caption{Beam-separation dependence, during a simulated vertical \vdM\ scan, of the beam--size-bias factors at the scanning IP associated with the full beam-beam effect for $\NSIP = 0$ using the \Lumz normalization ($\sigma_{y1}/\sigma^0$, open squares), and for $\NSIP = 1$ and using the \Lumu normalization ($\sigma_{y1}/\sigma^u_{y1}$, filled green lozenges). For the sake of clarity, only beam-1 results are displayed; the beam-2 points would be indistinguishable. The  horizontal axis is the beam separation in units of the nominal transverse beam size $\sigma_0$. The span of the vertical scale is identical to that of Fig.~\ref{fig:sYmu-o-s0_1wIP}.}
    \label{fig:sYm-o-sYu0_1n0wIP}
\end{figure}

This same luminosity-bias factor that is presented in Fig.~\ref{fig:Lmu-o-L0_1wIP}  for $\NSIP = 1$, is compared in Fig.~\ref{fig:Lm-o-Lu0_1n0wIP} to the equivalent quantity for $\NSIP = 0$.  Up to a small systematic difference that represents the impact
of the additional head-on collisions that occur at the non-scanning IP,  the two curves are essentially identical. This observation suggests that the beam-separation dependence of the beam--beam-induced orbit-shift and optical-distortion effects is encapsulated in that of \LoLuF, even in the presence of collisions at the non-scanning IP(s). This interpretation is confirmed by comparing the separation dependence of the actual single-beam transverse RMS sizes $\sigma_{iB}$ ($i = x,  y$; $B = 1,2$)  with, and without, additional collisions at the non-scanning IP. When $\NSIP = 1$ (Fig.~\ref{fig:sYmu-o-s0_1wIP}),  $\sigma_{y1} > \sigma_{y2} $; these two quantities exhibit  a very similar beam-separation dependence with respect to the corresponding unperturbed beam size $\sigma^u_{yB}$. A direct comparison of the $\sigma_{yB}/\sigma^u_{yB}$ ratio with and without a non-scanning IP (Fig.~\ref{fig:sYm-o-sYu0_1n0wIP}) shows that the two configurations result in an almost identical beam-separation dependence, the shape of which closely parallels that computed~\cite{Herr} in the linear approximation using MADX, once the qualitative impact of amplitude detuning is taken into account.
%

\par
\begin{figure}
    \centering
    \includegraphics[width=0.48\textwidth]{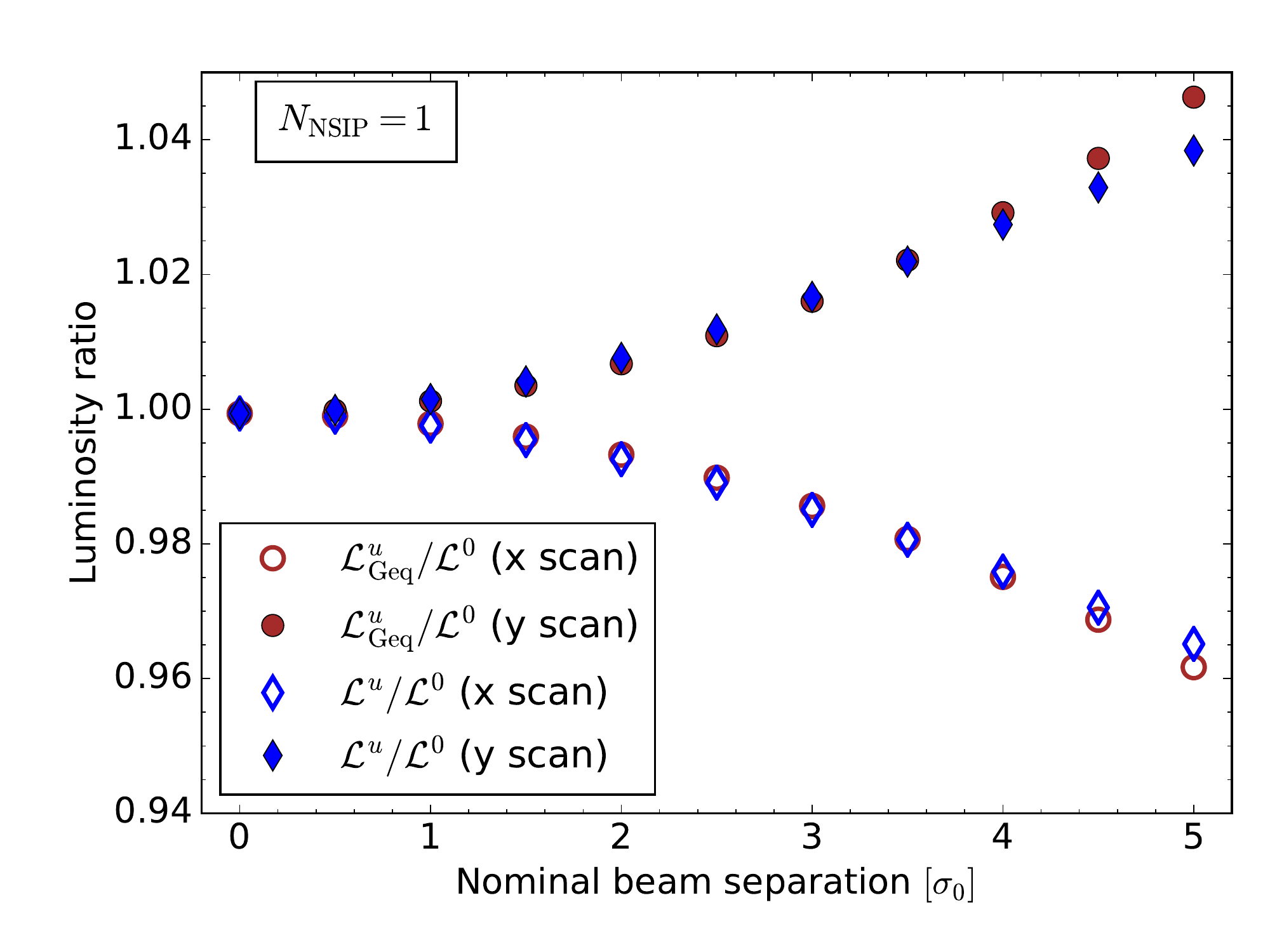}
    \caption{Beam-separation dependence, during \vdM\ scans simulated in the Run-2 configuration with one non-scanning IP, of the luminosity-bias factors at the scanning IP associated with the unperturbed luminosity ($\Lumu/\Lumz$, lozenges), and with the Gaussian-equivalent unperturbed luminosity ($\LuGeq / \Lumz$, circles). The  horizontal axis is the beam separation in units of the nominal transverse beam size $\sigma_0$. The filled (open) symbols corresponds to a vertical (horizontal) scan.
}
    \label{fig:LGeqVsLu_xNy}
\end{figure}

The contrast in beam-separation dependence between the nominal (\Lumz) and the unperturbed (\Lumu) luminosity-bias curves apparent in Fig.~\ref{fig:Lmu-o-L0_1wIP}, and the ``universality'' of the \LoLu luminosity-bias curves (Fig.~\ref{fig:Lm-o-Lu0_1n0wIP}), can be further characterized by comparing the separation dependence of the unperturbed luminosity \Lumu, computed numerically as the overlap integral over the unperturbed macroparticle distributions, with that of its Gaussian-equivalent counterpart \LuGeq. The latter is defined as\footnote{Some of the constants appearing in the complete formulas (Eqs.~(\ref{eqn:lumifin})--(\ref{eqn:sigmaVis})) are irrelevant here, and omitted for the sake of clarity.}:
\begin{equation}
\LuGeq(\delta_x) =  \frac{1}{2\pi \CSxuGeq  \CSxuGeq}  
		\exp\left[ - \frac{1}{2} \left( \frac{\delta_x}{ \CSxuGeq} \right)^2 \right]\, ,
		\nonumber
\end{equation}
and similarly for $\LuGeq(\delta_y)$. It is calculated under the assumption that the unperturbed macroparticle distributions are Gaussian, and adequately described (for this particular purpose) by the transverse RMS single-beam sizes $\sigma^u_{iB}$ ($i = x,y$; $B = 1,2$):
\begin{equation}
\Sigma^u_{i\mathrm{,Geq}}  = \sqrt{(\sigma^u_{i1})^2 + (\sigma^u_{i2})^2}\, .
\nonumber
\end{equation}
The comparison between \Lumu and \LuGeq is presented in Fig.~\ref{fig:LGeqVsLu_xNy}. During the vertical scan (filled symbols), the unperturbed luminosity (already shown in Fig.~\ref{fig:Lmu-o-L0_1wIP}) and its Gaussian-equivalent counterpart are almost exactly equal, except at large separation ($\delta_y/\sigma^0 > 4$); both increase with separation relative to \Lumz. The same occurs during the horizontal scan (open symbols), except that here both curves drop with increasing separation, because the horizontal phase advances are such that the dynamic-$\beta$ effect at IP5 translates into smaller horizontal beam sizes at IP1. As already documented~\cite{Herr} in the linear approximation, the dynamic-$\beta$ effect at the non-scanning IP(s) propagates to the scanning IP, resulting in either larger or smaller unperturbed transverse beam-sizes depending on the phase advance. 

\par
The fundamental observable of the van der Meer method is the {\em beam-separation dependence} of the measured interaction rate. Intuitively therefore, one can argue that a separation-independent change in unperturbed transverse beam size at the scanning IP (Fig.~\ref{fig:sYmu-o-s0_1wIP}, open symbols) should not bias the \svis measurement, provided the shift in unperturbed tunes is properly accounted for. The same reasoning is offered in Ref.~\cite{Herr}, where the argument is more straightforward: in the linear approximation (Sec.~\ref{subsec:MADXdescr}), additional collisions are tantamount to inserting a quadrupole-like perturbation at the non-scanning IP(s), that in turn results in a small tune shift accompanied by a $\beta$-beating wave around each ring. In the more general case treated here, the excellent agreement between \Lumu and \LuGeq  over most of the scanning range (Fig.~\ref{fig:LGeqVsLu_xNy})  indicates that the dominant effect is that of a beam- and plane-dependent shift of the effective \bst value, and of the corresponding unperturbed RMS single-beam sizes. Non-Gaussian tails associated with the non-linearity of the beam-beam force at the non-scanning IP(s) have but a very minor impact on the unperturbed transverse-density distributions at the scanning IP.

\par
As a first illustration of the magnitude of multi-IP effects and of the quantitative impact, on beam-beam corrections, of the difference between the green circles and the red squares in Fig.~\ref{fig:Lmu-o-L0_1wIP}, Table~\ref{tab:FoM_LzVu} compares the separation-integrated FoMs (Sec.~\ref{subsubsec:FoMs}) extracted from a simulated \vdM-scan pair at IP1, with and without additional head-on collisions at IP5. 
\begin{itemize}
\item
In the \Lumu normalization, and as expected from Fig.~\ref{fig:Lm-o-Lu0_1n0wIP}, the hierarchy of the FoMs is very similar with ($\NSIP = 1$) and without ($\NSIP = 0$) collisions at the non-scanning IP. The beam-beam bias is  systematically more negative for $\NSIP = 1$: this is qualitatively consistent with the combination of a decrease of the mean unperturbed tunes (Sec.~\ref{subsubsec:mIPtunes}), and of  the tune-dependence of the luminosity-bias curves (Fig.~\ref{fig:FullBB-ll0-Q-dependence}). As a result, the magnitude of the overall beam-beam bias on \svis (row 7, columns 3 and 4) increases substantially in the presence of a non-scanning IP.
\item
For $\NSIP = 1$, the hierarchy of the FoMs is very different in the \Lumz and \Lumu normalizations, because the dynamic-$\beta$ effect at the non-scanning IP has a significant impact on the transverse unperturbed beam sizes at the scanning IP (Fig.~\ref{fig:sYmu-o-s0_1wIP}). The fact that the \CSx (\CSy) bias is more negative (positive) in the \Lumz normalization reflects the fact that the unperturbed horizontal (vertical) beam size at the scanning IP is smaller (larger) than its nominal value (Fig.~\ref{fig:LGeqVsLu_xNy}).
\item
In contrast, the \svis-bias values for $\NSIP = 1$ agree to better than 0.005\% between the \Lumz and \Lumu normalizations, indicating that in practice, the two normalizations yield equivalent beam-beam corrections. This can be understood by examining the bottom half of Table~\ref{tab:FoM_LzVu}, that summarizes the impact of collisions at IP5 on the unperturbed scan variables at IP1: $\Sigma^u_x$ decreases, $\Sigma^u_y$ increases by a little more, resulting in a small decrease of the unperturbed peak interaction rate $ \mu^u_{pk}$; the net change in \svis is at the 0.004\% level. Physically, this expresses the fact that the changes in the unperturbed transverse convolved beam sizes and the unperturbed head-on luminosity are correlated such that the visible-cross-section 
remains invariant,
 at least in the configuration simulated here.
\end{itemize}

\begin{table*} [htb]
\centering
\begin{tabular}{|c|cc|c|} 
\hline
										& \multicolumn{2}{c|}{$\NSIP =1$}						& $\NSIP =0$			\\
\hline
										& \multicolumn{3}{c|}{Normalization:}											\\
FoM										& \multicolumn{2}{c|}{\Lnobb =}							& \Lnobb = \Lumz  		\\
										&		\Lumz			&	\Lumu				& (\Lumu = \Lumz)		\\				
\hline \hline
										 \multicolumn{4}{|c|}{Beam-beam bias on measured scan variables [\%]}				\\
\hline
$ \mu_{pk} / \mu^{\mathrm no-bb}_{pk} - 1$		& 	 0.346	& 	 0.410	& 	 0.431		  		\\
$\Sigma_x / \Sigma^{\mathrm no-bb}_x - 1$		&	-0.784	&	-0.454	&	-0.405				\\
$\Sigma_y / \Sigma^{\mathrm no-bb}_y - 1$		&	0.076	&	-0.323	&	-0.239				\\
\hline
$\sigma_{vis} / \sigma^{\mathrm no-bb}_{vis} - 1$	&	-0.365	&	-0.369	&	-0.215				\\ 
\hline \hline
										 \multicolumn{4}{|c|}{Beam-beam bias on unperturbed scan variables [\%]}			\\
\hline
$ \mu^u_{pk} / \mu^{\mathrm no-bb}_{pk} -1 $		&	-0.064	&	-		&	0					\\
$\Sigma^u_x / \Sigma^{\mathrm no-bb}_x - 1$		&	-0.331	&	-		&	0					\\
$\Sigma^u_y / \Sigma^{\mathrm no-bb}_y - 1$		&	 0.401	&	-		&	0					\\
\hline
$\sigma^u_{vis} / \sigma^{\mathrm no-bb}_{vis} - 1$	&	0.004	&	-		&	0					\\
\hline
\end{tabular}
\caption{Fractional peak-rate, beam-size and \svis beam--beam-induced bias on measured (rows 3-7) and unperturbed (rows 8-12) scan variables, for simulated \vdM scans at IP1, with (columns 2-3) and without (column 4) head-on collisions at IP5, using the nominal bunch parameters shown in Table~\ref{tab:mIPparms} and the phase-advance settings listed in Table~\ref{tab:R2PhasAdv}. 
}
\label{tab:FoM_LzVu}
\end{table*}

	\subsubsection{Sensitivity of beam-beam corrections to the phase advance between IPs}
	\label{subsubsec:mIPambig}

The results presented in the preceding paragraphs indicate that even though the separation dependence of the luminosity-bias factors are very different in the \Lumz and \Lumu normalizations (Fig.~\ref{fig:Lmu-o-L0_1wIP}), these two reference configurations appear to yield almost identical beam-beam corrections to \svis (Table~\ref{tab:FoM_LzVu}). What remains to be quantified is how robust this conclusion is against large variations in the assumed phase advance between consecutive IPs, as well as in the presence of more than one non-scanning IP.

\par
The study reported in Sec.~\ref{subsubsec:LzVsLu} assumes the phase-advance values between IP1 and IP5 that are listed in the top two rows of Table~\ref{tab:R2PhasAdv}; these correspond to one point in four-dimensional phase-advance space ($\mu_x^{\mathrm{B}1}, \mu_y^{\mathrm{B}1}, \mu_x^{\mathrm{B}2}, \mu_y^{\mathrm{B}2}$). To address the first question above, the analysis detailed in Sec.~\ref{subsubsec:LzVsLu} was repeated while varying the horizontal and vertical phase advance between IP1 and IP5 over most of the $[0, 2\pi]$ range, while keeping the nominal tunes constant; in order to keep the parameter space manageable, the B1 and B2 phase-advance values were in addition constrained to be the same, separately in the horizontal and the vertical plane ($\mu_x^{\mathrm{B}1} = \mu_x^{\mathrm{B}2} = \mu_x$, $\mu_y^{\mathrm{B}1} = \mu_y^{\mathrm{B}2} = \mu_y$). \vdM scans were simulated at IP1 over a grid in ($\mu_x, \mu_y$) space, with the bunches colliding head-on at IP5, and using the bunch parameters listed in Table~\ref{tab:mIPparms}. 

\par
\begin{figure}
    \centering
    \includegraphics[width=0.48\textwidth]{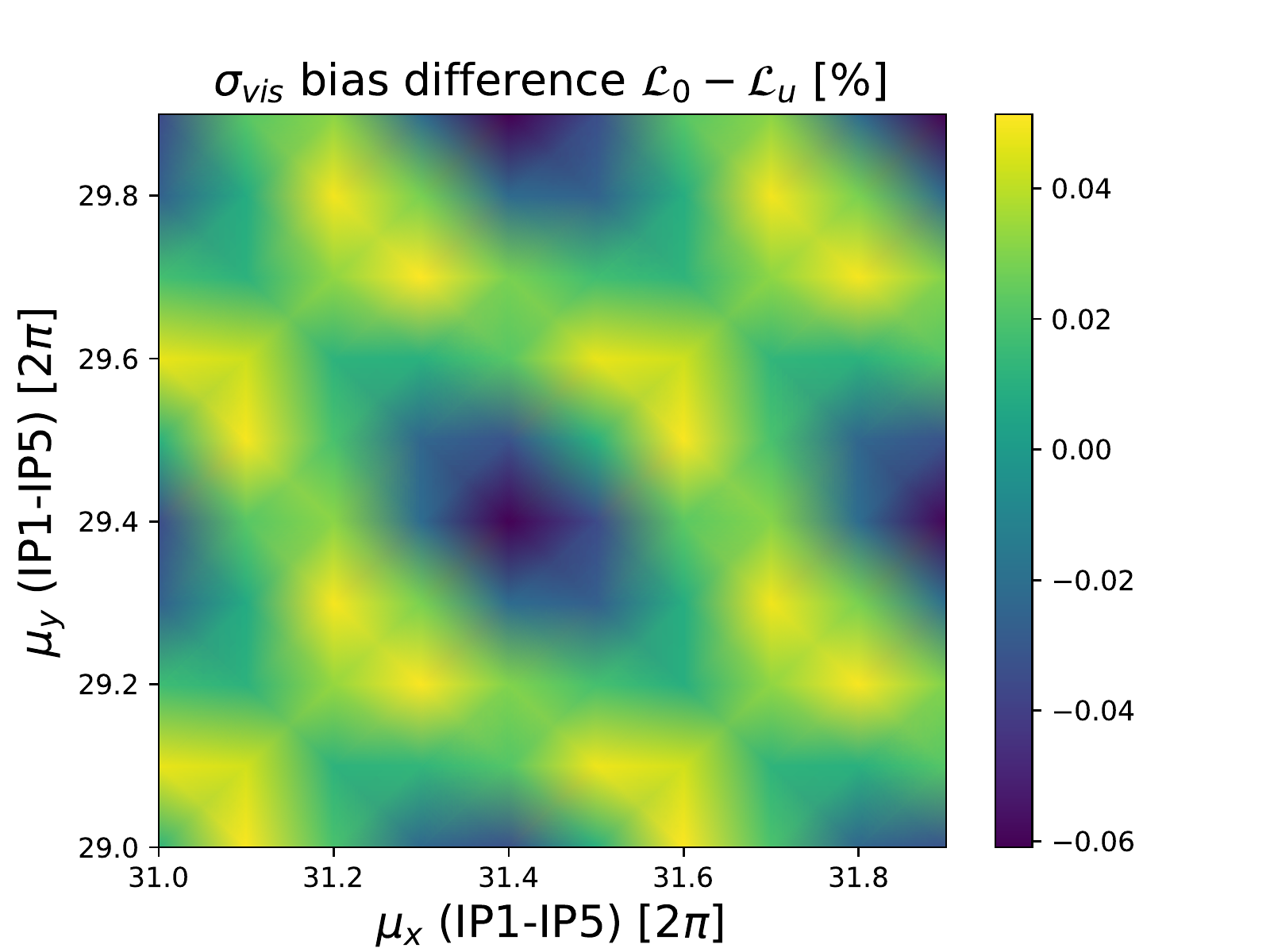}
    \caption{Phase-advance dependence of the difference, between the \Lumz and the \Lumu normalization, of the beam-beam induced \svis bias. The horizontal and vertical axes represent the betatron phase advance between IP1 and IP5, here assumed to be the same for B1 and B2. The color scale quantifies the fractional inconsistency in \svis bias between the two reference configurations; it ranges from $-0.06$ to $+ 0.05$\% of \svis.
}
    \label{fig:LsvBias_Lu-Lz_1wIP}
\end{figure}

At each point in the ($\mu_x, \mu_y$) space, the luminosity-bias curves and the corresponding separation-integrated FoMs were computed in both the \Lumz and the \Lumu normalization. The beam--beam-induced \svis bias, or equivalently the magnitude of the overall beam-beam correction to the \vdM-based absolute luminosity scale, was found to depend slightly on the choice of the reference ``no beam-beam'' configuration, \ie on whether the beam-beam correction is calculated in the \Lumz or in the \Lumu normalization. The difference is phase-advance dependent, and exhibits a clear periodic structure (Fig.~\ref{fig:LsvBias_Lu-Lz_1wIP}); periodic structures are also observed in such two-dimensional maps of the luminosity-bias factor (not shown), with different patterns for  $\LoLz$ and $\LoLu$.

\par
These periodicities have been verified to arise neither from statistical fluctuations, nor from numerical effects in the simulation.\footnote{It is interesting to note that the interplay between beam-beam effects at different IPs, and its sensitivity to the phase advance between IPs, was observed at LEP-II already, where they could strongly affect the instantaneous-luminosity balance between the four experimental detectors~\cite{bib:LEPmIP}.} They are attributed to cross-talk between collision points: beam-beam kicks at one IP induce separation- and amplitude-dependent $\beta$-beating that propagates in a phase-dependent manner to the next IP, where it slightly modifies the transverse-density distributions and the particle-averaged $\beta$ functions; the tunes are affected as well. The superposition of these effects across multiple IPs, in turn, modulates the associated beam--beam-induced biases. This interpretation is supported by the fact that the detailed shape of the simulated tune spectra, that of the transverse luminosity profiles at the witness IP, as well as that of the beam-separation dependence of the $\Lum/\Lumu$ ratio at the scanning IP, are all observed to depend on the phase advance between IPs. A deeper understanding of these mechanisms may provide the key for resolving the ambiguity between the \Lumz and  \Lumu normalizations; as such, they  remain under active study.

\par 
Fortunately, these subtle effects have only a minor impact on beam-beam corrections to \vdM calibrations. The largest amplitude of the \Lumz-\Lumu discrepancy over the entire plane amounts  to 0.06\% of \svis, which is smaller than, or at most comparable to, the absolute accuracy of the B*B or COMBI simulations. The reason why in Table~\ref{tab:FoM_LzVu}  the \Lumz and \Lumu normalizations yield essentially identical results, is that the Run-2 phase-advance between IP1 and IP5 happens to lie in a region of the  $\mu_x - \mu_y$ plane where the inconsistency is the smallest. In this case therefore, and for bunches that collide only in these two IPs, the difference in \svis bias between the \Lumz and \Lumu normalizations is negligible.

\par
To assess the impact of more than one non-scanning IP, the above phase scans were repeated for collision patterns with up to three non-scanning IPs. To deal with the increased dimensionality of the problem, the fractional phase advance from IP1 to IP5 was fixed to $\Delta\mu_x = \Delta\mu_y = 0.405$, a choice that in Fig.~\ref{fig:LsvBias_Lu-Lz_1wIP}  corresponds to the largest absolute difference in \svis bias between the \Lumu  and \Lumz normalizations. With that constraint, and with the overall tunes and the IP5-IP8 phase advances fixed at their nominal values, the horizontal and vertical fractional phase-advance values from IP1 to IP2 were scanned over the range $[0, 2\pi]$. The study was then repeated with the roles of IP2 and IP8 interchanged. In all these configurations, the difference in \svis bias between the \Lumu and \Lumz normalizations displays the same periodic pattern as that in Fig.~\ref{fig:LsvBias_Lu-Lz_1wIP}, and the largest absolute difference in \svis bias between the two reference configurations does not exceed 0.15\% (for $\xi = 0.006$). While strictly speaking, the ensemble of these scans does not constitute an exhaustive sampling of the 12-dimensional phase-advance space (2 beams $ \times$ 2 planes $\times$ 3 IP combinations), it suggests that the largest possible disagreement between the  \Lumz and \Lumu normalizations amounts to 0.1-0.2\% of \svis when considering the full parameter space. Since the phase advance between LHC IPs can be measured to a few degrees ($\sim 0.015 \times 2\pi$) per beam and per plane, the ``operating point'' in phase-advance space is known to very good precision. In some cases therefore, the ambiguity on the \svis bias associated with the choice of reference configuration can be significantly reduced, as already demonstrated for bunches that collide in IP1 and IP5 only.  

\par

	\subsubsection{Parameterization of multi-IP effects}
	\label{subsubsec:mIPparam}

	To gain further insight into multi-IP effects and account, in  a pragmatic fashion, for their impact on beam-beam corrections to luminosity calibrations, COMBI was used to simulate scans over the full range of beam-beam parameter values typical of \vdM sessions at $\sqrt{s} = 13$\,TeV, using the Run-2 phase-advance settings (Table~\ref{tab:R2PhasAdv}), both without  ($\NSIP =0 $) and with ($1 \le \NSIP \le 3$) head-on collisions at one or more non-scanning IP(s).

\par
\begin{figure}[!htbp]
	\centering
	\includegraphics[width=0.47\textwidth]{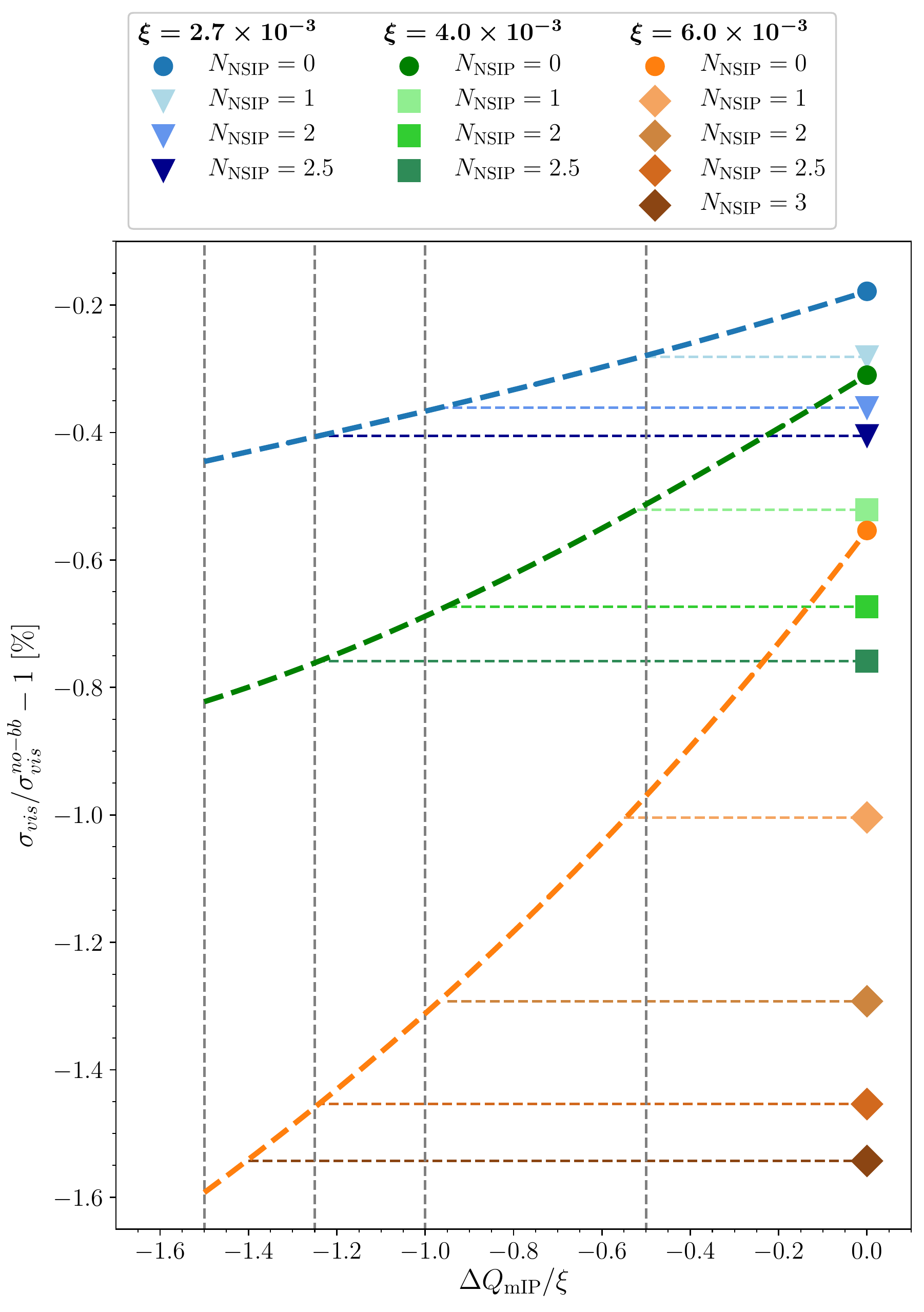}
	\caption{Horizontal lines, and markers at $\DQmIP = 0$: visible cross-section bias in the \Lumu normalization, associated with \vdM scans at IP1, for bunch pairs that in addition collide head-on at a fixed number of non-scanning IPs, and with the tunes set to their nominal value. The three sampled values of the single-IP beam-beam parameter $\xi$ are identified by the blue triangles ($\xi = 2.7\times 10^{-3}$), green squares ($\xi = 4.0\times 10^{-3}$) and brown lozenges ($\xi = 6.0\times 10^{-3}$). For a given value of $\xi$, the lighter, medium and darker colors correspond to additional collisions at $\NSIP = 1$, 2, 2.5 and 3 non-scanning IPs respectively (see text). Inclined curves and circular markers: visible cross-section bias associated with \vdM scans at IP1 with beams fully separated at all other IPs ($\NSIP = 0$), as a function of the multi-IP equivalent mean tune shift \DQmIP expressed in units of $\xi$.
The circular markers at $\DQmIP/\xi = 0$ show the value of the \svis bias corresponding to beams colliding only at the scanning IP, with the unperturbed tunes set to their nominal value. The vertical dashed lines correspond, from right to left,  to  $\DQmIP/\xi = -0.5$, $-1.0$, $-1.25$ and $-1.5$, and approximately correspond to the values of the effective tune shift \DQmIP where an inclined curve associated with a given value of $\xi$ intersects the corresponding horizontal lines.
}
	\label{fig:relDiff_multiIP}
	\end{figure}

	
	The results are summarized in Fig.~\ref{fig:relDiff_multiIP} for three representative values of the single-IP beam-beam parameter $\xi$, using the \Lumu normalization. Since both the orbit-shift and the optical-distortion bias are sensitive to the unperturbed-tune values (Sec.~\ref{subsubsec:xiNTuneDep}), the overall beam-beam bias on the visible cross-section associated with a given value of $\xi$ is significantly affected by collisions at non-scanning IPs. Taking as an example the case of $\xi = 6\times 10^{-3}$, the \svis bias when colliding only at the scanning IP amounts to $-0.55$\% (orange circle). It grows to $-1.0$\% for one non-scanning IP (pale-brown lozenge), to $-1.3$\% for two non-scanning IPs (medium-brown lozenge), and to a little over $-1.5$\% for the worst case of $\NSIP = 3$ (darkest-brown lozenge).\footnote{Half-integer values such as $\NSIP = 2.5$ correspond to a bunch pair that collides at IP1 and in which one of the bunches experiences three additional collisions (IP5, plus both IP2 and IP8), while the other one is involved in only two (IP5, plus either IP2 or IP8).}
As \NSIP increases by one unit, both the horizontal and the vertical mean unperturbed tune decrease by approximately $\xi/2$; as a consequence, the luminosity-bias curves shift downwards (Figs.~\ref{fig:Opt-ll0-Q-dependence}, \ref{fig:FullBB-ll0-Q-dependence}) and the \svis bias becomes accordingly more negative. The impact of this ``multi-IP tune shift'' is comparable in magnitude to the \svis bias for $\NSIP = 0$; it varies from one bunch pair to another in a manner that depends on the single-bunch parameters and on the value of \NSIP associated with each-colliding bunch pair at the IP where the scans are taking place. Fortunately however, this effect lends itself to a simple 
parameterization.

\par
	The diagonal curves in Fig.~\ref{fig:relDiff_multiIP} display the visible cross-section bias for scans simulated at IP1 in the absence of collisions at any non-scanning IP, but with both the horizontal and the vertical unperturbed tune shifted by a quantity \DQmIP that scales with $\xi$. For a given value of $\xi$, the intersections of the corresponding diagonal curve (orange in the example above) with the horizontal lines of the same hue corresponding to $\NSIP = 1, 2,...$, determine the magnitude of the multi-IP equivalent tune shift \DQmIP that needs to be applied simultaneously to the unperturbed horizontal and vertical tunes input to the COMBI simulation to mimic the effect of collisions at non-scanning IPs. The corresponding values of $\DQmIP / \xi$ are indicated by the vertical dashed lines in the figure.
 
\par
\begin{figure}
    \centering
    \includegraphics[width=0.47\textwidth]{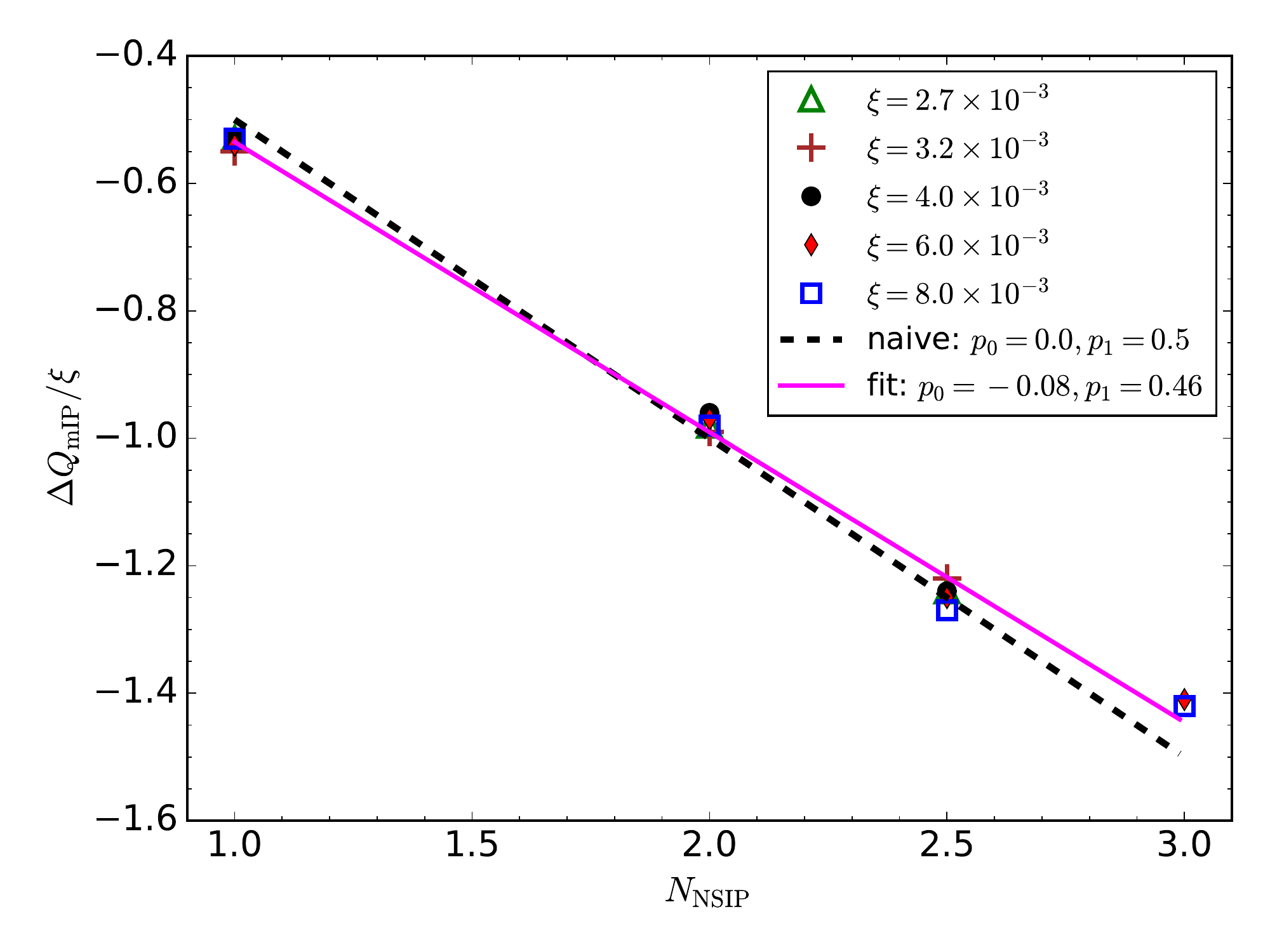}
    \caption{Dependence of the normalized  multi-IP equivalent tune shift  $\DQmIP / \xi$ on the number of non-scanning IPs, for several values of the single-IP beam-beam parameter $\xi$. The scanning IP is set to IP1, and the \svis bias is calculated using the \Lumu normalization. The solid magenta line represents a two-parameter linear fit to the points; the black dashed line represents a naive parameterization of the form  $\DQmIP / \xi = -0.5 \times \NSIP$.
}
    \label{fig:mIP-xiScaling}
\end{figure}

A more telling representation of the same results is offered in Fig.~\ref{fig:mIP-xiScaling}, which displays the  multi-IP equivalent tune shift as a function of the number of non-scanning IPs. The results are consistent with a linear dependence that differs only slightly from the naive scaling law
\begin{equation}
    \label{eqn:multIPShift}
        \DQmIP  = -0.5 \, \xi  \NSIP = -p_1 \, \xi  \NSIP
\end{equation}
inspired by the qualitative evolution of the tune spectra. The purely empirical, two-parameter linear fit represented by the purple solid line:
\begin{equation}
       \DQmIP = p_0 - p_1 \,  \xi \, \NSIP
\nonumber
\end{equation}
provides a slightly better parameterization than Eq.~(\ref{eqn:multIPShift}), and appears accurate at the level of $\DQmIP  \sim \pm 0.05\, \xi$.

\par
\begin{figure}
    \centering
    \includegraphics[width=0.48\textwidth]{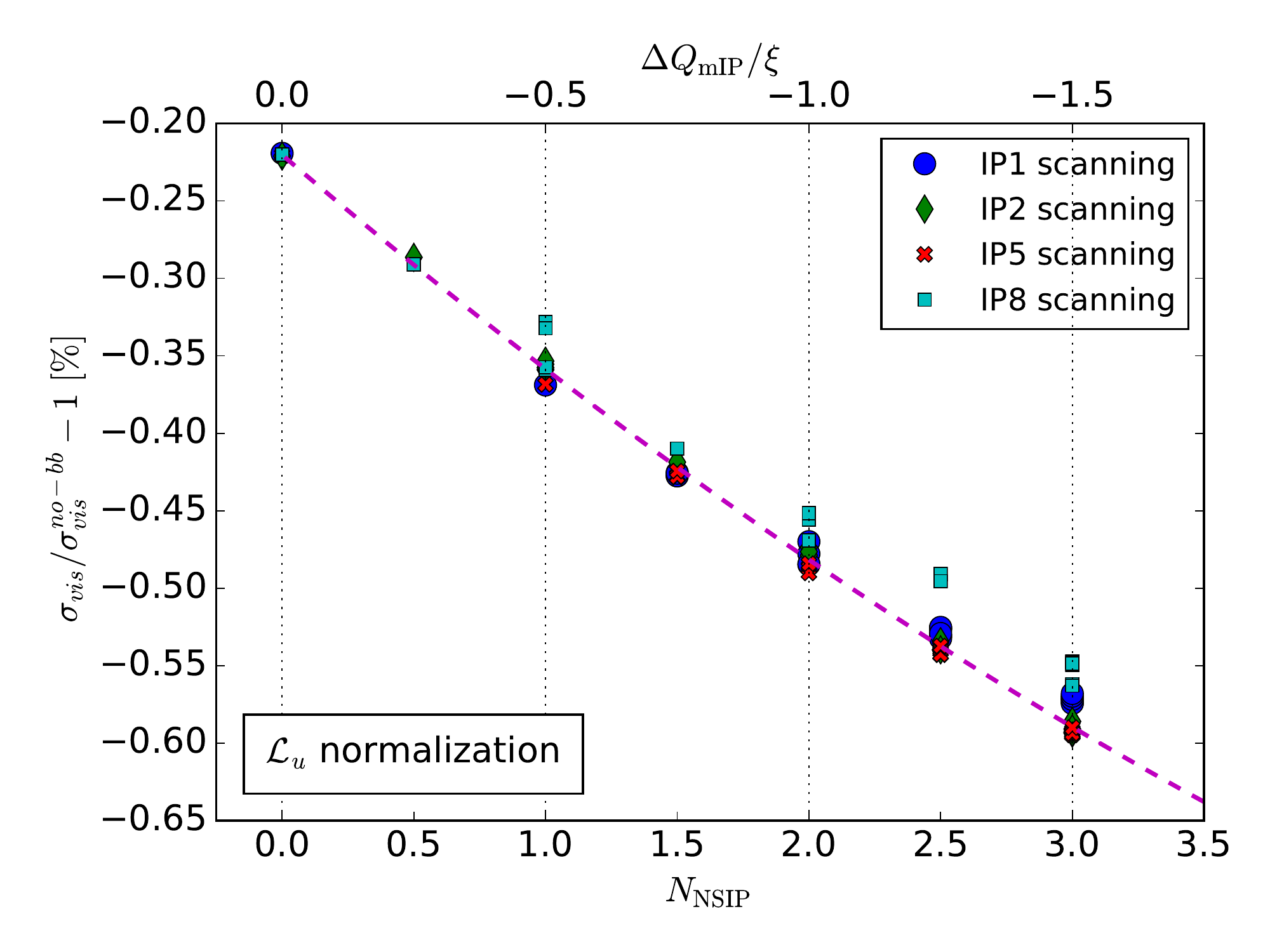}
    \caption{Visible cross-section bias in the \Lumu normalization as a function of the number of non-scanning IPs (lower horizontal axis), for the four possible choices of scanning IP. The curve represents the dependence of the \svis bias in the $\NSIP = 0$ case, but with the horizontal and vertical unperturbed tunes shifted downwards by $\DQmIP / \xi = -0.5 \times \NSIP$ (upper horizontal axis). For a given scanning IP, there exist several B1-B2 bunch-collision patterns that yield the same value of \NSIP. The beam-beam parameter and phase-advance values are listed in Tables ~\ref{tab:mIPparms} and \ref{tab:R2PhasAdv} respectively. 
 }
    \label{fig:sIPScaling_Lu}
\end{figure}

Figure~\ref{fig:mIP-xiScaling} demonstrates that for a given scanning IP and  for all values of $\xi$, the normalized multi-IP equivalent tune shift scales linearly with the number of non-scanning IPs. To what extent this scaling is ``universal'', \ie how strongly it depends on the choice of the IP where the scans are taking place, is addressed by Fig.~\ref{fig:sIPScaling_Lu}. Irrespective of the scanning IP considered, the overall dependence of the \svis bias on the number of scanning IPs is reasonably well represented by the naive scaling law of Eq.~(\ref{eqn:multIPShift}), demonstrating that the dominant effect is the downward shift of the mean unperturbed tunes. There are, however, significant differences between marker families, most notably between the red crosses (IP5) and the cyan squares (IP8): these reflect the variety of phase-advance patterns between the scanning IP and the other collision points. Similarly, within a given family, there exist several bunch pairings between B1 and B2 that yield the same number of additional collisions, but that correspond to different phase-advance configurations. When IP5 is the scanning IP, for instance, the configuration $\NSIP = 2$ can be achieved with additional collisions at IP1 and at either IP2 or IP8; these two cases can therefore yield slightly different \svis biases. To minimize modeling errors, therefore, it is advisable that scaling parameterizations such as that illustrated by Fig.~\ref{fig:mIP-xiScaling} be developed for each scanning IP separately,  preferably taking into account the actual collision pattern experienced by each bunch that collides at the scanning IP under consideration.
\begin{figure}
    \centering
    \includegraphics[width=0.48\textwidth]{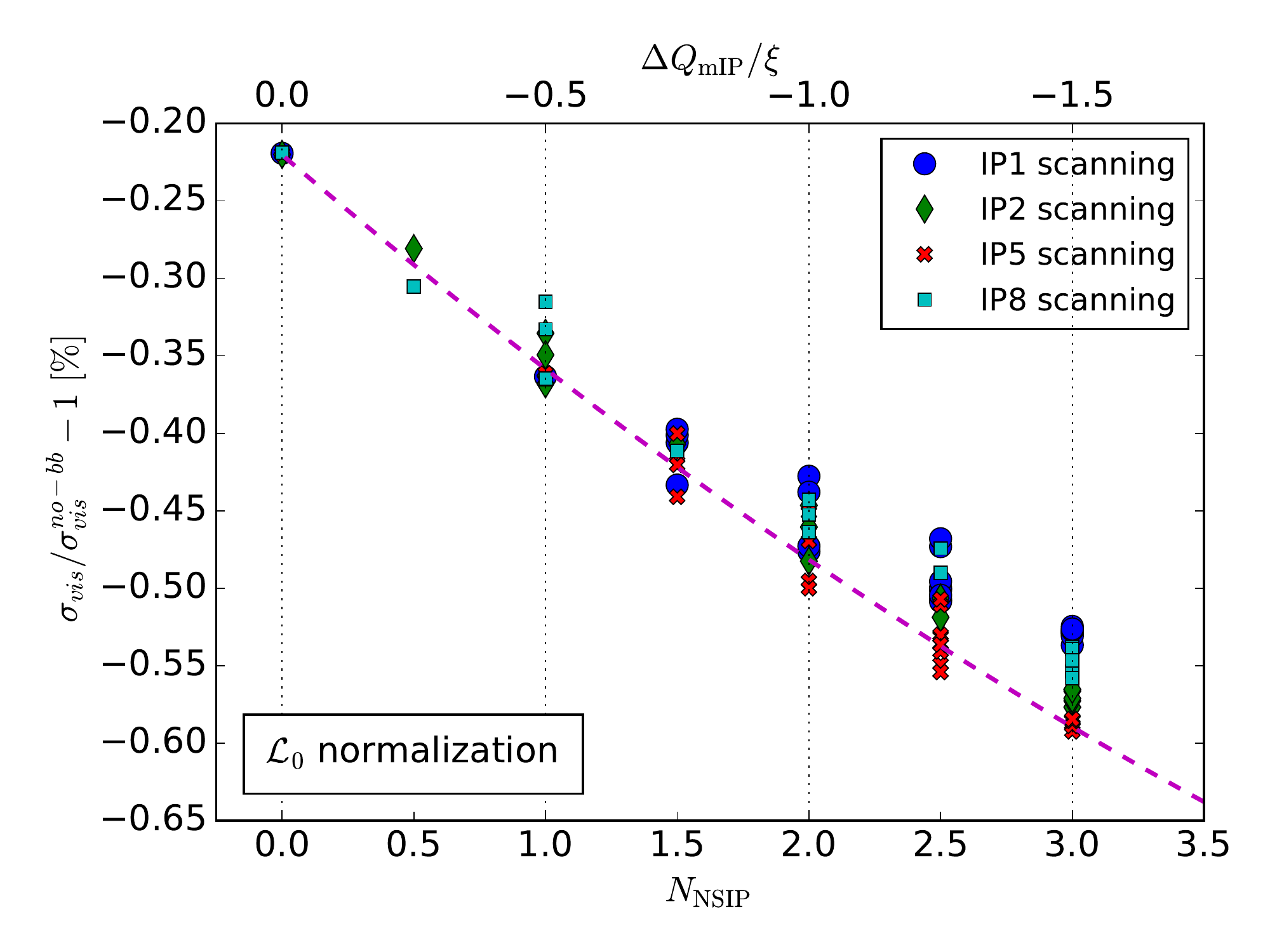}
    \caption{Visible cross-section bias in the \Lumz normalization as a function of the number of non-scanning IPs, for the four possible choices of scanning IP. The results are extracted from the same simulations as those in Fig.~\ref{fig:sIPScaling_Lu}; only the reference configuration differs. The curve, the axis scales and the legend are identical.
}
    \label{fig:sIPScaling_L0}
\end{figure} 
%


\par
Pursuing the same strategy in the \Lumz normalization meets with limited success (Fig.~\ref{fig:sIPScaling_L0}): for a given value of \NSIP, the spread across different collision patterns is noticeably larger than in the \Lumu normalization, even when restricted to one given scanning IP. This observation reflects the same underlying physics as the periodic structure in Fig.~\ref{fig:LsvBias_Lu-Lz_1wIP}; its implications are addressed below.

	\subsubsection{Beam-beam correction strategies in the presence of multi-IP collisions}
	\label{subsubsec:mIPSmry}

	In summary, collisions at non-scanning IPs manifest themselves in two main ways. The widening of the tune spectra and the resulting downward shift of the mean unperturbed tunes (Sec.~\ref{subsubsec:mIPtunes}) significantly increase the beam--beam-induced bias on the absolute luminosity scale (Fig.~\ref{fig:sIPScaling_Lu}), mainly because of the reduced amplitude of the optical-distortion bias (Sec.~\ref{subsubsec:xiNTuneDep}). This tune shift is accompanied by a beam- and plane-dependent $\beta$-beating wave around each ring~\cite{Herr}, that translates into a sub-percent rescaling of the unperturbed transverse single-beam sizes at the scanning IP (Fig.~\ref{fig:sYmu-o-s0_1wIP}). The latter causes significant distortions of the luminosity-bias curves when using the \Lumz normalization (Fig.~\ref{fig:Lmu-o-L0_1wIP}), that depend strongly on the phase advance between consecutive collision points.
	
\par
The shift in unperturbed tunes (Sec.~\ref{subsubsec:mIPtunes}) has, by far, the dominant impact on beam-beam corrections to \vdM calibrations. It can be accounted for by simulating collisions simultaneously at all relevant IPs, with the input tunes set to their nominal value, and by correcting the luminosity-scan curves to ``no beam-beam anywhere''. This \Lumz normalization is the approach adopted in Ref.~\cite{Balagura_2021}. The alternative is to use the \Lumu normalization. It is inspired by the fact that the fundamental observable of the \vdM method is the beam-separation dependence of the measured interaction rate. What should matter, therefore, is the variation of the beam-beam force {\em during the scan}: underlying, separation-independent modifications of the unperturbed transverse-density distributions are expected not to affect the visible cross-section. 
In the \Lumu approach, the luminosity-scan curves extracted from the same multi-IP simulation as above are corrected to ``no beam-beam at the scanning IP only'' (Sec.~\ref{subsubsec:LzVsLu}); the resulting luminosity-bias functions become much less sensitive to the presence of non-scanning IPs (Fig.~\ref{fig:Lm-o-Lu0_1n0wIP}). 

\par
The \Lumz and \Lumu normalizations yield beam-beam corrections that, depending on the phase advance between consecutive IPs, either are indistinguishable (Table~\ref{tab:FoM_LzVu}), or in the worst case differ by at most 0.1-0.2\% of \svis (Sec.~\ref{subsubsec:mIPambig}). These observations demonstrate that no phase-advance configuration exists in which beam-beam induced $\beta$-beating would significantly reduce the beam-beam bias on \svis.

\par
The results of the full multi-IP simulations can be very closely approximated by those of the much faster single-IP simulation, with as input the same value of the beam-beam parameter $\xi$, but with the unperturbed tunes shifted downwards by an amount \DQmIP that in the \Lumu normalization and for a given scanning IP, scales with $\xi$ and depends linearly on \NSIP (Fig.~\ref{fig:mIP-xiScaling}). The scaling parameters are not fully universal, in the sense that they depend slightly on the choice of scanning IP (Fig.~\ref{fig:sIPScaling_Lu}), or more fundamentally on the phase advance between the scanning and non-scanning IPs. 
%
%
%

\par
The scaling prescriptions offered in Sec.~\ref{subsubsec:mIPparam} in the \Lumu normalization account for collisions at non-scanning IPs to better than 0.1-0.2\% on \svis. It should be pointed out, however, that these simple-minded models reproduce only the \svis bias extracted from the multi-IP simulation (to an accuracy limited by the phase--advance-dependent inconsistency between the \Lumz and \Lumu normalizations), but that they are not meant to reproduce the corresponding tune spectra. An additional minor limitation is that they assume that collisions at non-scanning IPs occur with zero transverse separation, and with zero crossing angle. These {\it ad hoc} recipes, therefore, offer but a partial substitute for full-fledged simulations that account in detail for every distinct multi-IP collision pattern, separately for each member of every colliding-bunch pair and for each scanning IP. For many purposes, however, these simple scaling laws appear sufficiently accurate to cover most beam conditions encountered so far during \vdM sessions at the LHC. 

\par
\par
On the practical side, the \Lumu normalization offers several advantages. First, by reducing the multi-IP problem to that of a single collision point, it greatly simplifies the interpretation of simulation results and the estimate of systematic uncertainties. The \Lumu approach also automatically accounts for the static (in the sense of separation-independent) change in unperturbed beam size associated with the dynamic-$\beta$ effect at the non-scanning IPs. Finally, combining the parameterization of single-IP beam-beam biases offered in Sec.~\ref{subsubsec:CorImplement} with the scaling prescriptions of Sec.~\ref{subsubsec:mIPparam} provides a beam--beam-correction procedure that is straightforward to implement, requires no or little investment by individual users in complex, CPU-intensive simulations, and is in many cases sufficient to calculate absolute bunch-by-bunch corrections to \vdM calibrations, or to estimate systematic uncertainties. This is particularly valuable when dealing with five to ten \vdM-scan pairs that involve up to 150 colliding bunches with different populations and emittances, arranged in multiple collision patterns that each require a different multi-IP correction~\cite{bib:ATLR2Lum}: the computing overhead that would be associated with simulating each of those configurations in detail is considerable.

\par
At the present level of understanding, there appears to be no factual argument in favor of the \Lumz normalization. Even if it eventually turns out that correcting to ``no beam-beam anywhere'' provides a more accurate correction in principle, the potential reduction in systematic uncertainty remains moderate ($< 0.2$\%) when compared in quadrature to the overall luminosity-calibration uncertainty (around 1\%). In addition, and because of the larger phase-advance sensitivity of the \Lumz normalization, this hoped-for improvement could only be secured by simulating separately each and every collision pattern associated with the scanning IP under consideration. Finally, the cumulative impact of the dynamic-$\beta$ effect at the non-scanning IPs, that can shift (Sec.~\ref{subsubsec:LzVsLu}) the underlying unperturbed beam sizes $\sigma^u_{iB}$ at the scanning IP by $\mathcal{O}$(0.5-1\%) per plane, per beam and per non-scanning IP, totally blurs both the conceptual and the operational distinction between the orbit-shift and the optical-distortion correction. This is an additional reason why in the \Lumz normalization, a parameterization approach would be inapplicable: one must resort to a combined, ``full beam-beam'' multiplicative correction to the measured collision rate that can only be extracted from a dedicated simulation that has been tailored, separately for each colliding bunch, to the actual conditions of the \vdM-calibration session.

\par
In conclusion, clarifying the origin of periodic structures such as that in Fig.~\ref{fig:LsvBias_Lu-Lz_1wIP},  and explaining why beam-beam corrections are less sensitive to phase-advance effects in the \Lumu (Fig.~\ref{fig:sIPScaling_Lu})  than in the \Lumz   (Fig.~\ref{fig:sIPScaling_L0}) normalization, remain topics of active study. The ambiguity stands, therefore, as to which of the two reference ``no beam-beam'' configurations (\ie which of the \Lumz or the \Lumu normalization) provides the most accurate beam-beam correction in absolute terms. Pragmatically, the practical advantages of the \Lumu normalization, and the larger complexity in implementing the \Lumz normalization in the \vdM-data analysis, tilt the balance in favor of the former - at least for now. Until the \Lumz-\Lumu ambiguity is lifted, and irrespective of the choice of normalization, it seems prudent to cover the difference by assigning a phase--dependence-related systematic uncertainty to the absolute magnitude of the beam-beam corrections; an explicit prescription to this effect will be offered in Sec.~\ref{subsec:multIPSyst}.

	\subsection{Impact of beam-beam asymmetry}
	\label{subsec:ImbalImpact}

The results presented so far in this chapter, with the exception of those in Sec.~\ref{subsec:EllipImpact}, assume beams that are round in the transverse plane ($\beta^*_{xB} = \beta^*_{yB} =  \beta^*_B$ and $\sigma_{xB} = \sigma_{yB} = \sigma_B$, where $B = 1,2$). In such a case, the distinction between horizontal and vertical beam-beam parameters disappears, and the beam-beam parameter of beam 2, defined in Eq.~(\ref{eqn:xiGDef}), simplifies to
\begin{equation}
\label{eqn:xiAsymDef}
\xi_2 = \frac{n_1 \,  r_0 \, \beta^*_2}{4 \pi \, \gamma_2 \, \sigma_1^2} ,
\end{equation}	
where $\sigma_1$ is the transverse RMS size of beam 1. The same formula, {\it mutatis mutandis}, is used to define $\xi_1$. 

\par
The parameterization described in Sec.~\ref{subsec:corProcRefConf}, and the results derived therefrom, further assume that the beam-beam forces are exactly balanced between the two beams, \ie that $\xi_1 = \xi_2 = \xiR$ (Eq.~(\ref{eqn:xiRDef2})). Beam-beam asymmetry occurs when this assumption is violated, \ie when $\xi_1 \neq \xi_2$, which can arise under three different scenarios (Sec.~\ref{subsubsec:ImbalScen}). Even though a parametric approach can provide physically intuitive insight (Sec.~\ref{subsubsec:ImbalParam}), dedicated simulations are required to quantify the impact of such asymmetries on the accuracy of the beam-beam correction procedure (Sec.~\ref{subsubsec:BeamSizImbal}).

	\subsubsection{Beam-beam asymmetry scenarios}
	\label{subsubsec:ImbalScen}
	
Consider first the case of a bunch-population asymmetry ($n_1 \neq n_2$, with $\beta^*_1 = \beta^*_2$ and $\sigma_1 = \sigma_2$), which can be simulated in B*B by assigning different populations to the source beam and to the witness beam. Since the source beam in B*B remains unaffected by the beam-beam interaction, and since the two beams are subject to different beam-beam parameters, the simulation must be run twice, first with $n_1$ and $n_2$ assigned to, respectively, the source  and the witness beam, and then with the inverse assignment. The predicted luminosity-bias curves, and the corresponding corrections to the FoMs, are then given by adding the two sets of results\footnote{When B*B is run on a symmetric configuration, the scan is simulated only once, and the luminosity-bias result is simply doubled to account for the conceptual swap between the source and the witness beam.},
implicitly assuming that strong-strong effects do not play a significant role even in the presence of a large bunch-population asymmetry (at least in the \vdM regime). This procedure was validated by confronting the B*B results with those from COMBI for several asymmetric configurations: even in the worst case considered ($n_2/n_1 = 2$, $\xi_1 = 5.30 \times 10^{-3}$, $\xi_2 = 2.65 \times 10^{-3}$), the beam-beam corrections calculated by the two packages agreed to better than one part in 1600.

\par
The B*B and COMBI results above are accurately reproduced by the parameterization detailed in Sec.~\ref{subsubsec:CorImplement}, with as input the beam-averaged beam-beam parameter:
\begin{equation}
 \xiAv = \frac{\xi_1 + \xi_2}{2}~ . 
 \label{eqn:xiAvDef}
 \end{equation}
This is a consequence of the fact that the $\xi$-dependence of the FoMs is close enough to linear (Figs.\,\ref{fig:FoM-mu}--\ref{fig:FoM-sigvis}) for the averaging of the FoM values at two different $\xi$ settings to be tantamount to computing the FoM once, with \xiAv as input.

\par
The second scenario is that of a $\beta$-function asymmetry ($\beta^*_1 \neq\beta^*_2$, with $n_1 =n_2$ and $\sigma_1 = \sigma_2$). This case is equivalent to that of the bunch-population asymmetry above, since only the product $n_k \beta^*_l$ ($k, l = 1, 2$ with $l \neq k$) plays a role in Eq.~(\ref{eqn:xiAsymDef}). The bunch population $n_k$ of beam $k$ controls the scale of the beam-beam force beam $k$ exerts on beam $l$, while $\beta^*_l$ determines how sensitive beam $l$ is to this force.

\par
In contrast to the above two scenarios, which involve only scale effects, a beam-size or emittance asymmetry ($\sigma_1 \neq \sigma_2$, with $n_1 =n_2 = n$ and $\beta^*_1 = \beta^*_2 = \bst$) modifies in different ways the spatial distributions of the source field and of the particles in the witness bunch, the trajectories of which are affected by this field.

	\subsubsection{Parametric characterization of beam-beam asymmetries}
	\label{subsubsec:ImbalParam}

	A full characterization of beam-beam asymmetries requires the two parameters $\xi_1$ and $\xi_2$.  A physically intuitive parameterization can however  be provided by a single, ``effective'' beam-beam parameter \xibar, defined as
 	\begin{equation}
	\xibar =  \frac{\xi_1 \sigma_1^2  +  \xi_2 \sigma_2^2}{\sigma_1^2+\sigma_2^2} ~,
	\label{eqn:xibarDef}
	\end{equation}	
on the basis of the following argument.

\par
Under the naive assumption that the optical distortion of B1 induced by the field of B2 is proportional to $\xi_1$, and that it can be adequately represented by a small fractional change $\lambda_1$ in the value of $\sigma_1$:
	\begin{equation}
	\sigma_1 \rightarrow \sigma_1 (1+\lambda_1)\, ,
	\nonumber
	\end{equation}
where $\lambda_1 \propto \xi_1$, one can show, using Eqs.~(\ref{eqn:lumifin}) and (\ref{eqn:capSigDef}) and after a bit of algebra, that to first order in $\xi_1$,  the luminosity-bias factor $[\Lum / \Lumz]_{Opt, 1}$ associated with the optical distortion of B1 alone obeys
 	\begin{equation}
	[\Lum / \Lumz]_{Opt, 1}  - 1    \propto \frac{\lambda_1 \sigma_1^2}{\sigma_1^2+\sigma_2^2}   \propto \frac{\xi_1 \sigma_1^2}{\sigma_1^2+\sigma_2^2}~.
	\nonumber
	\end{equation}
Applying the same argument to the optical distortion of B2 induced by the field of B1, and combining the impact of the optical distortions of the two beams, it follows that
 	\begin{equation}
	[\Lum / \Lumz]_{Opt}  - 1  \propto \frac{\xi_1 \sigma_1^2  + \xi_2 \sigma_2^2}{\sigma_1^2+\sigma_2^2} = \xibar~ .
		\nonumber
	\end{equation}
	
\par
	In the limit of equal beam sizes ($\sigma_1=\sigma_2$),  \xibar reduces to \xiAv, 
and as such properly characterizes bunch-population or $\beta^*$ asymmetries. If instead the bunch populations are balanced ($n_1 = n_2 = n$) but the transverse beam sizes differ ($\sigma_2 \neq \sigma_1$), then for a given value of $\CSR = \sqrt{\sigma_1^2 + \sigma_2^2} $, the effective beam-beam parameter can be written as
	\begin{equation}
	\xibar =  \frac{1}{2}  \left[ (\sigma_2 / \sigma_1)^2 + (\sigma_1 / \sigma_2)^2 \right] \, \frac{n r_0 \bst}{2 \pi \gamma \CSR^2} ~,
	\label{eqn:xibarAsym}
	\end{equation} 
thereby incorporating the effect of beam-size asymmetries in a single-parameter, but approximate, characterization of the strength of the beam-beam interaction. In the limit of $\sigma_2 / \sigma_1 = 1$, \xibar reduces to the symmetric, round-beam equivalent beam-beam parameter \xiR (Eq.~(\ref{eqn:xiRDef2})).

	\subsubsection{Predicted impact of transverse beam-size asymmetries}
	\label{subsubsec:BeamSizImbal}
	
Since the beam-beam induced orbit shift depends only on the convolved beam size \CSR (Sec.~\ref{subsubsec:orbShift}), but not on the $\sigma_2 / \sigma_1$ ratio, only the beam--beam-induced optical distortion may be affected by a beam-size asymmetry. It is natural, therefore, to characterize this effect by varying, in the simulation, the common bunch intensity and the beam-size ratio, while keeping \CSR constant.  When using COMBI, this can be done in a single step for each parameter set, since the spatial density distributions of both bunches evolve simultaneously. When using B*B, however, the source bunch remains unaffected. Therefore the simulation must  be run twice, first with $\sigma_1$ and $\sigma_2$ assigned to, respectively, the source  and the witness beam, and then with the inverse assignment, after which the contributions of the two passes are combined. 

\par
\begin{figure}
	\centering
	\includegraphics[width=0.45\textwidth]{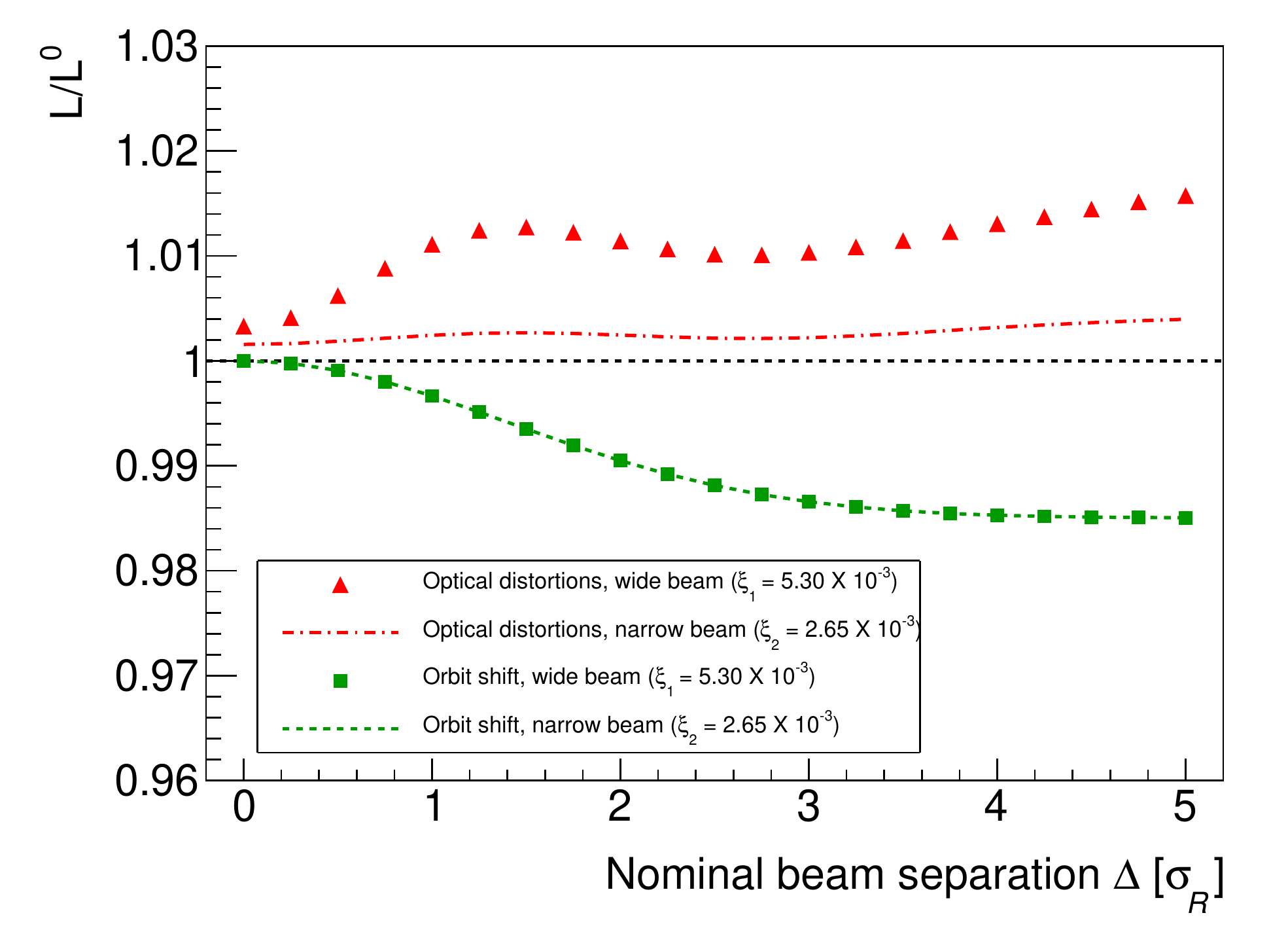}
	\caption{Single-beam contributions to the beam-separation dependence, during a horizontal \vdM\ scan simulated using B*B, of the luminosity-bias factor associated with the optical distortions (\LoLzOpt, red curve and triangles), and of that induced by the beam-beam orbit shift (\LoLzOrb, green curve and squares). The unperturbed beams are assumed to be round and perfectly Gaussian. Their bunch populations are assumed equal ($n_1 = n_2 = 0.78 \times 10^{11}$ p/bunch), and their transverse beam sizes differ 
by a factor of $\sqrt{2}$ ($\sigma_1 = 99.6\,\mu$m, $\sigma_2 = 70.5\,\mu$m). 
The wide-beam contributions are shown by the filled markers, the narrow-beam contributions  by the dashed curves. The beams collide only at the IP where the scan is taking place. The  horizontal axis is the nominal beam separation in units of the effective transverse beam size $\sigma_R = \CSR/\sqrt{2}$. 
}
	\label{fig:LoLzImSep}
	\end{figure}
\begin{figure}
	\centering
	\includegraphics[width=0.45\textwidth]{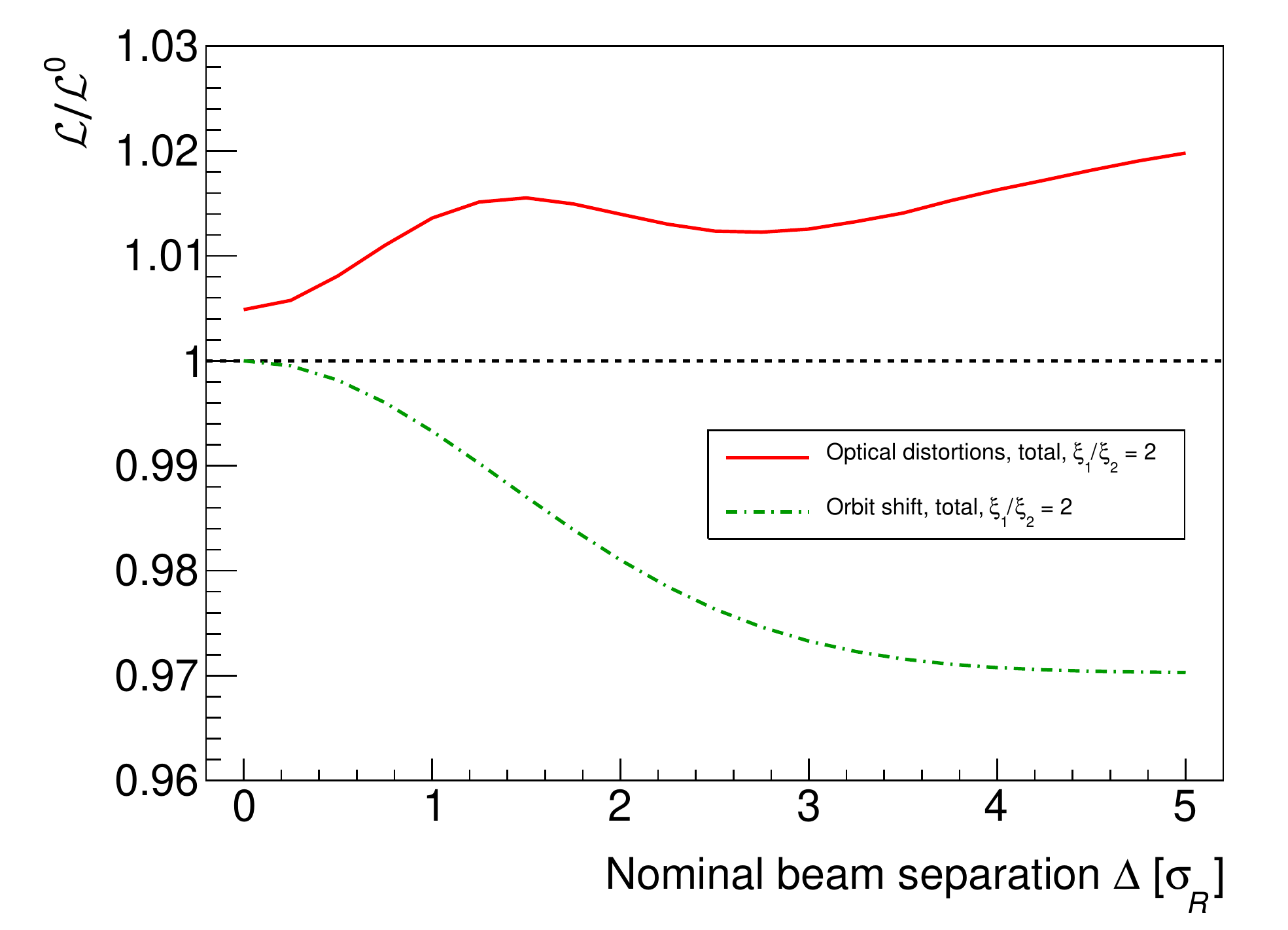}
	\caption{Beam-separation dependence, during a horizontal \vdM\ scan simulated using B*B, of the luminosity-bias factor associated with the optical distortions (\LoLzOpt, red solid curve), and of that induced by the beam-beam orbit shift (\LoLzOrb, green dot-dashed curve), in the presence of  a beam-size asymmetry corresponding to $\xi_1/\xi_2 = 2$. The single-beam parameters are listed in the caption of Fig.~\ref{fig:LoLzImSep}. The beams collide only at the IP where the scan is taking place. The  horizontal axis is the nominal beam separation in units of the effective transverse beam size $\sigma_R = \CSR/\sqrt{2}$. 
}
	\label{fig:LoLzImComb}
	\end{figure}

Such a study is presented in Fig.~\ref{fig:LoLzImSep}, for a highly asymmetric example in which 
$\xi_1/\xi_2 = 2$ and $\xibar = 4.42 \times 10^{-3}$. When the wider of the two bunches ($\xi_1 = 5.30 \times 10^{-3}$)
plays the role of the witness beam (triangles and squares), the orbit-shift and optical-distortion contributions are of opposite 
sign and of comparable magnitude; when the narrower bunch ($\xi_2 = 2.65 \times 10^{-3}$) 
becomes the witness beam (dashed curves), the beam-beam parameter drops by a factor of 
two,
the orbit-shift contribution remains unchanged as it should, but the optical-distortion effect becomes almost negligible. The combination of the two contributions, obtained by multiplying the luminosity-distortion factors associated with the two witness beams, is presented in Fig.~\ref{fig:LoLzImComb}. The two beams contribute equally to the orbit effect, but the optical-distortion is dominated by that experienced by the wider (\ie the weaker) of the two beams.  

\par
\begin{figure}
	\centering
	\includegraphics[width=0.45\textwidth]{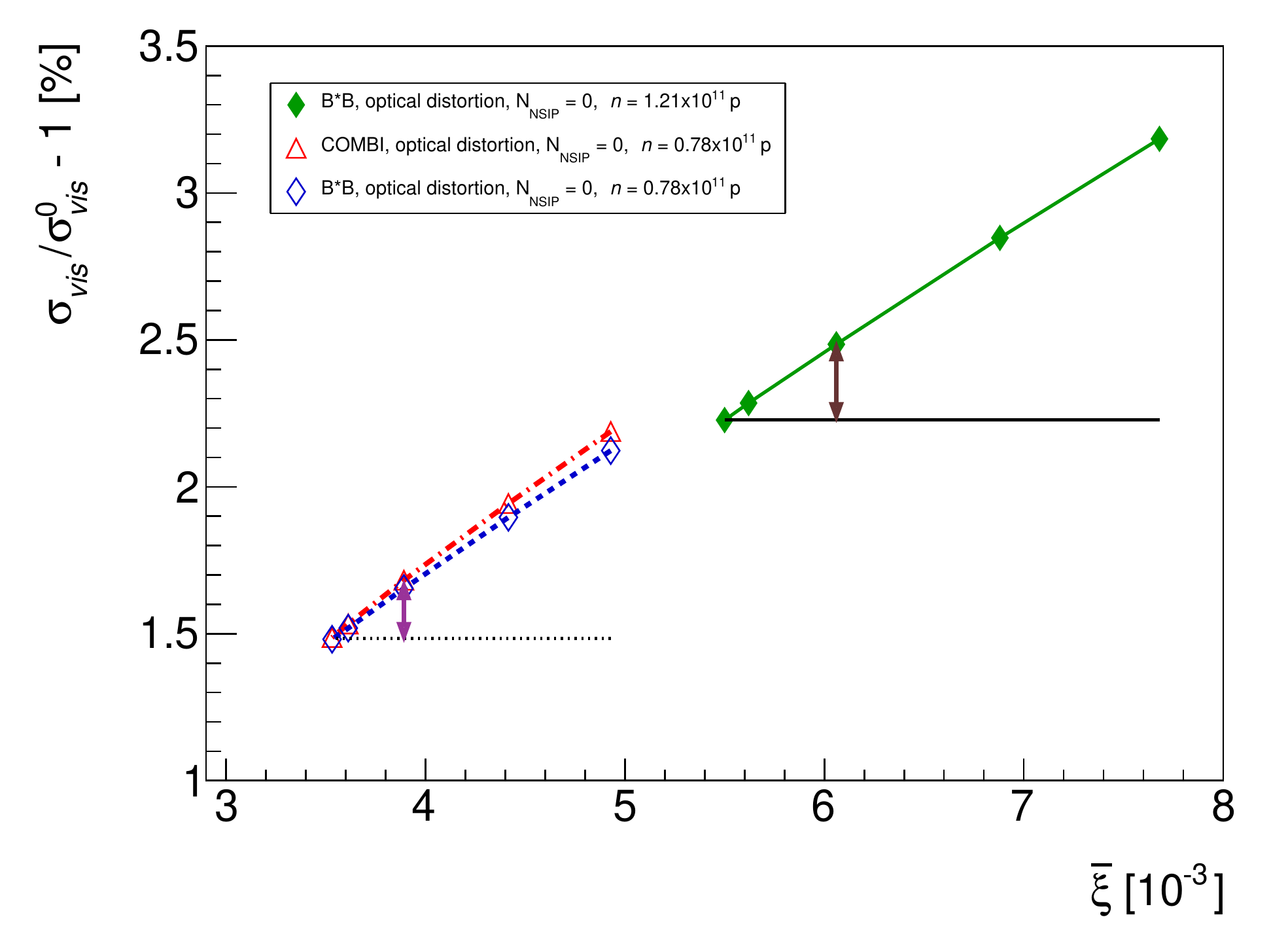}
	\caption{Dependence on the effective beam-beam parameter \xibar of the visible cross-section bias $\svis/\svisz-1$ associated with optical distortions, as predicted by B*B (diamonds) or COMBI (triangles) simulations, for low-intensity (open blue and red markers, bottom left) and high-intensity (filled green markers, top right) colliding-bunch configurations that cover a wide range of beam-size asymmetries at constant \CSR (see text). The beams collide at the scanning IP only. The \svis bias in the fully symmetric configuration ($\sigma_1 = \sigma_2$) corresponds to the lowest value of \xibar in each group, and is indicated by a horizontal black line. The curves are second-order polynomial fits to the points, and are intended to guide the eye. The arrows indicate the impact on the absolute luminosity scale of  a 20\% beam-size imbalance ($\sigma_2/\sigma_1 = 0.8$). 
}
	\label{fig:svisbVxib}
	\end{figure}
\begin{figure}
	\centering
	\includegraphics[width=0.45\textwidth]{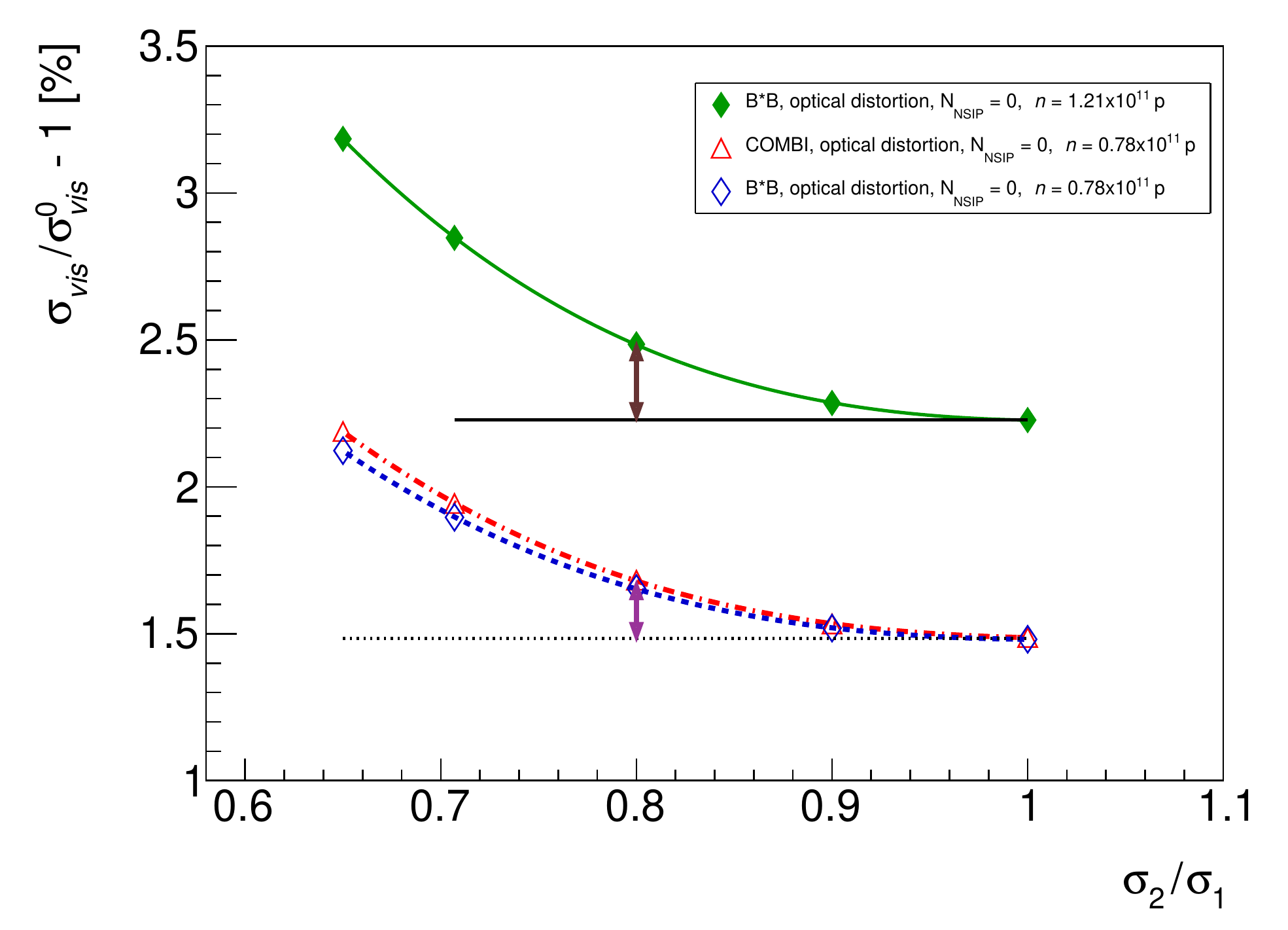}
	\caption{Dependence on the transverse beam-size ratio $\sigma_2/\sigma_1$ of the visible cross-section bias $\svis/\svisz-1$ associated with optical distortions, as predicted by B*B (diamonds) or COMBI (triangles) simulations, for low-intensity (red and blue bottom curves) and high-intensity (green top curve) colliding-bunch configurations that cover a wide range of beam-size asymmetries at constant \CSR (see text). The beams collide at the scanning IP only. The \svis bias in the fully symmetric configuration ($\sigma_1 = \sigma_2$)  is indicated by a horizontal black line in each group. The curves are third-order polynomial fits to the points, and are intended to guide the eye. The arrows indicate the impact on the absolute luminosity scale of  a 20\% beam-size imbalance ($\sigma_2/\sigma_1 = 0.8$).
}
	\label{fig:svisbVs2os1}
	\end{figure}

To quantify the impact of a transverse beam-size asymmetry on beam-beam corrections to \vdM calibrations, the simulation protocol detailed above was repeated for several configurations in which the bunch populations were  symmetric ($n_1 = n_2 = n$), the beam-size ratio $\sigma_2 / \sigma_1$ was varied from 1.0 down to 0.65, and the convolved beam size was kept constant at $\CSR = 122\, \mu$m. Two values of the common bunch population were considered ($n= 0.78 \times 10^{11}$ and $n = 1.21 \times 10^{11}$ p/bunch), near the lower end, and beyond the upper end, of the bunch-intensity range used during \vdM scans; these translate into round-beam equivalent beam-beam parameter values of, respectively $\xiR = 3.53\times 10^{-3}$ and $5.50\times 10^{-3}$. The visible cross-section bias $\svis / \svisz - 1$ corresponding to each of these configurations, and computed from luminosity-bias curves such as those displayed in Fig.~\ref{fig:LoLzImComb}, is presented in Figs.~\ref{fig:svisbVxib} and \ref{fig:svisbVs2os1}.

\par
As the beam-size ratio deviates more and more from unity, \xibar grows (see Eq.~(\ref{eqn:xibarAsym})), and so does the \svis bias. Its dependence on \xibar  is almost linear (Fig.~\ref{fig:svisbVxib}), with only a small quadratic component. The scaling law, however, is approximate only, as evidenced by the slight relative misalignment of the B*B curves at the interface between the two groups ($\xiR \sim 5.2$). The dependence on the beam-size ratio (Fig.~\ref{fig:svisbVs2os1}) is well modeled by a third-order polynomial of $\sigma_2/\sigma_1$, the coefficients of which depend on the $n \bst$ product. A systematic difference between COMBI ands B*B becomes apparent at large asymmetry, suggesting that coherent oscillations may start to play a role in that regime. Quantitatively however, this discrepancy is small enough in the asymmetry range of interest at LHC (at most 0.01\% on \svis for $\sigma_2/\sigma_1 > 0.9$, \ie for an emittance imbalance of 20\%) that it can be absorbed in a systematic uncertainty.	

\par
\begin{figure}
	\centering
	\includegraphics[width=0.45\textwidth]{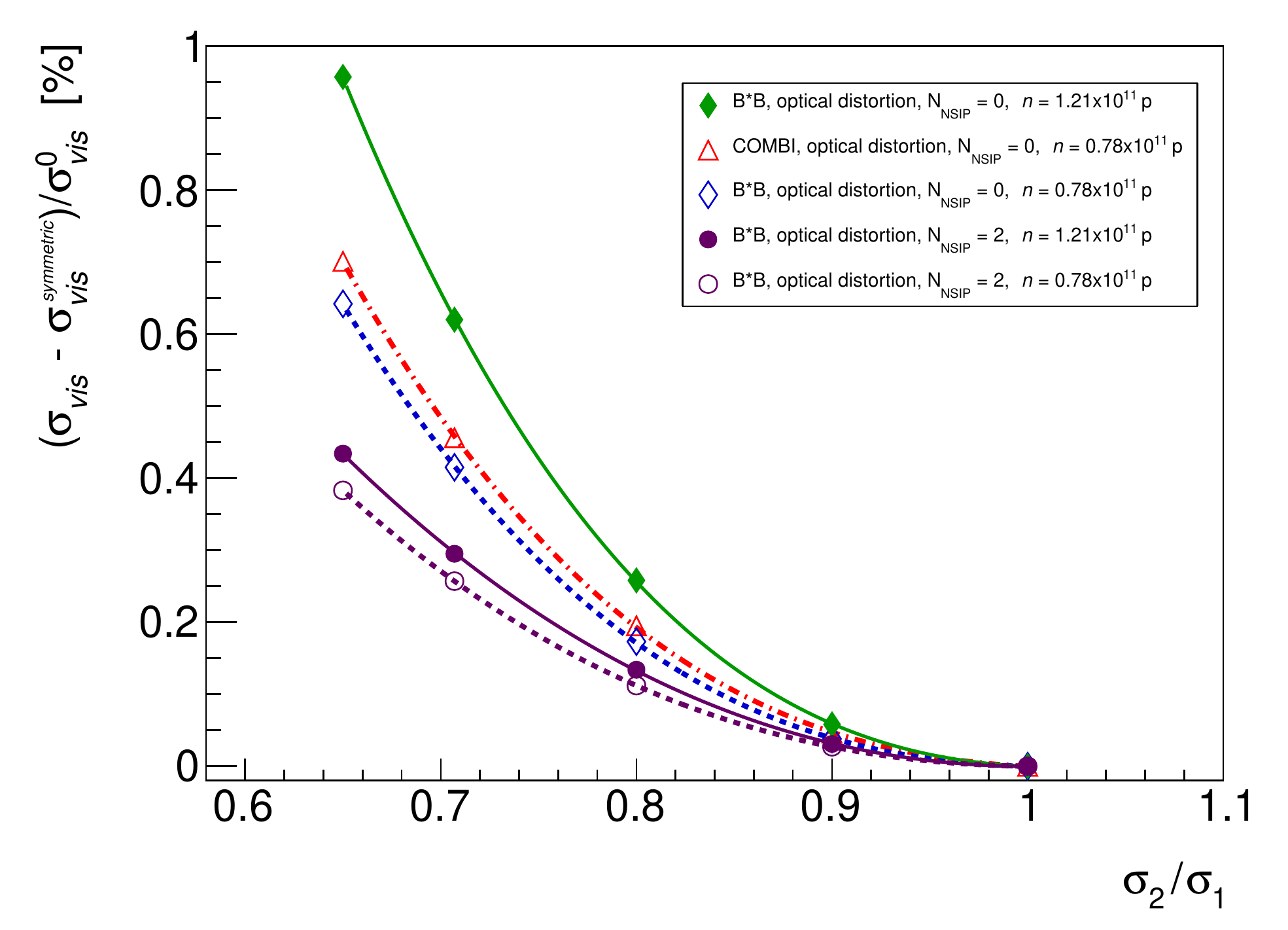}
	\caption{Impact of the beam-size asymmetry on the visible cross-section bias $\svis/\svisz-1$ associated with optical distortions, as simulated using either B*B (diamonds and circles) or COMBI (triangles). Shown is the change is \svis bias between an asymmetric beam-beam configuration characterized by a given setting of $\sigma_2/\sigma_1 < 1$, and the corresponding symmetric configuration ($\sigma_2 = \sigma_1$). The top three curves (green diamonds, red triangles and blue diamonds) correspond to the data displayed in Fig.~\ref{fig:svisbVs2os1}, when the beams collide at the scanning IP only. The bottom two curves (circles) show the same results when the beams also collide head-on at two non-scanning IPs. The open (filled) symbols correspond to the low- (high-) intensity bunches (see text). The curves are third-order polynomial fits to the points, and are intended to guide the eye.}
	\label{fig:svisbAsMS}
	\end{figure}

The impact of the beam-size asymmetry on the  absolute luminosity scale, \ie the difference in the \svis bias between a given asymmetric configuration (in this example $\sigma_2/\sigma_1 = 0.8$) and the fully symmetric case ($\sigma_1 = \sigma_2$), is indicated by the arrows. The same results are presented in Fig.~\ref{fig:svisbAsMS} (top three curves), on an expanded scale and after subtracting the \svis bias corresponding to the symmetric configuration.
	
\par
The results displayed in Figs.~\ref{fig:LoLzImSep}--\ref{fig:svisbVs2os1} assume that the beams collide at the scanning IP only. Additional collisions at other IPs lower the unperturbed tunes  (Sec.~\ref{subsubsec:mIPtunes}): as a result, the impact of the beam--beam-induced optical distortion is significantly reduced (Fig.~\ref{fig:Opt-ll0-Q-dependence}), and so is the impact of the transverse beam-size asymmetry. This is illustrated by the two bottom curves in Fig.~\ref{fig:svisbAsMS}, for the  case of bunches colliding at two non-scanning IPs ($\NSIP = 2$), that was modeled using B*B to simulate single-IP collisions but with the unperturbed tunes shifted as prescribed by Eq.~(\ref{eqn:multIPShift}). At the largest asymmetry considered ($\sigma_2/\sigma_1 = 0.65$), the sensitivity of the \svis bias to the beam-size asymmetry drops by more than a factor of two. In addition, the bunch-current dependence of this bias, manifested by the difference between the filled and the open circles, is several times smaller for the $\NSIP = 2$ case. This occurs because the shift \DQmIP in unperturbed tunes is proportional to both $\xi$ and $\NSIP$: when both are large enough, this tune shift more than compensates the increase in optical distortion at the scanning IP associated with the increased beam-beam parameter experienced by the weaker of the two beams. 

\par	
During Run-2 $pp$ \vdM scans at $\sqrt{s} = 13$\, TeV, the bunch population $n_p$ rarely exceeded $10^{11}$ (Table~\ref{tab:vdMBeamParms}), and the typical beam-size asymmetry $\sigma_2/\sigma_1$ hovered around 0.94, as extracted from one of the non-factorization analyses outlined in Sec.~\ref{subsubsec:2GBunchParms}. In practice therefore, for these \vdM sessions the impact of beam-beam asymmetries on beam-beam corrections to \vdM scans does not exceed 0.02\% (Fig.~\ref{fig:svisbAsMS}); such a small number can be safely accounted for as a systematic uncertainty in the absolute luminosity scale.
	
\section{Systematic uncertainties}
\label{sec:bbsysts}

\begin{sidewaystable*}[htbp]
\centering
\begin{tabular}{|c|c|c|c|c|c|c|c|c}
\hline
Beam-beam (b-b) 
 & $\xi_{\mathrm sim}$					& Uncertainty-determination	
 & \multicolumn{3}{c|}{\svis uncertainty [\%]} 	& Comments														& See					\\
uncertainty
 & $[10^{-3}]$						& procedure			
 &\multicolumn{3}{c|}{for \NSIP =}		&																	& Sec.					\\	
source 	&		 				& 		& 0	   	& 1		&  2			& 								&						\\	
\hline 
\hline
Absolute $\xi$ scale:
  & 5.60					& Vary \bst by $\pm 10\%$ in the simulation 			
  & 0.06		& 0.10		&  0.13										& \bst uncertainty assumed			& \ref{subsubsec:scalHypVal}	\\	
\bst uncertainty 
  &   			 			& or parameterization (Sec.~\ref{subsubsec:CorImplement}),			
  & \multicolumn{3}{c|}{(total for both beams}  									& uncorrelated between beams,		&		+				\\
at the scanning IP
  &   			 			& for each beam and in each plane			
  & \multicolumn{3}{c|}{and both planes)} 										& correlated between planes			& \ref{subsubsec:bstSyst}		\\
 \hline 
Nominal
  & 5.60					& Vary \qx, \qy by $\pm 0.002$ 			
  & 0.26		& 0.23		&  0.20										& Tune uncertainty assumed 			& \ref{subsubsec:xiNTuneDep} 	\\	
collision tunes 
  &   			 			& in the simulation or parameterization,			
  & \multicolumn{3}{c|}{(total for both beams}									& correlated between beams			&		+				\\	

  &   			 			& for each beam 			
  & \multicolumn{3}{c|}{and both planes)} 										& and between planes				& \ref{subsubsec:tuneSyst}	\\   
\hline
Non-Gaussian
  & 5.60 			 		& B*B (or COMBI) simulations			
  & 0.13		& 0.22		&  0.30										& Simulated for $\NSIP = 0$,			& \ref{subsec:nonGausImpact} 	\\	
transverse-density
  &   			 			& 			
  & \multicolumn{3}{c|}{ } 													& extrapolated to $\NSIP \ge 1$ 		&		+				\\    
distributions
  &   			 			& 			
  & \multicolumn{3}{c|}{ }  				 									& using Eq.~(\ref{eqn:multIPShift})		& \ref{subsubsec:nonGSyst}	\\    
\hline
Beam ellipticity
  & 5.60					&  B*B (or COMBI) simulations. 			
  & \multicolumn{3}{c|}{ $0.03$}										   		& Simulated for 				 	&\ref{subsec:EllipImpact} 		\\	
at the scanning IP
  &   			 			& Uncertainty scaled linearly			
  &  \multicolumn{3}{c|}{ }													& $\xiR \le 4.2 \times 10^{-3}$,			&		+				\\ 
  &   			 			& from \xiR to $ \xi_{\mathrm sim}$			
  &  \multicolumn{3}{c|}{ }				 									& $0.7 < \CSy/\CSx < 1.4$			& \ref{subsubsec:elliptSyst}	\\  
\hline
Non-zero
  & $\le 5.60$				& COMBI simulations			
   & \multicolumn{3}{c|}{$< 0.01^*$}											&  for $\theta_c \le 10\,\mu$rad$^*$			& \ref{subsec:XingAngImpact} +  \\	
crossing angle
  &   			 			& 			
  & \multicolumn{3}{c|}{$< 0.02~$} 											& for $\theta_c \le 150\,\mu$rad 		& \ref{subsubsec:longitSyst}	\\ 
\hline
Beam-beam 
  & 5.60			 		& B*B and COMBI simulations			
  & 0.016$^*$	& 0.012$^*$	&  0.008$^*$									& for $\sigma_2/\sigma_1 > 0.95^*$		& \ref{subsec:ImbalImpact} 	\\	
imbalance
  &   			 			& 			
  & 0.059~		& 0.045~		&  0.032~		 								& for $\sigma_2/\sigma_1 > 0.90~$		& 		+				\\   

  &   			 			& 			
  & 0.136~		& 0.104~		&  0.072~		 								& for $\sigma_2/\sigma_1 > 0.85~$		& \ref{subsubsec:BBimbalSyst}	\\     
\hline
Multiple IPs:~~~~~~~~~
  & 			 			& 			
  &  \multicolumn{3}{c|}{ }													& 								& 						\\	
~phase advance~~~~~~
  & 5.60  			 		& COMBI (or B*B) simulations			
  &  0			& \multicolumn{2}{c|}{$< 0.20$ }								& Worst case: arbitrary				&\ref{subsubsec:mIPambig} 	\\  

  &   			 			& 			
  & \multicolumn{3}{c|}{ } 													&  phase advances between IPs		& 		+				\\       
~~multi-IP tune shift
  & 5.60  			 		& Vary $p_1$ in Eq.(\ref{eqn:multIPShift}) by $\pm 15\%$ 
  &  0			& 0.05		& 0.09										&  								&\ref{subsubsec:mIPparam} 	\\ 
  
  &   						& in single-IP simulations.			
  & \multicolumn{3}{c|}{ } 													& 								& 		+				\\ 
  &   			 			& Ignore if using multi-IP simulation			
  &  \multicolumn{3}{c|}{ }		 											& 								&  \ref{subsec:multIPSyst}		\\ 
\hline
Long-range 
  & -						& None at the scanning IP		
  &  \multicolumn{3}{c|}{-}													&					 			&\ref{subsubsec:LRSyst} 		\\	
encounters
  &   			 			&  during $pp$ \vdM scans at the LHC			
  &  \multicolumn{3}{c|}{ }		 											& 								& 						\\ 
\hline
Lattice 
  & -						& COMBI simulations, with		
  & \multicolumn{3}{c|}{0.01$^*$}											& for $E_B \ge 6.5$\,TeV$^*$				&\ref{subsubsec:nonLinSyst} 	\\	
non-linearities
  &   			 			& sextupoles and octupoles included			
  & \multicolumn{3}{c|}{0.03~~}												& at lower beam energies				&						\\
\hline
Numerical accuracy
  & -						& 		
    &  \multicolumn{3}{c|}{$< 0.10$}											& Ignore if using simulation			&\ref{subsubsec:NumParztn} \\	
of parameterization
  &   						& 			
  &  \multicolumn{3}{c|}{  }		 											& rather than parameterization			& 						\\       
\hline\hline
Total uncertainty
  & 5.60						& Uncertainties summed in quadrature				
  & $\pm 0.32$	& $\pm 0.41$	&  $\pm 0.46$									&  \% of \svis					 	& \ref{subsec:SystSmry}		\\	
\hline
Total b-b correction
  & 5.60						& Parameterization (Secs.~\ref{subsubsec:CorImplement} \& \ref{subsubsec:mIPparam})		
  &  +0.52		& +0.86		&  +1.17										&  \% of \svis					 	& \ref{subsec:SystSmry}						\\	  
\hline        
\end{tabular}
\caption{Typical systematic uncertainties affecting beam-beam corrections to a hypothetical $pp$ \vdM calibration in a fully symmetric Gaussian-beam configuration, with the round--beam-equivalent beam-beam parameter set equal to $\xi_{\mathrm sim}$, for three values of \NSIP. For each source, the uncertainty is either evaluated at, or scaled linearly to, the value of $\xi_{\mathrm sim}$ indicated in the second column; if no value of $\xi_{\mathrm sim}$ is specified, the uncertainty listed covers the full range of $\xi$ values encountered during $pp$ \vdM scans at the LHC. When an uncertainty is assumption-dependent, the value flagged by an asterisk is that used in computing the total uncertainty; the latter is compared to the overall beam-beam correction itself in the bottom two rows of the Table. The rightmost column indicates the chapter(s) where the corresponding issues are discussed in detail.
}
\label{tab:SystSmry}
\end{sidewaystable*}

	The sources of systematic uncertainty that may affect the absolute magnitude of beam-beam corrections during \vdM scans, and that are detailed in Sec.~\ref{sec:bbImpact}, can be regrouped in three categories: optical configuration of the LHC rings, deviations from the fully symmetric Gaussian-beam configuration, and collisions at multiple IPs. The associated uncertainties are briefly elaborated upon in Secs.~\ref{subsec:ringOptics}, \ref{subsec:devFromGauss} and \ref{subsec:multIPSyst} below; additional sources of potential bias are addressed in Sec.~\ref{subsec:otherSysts}.  All the uncertainties are then added in quadrature, and their combined magnitude compared in Sec.~\ref{subsec:SystSmry} to that of the overall beam-beam correction itself.

\par
The quantitative impact of each uncertainty on the \svis bias, or equivalently on the magnitude of the beam-beam correction, is summarized in Table~\ref{tab:SystSmry} for scans simulated using three different collision patterns ($\NSIP = 0, 1, 2$). For each uncertainty source, the beam conditions input to the simulation are detailed in the chapters listed in the last column of the table. Not all simulations, however, use exactly the same beam conditions. In order to provide a self-consistent picture of the relative magnitude of the various uncertainties, and since the magnitude of the beam-beam bias, as well as most of the uncertainties, scale with $\xi$, the uncertainties listed in Table~\ref{tab:SystSmry} are all expressed at a common value $\xi_{\mathrm sim}$ of the beam-beam parameter; the details of this procedure can be found either in the table itself, or in Secs.~\ref{subsec:ringOptics} to \ref{subsec:multIPSyst}. Because the chosen value of $\xi_{\mathrm sim}$  lies at the upper end of the range covered during Run-2 vdM scans, Table~\ref{tab:SystSmry} illustrates something of a ``worst-case'' scenario with respect to the magnitude of the beam-beam correction and of the associated systematic uncertainty. The numerical results listed in that table should therefore not be applied blindly to estimating actual \vdM-calibration uncertainties, but rather used as guidance for a case-by-case and bunch-by-bunch error analysis of actual \vdM-calibration data.


\subsection{Optical configuration}
\label{subsec:ringOptics}

	\subsubsection{\bst uncertainty at the scanning IP}
	\label{subsubsec:bstSyst}

	Since the \svis bias scales with the beam-beam parameter (Sec.~\ref{subsec:corProcRefConf}), any measurement error on $\xi$ directly translates into an error of similar relative magnitude on beam-beam corrections to the luminosity scale. Existing accelerator instrumentation unfortunately does not allow accurate enough a determination of the transverse single-beam sizes that enter the definition of $\xi$ (Eq.~(\ref{eqn:xiGDef})). The horizontal and vertical beam-beam parameters must therefore be redefined in terms of two-beam observables that can be reliably measured:
\begin{eqnarray}
\xi_{xAv} 	& =	& \frac{n \,  r_0 \, \beta^*_{xAv}}{ \pi \, \gamma \, \CSx (\CSx + \CSy)} 	\label{eqn:xixAv}		\\ 
\xi_{yAv} 	& =	& \frac{n \,  r_0 \, \beta^*_{yAv}}{ \pi \, \gamma \, \CSy (\CSx + \CSy)} \, . 	\label{eqn:xiyAv}
\end{eqnarray}	
Here, $\beta^*_{xAv} = (\beta^*_{x1} + \beta^*_{x2})/2$  is the beam-averaged horizontal $\beta$-function at the scanning IP, and similarly for $\beta^*_{yAv}$. The quantity $n = (n_1 + n_2)/2$ refers to the beam-averaged bunch population associated with the colliding-bunch pair under study, $\gamma = \gamma_1 = \gamma_2$ is the common relativistic factor (assumed here to be the same for B1 and B2), and \CSx and \CSy are the convolved transverse bunch sizes. During \vdM scans, the observables $n$, \CSx  and \CSy are measured on a bunch-by-bunch basis to sub-percent accuracy~\cite{bib:ATLR2Lum, bib:LHCbLumPap2, bib:CMS201516Lum, bib:ALICE2016-18Lum}, and the uncertainty on the absolute beam energy is totally negligible for the purposes of this paper. The main uncertainty affecting the absolute scale of the beam-beam parameter, therefore, is that associated with potential deviations of \bst from its nominal or assumed value.

\par
The facts that the $\xi$ dependence of the \svis bias is roughly linear (Fig.~\ref{fig:FoM-sigvis}), and that the measured value of $\xi_{xj}$  (Eq.~(\ref{eqn:xiGDef})) depends linearly on $n_k$ and $\beta^*_{xj}$ ($j, k = 1,2; j \ne k$), justify the use of the beam-averaged variables defined in Eqs.~(\ref{eqn:xixAv})-(\ref{eqn:xiyAv}) above. Furthermore, the very weak dependence of the \svis bias on the $\CSyz/\CSxz$ aspect ratio (Fig.~\ref{fig:svisBias_EmR-vs-AR})  indicates that ``averaging'' over $x$ and $y$, in the sense of substituting the round-beam equivalent parameter \xiR (Eq.~(\ref{eqn:xiRDef2})) for the pair of horizontal and vertical beam-beam parameters ($\xi_{xAv}, \xi_{yAv}$), results in an acceptably small bias. What finally matters, therefore, is the uncertainty on the beam- and plane-averaged value of \bst at the scanning IP.

\par
The \bst values at the LHC IPs are extracted from optical-function measurements that are carried out for every colliding-beam configuration as part of routine accelerator commissioning. During the first few years of LHC operation, these values were typically obtained using the phase-advance method; k-modulation is now considered a more promising approach~\cite{bib:bstKmod}. It targets an absolute accuracy of a few percent on $\beta^*_{iB}$ ($i = x, y$; $B =1, 2$), provided sufficient beam time can be dedicated to the measurement.

\par
The absolute \bst-accuracy requirements associated with beam--beam-correction uncertainties are rather modest. When beams collide at the scanning IP only, and for $\xi = 5.6 \cdot 10^{-3}$, a $\pm 10\%$ uncertainty in \bst at the scanning IP, uncorrelated between beams (because the optics of the two rings are almost completely independent) but correlated between planes (because a focusing error presumably affects both the horizontal and the vertical $\beta$-function) results in $\pm 0.06\%$ uncertainty in \svis (Table~\ref{tab:SystSmry}). Treating this uncertainty as fully uncorrelated (correlated) between beams and between planes would decrease (increase) the corresponding \svis uncertainty by a factor of $\sqrt{2}$.  The absolute uncertainty increases to 0.10\% (0.13\%) for $\NSIP = 1 (2)$, but its magnitude relative to the overall beam-beam correction remains about the same. 

\par
The $\pm 10\%$ \bst-uncertainty assumed in the example above is very conservative, and is representative of the typical difference between the nominal, \ie the dialed-in, \bst setting and the measured values. Smaller uncertainties can be achieved by calculating $\xi$ based on the measured \bst values themselves, while accounting rigorously for the systematic \bst-measurement uncertainties and for their correlation between beams and between planes.

	\subsubsection{Nominal collision tunes}
	\label{subsubsec:tuneSyst}

	Figures~\ref{fig:Opt-ll0-Q-dependence} and \ref{fig:FullBB-ll0-Q-dependence} illustrate the sensitivity of the optical distortion and of the full beam-beam bias to the unperturbed fractional tunes. Any deviation of (\qx, \qy) from their nominal setting results in an inaccurate prediction of the \svis bias. 
	
\par
	For the purposes of the present discussion, a conservative uncertainty of $\pm 0.002$ is assigned to each of \qx and \qy, and is assumed to be correlated  between beams (as could be caused by, for instance, instrumental effects) and between planes (because a focusing error presumably affects both the horizontal and the vertical betatron frequency). The magnitude of this uncertainty is based on LHC operational experience\footnote{While tunes can often be measured to 0.001 or better before beams are brought into collision, the beam-beam tune spread makes it quite challenging to monitor the unperturbed LHC tunes while the beams are colliding.}, and is meant to cover setting errors, as well as the stability of the unperturbed tunes, over the course of the \vdM-calibration fill(s). The corresponding uncertainty in \svis (Table~\ref{tab:SystSmry}) dominates the overall beam-beam uncertainty. It decreases slightly when the number of non-scanning IPs increases, because the additional collisions lower the unperturbed tunes, thereby reducing the sensitivity of the optical distortions to the exact values of the fractional tunes. The same effect is responsible for the \NSIP-dependent sensitivity of \svis to beam-beam asymmetries (Fig.~\ref{fig:svisbAsMS}).

\subsection{Deviations from the fully symmetric Gaussian-beam configuration}
\label{subsec:devFromGauss}

	Except where specifically stated otherwise, most of the results presented in this paper, and in particular the parameterizations detailed in Secs.~\ref{subsubsec:CorImplement} and \ref{subsubsec:mIPparam}, are based on the assumptions listed in Sec.~\ref{subsec:bbCorMeth}. Each of these will be lifted, one at a time, in the following Sections, and their individual impact on beam--beam-correction uncertainties estimated using the studies detailed in Secs.~\ref{subsec:nonGausImpact} to \ref{subsec:ImbalImpact}.

	\subsubsection{Non-Gaussian unperturbed transverse-density profiles}
	\label{subsubsec:nonGSyst}

Of the three single-bunch models described in Table~\ref{tab:2gModels}, the 2012 parameter set represents an extreme case, both because this is the \vdM session that revealed the largest non-factorization biases at LHC ever~\cite{bib:ATL2012Lum,bib:ATLR2Lum}, and because the parameters in that Table are deliberately chosen to represent the worst case among all scans and all bunches in a given \vdM session. The 2018 model emerges as the most appropriate to estimate the uncertainty associated with non-Gaussian tails in the unperturbed density distributions: not only is the kurtosis intermediate between that of the other two parameter sets, but it also reflects the performance of the injector chain in the fourth year of LHC Run 2, by which time procedures had long stabilized and beam conditions could be considered reasonably reproducible. 

\par	
	Since the factorizable double-Gaussian configuration systematically yields a larger deviation from the corresponding single-Gaussian model than its non-factorizable counterpart (Table~\ref{tab:2Gimpact}), it is favored for the purpose of estimating systematic errors. The uncertainty listed in Table~\ref{tab:SystSmry} for $\NSIP = 0$ is therefore taken from the top half of Table~\ref{tab:2Gimpact}. It is extrapolated to the cases of one and two non-scanning IPs by assuming that the fractional impact of additional collisions on the \svis bias is dominated by the multi-IP tune shift (Sec.~\ref{subsubsec:mIPtunes}), and thus insensitive to the shape of the unperturbed transverse-density distributions.
	
\par	
The choice of the 2018 factorizable model to estimate the impact of non-Gaussian tails is not devoid of some arbitrariness. The uncertainties listed in Table~\ref{tab:SystSmry}, therefore, should be considered as a first indication, albeit a prudent one. A refined evaluation would require a more extensive characterization of the transverse-density distributions extracted from non-factorization analyses, that could then be used to produce a wider palette of more realistic single-bunch models for input to B*B or COMBI. In order to explore potential correlations between systematic biases, one could in addition extend the existing studies to simulate bunches that are not purely Gaussian and that in addition collide at multiple interaction points.

	\subsubsection{Beam ellipticity at the scanning IP}
	\label{subsubsec:elliptSyst}

The dependence of the optical-distortion bias on the $\CSy / \CSx$ aspect ratio is summarized by Fig.~\ref{fig:svisBias_EmR-vs-AR}, and does not exceed $\pm 0.025\%$ for $\xiR \le 4.2 \cdot 10^{-3}$. Scaling this linearly to the value of $\xi_{\mathrm sim}$ shown in Table~\ref{tab:SystSmry} yields a beam--beam-correction uncertainty of 0.03\%, under the condition that the orbit-shift correction take the actual ellipticity into account, \ie that it be computed using the Bassetti-Erskine formalism~\cite{Bassetti} with, as input, the measured values of \CSx and \CSy.

	\subsubsection{Non-zero crossing angle}
	\label{subsubsec:longitSyst}

The crossing-angle dependence of the beam-beam bias in the \vdM regime is summarized by Fig.~\ref{fig:sigVis_vsXA}. The associated  systematic uncertainty is taken as the difference in \svis bias between $\theta_c = 0$ and the actual crossing-angle value; it is listed in Table~\ref{tab:SystSmry} for two values of the full crossing angle.

\par
It should be emphasized that this small uncertainty is applicable only at moderate crossing angles ($\theta_c \le 150\,\mu$rad), and only in the \vdM regime ($\xi \le 0.006$, $\bst \ge 10$\,m). For larger crossing angles in the \vdM regime, additional effects may come into play. This may also be the case for scans performed in the physics regime, even at moderate crossing angle, when the beam-beam parameter becomes large enough, and/or when \bst becomes small enough ($\bst \le 30$\,cm) for the hourglass effect to no longer be negligible. Of particular concern are the so-called ``emittance scans'',  that are \vdM-like beam-separation scans that are used to monitor luminometer performance and transverse emittances during routine LHC operation~\cite{bib:eScans}: the study of their sensitivity to beam-beam-induced biases remains mostly unexplored territory.

	\subsubsection{Beam-beam imbalance}
	\label{subsubsec:BBimbalSyst}	
	
	Of the three beam--beam-asymmetry scenarios considered in Sec.~\ref{subsec:ImbalImpact}, only the transverse-emittance imbalance (Sec.~\ref{subsubsec:BeamSizImbal}) requires special attention, since bunch-current and \bst-asymmetries are accounted for by the use of the beam-averaged beam-beam parameters defined in Eqs.~(\ref{eqn:xixAv})-(\ref{eqn:xiyAv}). The emittance-asymmetry dependence of the \svis bias is controlled by the combination of \xiR and of the transverse beam-size ratio $\sigma_2 / \sigma_1$ (Eq.~(\ref{eqn:xibarAsym})); it also is sensitive to the number of non-scanning IPs (Fig.~\ref{fig:svisbAsMS}). 
	
\par	
	The beam--beam-imbalance  uncertainty is taken as the difference in \svis bias between an asymmetric beam-beam configuration characterized by a given value of $\sigma_2/\sigma_1 < 1$, and the corresponding symmetric configuration ($\sigma_2 = \sigma_1$); it is listed in Table~\ref{tab:SystSmry} for three values of the transverse beam-size ratio. The ratio closest to unity ($\sigma_2 / \sigma_1 > 0.95$), that emerged as the most representative of typical beam conditions during the 2015-2018 $pp$ \vdM sessions~\cite{bib:ATLR2Lum}, is the one used when computing the total beam-beam uncertainty (penultimate row of Table~\ref{tab:SystSmry}). The \NSIP dependence  is discussed in Sec.~\ref{subsubsec:BeamSizImbal}, and is closely related to the tune-dependence of the optical-distortion bias (Sec.~\ref{subsubsec:tuneSyst}).

\subsection{Multiple interaction points}
\label{subsec:multIPSyst}
	
	Two sources of uncertainty affect the beam-beam correction strategies in the presence of multi-IP collisions, that are detailed in Sec.~\ref{subsubsec:mIPSmry}: the phase advance between consecutive IPs (Sec.~\ref{subsubsec:mIPambig}), and the parameterization of the multi-IP equivalent tune shift \DQmIP (Sec.~\ref{subsubsec:mIPparam}). These uncertainties affect only those colliding-bunch pairs in which at least one of the two opposing bunches experiences collisions at one or more IPs other than the scanning IP ($\NSIP > 0$).
	
\par
The phase-advance uncertainty arises from the currently unresolved ambiguity in the choice of the reference ``no beam-beam'' configuration,  or,  in more technical terms, from the ambiguity between the \Lumz and the \Lumu normalizations (Sec.~\ref{subsubsec:LzVsLu}). In the absence of any phase-advance information, \eg for a bunch pattern that mixes different collision configurations that are all treated on the same footing, a blanket systematic uncertainty of at most 0.2\% appears adequate (Sec.~\ref{subsubsec:mIPambig}). This uncertainty can, in some cases, be significantly lowered by simulating the actual collision pattern associated with each colliding-bunch pair, with as input the actual IP-to-IP phase advance per beam and per plane, and by then comparing the results between the two reference no-beam configurations. An illustration of the potential gain provided by such a refined analysis is offered by the results presented in Table~\ref{tab:FoM_LzVu} for the $\NSIP = 1$ case, a configuration that applies to the 13\,TeV $pp$ \vdM scans at IP1 and IP5 in 2015, 2016 and 2017~\cite{bib:ATLR2Lum}.
	
\par
The uncertainty associated with the equivalent multi-IP tune shift \DQmIP needs to be considered only when invoking the parameterization of single-IP simulations proposed in Sec.~\ref{subsubsec:mIPparam} and encapsulated in Eq.~(\ref{eqn:multIPShift}), to compute beam-beam corrections in the presence of additional collisions. This parametrization, that is illustrated by the magenta curve in Fig.~\ref{fig:sIPScaling_Lu}, is in good agreement with the simulation results when IP5 is the scanning IP. The discrepancies that are apparent for scans at some of the other IPs, reflect the variety of phase-advance patterns between the scanning IP and the other collision points. Varying the $p_1$ parameter in Eq.~(\ref{eqn:multIPShift}) by $\pm 15\%$ of its central value covers all the collision patterns shown in Fig.~\ref{fig:sIPScaling_Lu}, thereby providing a measure of the uncertainty associated with this simplified procedure (Table~\ref{tab:SystSmry}). More precise results, and a smaller uncertainty, could again be obtained by simulating separately each of the collision patterns associated with the B1 and B2 bunch strings circulating in the LHC ring during the \vdM-calibration session under analysis.

\subsection{Other potential sources of systematic uncertainty}
\label{subsec:otherSysts}

	\subsubsection{Long-range encounters}
	\label{subsubsec:LRSyst}
		
	During $pp$ \vdM scans at the LHC, the bunches are longitudinally isolated, in the sense that consecutive bunches are separated by more than 500\,ns, rather than grouped in trains with only 25\,ns spacing. In addition, the fill pattern is systematically optimized to avoid parasitic crossings, at least at the scanning IP. This strategy eliminates separation-dependent, long-range beam-beam kicks that might distort the luminosity-scan curves.\footnote{The situation is different during heavy-ion (HI) scans, which for operational reasons are carried out in bunch-train mode with a non-zero nominal crossing angle. However, the bunch charge is considerably smaller than during $pp$ \vdM scans, and the beam-beam parameter about an order of magnitude smaller. Parasitic crossings, therefore, are no issue from the viewpoint of beam-beam corrections to HI \vdM calibrations.}

\par
In contrast, during high-luminosity $pp$ physics running, consecutive bunches are typically separated longitudinally by only 25\,ns, the emittance is smaller than during \vdM scans, and the bunch current significantly larger. As a result, and in spite of the non-zero crossing angle, long-range beam-beam effects are known to significantly impact emittance scans in several ways, most of which remain to be understood.

	\subsubsection{Lattice non-linearities}
	\label{subsubsec:nonLinSyst}

The strongest non-linearity is introduced by the beam-beam interaction itself. However, non-linear magnetic elements in the LHC rings also affect beam dynamics, potentially leading to additional distortions of the transverse particle-density distributions. Because the LHC relies on the combination of high chromaticity ($Q' \approx 10-15$ units) and of high Landau-octupole currents to mitigate some strong impedance-driven instabilities, the next largest optical non-linearities at beam energies above 1\,TeV are associated with the corresponding sextupoles and octupoles~\cite{Elias}.  As demonstrated experimentally~\cite{Non-linearities, Non-linearities2, ewen}, machine imperfections contribute even less by comparison, and are therefore neglected in the present study. 

\par
\begin{figure}
\centering
\includegraphics[width=0.48\textwidth]{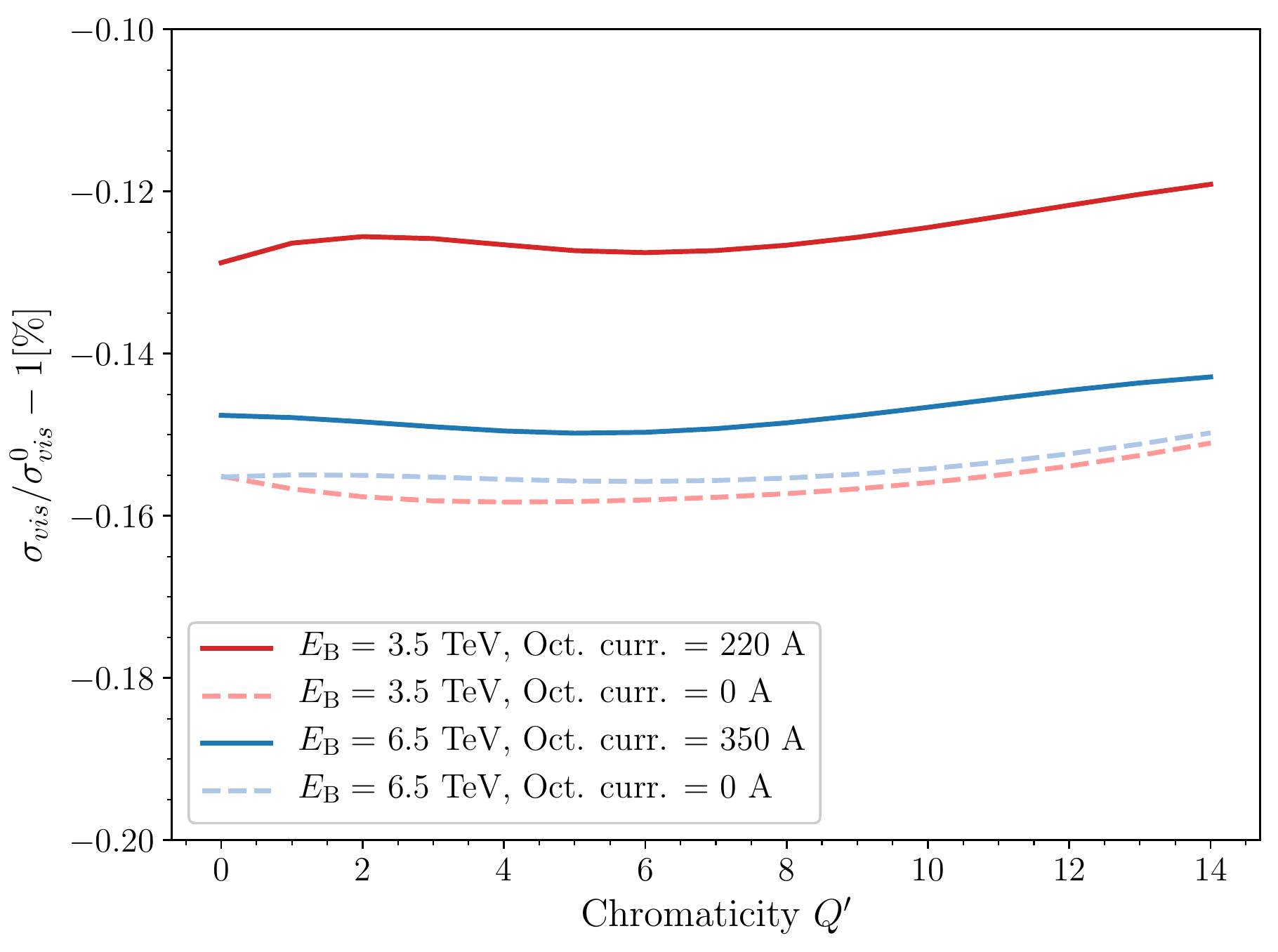}
\caption{Chromaticity dependence of the \svis bias predicted by COMBI for $pp$ \vdM scans, with octupoles either under power (solid curves) or left at zero current (dashed curves). The red curves correspond to the 2011 configuration ($E_B = 3.5$\,TeV), the blue curves to that in 2018 ($E_B = 6.5$\,TeV); the corresponding octupole currents are indicated in the legend.  The momentum spread and the synchrotron tune are set to, respectively,  $\num{1.12d-4}$ and $\num{2d-3}$. The beam-beam parameter is set to $\xi = \num{2.6 d-3}$, and the beams collide at the scanning IP only.
} 
\label{fig:chromaScan} 
\end{figure}

In order to quantify the impact of these magnets on the \svis bias, both the chromaticity and the linear detuning with amplitude introduced by the Landau octupoles were added to the  COMBI model. The chromaticity dependence of the beam-beam bias is presented in Fig. \ref{fig:chromaScan} for two different beam energies. As expected, the impact of the octupoles is more important in the 2011 case because of the lower beam energy: at zero chromaticity, powering the octupoles changes the beam-beam bias from $-0.155\%$ to  $-0.130\%$, a reduction in relative magnitude of about one sixth. In contrast, in the 2018 configuration the effect of the octupoles is three times smaller, and almost negligible. As for the chromaticity, its impact is negligible in both configurations.

\par
 The very low sensitivity of beam-beam corrections to these non-linear elements can be understood as follows. Octupoles act more strongly on larger-amplitude particles. These barely contribute to the luminosity when the beams collide head-on; the fractional contribution of the tails to the overlap integral is highest at large beam separation, but there the luminosity is the lowest. This is why modifying the trajectory of the tail particles barely influences the separation-integrated FoMs, and therefore the \svis bias.

\par 
 Similarly, high chromaticity potentially enhances synchro-betatron coupling; however since \vdM scans are carried out with relaxed \bst settings and at zero crossing angle, this is of no concern. 

\par 
 Even though lattice non-linearities are almost negligible in vdM scans (Table~\ref{tab:SystSmry}),  it remains important to check systematically the potential influence of the octupoles under operational scenarios where they are pushed to maximum power to ensure beam stability: their impact on beam-beam uncertainties may turn out to be no longer negligible, for instance during emittance scans.
 
	\subsubsection{Numerical accuracy of polynomial parameterization}
	\label{subsubsec:NumParztn}

	Propagating to the \svis bias the parameterization uncertainties that affect the separation-dependent luminosity-bias factors (Eq.~(\ref{eqn:ll0_xiqxqy})) and that are discussed in Sec.~\ref{subsubsec:CorImplement}, results in an upper limit of 0.1\% on the associated beam-beam uncertainty. This contribution should be ignored when resorting to a full B*B or COMBI simulation at each scan point, rather than to the parameterization.

\subsection{Summary of systematic uncertainties}
\label{subsec:SystSmry}

Adding in quadrature all the above uncertainties yields, for the example examined here ($\xi_{\mathrm sim} = 5.6 \cdot 10^{-3}$), a total beam-beam uncertainty of 0.32 to 0.46\%, depending on the number of non-scanning IPs (Table~\ref{tab:SystSmry}). The corresponding value of the beam-beam correction varies from 0.52 to 1.17\%. The absolute magnitude of the total uncertainty grows when \NSIP increases from 0 to 2, but its magnitude relative to the correction itself decreases from about 60\% to roughly 40\% through the mechanism dicussed in Secs.~\ref{subsubsec:tuneSyst} and \ref{subsubsec:BBimbalSyst}. 

\par
Under operating conditions typical of actual \vdM-calibration sessions, the number of non-scanning IPs is at least one.  For $\NSIP = 1$, the four dominant uncertainty contributions, in order of decreasing  importance, are associated with the nominal tunes, non-Gaussian tails, multi-IP effects (in the absence of collision--pattern-specific studies), and the \bst uncertainty. Estimating the uncertainty associated with non-Gaussian tails, and verifying whether beam-size asymmetries might contribute at a significant level, imperatively requires input from the non-factorization analysis of luminous-region data. 

\par
It is worth noting that the larger \NSIP, the larger the contribution of the orbit-shift correction relative to that of the optical distortion. Qualitatively therefore, estimating precisely the effects that contribute only to the optical-distortion uncertainty, viz. the beam ellipticity and the beam-size asymmetry, becomes less critical as the number of non-scanning IPs increases. In contrast, the \bst and tune uncertainties affect the magnitude of both the orbit shift and the optical distortion, in a fully correlated manner.

\par
The uncertainties detailed in Table~\ref{tab:SystSmry} have been deliberately estimated in a conservative fashion, and for the largest value of the beam-beam parameter encountered during Run-2 \vdM scans in $pp$ collisions at $\sqrt{s} = 13$\,TeV. Most uncertainties scale with $\xi$, either linearly or with a slightly positive quadratic component. For the purposes of a rough estimate, therefore, they can be safely extrapolated linearly to smaller $\xi$ values, that are more representative of typical \vdM conditions. A more precise determination of these uncertainties requires a case-by-case analysis, possibly augmented by some of the additional simulation studies suggested in some of the preceding sections in this chapter.

\section{Summary and outlook}
\label{sec:Concl}

Under conditions typical of \vdM luminosity-calibration scans in $pp$ collisions at the LHC, beam--beam-induced orbit shifts and optical distortions, if left uncorrected, bias the absolute luminosity scale by an amount comparable to, or larger than, the total luminosity-uncertainty budget. The magnitude of the correction, which ranges from approximately 0.2\% to 1.2\%, mainly depends on the beam-beam parameter $\xi$ and on the number of non-scanning IPs. 

\par
The contributions to the associated systematic uncertainty are listed in Table~\ref{tab:SystSmry} and elaborated upon in Sec.~\ref{sec:bbsysts}. The total uncertainty amounts to roughly half of the correction itself, and is dominated by tune-related effects, either directly (accuracy and reproducibility of nominal-tune settings) or indirectly (beam-beam tune shift at non-scanning IPs, phase advance between consecutive IPs). The next  most important source of uncertainty is the potential deviation of the unperturbed transverse bunch-density distributions from a perfectly Gaussian shape. Simulating the impact of non-Gaussian tails cannot be based on luminosity data alone: it requires, as input, single-bunch parameters that can only be extracted from the measured beam-separation dependence, during the scan, not only of the luminosity but also of the position, shape and orientation of the luminous region. These single-bunch parameters are also crucial to quantify transverse beam-size differences between the two beams, which, if large enough, may become a significant source of uncertainty.

\par
The scope of the present paper is limited to beam-separation scans in the \vdM regime ($\xi \le 6 \cdot 10^{-3}$), under beam conditions that deviate only moderately from round, initially Gaussian bunches of equal brightness that collide with a zero or small nominal crossing angle. In the course of investigating the impact of actual departures from that idealized limit, several areas of further study have been identified, most notably:
\begin{itemize}
\item
improved single-bunch models of non-Gaussian tails (Sec.~\ref{subsec:nonGausImpact}), both factorizable and non-factorizable, that are more sophisticated than the simple double-Gaussian functions considered in Eqs.~(\ref{eqn:FDGparmtn}) and (\ref{eqn:NFDGparmtn});
\item
the interplay between non-factorization, non-Gaussian tails, and beam-beam corrections, and in particular the application of beam-beam corrections to two-dimensional grid scans;
\item
the ambiguity affecting the choice of the reference ``no beam-beam'' configuration, \ie the phase-dependence of the difference between the \Lumz and the \Lumu normalizations (Sec.~\ref{subsec:multiIP});
\item
the characterization of optical distortions at large nominal crossing angle (Sec.~\ref{subsec:XingAngImpact}) in the \vdM regime;
\item
the potential interplay between the impact of non-Gaussian beams (Sec.~\ref{subsec:nonGausImpact}), beam ellipticity (Sec.~\ref{subsec:EllipImpact}), crossing angle (Sec.~\ref{subsec:XingAngImpact}), and multi-IP effects (Sec.~\ref{subsec:multiIP}), that have so far been treated as fully uncorrelated;
\item
the extension of the $\xi$-scaling laws (Sec.~\ref{subsec:corProcRefConf}) to the physics regime ($\xi \le 10^{-2}$), with emittance scans as the primary use case;
\item
the combined impact, during emittance scans at low \bst in the physics regime, of higher-beam-beam parameter values and of the hourglass effect on the crossing-angle dependence of beam-beam biases;
\item
the impact of parasitic crossings on the absolute accuracy of emittance scans during routine physics running;
\item
the impact of lattice non-linearities on the absolute accuracy of emittance scans during routine physics running.
\end{itemize}

\section{Acknowledgments}

We thank our colleagues in the LHC Luminosity Monitoring and Calibration Working Group (LLCMWG), and in  particular V. Balagura, X. Buffat, R. Hawkings, W. Herr and J. Wenninger, for many fruitful discussions and for valuable comments on part or all of the manuscript. We are also indebted to V. Balagura for teaching us how to run his B*B package, as well as for implementing additional functionalities needed by the studies reported above. Some of the authors have received support from either the Swiss Accelerator Research and Technology Institute (CHART), the U.S. Department of Energy (DOE Contract No. DE-AC02-76SF00515), or the Hungarian National Research, Development and Innovation Office (NKFIH Contract Nos. 124845 and 143460).

\appendix
	\section{Impact of non-Gaussian distortions of the source field}
	\label{sec:nonGausFieldImpact}

Because B*B is fundamentally a weak-strong model, beam--beam-induced shape-distortions of the witness bunch, \ie of the multiparticle distribution, influence neither the shape of the source bunch nor the spatial dependence of the electromagnetic field it generates. The latter is computed once and for all at initialization time, and remains static both turn-by-turn and as a function of beam separation.

\par
 COMBI, in contrast, is based on a strong-strong model: both bunches are represented by macroparticles; both play, with respect to their partner, the dual roles of source and witness bunch; and the electromagnetic fields they generate are updated on a turn-by-turn basis, and therefore also at each scan step. In order to estimate the impact of separation-dependent, non-Gaussian field distortions caused by the mutual interaction of the opposing bunches, the  \svis bias (Eq.~(\ref{eqn:FoM-svis})) has been computed using COMBI in two different modes:
\begin{itemize}
\item
for each witness bunch in turn, the field that its macropraticles are subjected to, is computed under the assumption that the transverse macroparticle distribution of the source bunch is a two-dimensional Gaussian, with the position of its centroid and its RMS widths computed on every turn from the actual spatial distribution of the ``source'' macropraticles. This procedure tracks the evolution of the relative position and of the width of the opposing multiparticle distributions, but the field calculation ignores potential deviations of the bunches from their initial, perfectly Gaussian shape. This is the mode in which all the COMBI results in this paper have been obtained. 
\item
for each witness bunch, the field it is subjected to is computed by the HFMM method~\cite{HFMM} from the actual charge distribution in the source bunch. In this mode, each macroparticle in the witness bunch is subjected to an electromagnetic field computed as the vector sum of the contributions of all the macroparticles in the source bunch. Since this computation must be repeated on every turn and for the opposing bunch, it is much more CPU-intensive than the Gaussian-field approximation above.
\end{itemize} 

\par

\begin{figure}[h]
	    \centering
	    \includegraphics[width=0.48\textwidth]{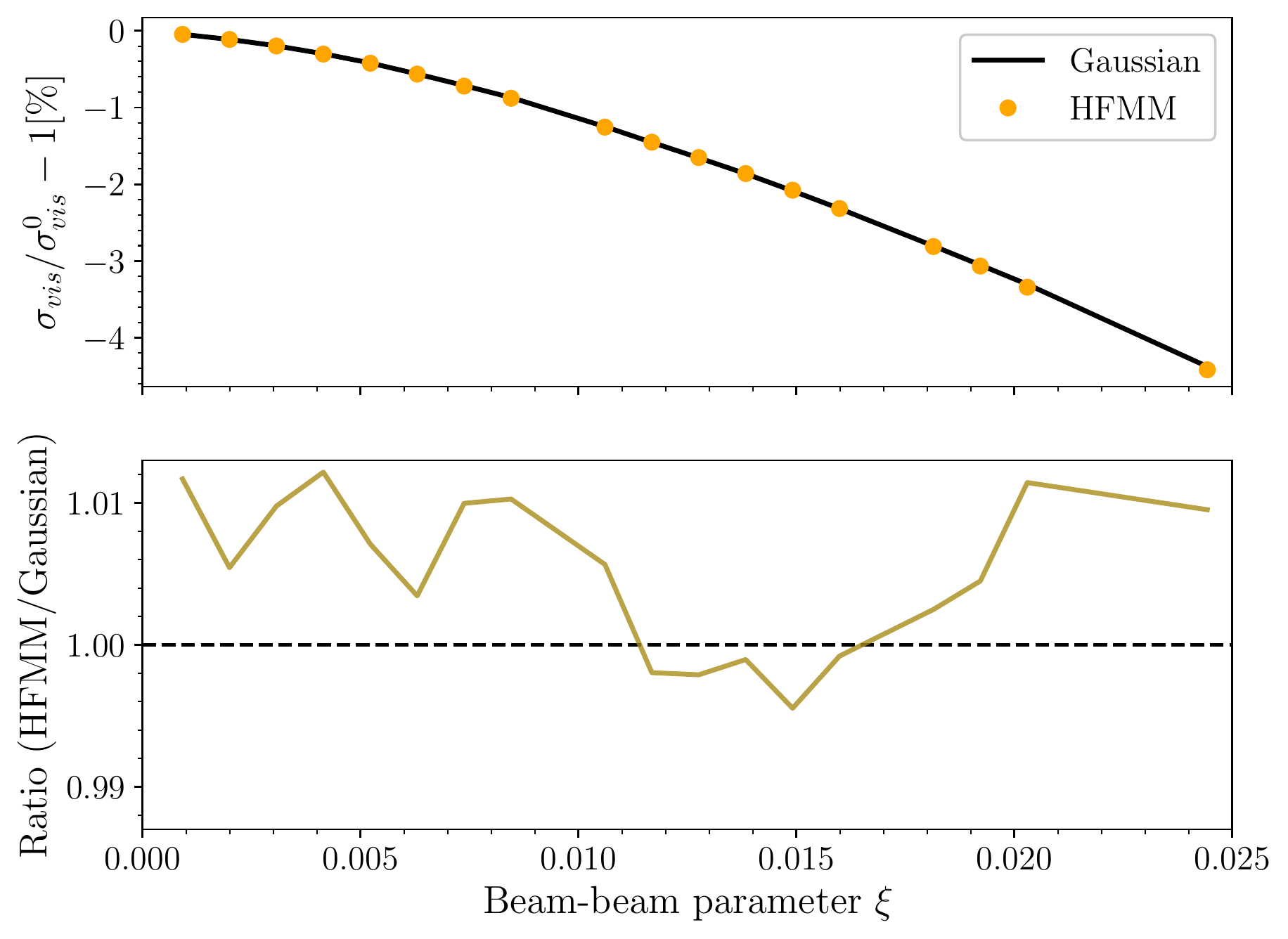}
	    \caption{Top: $\xi$ dependence of the beam-beam induced $\sigma_{vis}$ bias,  computed either in the Gaussian-kick approximation (solid black line), or with a self-consistent kick calculated using the HFMM field solver (filled orange circles). Bottom: $\xi$ dependence of the ratio between the \svis-bias values computed by the two methods. }
	    \label{fig:HFMM}
	\end{figure}

Figure~\ref{fig:HFMM} displays the beam--beam-parameter dependence of the \svis bias over a  $\xi$ range that extends all the way up to $25\times 10^{-3}$, more than four times the highest $\xi$ value encountered in \vdM scans, and significantly higher than the largest per-IP beam-beam parameter that is sustained operationally in the actual LHC rings. The two calculations agree within 1\% or better; for the $\xi$ range and the nominal tunes considered in Table~\ref{tab:vdMBeamParms}, this translates into a \svis difference of less than 0.005\%. Computing the electromagnetic field as that of an equivalent Gaussian, the parameters of which evolve turn by turn, therefore constitutes a fully justified approximation whose impact on beam-beam corrections can be safely neglected.

\section{Luminosity Integration in COMBI}
\label{sec:Integration}

In COMBI, the calculation of the luminosity, or more precisely of the overlap integral (Eq.~(\ref{eqn:lumi})), is based on a complete description of the macroparticle distribution in the two colliding bunches: no assumption is made about the shape of the density distributions when evaluating this integral. The latter is calculated on a turn-by-turn basis, both at the scanning IP and at the non-scanning IP(s), if any. The positions of all the macroparticles are updated at each IP on every consecutive turn, and each step in a beam-separation scan is initialized and simulated separately. 

\par
The transverse distribution of the macroparticles, called $H^B_{x,y}$,  is discretized, separately for the two beams $B$ ($B = 1, 2$), on a two-dimensional grid in the $x$-$y$ plane. The number of cells in each grid is given by $N_\mathrm{cells} = n \times m$, where $n$ and $m$ represent the number of bins in the $x$ and $y$ directions respectively. The grid boundaries are located at a distance $k_i \times \sigma^0_i$  ($i = x, y$) from the center of the grid, where $\sigma^0_i$ is the initial, unperturbed nominal  transverse beam size inferred from the input emittance and $\bst$ values in the $i$ plane; the scale factor $k_i$ is typically set to 12 in both the positive and the negative direction along the $x$ and $y$ axes. The cell area is thus given by 
\begin{equation}
\Delta S = \Delta x \times \Delta y = 2k_x \sigma^0_x/n \times 2k_y \sigma^0_y/m \,. 
\nonumber
\end{equation}
The separation between the two beams is taken into account when filling two-dimensional histograms of the macroparticle distributions.

\par
With these definitions, the discretized macroparticle density distribution is given by
\begin{equation}
h_{x,y}^B=\frac{H^B_{x,y}}{N_{part} \, \Delta S}\, ,
\nonumber
\end{equation}
where $N_{part}$ is the total number of tracked macroparticles.  The overlap density of the two bunches, \ie the product $\hat{\rho} _1 (x,y)\, \hat{\rho} _2(x,y)$ in Eq.~(\ref{eqn:lumi}), is therefore represented by:
\begin{equation}
\lambda_{x,y}=h^1_{x,y}  h^2_{x,y}\, .
\nonumber
\end{equation}
The bunch luminosity can be calculated from the overlap integral $I_\mathrm{ovlp}$ as:
\begin{equation}
\Lum_{\mathrm b} = f_{\mathrm r}\, n_1 n_2\, I_\mathrm{ovlp} ,
\label{eq: lumi1}
\end{equation}
where $n_1$ and $n_2$ are the bunch populations and $f_{\mathrm r}$ the revolution frequency.
The overlap integral is estimated by the two-dimensional trapezoidal method:
\begin{eqnarray}
\label{eq: trap}
 I_\mathrm{ovlp} & \simeq & 
  \frac{1}{4} \Delta x  \Delta y \Biggl(\lambda_{0,0} + \lambda_{m,0} + \lambda_{0, n} +\lambda_{m, n} + \nonumber
\\
&  & 2 \sum_{i=1}^{m-1} \lambda_{i,0} + 2 \sum_{i=1}^{m-1} \lambda_{i,n}+ 2  \sum_{j=1}^{n-1} \lambda_{0,j} + \nonumber\\ 
&  & 2 \sum_{j=1}^{n-1} \lambda_{m,j} + 4 \sum_{j=1}^{n-1} \Biggl( \sum_{i=1}^{m-1} \lambda_{i,j} \Biggr) \Biggr).        \nonumber
\end{eqnarray}
The instantaneous luminosity in Eq.~(\ref{eq: lumi1}) is calculated at each turn, and the result is averaged over the total number of selected turns. Typically a few hundred turns are necessary for the result to stabilize. The reliability of this method was confirmed by benchmarking it against analytical calculations, with the beam-beam effects turned off.

\par
\begin{figure}[ht]
\centering
\includegraphics[width=0.5\textwidth]{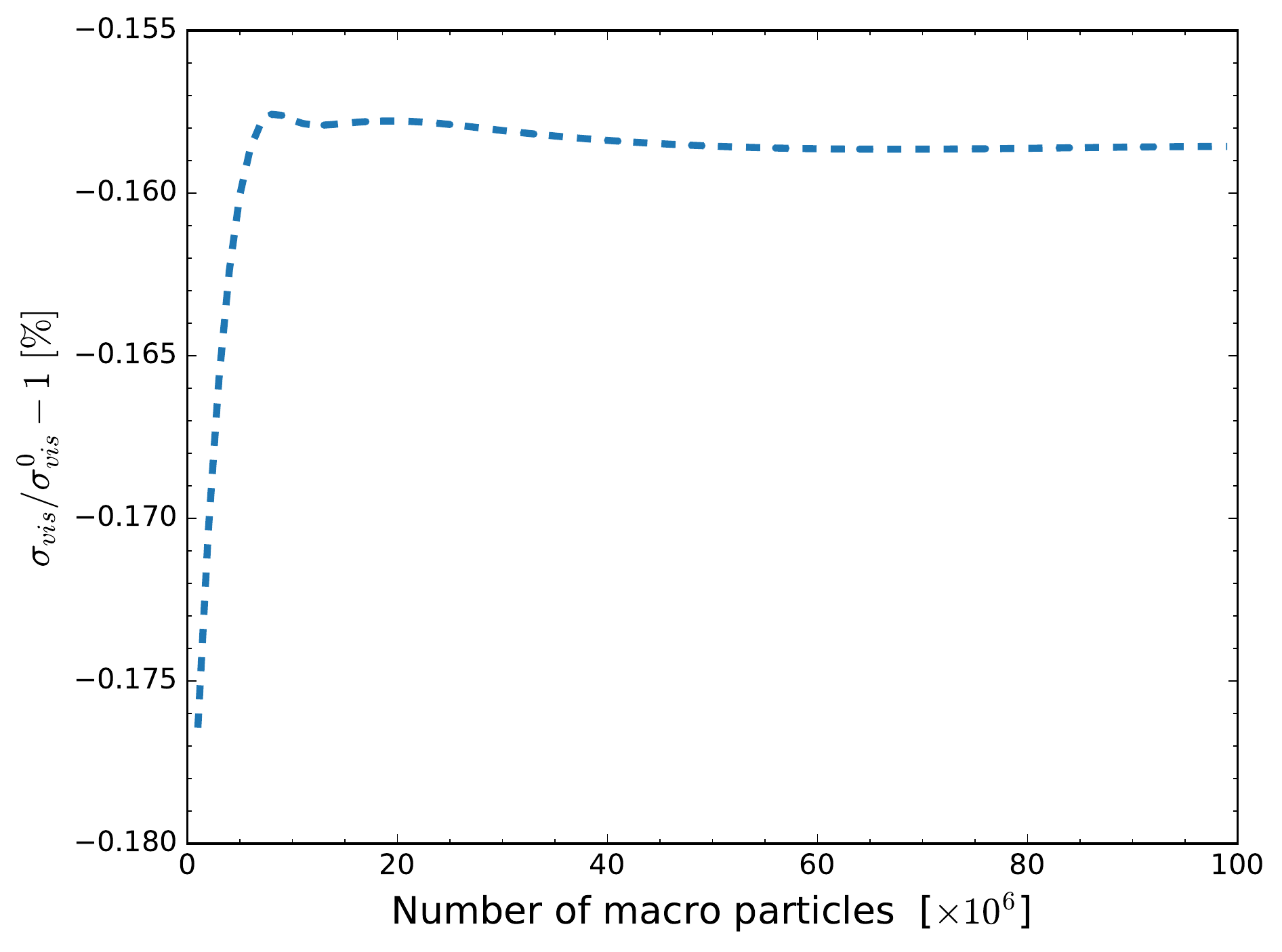}
\caption{Visible cross-section bias as a function of the number of macroparticles used to model the transverse-density distributions.}
\label{fig: mpart_scan} 
\end{figure}

Figure~\ref{fig: mpart_scan} shows the evolution of the \svis bias as a function of the number of macroparticles used in the simulation. The result converges to within 0.01\% of its asymptotic value for $5 \times 10^6$ macroparticles. Most of the simulation results presented in this paper are based on  $10\times 10^6$ macroparticles per bunch, implying that the results are numerically stable at the $0.001\%$ level. 

\par
The uncertainty associated with statistical fluctuations in the discretization of the transverse-density distributions was evaluated separately, and typically cancels out when computing luminosity ratios such as \LoLz. It becomes significant only at large beam separation, when the overlap integral is computed from a small number of macroparticles in the tails. At these scan points, however, the luminosity values are very small, and therefore have a negligible impact on the estimation of the beam-beam bias factors.

\section{Parameterization of optical-distortion corrections}
\label{subsec:OptCorrTables}

\begin{table*}[hbt]
\centering
\renewcommand{\arraystretch}{1.2}
\begin{tabular}{|c|c|}
\hline
Horizontal scans						& Vertical scans 					\\
\hline\hline
 \multicolumn{2}{|c|}{$0.002 \leq \xiR \leq 0.007$}								\\
 \hline
$0.2975 \leq \qx \leq 0.3100$  				& $\qy - 0.0125 \leq \qx \leq \qy - 0.0075$	\\
$\qx + 0.0075 \leq \qy \leq \qx + 0.0125$  		& $0.3075 \leq \qy \leq 0.3200$ 		\\
\hline
\end{tabular}
\caption{Parameter space used in the determination of the \LoLzOpt polynomial coefficients.}
\label{tab:xiqspace}
\end{table*}

The polynomial parameters $p_{0k},...,p_{9k}$ from which to calculate the optical-distortion luminosity-bias factor
\begin{align}
&\LoLzOpt(\xiR, \qx, \qy|\Delta_k) = p_{0k} + p_{1k}\, \xiR 							 \nonumber	\\
&+ p_{2k}\, \qx + p_{3k}\, \qy + p_{4k}\, \xiR^{2} + p_{5k}\, \qx^{2} + p_{6k}\, \qy^{2} 		 \nonumber 	\\
& + p_{7k} \xiR \qx + p_{8k}\, \xiR\, \qy +p_{9k}\, \qx\, \qy							
\label{eqn:LL0_xiqxqyA}
\end{align}
defined in Eq.~(\ref{eqn:ll0_xiqxqy}) of Sec.~\ref{subsubsec:CorImplement}, have been tabulated, for normalized nominal-separation values covering the range $0 < |\Delta_k / \sigma_0| < 6$ in steps of 0.25. Here the index $k$ ($k =$ 1, 25) labels the nominal-separation bin $\Delta_k$ in the scanning plane considered. This parameterization is valid in the fully symmetric Gaussian-beam configuration with zero non-scanning IPs, over the three-dimensional space of fractional tunes (\qx, \qy) and round--beam-equivalent beam-beam parameter $(\xiR)$ values defined in Table~\ref{tab:xiqspace}. The values of $\bst$, $\gamma$, $\sigma_0$ and $n$ in Eq.~(\ref{eqn:xiRDef}) were chosen such that they correspond to typical settings for $pp$ \vdM-calibration sessions during Run 2 of the LHC.

\par
The coefficients $p_{0k},...,p_{9k}$ are listed in Tables~\ref{tab:parhoriz} and  \ref{tab:parv} for horizontal and vertical \vdM scans, respectively; they are also  publicly accessible, in computer-readable form, in Ref.~\cite{bib:OptCorrTables}. In the unabridged tables, \ie for the studies reported in Secs.~\ref{subsubsec:CorImplement} and \ref{subsubsec:FoMs}, all parameters were stored with a number of digits corresponding to the precision of the {\it IEEE 754} double-precision floating-point data format. In the tables below, the coefficients are listed with enough precision to reproduce the 
values of \LoLzOpt  in the unabridged tables to at least five decimal digits. For verification purposes, $50\times 10^{3}$ points were randomly chosen across the $(\xiR,\qx,\qy)$ space defined in Table~\ref{tab:xiqspace}, and for each of these, 200 $\Delta$ values were randomly generated. The resulting \LoLzOpt values, obtained using the unabrigded tables or the ones below, were then compared. The maximum differences found in $10\times 10^{6}$ evaluations were $5.4\times 10^{-6}$ and $9.5\times 10^{-6}$ for horizontal and vertical scans, respectively.

\par
The procedure to correct an $x$-$y$ pair of experimentally measured \vdM-scan curves for beam--beam-induced optical distortions is detailed below. It is valid only for on-axis, one-dimensional \vdM scans with zero crossing angle; off-axis one-dimensional scans, as well as two-dimensional grid scans in the $(\delta_x, \delta_y)$ beam-separation plane, require dedicated simulations.
\begin{enumerate}
\item
For a given colliding-bunch pair, calculate the values of \sigR and \xiR from the measured bunch-current and uncorrected (\CSx, \CSy) values, following the prescriptions in Sec.~\ref{subsubsec:CorImplement};
\item
for that same bunch pair, calculate the values of the effective fractional tunes (\qx, \qy), following the prescription in Sec.~\ref{subsubsec:mIPparam} with, as input, the nominal tune values (\Qx, \Qy), the appropriate number \NSIP of non-scanning IPs, and the resulting multi-IP equivalent tune shift  \DQmIP;
\item
for each scanning plane ($x$ and $y$) and each tabulated value of the nominal separation $\Delta_k$, calculate the corresponding value of $\{\LoLzOpt\}_k =  \LoLzOpt(\xiR, \qx, \qy|\Delta_k)$ using Eq.~(\ref{eqn:LL0_xiqxqyA}), with the polynomial coefficients taken from the relevant row of Table~\ref{tab:parhoriz} or \ref{tab:parv}, as appropriate;
\item
use all pairs of numbers $(\Delta_k, \{\LoLzOpt\}_k)$ as input to a suitably chosen numerical-interpolation algorithm\footnote{Cubic-spline interpolation is recommended.} that can return the value of $\LoLzOpt(\Delta/\sigR)$ for any nominal separation $\Delta$;
\item
at each scan step, multiply the measured collision rate by $1/ \LoLzOpt(\Delta/\sigR)$, where $\Delta$ is the nominal separation for the scan step and in the scanning plane considered. This final step yields, for the bunch pair considered, the horizontal and vertical \vdM-scan curves corrected for optical distortions.
\end{enumerate}

%
%
\onecolumn
\begin{sidewaystable}
\centering
\renewcommand{\arraystretch}{1.2}
\renewcommand{\tabcolsep}{6pt}
\begin{tabular}{| c | c| c | c | c | c | c | c | c | c | c |}
\hline
$\Delta_{k} [\sigma_0]$	
& $p_{0k}$		& $p_{1k}$		& $p_{2k}$		& $p_{3k}$		& $p_{4k}$	
& $p_{5k}$		& $p_{6k}$		& $p_{7k}$		& $p_{8k}$		& $p_{9k}$		\\
\hline
%
%
0.00&1.00944&-9.9169&0.05163&-0.10946&-11.343&0.27849&0.51631&17.907&17.933&-0.70617 \\
0.25&1.0965&-11.2335&-0.37109&-0.25749&-15.156&0.8075&0.59757&21.37&19.207&-0.38121 \\
0.50&1.33057&-14.2614&-1.41804&-0.75044&-23.614&2.23701&1.09972&29.115&22.489&0.21938 \\
0.75&1.6732&-18.353&-3.14234&-1.28652&-32.777&4.71339&1.58581&37.476&28.876&0.9846 \\
1.00&2.09627&-23.4765&-4.63004&-2.55265&-38.221&3.68123&0.36501&50.199&34.146&7.71504 \\
1.25&2.35234&-28.0037&-5.75125&-3.10402&-38.905&3.96085&-0.21768&57.845&41.93&10.74776 \\
1.50&2.5998&-30.8986&-6.20145&-4.24446&-36.073&8.14783&4.83319&59.119&50.075&4.06889 \\
1.75&2.71221&-32.1903&-5.80054&-5.33954&-33.326&6.80135&5.92624&54.751&58.075&5.3906 \\
2.00&2.68384&-31.9924&-4.60368&-6.30519&-25.565&5.92361&8.49128&47.367&63.795&3.25386 \\
2.25&2.72577&-30.6931&-4.15275&-7.00444&-15.931&4.41796&8.97022&42.853&63.188&4.6426 \\
2.50&2.71357&-29.8333&-3.31831&-7.7255&-17.797&-0.48377&6.79757&35.712&67.066&11.47357 \\
2.75&2.55328&-28.5073&-2.93845&-7.07616&-19.558&2.2039&8.83346&28.805&69.588&5.10042 \\
3.00&2.63081&-28.984&-3.48143&-7.0339&-13.235&1.7513&7.59532&37.803&62.493&7.5962 \\
3.25&2.51794&-29.2876&-3.37658&-6.41585&-9.579&2.73257&7.63313&38.021&63.687&5.42281 \\
3.50&2.57552&-30.733&-3.87832&-6.30735&-5.79&3.30687&7.27443&49.126&58.156&5.87508 \\
3.75&2.7132&-34.1604&-4.72201&-6.34866&0.329&4.88899&7.5148&62.801&56.443&5.47604 \\
4.00&2.80473&-36.4992&-6.20828&-5.49987&-0.487&1.89557&1.09204&77.16&50.842&15.98952 \\
4.25&2.84402&-40.437&-7.01717&-4.94963&5.671&5.45919&2.26449&89.531&51.974&11.6835 \\
4.50&2.92941&-42.1662&-7.69945&-4.84568&7.414&12.10344&7.25898&100.023&47.916&1.02831 \\
4.75&3.00531&-44.2824&-8.94959&-4.12019&4.043&8.3604&0.8071&119.399&36.595&12.11431 \\
5.00&3.1174&-47.4622&-9.45872&-4.32462&10.289&7.49102&-0.57319&125.029&41.518&15.50293 \\
5.25&3.08992&-48.0346&-9.55933&-4.06527&9.796&14.24273&5.29776&139.651&29.582&2.65662 \\
5.50&3.22545&-51.1449&-10.7063&-3.79323&9.956&14.35667&3.04625&141.488&38.024&6.20924 \\
5.75&3.25358&-51.7947&-11.62684&-3.08922&7.899&12.51096&-1.17893&151.272&30.874&12.68129 \\
6.00&3.15344&-50.9016&-10.84895&-3.228&7.762&19.8657&7.03849&145.734&33.547&-3.89875 \\
\hline
\end{tabular}
\caption{Parameterization of the \LoLzOpt luminosity-bias factor as a function of the normalized nominal separation, for horizontal \vdM scans. The functional form is defined in Eq.~(\ref{eqn:LL0_xiqxqyA}).}
\label{tab:parhoriz}
\end{sidewaystable}

%
\begin{sidewaystable}
\centering
\renewcommand{\arraystretch}{1.2}
\renewcommand{\tabcolsep}{6pt}
\begin{tabular}{| c | c| c | c | c | c | c | c | c | c | c |}
\hline
$\Delta_{k} [\sigma_0]$	
& $p_{0k}$		& $p_{1k}$		& $p_{2k}$		& $p_{3k}$		& $p_{4k}$	
& $p_{5k}$		& $p_{6k}$		& $p_{7k}$		& $p_{8k}$		& $p_{9k}$		\\
\hline
0.00&1.00627&-9.8954&0.02224&-0.06076&-11.359&0.0568&0.18622&17.901&17.872&-0.18399 \\
0.25&1.20592&-12.0782&-0.02827&-1.29247&-17.815&0.1942&2.21186&19.01&24.368&-0.28862 \\
0.50&1.75762&-17.6722&-0.35287&-4.51951&-30.505&-0.9264&5.84403&22.471&40.37&2.91383 \\
0.75&2.47795&-24.582&-0.92069&-8.60293&-41.735&1.3187&13.56868&20.768&65.956&0.45508 \\
1.00&3.16343&-32.8235&-1.16788&-12.75235&-55.11&-0.3834&18.27023&29.422&85.709&4.49981 \\
1.25&3.74028&-38.7998&-2.36278&-15.28112&-48.5&0.3875&21.25607&39.51&95.711&6.74987 \\
1.50&3.96499&-42.0778&-3.06102&-16.01822&-37.845&-1.1981&19.73221&36.839&108.559&12.15308 \\
1.75&3.69522&-40.042&-3.62485&-13.73364&-31.695&-1.4007&14.95999&39.742&98.388&14.38483 \\
2.00&3.17695&-34.3599&-3.98479&-10.06481&-23.614&5.9537&15.52287&48.181&70.742&1.19056 \\
2.25&2.54203&-27.8488&-4.42917&-5.57026&-15.667&5.2247&6.89102&46.516&50.153&4.09602 \\
2.50&1.93646&-21.3641&-4.22219&-1.90718&-15.994&5.1956&1.34853&47.557&27.611&3.48476 \\
2.75&1.64684&-18.4268&-4.50556&0.22396&-11.197&5.0949&-2.57366&49.259&16.174&4.5606 \\
3.00&1.58403&-17.9608&-4.4421&0.56159&-7.635&6.066&-2.11206&48.03&16.126&2.48878 \\
3.25&1.84422&-21.9484&-4.56541&-0.96411&-2.223&3.7746&-2.00762&49.005&28.617&7.28473 \\
3.50&2.32149&-27.3088&-4.57251&-4.00249&1.528&4.2728&3.31839&48.531&47.316&6.31746 \\
3.75&2.86007&-35.4916&-3.79586&-8.17387&13.02&7.1856&13.85154&44.04&78.901&-1.77626 \\
4.00&3.52241&-43.4275&-3.74323&-12.44714&15.761&2.4115&16.29906&43.702&105.987&7.26694 \\
4.25&4.18491&-52.8584&-3.75654&-16.63881&17.478&2.9322&23.4521&43.277&137.933&6.26266 \\
4.50&4.82121&-61.3184&-4.12935&-20.31864&25.852&-1.7901&24.22835&37.512&171.614&16.6589 \\
4.75&5.33928&-68.7429&-4.41872&-23.34458&24.7&3.8166&33.91458&37.937&196.185&6.67136 \\
5.00&5.80246&-73.1287&-3.19177&-27.51882&20.495&18.7373&56.57897&38.918&210.435&-26.21613 \\
5.25&6.44223&-79.2346&-5.10161&-29.74714&33.649&17.5226&56.18781&50.413&219.306&-17.92498 \\
5.50&7.17335&-86.1747&-5.53665&-33.9533&35.037&6.9886&52.15683&35.086&257.011&3.99827 \\
5.75&6.8188&-87.1346&-4.86936&-32.37045&28.729&-1.12&43.18943&42.423&253.701&17.49413 \\
6.00&6.41896&-87.7396&-0.88027&-33.73032&32.935&1.0047&53.37414&28.725&269.298&0.87602 \\
\hline
\end{tabular}
\caption{Parameterization of the \LoLzOpt luminosity-bias factor as a function of the normalized nominal separation, for vertical \vdM scans. The functional form is defined in Eq.~(\ref{eqn:LL0_xiqxqyA}).}
\label{tab:parv}
\end{sidewaystable}
\twocolumn

\clearpage

\nocite{*}

\printbibliography

\end{document}